\documentclass[onecolumn,notitlepage,floats,floatfix,amssymb,aps,prd,showpacs,superscriptaddress,groupedaddress,nofootinbib]{revtex4-2}  

\usepackage[T1]{fontenc}
\usepackage{lmodern} 
\usepackage{graphicx}  
\usepackage{dcolumn}   
\usepackage{bm}        
\usepackage{amssymb}   
\usepackage{amsmath}
\usepackage{bbold}
\usepackage{array}
\usepackage{makecell}
\usepackage{braket}
\usepackage{longtable}
\usepackage{supertabular,booktabs}

\usepackage{titlesec} 
\usepackage[colorlinks,citecolor=blue,urlcolor=blue,hypertexnames=true]{hyperref}
\usepackage{subfigure}

\usepackage[usenames,dvipsnames,svgnames,table]{xcolor} 

\renewcommand{\arraystretch}{2}

\newcommand{\be}{\begin{equation}}
\newcommand{\ee}{\end{equation}}
\newcommand{\bea}{\begin{eqnarray}}
\newcommand{\eea}{\end{eqnarray}}
\newcommand{\bml}{\begin{subequations}}
\newcommand{\eml}{\end{subequations}}
\newcommand{\bfig}{\begin{figure}}
\newcommand{\efig}{\end{figure}}

\newcommand{\Del}{\Delta}

\newcommand{\slashed}{\hspace{-1.1ex}/}

\newcommand{\bmat}{\begin{pmatrix}}
\newcommand{\emat}{\end{pmatrix}}
\usepackage{graphicx,slashed,booktabs,xcolor,multirow,float,
amsfonts,bbold,mathtools,sidecap,tikz,bm,enumitem}
\usepackage{multirow}
\usepackage{bbding}
\usepackage{titlesec}
\usepackage{hyperref}
\usepackage{cancel}
\usepackage{wasysym}
\usepackage{amssymb}
\usepackage{pifont}
\newcommand{\grad}{\nabla}

\renewcommand{\leq}{\leqslant}
\renewcommand{\geq}{\geqslant}

\usepackage[dvipsnames, usenames]{xcolor}

\definecolor{linkcolor}{rgb}{0.55, 0.13, .32}

\definecolor{oucrimsonred}{rgb}{0.6, 0.0, 0.0}
\definecolor{persianblue}{rgb}{0.11, 0.22, 0.73}
\definecolor{forestgreen}{rgb}{0.13,0.35,0.13}
\definecolor{lightgray}{rgb}{0.83, 0.83, 0.83}
 \hypersetup{colorlinks, citecolor=oucrimsonred, linkcolor=persianblue, urlcolor=oucrimsonred}
\definecolor{cornellred}{rgb}{0.7, 0.11, 0.11}
\definecolor{navyblue}{rgb}{0.0, 0.0, 0.5}
\definecolor{amethyst}{rgb}{0.6, 0.4, 0.8}
\definecolor{yellow}{rgb}{1.0, 1.0, 0.0}
\definecolor{firebrick}{rgb}{0.7, 0.13, 0.13}
\definecolor{tangerineyellow}{rgb}{1.0, 0.8, 0.0}
\definecolor{deepfuchsia}{rgb}{0.76, 0.33, 0.76}
\definecolor{amber}{rgb}{1.0, 0.75, 0.0}
\definecolor{VioletRed4}{rgb}{0.55, 0.13, .32}
\definecolor{indiagreen}{rgb}{0.07, 0.53, 0.03}
\definecolor{VioletRed4}{rgb}{0.55, 0.13, .32}

\usepackage{hyperref}

\usepackage{graphics,appendix,afterpage,makecell}

\usepackage{bbold}
\usepackage{tikz}
\usepackage{adjustbox}
\usepackage{tikz-feynman}
\usepackage{tcolorbox}

\def\abs#1{\left| #1\right|}

\definecolor{oucrimsonred}{rgb}{0.6, 0.0, 0.0}
\definecolor{persianblue}{rgb}{0.11, 0.22, 0.73}
\definecolor{forestgreen}{rgb}{0.13,0.35,0.13}
\definecolor{lightgray}{rgb}{0.83, 0.83, 0.83}
 \hypersetup{colorlinks, citecolor=oucrimsonred, linkcolor=persianblue, urlcolor=oucrimsonred}
\definecolor{cornellred}{rgb}{0.7, 0.11, 0.11}
\definecolor{navyblue}{rgb}{0.0, 0.0, 0.5}
\definecolor{amethyst}{rgb}{0.6, 0.4, 0.8}
\definecolor{yellow}{rgb}{1.0, 1.0, 0.0}
\definecolor{firebrick}{rgb}{0.7, 0.13, 0.13}
\definecolor{tangerineyellow}{rgb}{1.0, 0.8, 0.0}
\definecolor{deepfuchsia}{rgb}{0.76, 0.33, 0.76}
\definecolor{amber}{rgb}{1.0, 0.75, 0.0}
\definecolor{VioletRed4}{rgb}{0.55, 0.13, .32}
\definecolor{indiagreen}{rgb}{0.07, 0.53, 0.03}
\definecolor{VioletRed4}{rgb}{0.55, 0.13, .32}

\definecolor{oucrimsonred}{rgb}{0.6, 0.0, 0.0}
\newcommand\vertarrowbox[3][6ex]{%
  \begin{array}[t]{@{}c@{}} #2 \\
  \left\uparrow\vcenter{\hrule height #1}\right.\kern-\nulldelimiterspace\\
  \makebox[0pt]{\scriptsize#3}
  \end{array}%
}

\definecolor{mtcolor}{rgb}{.8,.3,.1}

\definecolor{violachiaro}{rgb}{1,0.6,1}

\definecolor{gbcolor}{rgb}{.43,.22,.12}
 
\definecolor{gbcolor2}{rgb}{.9,.2,.6}
\definecolor{gbcolor3}{rgb}{.3,.2,.6}

\definecolor{verdechiaro}{rgb}{0.6,1,0.6}
\definecolor{giallochiaro}{rgb}{1,1,0.6}
\definecolor{bluscuro}{rgb}{0.15, 0.2, 0.9}
\definecolor{verdes}{rgb}{0.1, 0.5, 0.1}%
\definecolor{tangerineyellow}{rgb}{1.0, 0.8, 0.0}
\definecolor{smokyblack}{rgb}{0.06, 0.05, 0.03}

\definecolor{americanrose}{rgb}{1.0, 0.01, 0.24}
\definecolor{cobalt}{rgb}{0.0, 0.28, 0.67}
\definecolor{brandeisblue}{rgb}{0.0, 0.44, 1.0}
\definecolor{mycolor}{rgb}{0.0, 0.0, 0.5}
\definecolor{oxfordblue}{rgb}{0.0, 0.13, 0.28}
\definecolor{azure}{rgb}{0.0, 0.5, 1.0}
\definecolor{turquoiseblue}{rgb}{0.0, 1.0, 0.94}
\newtcolorbox{mynewbox}[1]{colback=white!5!white,colframe=azure!75!black,fonttitle=\bfseries,title=#1}
\newtcolorbox{mybox}{colback=mycolor!5!white,colframe=azure!75!black}
\newtcolorbox{mynamedbox}[1]{colback=mycolor!5!white,colframe=azure!75!black,title=#1}
\definecolor{venetianred}{rgb}{0.78, 0.03, 0.08}
\newtcolorbox{mynamedbox1}[1]{colback=venetianred!5!white,colframe=venetianred!80!black,title=#1}
\newtcolorbox{mynamedbox2}[1]{colback=azure!5!white,colframe=azure!80!black,title=#1}

\definecolor{rossocorsa}{rgb}{0.83, 0.0, 0.0}

\tikzset{->-/.style={decoration={
  markings,
  mark=at position #1 with {\arrow{>}}},postaction={decorate}}}
\tikzset{-<-/.style={decoration={
  markings,
  mark=at position #1 with {\arrow{<}}},postaction={decorate}}} 

\def\be{\begin{equation}}
\def\ee{\end{equation}}
\def\ba{\begin{eqnarray}}
\def\ea{\end{eqnarray}}

\def\v{{\cal V}}

\def\L*{{\cal L}_*}
\def\L{\mathcal{L}}
\def\({\left(}
\def\){\right)}

\def\<{\langle}
\def\>{\rangle}

 \def\neq {\not\equiv}

\def\cs2{c_{s}^{2}}

 \def\be   {\begin{equation}}   \def\ee   {\end{equation}}
 \def\ba   {\begin{array}}      \def\ea   {\end{array}}
 \def\bea  {\begin{eqnarray}}   \def\eea  {\end{eqnarray}}
 \def\bean {\begin{eqnarray*}}  \def\eean {\end{eqnarray*}}

\titleclass{\subsubsubsection}{straight}[\subsection]

\newcounter{subsubsubsection}[subsubsection]
\renewcommand\thesubsubsubsection{\thesubsubsection.\arabic{subsubsubsection}}

\titleformat{\subsubsubsection}
  {\normalfont\normalsize\bfseries}{\thesubsubsubsection}{1em}{}
\titlespacing*{\subsubsubsection}
{0pt}{3.25ex plus 1ex minus .2ex}{1.5ex plus .2ex}

\makeatletter
\renewcommand\paragraph{\@startsection{paragraph}{5}{\z@}%
  {3.25ex \@plus1ex \@minus.2ex}%
  {-1em}%
  {\normalfont\normalsize\bfseries}}
\renewcommand\subparagraph{\@startsection{subparagraph}{6}{\parindent}%
  {3.25ex \@plus1ex \@minus .2ex}%
  {-1em}%
  {\normalfont\normalsize\bfseries}}
\def\toclevel@subsubsubsection{4}
\def\toclevel@paragraph{5}
\def\toclevel@paragraph{6}
\def\l@subsubsubsection{\@dottedtocline{4}{7em}{4em}}
\def\l@paragraph{\@dottedtocline{5}{10em}{5em}}
\def\l@subparagraph{\@dottedtocline{6}{14em}{6em}}
\makeatother

\begin{document}


\definecolor{lime}{HTML}{A6CE39}
\DeclareRobustCommand{\orcidicon}{\hspace{-2.1mm}
\begin{tikzpicture}
\draw[lime,fill=lime] (0,0.0) circle [radius=0.13] node[white] {{\fontfamily{qag}\selectfont \tiny \,ID}}; \draw[white, fill=white] (-0.0525,0.095) circle [radius=0.007]; 
\end{tikzpicture} \hspace{-3.7mm} }
\foreach \x in {A, ..., Z}{\expandafter\xdef\csname orcid\x\endcsname{\noexpand\href{https://orcid.org/\csname orcidauthor\x\endcsname} {\noexpand\orcidicon}}}
\newcommand{\orcidauthorA}{0000-0002-0459-3873}
\newcommand{\orcidauthorD}{0009-0003-9227-8615}
\newcommand{\orcidauthorB}{0009-0003-2854-9708}
\newcommand{\orcidauthorC}{0000-0003-1081-0632}


\title{\textcolor{Sepia}{\textbf \Large\LARGE
Untangling PBH overproduction in $w$-SIGWs generated by Pulsar Timing Arrays for MST-EFT of single field inflation
}}


\author{{\large  Sayantan Choudhury\orcidA{}${}^{1}$}}
\email{sayanphysicsisi@gmail.com, 
sayantan_ccsp@sgtuniversity.org,  sayantan.choudhury@nanograv.org (Corresponding author)} 
\author{\large Kritartha Dey\orcidB{}${}^{1}$}
\email{kritartha09@gmail.com}
\author{{\large  Ahaskar Karde\orcidD{}${}^{1}$}}
\email{kardeahaskar@gmail.com}

\affiliation{ ${}^{1}$Centre For Cosmology and Science Popularization (CCSP),\\
        SGT University, Gurugram, Delhi- NCR, Haryana- 122505, India.}

\begin{abstract}

Our work highlights the crucial role played by the equation of state (EoS) parameter $w$ within the context of single field inflation with Multiple Sharp Transitions (MSTs) to untangle the current state of the PBH overproduction issue. We examine the situation for a broad interval of EoS parameter that remains most favourable to explain the recent data released by the pulsar timing array (PTA) collaboration. Our analysis yields the interval,  $0.2 \leq w \leq 1/3$, to be the most acceptable window from the SIGW interpretation of the PTA signal and where sizeable PBHs abundance,  $f_{\rm PBH} \in (10^{-3},1)$, is observed. We also obtain $w=1/3$, radiation-dominated era, to be the best scenario to explain the early stages of the Universe and address the overproduction problem. Within the range of $1 \leq c_{s} \leq 1.17$, we construct a regularized-renormalized-resummed scalar power spectrum whose amplitude obeys the perturbativity criterion while being substantial enough to generate EoS dependent scalar induced gravitational waves ($w$-SIGWs) consistent with NANOGrav-15 data. 
Working for both $c_{s} = 1\;{\rm and}\;1.17$, we find the $c_{s}=1.17$ case more favourable for generating large mass PBHs, $M_{\rm PBH}\sim {\cal O}(10^{-6}-10^{-3})M_{\odot}$, as potential dark matter candidates with substantial abundance after constraints coming from microlensing experiments. 

\end{abstract}

\maketitle
\tableofcontents
\newpage

\section{Introduction}
\label{s1}

The observational evidence for the existence of the stochastic gravitational wave background (SGWB) at the nHz frequency is recently confirmed by the Pulsar Timing Array collaboration (PTA), which includes the NANOGrav \cite{NANOGrav:2023gor, NANOGrav:2023hde, NANOGrav:2023ctt, NANOGrav:2023hvm, NANOGrav:2023hfp, NANOGrav:2023tcn, NANOGrav:2023pdq, NANOGrav:2023icp, Inomata:2023zup}, EPTA \cite{EPTA:2023fyk, EPTA:2023sfo, EPTA:2023akd, EPTA:2023gyr, EPTA:2023xxk, EPTA:2023xiy, Lozanov:2023rcd}, PPTA \cite{Reardon:2023gzh, Reardon:2023zen, Zic:2023gta}, and CPTA \cite{Xu:2023wog}. Since then, it has been of growing interest in the scientific community due to its potential to probe the physics of the early universe beyond the scales observable till the last scattering surface. There is a vast amount of literature that has since been devoted to exploring the possible sources of this SGWB, ranging from phase transitions, domain walls, cosmic strings, and gravitational waves induced by large primordial fluctuations, namely scalar-induced gravitational waves (SIGWs) \cite{Choudhury:2023fwk,Choudhury:2023hfm,Bhattacharya:2023ysp,Franciolini:2023pbf,Inomata:2023zup,Wang:2023ost,Balaji:2023ehk,HosseiniMansoori:2023mqh,Gorji:2023sil,DeLuca:2023tun,Choudhury:2023kam,Yi:2023mbm,Cai:2023dls,Cai:2023uhc,Huang:2023chx,Vagnozzi:2023lwo,Frosina:2023nxu,Zhu:2023faa,Jiang:2023gfe,Cheung:2023ihl,Oikonomou:2023qfz,Liu:2023pau,Liu:2023ymk,Wang:2023len,Zu:2023olm, Abe:2023yrw, Gouttenoire:2023bqy,Salvio:2023ynn, Xue:2021gyq, Nakai:2020oit, Athron:2023mer,Ben-Dayan:2023lwd, Madge:2023cak,Kitajima:2023cek, Babichev:2023pbf, Zhang:2023nrs, Zeng:2023jut, Ferreira:2022zzo, An:2023idh, Li:2023tdx,Blanco-Pillado:2021ygr,Buchmuller:2021mbb,Ellis:2020ena,Buchmuller:2020lbh,Blasi:2020mfx, Madge:2023cak, Liu:2023pau, Yi:2023npi,Gangopadhyay:2023qjr,Vagnozzi:2020gtf,Benetti:2021uea,Inomata:2023drn,Lozanov:2023rcd,Basilakos:2023jvp,Basilakos:2023xof,Li:2023xtl,Domenech:2021ztg,Yuan:2021qgz,Chen:2019xse,Cang:2023ysz,Cang:2022jyc,Konoplya:2023fmh,Heydari:2023rmq,Li:2023xtl,Bernardo:2023jhs,Choi:2023tun,Elizalde:2023rds,Chen:2023bms,Nojiri:2023mbo,Domenech:2023jve,Liu:2023hpw,Huang:2023zvs,Oikonomou:2023bli,Cyr:2023pgw,Fu:2023aab,Kawai:2023nqs,Kawasaki:2023rfx,Maji:2023fhv,Bhaumik:2023wmw,He:2023ado,An:2023jxf,Zhu:2023lbf,Das:2023nmm,Roshan:2024qnv,Chen:2024fir}.  

In this work, we will focus on understanding the earlier history using the SIGWs, which result from the second-order sourcing of the enhancements in the primordial curvature perturbations at the small scales. From the current status of the CMB observations, we can only probe the Universe at large scales, but this is ineffective in providing sufficient information about the physics of the Universe near the end of inflation. On the contrary, Gravitational Waves (GWs) do not interact with the intervening matter and thus can explore the primordial universe before the advent of Big Bang Nucleosynthesis (BBN). So, it is a crucial tool to probe the smaller scales during inflation. Furthermore, the induced SIGWs have the potential to explain the PBH formation scenarios in the early universe, which are considered to be possible candidates of the dark matter and thus have recently observed a renewed interest in their study \cite{ Kawaguchi:2023mgk,Altavista:2023zhw}. 
The formation of PBH requires large enhancement in the fluctuations at the small scales not probed by the CMB, and these disturbances later gravitationally collapse upon horizon re-entry. Incidentally, the regime of sensitivity of the SGWB signal lies between ${\cal O}(10^{-9}-10^{-6})\;{\rm Hz}$. This frequency range coincides with the domain where primordial black holes (PBH) forming perturbations can cause observable distortions in signals from various gravitational wave (GW) events, such as mergers and microlensing events. Consequently, the SGWB serves as a robust counterpart for detecting primordial black holes. 
The spectrum of the GW energy density is usually determined using data fitting by a power law index $\Omega_{\rm GW} \propto  f^{\beta}$. This index, however, other than the values numerically allowed due to the fitting $\beta \in (-1.5,0.5)$, has the feature to be dependent on the equation of state (EoS) parameter $w$ of the early universe when corresponding wavelengths of primordial fluctuations re-enter the Hubble horizon \cite{Cai:2019cdl}. 

Now, the recent analysis, as presented by the NANOGrav collaboration, has mainly focused on the generation of these gravitational waves as being primarily sourced during the radiation-dominated era with an equation of state $w=1/3$. While there are proper motivations to consider such a scenario, it has recently shown to be worthwhile to explore the primordial universe with a constant EoS $w$ and propagation speed $c_{s}^{2}$ \cite{Liu:2023pau,Liu:2023hpw,Balaji:2023ehk,Domenech:2021ztg,Domenech:2020ers,Domenech:2019quo}. This effort promises methods to address scenarios that could have occurred before the onset of BBN. 



Our paper explores the scenario where the background of the early Universe gets dominated by a non-adiabatic perfect fluid with a constant $w$ and a constant propagation speed $c_{s}^{2}$ parameters.
Under the conditions of having a phase of relatively short duration, in addition to the conventional slow-roll, which violates the slow-roll conditions, we create a possibility to study the generation of primordial black holes (PBHs) \cite{Zeldovich:1967lct,Hawking:1974rv,Carr:1974nx,Carr:1975qj,Chapline:1975ojl,Carr:1993aq,Choudhury:2012whm,Choudhury:2012yh,Choudhury:2013woa,Yokoyama:1998pt,Kawasaki:1998vx,Rubin:2001yw,Khlopov:2002yi,Khlopov:2004sc,Saito:2008em,Khlopov:2008qy,Carr:2009jm,Choudhury:2011jt,Lyth:2011kj,Drees:2011yz,Drees:2011hb,Ezquiaga:2017fvi,Kannike:2017bxn,Hertzberg:2017dkh,Pi:2017gih,Gao:2018pvq,Dalianis:2018frf,Cicoli:2018asa,Ozsoy:2018flq,Byrnes:2018txb,Ballesteros:2018wlw,Belotsky:2018wph,Martin:2019nuw,Ezquiaga:2019ftu,Motohashi:2019rhu,Fu:2019ttf,Ashoorioon:2019xqc,Auclair:2020csm,Vennin:2020kng,Nanopoulos:2020nnh,Inomata:2021uqj,Stamou:2021qdk,Ng:2021hll,Wang:2021kbh,Kawai:2021edk,Solbi:2021rse,Ballesteros:2021fsp,Rigopoulos:2021nhv,Animali:2022otk,Correa:2022ngq,Frolovsky:2022ewg,Escriva:2022duf,Ozsoy:2023ryl,Ivanov:1994pa,Afshordi:2003zb,Frampton:2010sw,Carr:2016drx,Kawasaki:2016pql,Inomata:2017okj,Espinosa:2017sgp,Ballesteros:2017fsr,Sasaki:2018dmp,Ballesteros:2019hus,Dalianis:2019asr,Cheong:2019vzl,Green:2020jor,Carr:2020xqk,Ballesteros:2020qam,Carr:2020gox,Ozsoy:2020kat,Baumann:2007zm,Saito:2008jc,Saito:2009jt,Choudhury:2013woa,Sasaki:2016jop,Raidal:2017mfl,Papanikolaou:2020qtd,Ali-Haimoud:2017rtz,Di:2017ndc,Raidal:2018bbj,Cheng:2018yyr,Vaskonen:2019jpv,Drees:2019xpp,Hall:2020daa,Ballesteros:2020qam,Carr:2020gox,Ozsoy:2020kat,Ashoorioon:2020hln,Papanikolaou:2020qtd,Wu:2021zta,Kimura:2021sqz,Solbi:2021wbo,Teimoori:2021pte,Cicoli:2022sih,Ashoorioon:2022raz,Papanikolaou:2022chm,Papanikolaou:2023crz,Wang:2022nml,ZhengRuiFeng:2021zoz,Cohen:2022clv,Arya:2019wck,Correa:2022ngq,Cicoli:2022sih,Brown:2017osf,Palma:2020ejf,Geller:2022nkr,Braglia:2022phb,Frolovsky:2023xid,Aldabergenov:2023yrk,Aoki:2022bvj,Frolovsky:2022qpg,Aldabergenov:2022rfc,Ishikawa:2021xya,Gundhi:2020kzm,Aldabergenov:2020bpt,Cai:2018dig,Cheng:2021lif,Balaji:2022rsy,Qin:2023lgo,Riotto:2023hoz,Ragavendra:2020vud,Ragavendra:2021qdu,Ragavendra:2020sop,Gangopadhyay:2021kmf,Papanikolaou:2022did,Harada:2013epa,Harada:2017fjm,Kokubu:2018fxy,Gu:2023mmd,Saburov:2023buy,Stamou:2023vxu,Libanore:2023ovr,Friedlander:2023qmc,Chen:2023lou,Cai:2023uhc,Karam:2023haj,Iacconi:2023slv,Gehrman:2023esa,Padilla:2023lbv,Xie:2023cwi,Meng:2022low,Qiu:2022klm,Mu:2022dku,Fu:2022ypp,Davies:2023hhn,Raatikainen:2023bzk,Chen:2024gqn,Choudhury:2025kxg,Choudhury:2024aji,Choudhury:2024dzw,Choudhury:2024dei,Choudhury:2024jlz,Choudhury:2024ybk,Choudhury:2024one,Choudhury:2024kjj,Choudhury:2024ezx,Choudhury:2023kam}. This newly introduced ultra-slow roll phase further follows another slow-roll phase, which then lasts until the end of inflation. See refs. \cite{Kristiano:2022maq,Riotto:2023hoz,Choudhury:2023vuj,Choudhury:2023jlt,Choudhury:2023rks,Choudhury:2023hvf,Choudhury:2023kdb,Choudhury:2023hfm,Choudhury:2024one,Kristiano:2023scm,Riotto:2023gpm,Firouzjahi:2023ahg,Firouzjahi:2023aum,Franciolini:2023agm,Cheng:2023ikq,Tasinato:2023ukp,Motohashi:2023syh, Banerjee:2021lqu} for more details. One can treat the evolution of the curvature perturbation modes in a semi-classical manner as these exit the horizon, but there are also necessary quantum loop effects that one must incorporate to gain a more robust understanding of the primordial scalar power spectrum. It has recently been shown through a rigorous analysis in refs.\cite{Choudhury:2023vuj,Choudhury:2023jlt,Choudhury:2023rks} that a proper accounting of the one-loop corrections in deriving what becomes the final one-loop renormalized and further resummed scalar power spectrum warrants strict constraints on the duration of inflation, precisely $\Delta{\cal N}_{\rm Total} \sim {\cal O}(20-25)$, if one demands the production of near solar mass PBH $M_{\rm PBH} \sim {\cal O}(M_{\odot})$. A new and promising alternative, which can bypass the constraints coming from the said one-loop corrections and facilitate the formation of ${\cal O}(M_{\odot})$ PBH, comes through the implementation of \textit{Multiple Sharp Transitions} (MST) during inflation \cite{Bhattacharya:2023ysp}. We would be using the underlying construction of this theory in the present work to study the generation of large mass PBHs which corresponds to the frequency regime of the NANOGrav signal $(10^{-6}-10^{-9}){\rm Hz}$. 

Additionally, in this context, it is crucial to mention that earlier works attempting to explain the PBH overproduction issue, either utilize a Dirac delta power spectrum profile or a log-normal spectrum. Now, this kind of profile is pretty hard to construct from the dynamics of the primordial universe, if not impossible. Therefore, to realise a realistic scenario, we have adopted the MST set-up which provides us with a scalar power spectrum that is derived from a methodical and robust framework supported by rigorous renormalization and resummation procedures.

    \begin{figure}[ht!]
    	\centering
   {
   \includegraphics[width=18.5cm,height=13cm] {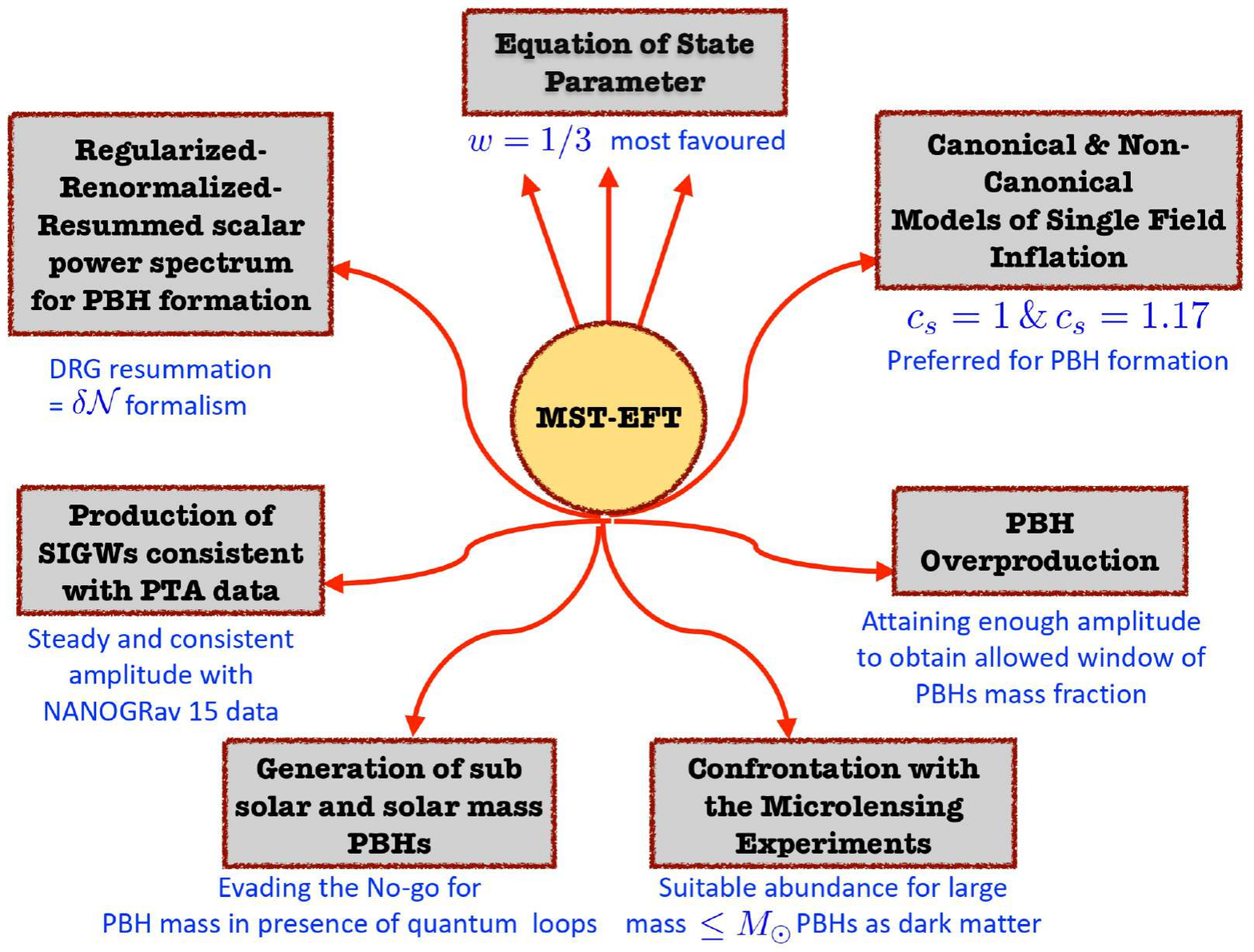}
    }
    	\caption[Optional caption for list of figures]{This branching diagram illustrates the crucial issues addressed in our research, showcasing the interconnected relationships and various dimensions of the set-up of Multiple Sharp Transitions. Each branch delves into specific aspects, contributing to a holistic understanding while discussing these prime subjects in detail.} 
    	\label{MST-EFT Branch}
    \end{figure}

Now the recently raised issue regarding PBH overproduction within the range of frequencies accessible by the NANOGrav is a severe concern, resolving which demands the utmost care. In this work, we explore the fact that the formation threshold for PBHs is an EoS-dependent quantity, and since we have the freedom not to know the exact state of the universe before it became radiation-dominated, exploring a range of EoS values where $w \ne 1/3$ enables us to identify the nature of the EoS and constrain the mass fraction of the PBH to give us a sizeable abundance thus avoiding overproduction. We have not yet incorporated non-linear effects in our approach. For a complete analysis to address the overproduction issue, it is necessary to investigate the effect of incorporating the equation of state in the non-linear regime using the gradient expansion approach. We plan to delve into this work shortly, despite the challenges it brings with it. But we hope that it will not bring vast changes in the results that we have obtained here. 

The branching fig.(\ref{MST-EFT Branch}) depicts the plethora of crucial topics that this paper is set out to address. The highlight of the paper is the Equation of state parameter that sits at the top, which when integrated within the EFT of MST through the arsenal of regularization, renormalization, and resummation can address the overproduction of PBHs, produce SIGWs consistent with the observational PTA  data, generate a spectrum of low mass and high mass PBHs which gets reassured by cross-referring with the Microlensing experiments. All of these have been done keeping in mind both canonical($c_s=1$), and non-canonical($c_s=1.17$) models of inflation. The sections in this paper are organized as follows: In section \ref{s2}, we present the underlying theory behind the setup of MSTs within the EFT of inflation framework. We discuss in detail the theoretical construct of MST, including the need to incorporate quantum one-loop effects and further details regarding the various renormalization and resummation procedures to address such quantum contributions adequately. \textcolor{black}{To ensure foundational robustness in our application of the renormalization techniques, we have detailed the formulation of the renormalized Lagrangian and derived the necessary counterterm contributions explicitly that later connect to the removal of harmful UV divergences and smoothening of the IR divergences. Further, we elaborate on the impact of the renormalization technique on the scalar power spectrum and cosmological beta functions.  } In section \ref{s3}, we provide an overview of the SIGW generation in the presence of a general EoS $(w)$ background. There, we also highlight specific cases of $w$ values explicitly. The section \ref{s4} discusses PBH formation and its mass fraction by incorporating proper changes from having a constant $w$ parameter. Section \ref{s5} presents a comprehensive analysis of our results, which deal with avoiding the overproduction issue for changing values of both $w$ and effective sound speed $c_{s}$. We also discuss the behaviour of abundance as a function of the PBH mass and further investigate the SIGW spectra for various $w$ and $c_{s}$ values. The results obtained from our theoretical setup are confronted with the NANOGrav-15 and EPTA data. Section \ref{s6} is devoted to highlighting the establishment of smooth transitions so far, stating contrasts with sharp transitions, and discussing further possibilities. Finally, section \ref{s7} ends with us presenting the conclusions drawn from our present work.

\section{Overview on the underlying theory of Multiple Sharp Transitions in EFT (MST-EFT)} \label{s2}

Amidst all the attempts to accommodate a diverse range of physics concerning inflationary scenarios, NANOGrav-15 data, and PBH formation, \textit{Multiple Sharp Transitions} (MSTs) can quite possibly be a boon to resolve existing concerns. The following sections of this paper will utilize the scalar power spectrum that has been obtained through a rigorous approach and in this section, we will briefly discuss the underlying construction behind the theory of MSTs.

\subsection{Motivation and Approach}

The main motivation and the approach for Multiple Sharp Transitions are appended in the points described below in the following subsubsections.
\subsubsection{Generation of large mass PBHs}

    Generating large mass PBHs while properly considering the one-loop corrections, particularly in the USR phase of the EFT of single field inflation, has been a debatable issue at the forefront for quite some time. The ongoing debate revolves around identifying the more favorable scenario in the presence of the aforementioned one-loop corrections - sharp transition or smooth transition or if both are equally probable. Noteworthy is the observation by the authors in \cite{Riotto:2023gpm,Firouzjahi:2023ahg,Firouzjahi:2023aum}, emphasizing that the utilization of a single smooth transition could provide a solution to circumvent challenges arising from substantial quantum fluctuations. This approach is suggested as a means to generate significant PBHs masses while concurrently addressing the production of SIGWs. However, due to the absence of robust renormalization procedures, there remains to be an idea in debate among the cosmology community regarding the proper implementation of smooth transitions and the accuracy of the results and conclusion. In contrast, the authors in ref. \cite{Choudhury:2023vuj,Choudhury:2023rks,Choudhury:2023jlt} have examined utilizing sharp Heaviside-like transitions that the one-loop corrections in the USR phase become highly constrained due to renormalization and resummation procedures. 
In light of these considerations, MST stands out as a set-up that has the potential to span across existing concerns and successfully resolve them. Its superiority is indebted to a rigorous procedure of renormalization and resummation which supports this theory. The prime advantage of MST lies in its ability to facilitate the generation of a wide range of PBH masses, ranging from ${\cal O}(10^{4}-10^{-36})M_{\odot}$ and realizing a successful inflationary setup in the presence of the constraints coming from the one-loop quantum corrections to the scalar power spectrum \cite{Choudhury:2023rks}. Not only is this setup able to produce a spectrum of PBH masses but also generates SIGWs that are consistent with the frequencies sensitive to the NANOGrav-15 signal.


\subsubsection{Preserving Perturbative Approximations}
Previous endeavours utilized a single field inflation framework to generate Solar mass Black Holes (${\cal O}(M_{\odot})$) with a single transition but were unable to accomplish this due to strong restrictions coming from one-loop corrections that disallowed a prolonged SR phase that succeeds a USR phase. This led to failing the necessary condition for the number of e-foldings to satisfy inflation. Keeping this in mind, the motivation to introduce multiple sharp transitions is rooted in the fact that one must give importance to the proper accounting of the one-loop contributions to the primordial scalar power spectrum. When performing this procedure, prompt identification of an issue arises, especially aiming for the ultimate expected output of generation of PBHs with mass $M_{\rm PBH} \sim {\cal O}(M_{\odot})$. Since it has been shown through robust analysis that the number of e-foldings for such a situation only turns out to satisfy $\Delta{\cal N}_{\rm Total} \sim {\cal O}(20-25)$ for a single transition, one must find an alternative to complete inflation for a single transition by achieving at least $\Delta{\cal N}_{\rm Total} \sim {\cal O}(60)$ and keep open the possibility to generate large mass PBH. A promising remedy for this situation lies in the introduction of multiple USR phases in our setup, with each following similar perturbativity approximations which restricts their duration in terms of the e-foldings, $\Delta{\cal N}_{\rm USR}\sim {\cal O}(2)$, and where the different transition scales allow for the possibility to observe the generation of PBH ranging from near solar mass to sub-solar mass in nature.

\subsubsection{Is six the requisite? - Light of the Horizon Problem}
The analysis developed in \cite{Bhattacharya:2023ysp} includes a setup where $6$ USR phases are incorporated to give us a scenario of successful inflation and keep the compulsory requirements of renormalization and resummation procedures intact. The rest of the e-foldings get occupied by multiple slow-roll phases and the initial slow-roll phase from where the first sharp transition commences. There are no stringent constraints on the number of USR phases or sharp transitions that can be allowed. We have taken the number $6$ to satisfy the number of e-foldings to be $\Delta{\cal N} \sim {\cal O}(60-65)$ which is necessarily required to solve the Horizon problem. Also noteworthy is that the SIGW generated with our present set-up is poised to explain the NANOGrav-15 frequency data and other GW observations, including from LISA, LIGO, VIRGO, and KAGRA, to name a few. Notably, we show the NANOGrav signal explained by the first transition in the GW spectrum. The relevance of this mentioned discussion is as follows. Now, the minimum criterion required to achieve inflation requires $6$ sharp transitions, below which the Horizon problem emerges as an issue. However, no proven higher bounds of $\Delta{\cal N}$ exist. So, one can go beyond six transitions, which also reassures satisfaction of obeying the necessary inflation conditions while automatically solving the Horizon problem without causing any complications for the inflation parameters. Additionally, a large number of sharp transitions exactly predicts the $\Omega_k=0$. This sort of physics can be explained easily with the conformal time diagram. The resulting SIGW spectrum that will be produced may also explain signals from future observations from distinguished observatories in the scope of near-future research. 

\textcolor{black}{Below we highlight some crucial developments point-wise which lay out the necessities to incorporate six sharp transitions for our set-up of MSTs:}
\begin{itemize}
    \item  \textcolor{black}{The case with a single sharp transition was originally introduced in the scenario of single-field inflation as a means to generate large mass PBHs, at least $M_{\rm PBH}\sim {\cal O}(M_{\odot})$, by generating enough enhancement in the curvature perturbation amplitude. This, however, brought along significant quantum loop effects which needs proper care to ensure perturbativity within the underlying arguments.}
    \item \textcolor{black}{The successful application of the appropriate renormalization and resummation conditions meant preserving the perturbative approximations but at the same time it was shown to not allow for successful inflation, if $M_{\rm PBH}\sim {\cal O}(M_{\odot})$ is the goal from our setup, leaving only $M_{\rm PBH}\sim {\cal O}(10^{2}{\rm gms})$ as a possible outcome for the PBHs. }
    \item \textcolor{black}{In order to witness complete inflation and concurrently produce large mass PBH, the idea of MST was put forward in \cite{Bhattacharya:2023ysp}. The conditions, albeit, remained the same where the moment between the start, at $k_{s}$ and end, at $k_{e}$, of a single USR phase must satisfy the ratio in wavenumbers, $k_{e}/k_{s}\sim {\cal O}(10).$ To satisfy this condition and the need for an enhancement at the suitable scales relevant for considerations of the NANOGrav 15 signal, we develop the remaining sharp transitions until the end of inflation.} 
    \item \textcolor{black}{Some important conditions for the model include successfully joining the different sharp transitions together such that progression between each phase retains continuity, the power spectrum amplitude right after the fall from the transition, and before entering into the next, does not reaches values as below as at the pivot scale, and the amplitude must also not climb to greater than ${\cal O}(10^{-2})$ after each transition. Further, our set-up is built in such a way that the amplitude of the power spectrum falls rapidly and sharply at the transition points. This does not allow the $\Delta{\cal N}$ to increase much during subsequent phases. Therefore, any number less than 6 would automatically lead to a lesser number of e-foldings to complete inflation. These altogether determine the number of MSTs possible.}
    \item \textcolor{black}{Given satisfying the above conditions, a total of six remain as the required MSTs to ultimately reach the end of inflation under sufficient total e-foldings of $\Delta{\cal N}\sim {\cal O}(60-65)$. Therefore $6$ now characterizes the minimum for our MST model to retain all the necessary conditions, including perturbativity in the underlying arguments, and solving the Horizon problem. Reiterating the fact that going beyond this number would also satisfy all the necessary conditions of perturbativity and inflation. For eg: one can as well perform the analysis presented in this paper with a number of sharp transitions N$ \ge 6$. }
\end{itemize}

\subsubsection{The necessity for Renormalization and Resummation} 
The plinth of the following discussion is based on the need to perform the renormalization and DRG resummation procedures to remove the harmful quadratic UV divergences coming from the one-loop integrals, soften the irremovable logarithmic IR divergences, and, lastly, obtain a finite output for our final expression of the scalar power spectrum including quantum loop corrections on the tree level contribution. For a complete analysis, consider \cite{Choudhury:2023rks, Choudhury:2023vuj}. The renormalized version of the scalar power spectrum is known to be independent of the chosen renormalization procedures, late-time or adiabatic/wave-function renormalization. The DRG approach ensures that the finite output from the resummation over the logarithmically divergent contributions mirrors respecting the cosmological perturbation theory in the horizon-crossing and the superhorizon scales. All the mentioned procedures do not allow for a prolonged USR phase and show that the amplitude of the one-loop renormalized and DRG-resummed scalar power spectrum falls sharply right after exiting from the USR before perturbativity is violated. Since we are interested in the use of this USR phase to provide a means to generate PBHs, the best-allowed scenario in terms of the necessary amplitude of the final power spectrum is shown to arise for the effective sound speed, $1 \leq c_{s} \leq 1.17$, in light of the above-mentioned procedures.

\subsubsection{Controlling the amplitude at each sharp transition - Is it a miraculous enigma ?}

In this setup, we try to comprehend the impact of the SIGW profile as it is a significant component in probing various observational data coming from EPTA, NANOGrav-15, and other gravitational wave observations. SIGW is sourced from the scalar perturbation in the second order within the cosmological perturbation theory, and the equation of state ($w$) particularly is of utmost importance in this context which we will discuss in more detail in section \ref{s5}. Interestingly, the case of $w=1/3$ is sort of a miracle that can preserve all the necessarily required theoretical constraints while also explaining physically consistent observations. Particularly, even after performing DRG resummation at each sharp transition scale over the existing loop corrections, the amplitude of neither the scalar power spectrum nor the tensor power spectrum keeps on rising breaking the perturbative approximations. Rather, the first sharp transition exactly complies with the NANOGrav-15 signal and the subsequent transitions have the potential to be viable in the ranges of other observational data. To emphasize, all of this is made possible without the requirement to add any coarse-graining factor to the final quantum loop corrected renormalized DRG resumed scalar power spectrum. However, for the other values of the equation of state, as will be evident from the section \ref{s5}, steering above $w=1/3$ will make the amplitude go uncontrollably higher at each sharp transition scale thus violating the perturbativity constraints. Therefore, to maintain a steady amplitude valid within the said approximation, there is a compulsion to incorporate coarse-graining factors phenomenologically. The number of coarse-graining factors will be one less than the number of sharp transitions as the initial sharp transition scale has been already set by us. The subsequent transitions would thus require the incorporation of this factor. For our analysis, a total of five coarse-graining factors are required. This coarse-graining factor will be different for different sharp transitions with each subsequent factor much lesser than the previous in magnitude since the peak amplitude now at succeeding transitions will effortlessly shoot much higher than its predecessor. Also, going below $w=1/3$ down to $w=0$, there is a rapid decrement in the peak amplitudes after each subsequent sharp transitions. Hence, the amplitude of the resulting spectrum becomes increasingly small at each transition scale. Even though this does not violate the perturbative approximation, the resulting fractional abundance generated is much smaller than $(10^{-3}-1)$, and can never explain the generation of high-mass PBHs. Hence, the resolution of this too requires the addition of coarse-graining factors just like previously except this time in the opposite sense of addition i.e. the factor serves to increase the amplitude to make it substantially important with regards to obtaining desired results. Further, going below $w=0$ till $w\sim-0.05$ produces a spectrum that still has a considerable amplitude, even higher than for $w=0$. This rise is however only till the EoS reaches $w\sim-0.05$. Seeking any lower value than this results in a spectrum with much-diminished amplitudes which is not very useful from the perspective of addressing the overproduction issue. The only logical conclusion that can be derived is that $w=1/3$ somehow serves as a miracle number as far as the inflation scenarios are concerned. This may be indebted to the underlying quantum fields starting to behave as radiations.

\subsection{The EFT Setup}

The general setup involves us working within the paradigm of the effective field theory (EFT) of inflation, where we tend to conduct our analysis regarding the theory of perturbations in the metric using an effective action valid below some fixed UV cut-off scale (usually in practice taken as the reduced Planck scale $M_{p}$). Starting with only a scalar field $\phi(t,\mathbf{x})$ driving inflation, its perturbations about a quasi de-Sitter background break the time-diffeomorphism symmetry but not the spatial diffeomorphism symmetry. The scalar perturbations under the action of time-diffeomorphism transform as- $\delta\phi \rightarrow \delta\phi + \dot{\phi_{0}}\xi^{0}(\mathbf{x},t)$, for $t \rightarrow t + \xi^{0}(\mathbf{x},t)$. At this stage, we invoke the unitary gauge, which enables the identification $\delta\phi(t,\mathbf{x}) = 0$. This step moves the scalar degree of freedom, $\delta\phi$, into the metric, thereby increasing its total degree of freedom to equal three: two helicities and one scalar mode.
Now neglecting the suppressed contributions, we write the effective action in terms of two derivatives in the metric as follows: 
 \bea \label{action}
  S &=& \int d^4 x \sqrt{-g} \Bigg(\frac{M_p^2}{2}R+M_{p}^2 \dot H g^{00}-M_p ^2 (3H^2 +\dot H)+\sum_{n=2,3}\frac{M_{n}^{4}(t)}{n!}(g^{00}+1)^{n}-\frac{\overline{M}_{1}^3 (t)}{2} (g^{00}+1) \delta K_{\mu}^{\mu} \nonumber\\ 
&&\quad\quad\quad\quad\quad\quad\quad\quad\quad\quad\quad\quad\quad\quad\quad\quad\quad\quad\quad\quad\quad\quad\quad\quad\quad\quad\quad\quad-\frac{\overline{M}_{2}^2 (t)}{2} (\delta K_{\mu}^{\mu})^2 - \frac{\overline{M}_{3}^2 (t)}{2} \delta K_{\nu}^{\mu}\delta K_{\mu}^{\nu}+\cdots\Bigg). 
    \eea   
For details regarding the construction of this EFT, refer to \cite{Choudhury:2017glj, Cheung:2007st}. The ellipsis represents the higher-order terms in the action. $M_{2}(t)$, $M_{3}(t)$, $\overline{M}_{1}(t)$, $\overline{M}_{2}(t)$, $\overline{M}_{3}(t)$  represent Wilson's coefficients, while $K_{\mu \nu}$ is the extrinsic curvature, described in the spatially flat FLRW background metric:
\bea
K_{\mu \nu} &=& h_{\mu}^{\alpha}\grad_{\alpha} n_{\nu}, \\
h_{\mu \nu} &=& g_{\mu \nu} + n_{\mu} n_{\nu},\\
n_{\mu} &=&
\frac{\delta_{\mu}t}{\sqrt{-g^{\mu \nu}\partial_{\mu}t \partial_{\nu}t}}.
\eea
The above expressions includes the induced metric $h_{\mu \nu}$ and the unit normal $n_{\mu}$. Additionally, $\delta K_{\mu \nu}$ represents the perturbed component of the extrinsic curvature given by: 
\bea
\delta K_{\mu \nu} = K_{\mu \nu} - a^2 H h_{\mu \nu},
\eea
where $H$ is the Hubble parameter. Now, this action of (\ref{action}) has a reduced symmetry and to restore the gauge invariance in the effective action, the St\"{u}ckelberg mechanism comes into play to make the scalar mode appear explicitly. The idea of the mechanism is to introduce a Goldstone mode that non-linearly realizes the broken time-diffeomorphism symmetry. This can be visualized as the following effect under the unitary gauge: $-\xi^{0}(t,\mathbf{x}) \rightarrow \tilde{\pi}(t,\mathbf{x})$, where $\tilde{\pi}(t,\mathbf{x})$ represents the shifted version of the Goldstone mode $\pi(t,\mathbf{x})$, or the scalar perturbation around the background metric. This method is similar to spontaneous symmetry breaking in $SU(N)$ non-abelian gauge theories.

\subsection{Mode functions for comoving scalar curvature perturbation}

Finally, after using the decoupling limit, $E_{\rm mix} = \sqrt{\dot{H}}$, one can neglect terms in the action, giving mixing contributions between the Goldstone mode and the metric fluctuations.    
Now, upon utilizing the one-to-one mapping of the Goldstone mode with the comoving curvature perturbation, that is $\zeta(t,\mathbf{x}) = -H\pi(t,\mathbf{x})$, we can write the perturbed version of the action up to the second order as follows:
\bea
S^{(2)}_{\zeta} = M^{2}_{p}\int d\tau\;d^{3}x\;a^{2}\frac{\epsilon}{c_{s}^{2}}\;\left[\zeta'^{2}-c_{s}^{2}(\partial_{i}\zeta)^{2}\right].
\eea

Varying the above action gives us a second-order differential equation called the Mukhanov Sasaki (MS) equation from which the mode solutions for different phases are obtained. We thus write the MS equation in Fourier space as:
\bea
\zeta_{\bf k}^{''}(\tau)+2\frac{z^{'}(\tau)}{z(\tau)}\zeta_{\bf k}^{'}(\tau)+c_s ^2 k^2 \zeta_{\bf k}(\tau) = 0.
\eea

Now in this set-up of 6 consecutive sharp transitions, the general $n$th USR phase is denoted as USR$_{n}$ and similarly the $(n+1)$th SR phase is denoted as SR$_{n+1}$ where the index $n$ runs from $1\to 6$ and the phase SR$_{1}$ is the first slow-roll, SRI, phase that satisfies $\Delta{\cal N}_{\rm SRI}=\ln{(k_{s_{1}}/k_{*})}\sim{\cal O}(12)$ for the fixed value of the pivot scale ($k_{*}=0.02{\rm Mpc^{-1}}$). Therefore, the corresponding mode solutions  for these SR and the USR phases that are obtained after solving for the MS equation, read as: 
\bea
\zeta_{\mathbf{k}, \rm SR_{1}}(\tau) &=& \bigg[\frac{ic_s H}{2 M_p \sqrt{\epsilon}}\bigg]_{*}\times \frac{1}{(c_s k)^{3/2}} \bigg[\alpha_{\mathbf{k}}^{(1)}(1+ikc_s \tau)\exp{(-ikc_s \tau)}-\beta_{\mathbf{k}}^{(1)}(1-ikc_s \tau)\exp{(ikc_s\tau)}\bigg], \nonumber \\
\zeta_{\mathbf{k}, \rm USR_n}(\tau) &=& \bigg[\frac{ic_s H}{2 M_p \sqrt{\epsilon}}\bigg]_{*} \times \bigg(\frac{\tau_{s_{n}}}{\tau}\bigg)^3 \bigg[\alpha_{\mathbf{k}}^{(2n)}(1+ikc_s \tau)\exp{(-ikc_s \tau)}-\beta_{\mathbf{k}}^{(2n)}(1-ikc_s \tau)\exp{(ikc_s\tau)}\bigg], \nonumber \\
\zeta_{\mathbf{k}, \rm SR_{n+1}}(\tau) &=& \bigg[\frac{ic_s H}{2 M_p \sqrt{\epsilon}}\bigg]_{*} \times \bigg(\frac{\tau_{s_{n}}}{\tau_{e_{n}}}\bigg)^3 \bigg[\alpha_{\mathbf{k}} ^{(2n+1)}(1+ikc_s \tau)\exp{(-ikc_s \tau)}-\beta_{\mathbf{k}}^{(2n+1)}(1-ikc_s \tau)\exp{(ikc_s\tau)}\bigg].
\eea

The expression in the parenthesis gets evaluated at the pivot scale. $\alpha_{\mathbf{k}} ^{(i)}$ and $\beta_{\mathbf{k}} ^{(i)}$ (where $i\in 1,2,3,\cdots$) are the Bogoliubov coefficients that arise as a result of implementing the Israel junction conditions at the boundaries over the conformal times $\tau=\tau_{s_{n}}$ and $\tau=\tau_{e_{n}}$. We will now discuss how the above mode solutions provide us with the means to incorporate the MST theory. For this, we begin with an initial SR$_{1}$ phase, which involves the pivot scale $k_{*}$ and where the first SR parameter $\epsilon$ is almost a constant while the second SR parameter $\eta = \epsilon'/\epsilon{\cal H}$ is a very small quantity and treated as constant. This phase persists until a sharp transition gets encountered at the conformal time $\tau=\tau_{s_{1}}$. We now enter into the first USR phase, USR$_{1}$, which follows the conformal time interval $\tau_{s_{1}} \leq \tau \leq \tau_{e_{1}}$. Importantly, during the transition into this phase, the $\eta$ parameter experiences a sudden jump from $\eta \rightarrow 0$ to $\eta \rightarrow -6$. This jump in value generates a drastic effect, which propagates into giving us significant one-loop contributions, which we will soon show as the case. This phase respects the necessary bound on the e-foldings: $\Delta{\cal N}_{\rm USR_{1}} \sim {\cal O}(2)$. From USR$_{1}$ at $\tau_{e_{1}}$, we experience another sharp transition into the new SR$_{2}$ phase which lasts within the interval $\tau_{e_{1}} \leq \tau \leq \tau_{s_{2}}$. During this new SR phase, the $\epsilon$ parameter again goes to being an almost constant with the value $\epsilon(\tau) = \epsilon(\tau_{e_{1}}/\tau_{s_{1}})^{6}$, with $\epsilon$ as in the SR$_{1}$, and the $\eta$ parameter also drops to its initial value of $\eta \rightarrow 0$. Thereafter, we witness another sharp transition at $\tau = \tau_{s_{2}}$ into a new USR$_{2}$ phase respecting the same constraint on its interval of $\Delta{\cal N}_{\rm USR_{2}} \sim {\cal O}(2)$ until $\tau = \tau_{e_{2}}$ occurs. This trend is continued similarly till we finally meet the condition of having the total e-foldings satisfy $\Delta{\cal N}_{\rm Total} \sim {\cal O}(60-70)$. This is mainly the reason why we chose to consider six sharp transitions in our analysis.\footnote{Note that the minimum number of sharp transitions required to satisfy the criterion for resolving the Horizon Problem is six. In practice, you can take more sharp transitions that will also obey the number of efoldings criterion for inflation while automatically resolving the Horizon problem.} We now turn to discuss the Bogoliubov coefficients ($\alpha_{\textbf{k}}$, $\beta_{\textbf{k}}$) for different phases and provide their expressions. The Bogoliubov coefficients for any arbitrary SR and USR phases are given as : 
\bea
\alpha^{(2n)}_{\textbf{k}}&=&-\frac{k_{e_{n-1}}^{3}k_{s_{n}}^{3}}{2k^{3}k_{s_{n-1}}^{3}}
\bigg[\bigg(3i-\frac{2k^{3}}{k_{s_{n}}^{3}}+\frac{3ik^{2}}{k_{s_{n}}^{2}}\bigg)\alpha^{(2n-1)}_{\textbf{k}}+3i\exp{\left(-\frac{2ik}{k_{s_{n}}}\right)}\bigg(i-\frac{k}{k_{s_{n}}}\bigg)^{2}\beta^{(2n-1)}_{\textbf{k}}\bigg],\\
\beta^{(2n)}_{\textbf{k}}&=&
\exp{\left(\frac{2ik}{k_{s_{n}}}\right)}\frac{k_{e_{n-1}}^{3}k_{s_{n}}^{3}}{2k^{3}k_{s_{n-1}}^{3}}
\bigg[3i\bigg(i+\frac{k}{k_{s_{n}}}\bigg)^{2}\alpha^{(2n-1)}_{\textbf{k}}+\exp{\left(-\frac{2ik}{k_{s_{n}}}\right)}\bigg(3i+\frac{2k^{3}}{k_{s_{n}}^{3}}+\frac{3ik^{2}}{k_{s_{n}}^{2}}\bigg)
\beta^{(2n-1)}_{\textbf{k}}\bigg],\\
\alpha^{(2n+1)}_{\textbf{k}}&=&\frac{k_{e_{n}}^{3}}{2k^{3}}
\bigg[\bigg(3i+\frac{2k^{3}}{k_{e_{n}}^{3}}+\frac{3ik^{2}}{k_{e_{n}}^{2}}\bigg)\alpha^{(2n)}_{\textbf{k}}+3i\exp{\left(-\frac{2ik}{k{e_{n}}}\right)}\bigg(i-\frac{k}{k_{e_{n}}}\bigg)^{2}\beta^{(2n)}_{\textbf{k}}\bigg],\\
\beta^{(2n+1)}_{\textbf{k}} &=& -\exp{\left(\frac{2ik}{k_{e_{n}}}\right)}\frac{k_{e_{n}}^{3}}{2k^{3}}
\bigg[3i\bigg(i+\frac{k}{k_{e_{n}}}\bigg)^{2}\alpha^{(2n)}_{\textbf{k}}+\exp{\left(-\frac{2ik}{k_{e_{n}}}\right)}\bigg(3i-\frac{2k^{3}}{k_{e_{n}}^{3}}+\frac{3ik^{2}}{k_{e_{n}}^{2}}\bigg)
\beta^{(2n)}_{\textbf{k}}\bigg],
\eea
these expressions are simplified from the use of the Horizon crossing conditions: $-k_{s_{n}}c_{s}\tau_{s_{n}} = -k_{e_{n}}c_{s}\tau_{e_{n}} = 1$. The effective sound speed $c_s$\footnote{Notice that this is different from the adiabatic fluid speed used in refs.\cite{Balaji:2023ehk,Domenech:2021ztg}} plays a crucial role in the present setup of sharp transitions through its specific parameterization. Its value at the pivot scale is previously established as $c_{s}(\tau_{*})=c_{s}$. At the moments for the occurrence of the sharp transition, its value, however, takes a new form of $c_{s}(\tau_{s_{n}}) = c_{s}(\tau_{e_{n}}) = \tilde{c_{s}} = 1 \pm \delta$, where $\delta \ll 1$. Here $\tilde{c_{s}} \ne c_{s}$, and the value of  $\tilde{c_{s}}$ remains the same at each successive conformal time of the transition $(\tau_{s_{n}},\tau_{e_{n}})$ for $n=1\rightarrow 6$. This design for the behaviour of the sound speed enables the generation of PBH from our setup by providing for the necessary amplification to the scalar perturbations and will later prove to be important also from the perspective of the one-loop contributions. 

\subsection{Tree level scalar power spectrum from MST-EFT}

Since the mode solutions are different for different regions of SR$_1$, USR$_n$, and SR$_{n+1}$, the dimensionless scalar power spectrum will have different values at the respective regions. However, there will be one common general expression valid for all the scales involved (subhorizon, superhorizon, and at horizon re-entry) given by :
\bea
\bigg[\Delta^{2}_{\zeta,{\bf Tree}}(k)\bigg] &=& \bigg[\frac{H^2}{8\pi ^2 M_{p}^2 \epsilon c_s}\bigg]_{*} \nonumber \\ 
\quad\quad\quad  &\times&  \begin{cases} (1+k^2c_s^2\tau^2)
& \mbox{if } k\ll k_{s_1} (\rm SR_{1})\\
    \bigg(\displaystyle{\frac{k_{e_n}}{k_{s_n}}}\bigg)^{6} \bigg|\alpha_{\bf k}^{(2n)}(1+ikc_s\tau)\exp{(-ikc_s\tau)} \\              \quad\quad\quad\quad\quad\quad\quad\quad\quad\quad\quad - \beta_{\bf k}^{(2n)}(1-ikc_s\tau)\exp{(ikc_s\tau)}\bigg|^{2} &\mbox{if } k_{s_n} \le k\le k_{e_n} (\rm USR_{n}) \\
      \bigg(\displaystyle{\frac{k_{e_n}}{k_{s_n}}}\bigg)^{6} \bigg|\alpha_{\bf k}^{(2n+1)}(1+ikc_s\tau) \exp{(-ikc_s\tau)}  \\ \quad\quad\quad\quad\quad\quad\quad\quad\quad\quad\quad - \beta_{\bf k}^{(2n+1)}(1-ikc_s\tau) \exp{(ikc_s\tau)} \bigg|^{2} & \mbox{if } k_{e_n}\le k \le k_{s_{n+1}} (\rm SR_{n+1})
    
    \end{cases}
\quad \eea

The expression in the parenthesis outside all the regions is evaluated in the pivot scale $k_{*}$. Now, implementing late time scale $\tau \to 0$, we write the scalar power spectrum as :
\begin{equation}
    \bigg[\Delta^{2}_{\zeta,{\bf Tree}}(k)\bigg] = \bigg[\frac{H^2}{8\pi ^2 M_{p}^2 \epsilon c_s}\bigg]_{*} \times \begin{cases} 1  &\mbox{if } k\ll k_{s_1} (\rm SR_{1})\\
    \bigg(\displaystyle{\frac{k_{e_n}}{k_{s_n}}}\bigg)^{6}\times \bigg|\alpha_{\bf k}^{(2n)}-\beta_{\bf k}^{(2n)}\bigg|^{2} & \mbox{if } k_{s_n} \le k\le k_{e_n} (\rm USR_{n}) \\
      \bigg(\displaystyle{\frac{k_{e_n}}{k_{s_n}}}\bigg)^{6} \times \bigg|\alpha_{\bf k}^{(2n+1)} - \beta_{\bf k}^{(2n+1)}\bigg|^{2} & \mbox{if } k_{e_n}\le k \le k_{s_{n+1}} (\rm SR_{n+1})
    
    \end{cases}
\end{equation}

Adding up the power spectrum for all the regions, we obtain the total tree-level scalar power spectrum at the superhorizon scale ($-kc_s\tau \ll 1$), in terms of the Bogoliubov coefficients as obtained previously, 
\bea
   \label{treeps}
\bigg[\Delta^{2}_{\zeta,{\bf Tree}}(k)\bigg]_{\bf Total} &=&\bigg[\Delta^{2}_{\zeta,{\bf Tree}}(k)\bigg]_{\textbf{SR}_{1}} \times \bigg[1+\sum^{6}_{n=1}\Theta(k-k_{s_{n}}) \left(\frac{k_{e_{n}}}{k_{s_{n}}}\right)^{6}\left|\alpha^{(2n)}_{\bf k}-\beta^{(2n)}_{\bf k}\right|^2\nonumber\\
&&\quad\quad\quad\quad\quad\quad\quad\quad\quad\quad\quad\quad\quad\quad\quad+\sum^{6}_{n=1}\Theta(k-k_{e_{n}})\times \left(\frac{k_{e_{n}}}{k_{s_{n}}}\right)^{6}\left|\alpha^{(2n+1)}_{\bf k}-\beta^{(2n+1)}_{\bf k}\right|^2\bigg].\quad\quad
\eea
The common factor for the total power spectrum valid for all the scales (subhorizon, superhorizon, and horizon re-entry) represents the SR$_1$ contribution and is given by :
\bea \bigg[\Delta^{2}_{\zeta,{\bf Tree}}(k)\bigg]_{\textbf{SR}_{1}} = \displaystyle{ \bigg[\frac{H^2}{8\pi ^2 M_{p}^2 \epsilon c_s}\bigg]_{*}}\left[1+(k/k_{s_{1}})^{2}\right] \xrightarrow{\rm Super-horizon \; scale \; k\ll k_{s_1}} \displaystyle{ \bigg[\frac{H^2}{8\pi ^2 M_{p}^2 \epsilon c_s}\bigg]_{*}}. \eea
Also, notice that the Heaviside theta function in eqn.(\ref{treeps}) is implemented to enforce the requirement of sharp transition at the scales where PBH formation occurs.

\subsection{Cut-off regularized one-loop correction to the scalar power spectrum from MST-EFT}

 As mentioned earlier, the quantum loop corrections to the power spectrum are significant due to the presence of multiple USR phases. To incorporate one-loop corrections, we need to first present the third-order action as follows:
\bea
\label{s3action}
         S_{\zeta}^{(3)} &=& \int d\tau\;d^3x\;M_p ^2\;a^2\bigg [\bigg(3(c_s ^2 -1)\epsilon + \epsilon ^2 - \frac{\epsilon ^3}{2}\bigg )\; \zeta^{\prime} {^2} \zeta + \frac{\epsilon}{c_s ^2}\bigg(\epsilon - 2s +1 -c_s ^2 \bigg)(\partial_i \zeta)^2 \zeta  - \frac{2 \epsilon}{c_s ^2}\zeta^{\prime} (\partial_i \zeta) \bigg (\partial_i \partial ^{-2}\bigg(\frac{\epsilon \zeta^{\prime}}{c_s ^2}\bigg)\bigg) \nonumber \\ 
        && \quad \quad \quad \quad \quad \quad \quad \quad -\frac{\epsilon}{aH}\bigg (1-\frac{1}{c_s ^2}\bigg) \left(\zeta^{\prime} {^3}+\zeta^{\prime}(\partial_i \zeta)^2 \right) + \frac{\epsilon}{2}\zeta \bigg(\partial_i \partial_j \partial^{-2}\bigg (\frac{\epsilon \zeta^{\prime}}{c_s ^2}\bigg)\bigg)^2 + \underbrace{\frac{\epsilon}{2c_s ^2}\partial_{\tau} \bigg(\frac{\eta}{c_s ^2}\bigg)\zeta^{\prime} \zeta^2}_{\textbf{Dominant term}}+.....\bigg].
   \eea
Notice that the ellipsis after the dominant term represent the other sub-dominant contributions that are suppressed. 
Here we focus our view on the third-order perturbed action $S_{\zeta}^{(3)}$ and find out that the final term in this action contributes as ${\cal O}(\epsilon^3)$ in the SR$_1$ and SR$_{n+1}$ phase and as ${\cal O}(\epsilon)$ in the USR$_n$ phase. It involves the conformal time-derivative of the combination $\eta/c_{s}^{2}$, which is the prime reason behind incorporating the abrupt changes in the $\eta$ parameter and the specific parameterization of $c_{s}$ which become significant at the sharp transitions. Specifically, the term $\partial_{\tau}(\eta/c_{s}^{2}) \approx 0$ during the conformal time interval of each SR$_{n}$ and USR$_{n}$ regimes, except at the transition moments. At the moments labelled by $(\tau_{s_{n}},\tau_{e_{n}})$ for $n=1\rightarrow 6$, the dominant term takes the values $\partial_{\tau}(\eta/c_{s}^{2}) \approx -\Delta\eta(\tau_{s_{n}})/\tilde{c_{s}}^{2}$ and $\partial_{\tau}(\eta/c_{s}^{2}) \approx \Delta\eta(\tau_{e_{n}})/\tilde{c_{s}}^{2}$. Now that we made clear how the parameterization of the two parameters $\eta$ and $c_{s}$ are important for our analysis of the enhancements in fluctuations and PBH formation, we can discuss further the leading order contributions to the scalar power spectrum from the correlation functions calculated using $S^{(3)}_{\zeta}$.

\textcolor{black}{We will now examine the contributions of the loop corrections by employing the Schwinger-Keldysh (In-In) formalism. With this, the two-point correlation function of $\langle\hat{\zeta_{\bf p}}\hat{\zeta_{\bf -p}}\rangle$ at the late time scale is written as:}
\bea
\big<\hat{\zeta}_{\bf p}\hat{\zeta}_{\bf -p}\big> &=& \big< \hat{\zeta}_{\bf p}(\tau)\hat{\zeta}_{\bf -p}(\tau)\big>_{\tau \rightarrow 0} \nonumber \\
&=& \bigg<\bigg[\bar{T} \exp{\bigg(i \int_{-\infty(1-i\epsilon)}^{\tau}d\tau' H_{\rm int}(\tau')\bigg)}\bigg]\hat{\zeta}_{\bf p}(\tau)\hat{\zeta}_{\bf -p}(\tau)\bigg[T \exp{\bigg(-i \int_{-\infty(1+i\epsilon)}^{\tau}d\tau'' H_{\rm int}(\tau'')\bigg)}\bigg]_{\tau \rightarrow 0}
\eea
\textcolor{black}{Here $T$ and $\bar{T}$ represent the time and anti-time ordering operators. The interaction Hamiltonian $H_{\rm int}(\tau)$ can be computed from the eqn.(\ref{s3action}) as follows:}
\begin{widetext}
\bea && H_{\rm int}(\tau)=-\int d^3x\;  M^2_{ pl}a^2\; \bigg[\left(3\left(c^2_s-1\right)\epsilon+\epsilon^2-\frac{1}{2}\epsilon^3\right)\zeta^{'2}\zeta+\frac{\epsilon}{c^2_s}\bigg(\epsilon-2s+1-c^2_s\bigg)\left(\partial_i\zeta\right)^2\zeta\nonumber\\ 
&&\quad\quad\quad\quad\quad\quad\quad\quad\quad\quad\quad-\frac{2\epsilon}{c^2_s}\zeta^{'}\left(\partial_i\zeta\right)\left(\partial_i\partial^{-2}\left(\frac{\epsilon\zeta^{'}}{c^2_s}\right)\right)-\frac{1}{aH}\left(1-\frac{1}{c^2_{s}}\right)\epsilon \bigg(\zeta^{'3}+\zeta^{'}(\partial_{i}\zeta)^2\bigg)
     \nonumber\\
&& \quad\quad\quad\quad\quad\quad\quad\quad\quad\quad\quad+\frac{1}{2}\epsilon\zeta\left(\partial_i\partial_j\partial^{-2}\left(\frac{\epsilon\zeta^{'}}{c^2_s}\right)\right)^2
+\underbrace{\frac{1}{2c^2_s}\epsilon\partial_{\tau}\left(\frac{\eta}{c^2_s}\right)\zeta^{'}\zeta^{2}}_{\bf Most~dominant ~term~in~USR}+\cdots
  \bigg],\quad\quad\eea
\end{widetext}
\textcolor{black}{ which is simply ${\cal H} = -{\cal L}^{(3)}_{\rm int}$. We can now further write the two-point correlation function of the scalar modes from the contributions up to the one-loop corrections in Dyson Swinger series.}
\begin{widetext}
    \bea  &&\label{g}\langle\hat{\zeta}_{\bf p}\hat{\zeta}_{-{\bf p}}\rangle= \underbrace{\langle\hat{\zeta}_{\bf p}\hat{\zeta}_{-{\bf p}}\rangle_{(0,0)}}_{\bf Tree\;level\;result}+\underbrace{\langle\hat{\zeta}_{\bf p}\hat{\zeta}_{-{\bf p}}\rangle_{(0,1)}+\langle\hat{\zeta}_{\bf p}\hat{\zeta}_{-{\bf p}}\rangle^{\dagger}_{(0,1)}+\langle\hat{\zeta}_{\bf p}\hat{\zeta}_{-{\bf p}}\rangle_{(0,2)}+\langle\hat{\zeta}_{\bf p}\hat{\zeta}_{-{\bf p}}\rangle^{\dagger}_{(0,2)}+\langle\hat{\zeta}_{\bf p}\hat{\zeta}_{-{\bf p}}\rangle_{(1,1)}}_{\bf One-loop\;level\;result},
\eea
\end{widetext}
\textcolor{black}{The contributions are given as:}
\bea
     &&\label{g0}\langle\hat{\zeta}_{\bf p}\hat{\zeta}_{-{\bf p}}\rangle_{(0,0)}=\left[\langle \hat{\zeta}_{\bf p}(\tau)\hat{\zeta}_{-{\bf p}}(\tau)\rangle\right]_{\tau\rightarrow 0},\\
    &&\label{g1}\langle\hat{\zeta}_{\bf p}\hat{\zeta}_{-{\bf p}}\rangle_{(0,1)}=\left[-i\int^{\tau}_{-\infty}d\tau_1\;\langle \hat{\zeta}_{\bf p}(\tau)\hat{\zeta}_{-{\bf p}}(\tau)H_{\rm int}(\tau_1)\rangle\right]_{\tau\rightarrow 0},\\
 &&\label{g2}\langle\hat{\zeta}_{\bf p}\hat{\zeta}_{-{\bf p}}\rangle^{\dagger}_{(0,1)}=\left[-i\int^{\tau}_{-\infty}d\tau_1\;\langle \hat{\zeta}_{\bf p}(\tau)\hat{\zeta}_{-{\bf p}}(\tau)H_{\rm int}(\tau_1)\rangle^{\dagger}\right]_{\tau\rightarrow 0},\\
 &&\label{g3}\langle\hat{\zeta}_{\bf p}\hat{\zeta}_{-{\bf p}}\rangle_{(0,2)}=\left[\int^{\tau}_{-\infty}d\tau_1\;\int^{\tau}_{-\infty}d\tau_1\;\langle \hat{\zeta}_{\bf p}(\tau)\hat{\zeta}_{-{\bf p}}(\tau)H_{\rm int}(\tau_1)H_{\rm int}(\tau_2)\rangle\right]_{\tau\rightarrow 0},\\
 &&\label{g4}\langle\hat{\zeta}_{\bf p}\hat{\zeta}_{-{\bf p}}\rangle^{\dagger}_{(0,2)}=\left[\int^{\tau}_{-\infty}d\tau_1\;\int^{\tau}_{-\infty}d\tau_1\;\langle \hat{\zeta}_{\bf p}(\tau)\hat{\zeta}_{-{\bf p}}(\tau)H_{\rm int}(\tau_1)H_{\rm int}(\tau_2)\rangle^{\dagger}\right]_{\tau\rightarrow 0},\\
  &&\label{g5}\langle\hat{\zeta}_{\bf p}\hat{\zeta}_{-{\bf p}}\rangle^{\dagger}_{(1,1)}=\left[\int^{\tau}_{-\infty}d\tau_1\;\int^{\tau}_{-\infty}d\tau_1\;\langle H_{\rm int}(\tau_1)\hat{\zeta}_{\bf p}(\tau)\hat{\zeta}_{-{\bf p}}(\tau)H_{\rm int}(\tau_2)\rangle^{\dagger}\right]_{\tau\rightarrow 0}.\eea
\textcolor{black}{Our task now remains to explicitly evaluate the equations (\ref{g1})-(\ref{g5}) both in the SR as well as in the USR region. The contribution from equation (\ref{g0}) representing the tree level effect is already computed explicitly in the previous section.}

\textcolor{black}{Let us now look into the highlighted dominant cubic self interaction of the Hamiltonian, which will further contribute to the two-point correlation function of the scalar modes at the one-loop level in the USR period, and quantified as follows:}
\begin{widetext}
     \bea   \langle\hat{\zeta}_{\bf p}\hat{\zeta}_{-{\bf p}}\rangle_{(0,1)}&=& -\frac{iM^2_{ pl}}{2}\int^{0}_{-\infty}d\tau\frac{a^2(\tau)}{c^2_s(\tau)}\epsilon(\tau)\partial_{\tau}\left(\frac{\eta(\tau)}{c^2_s(\tau)}\right)\nonumber\\
  &&\times\int \frac{d^{3}{\bf k}_1}{(2\pi)^3} \int \frac{d^{3}{\bf k}_2}{(2\pi)^3} \int \frac{d^{3}{\bf k}_3}{(2\pi)^3} \nonumber\\
  && \times\delta^3\bigg({\bf k}_1+{\bf k}_2+{\bf k}_3\bigg) \times \langle \hat{\zeta}_{\bf p}\hat{\zeta}_{-{\bf p}}\hat{\zeta}^{'}_{{\bf k}_1}(\tau)\hat{\zeta}_{{\bf k}_2}(\tau)\hat{\zeta}_{{\bf k}_3}(\tau)\rangle,\\
   \langle\hat{\zeta}_{\bf p}\hat{\zeta}_{-{\bf p}}\rangle_{(0,1)}&=& -\frac{iM^2_{ pl}}{2}\int^{0}_{-\infty}d\tau\frac{a^2(\tau)}{c^2_s(\tau)}\epsilon(\tau)\partial_{\tau}\left(\frac{\eta(\tau)}{c^2_s(\tau)}\right)\nonumber\\
  &&\times\int \frac{d^{3}{\bf k}_1}{(2\pi)^3} \int \frac{d^{3}{\bf k}_2}{(2\pi)^3} \int \frac{d^{3}{\bf k}_3}{(2\pi)^3} \nonumber\\
  && \times\delta^3\bigg({\bf k}_1+{\bf k}_2+{\bf k}_3\bigg) \times \langle \hat{\zeta}_{\bf p}\hat{\zeta}_{-{\bf p}}\hat{\zeta}^{'}_{{\bf k}_1}(\tau)\hat{\zeta}_{{\bf k}_2}(\tau)\hat{\zeta}_{{\bf k}_3}(\tau)\rangle^{\dagger},\\
     \langle\hat{\zeta}_{\bf p}\hat{\zeta}_{-{\bf p}}\rangle_{(0,2)}&=& -\frac{M^4_{ pl}}{4}\int^{0}_{-\infty}d\tau_1\frac{a^2(\tau_1)}{c^2_s(\tau_1)}\epsilon(\tau_1)\partial_{\tau_1}\left(\frac{\eta(\tau_1)}{c^2_s(\tau_1)}\right)\;\int^{0}_{-\infty}d\tau_2\;\frac{a^2(\tau_2)}{c^2_s(\tau_2)}\epsilon(\tau_2)\partial_{\tau_2}\left(\frac{\eta(\tau_2)}{c^2_s(\tau_2)}\right)\nonumber\\
  &&\times\int \frac{d^{3}{\bf k}_1}{(2\pi)^3} \int \frac{d^{3}{\bf k}_2}{(2\pi)^3} \int \frac{d^{3}{\bf k}_3}{(2\pi)^3} \int \frac{d^{3}{\bf k}_4}{(2\pi)^3} \int \frac{d^{3}{\bf k}_5}{(2\pi)^3} \int \frac{d^{3}{\bf k}_6}{(2\pi)^3}\nonumber\\
  &&\times \delta^3\bigg({\bf k}_1+{\bf k}_2+{\bf k}_3\bigg) \delta^3\bigg({\bf k}_4+{\bf k}_5+{\bf k}_6\bigg)\nonumber\\
  &&\times \langle \hat{\zeta}_{\bf p}\hat{\zeta}_{-{\bf p}}\hat{\zeta}^{'}_{{\bf k}_1}(\tau_1)\hat{\zeta}_{{\bf k}_2}(\tau_1)\hat{\zeta}_{{\bf k}_3}(\tau_1)\hat{\zeta}^{'}_{{\bf k}_4}(\tau_2)\hat{\zeta}_{{\bf k}_5}(\tau_2)\hat{\zeta}_{{\bf k}_6}(\tau_2)\rangle,\eea\bea
  \langle\hat{\zeta}_{\bf p}\hat{\zeta}_{-{\bf p}}\rangle^{\dagger}_{(0,2)}&=& -\frac{M^4_{ pl}}{4}\int^{0}_{-\infty}d\tau_1\frac{a^2(\tau_1)}{c^2_s(\tau_1)}\epsilon(\tau_1)\partial_{\tau_1}\left(\frac{\eta(\tau_1)}{c^2_s(\tau_1)}\right)\;\int^{0}_{-\infty}d\tau_2\;\frac{a^2(\tau_2)}{c^2_s(\tau_2)}\epsilon(\tau_2)\partial_{\tau_2}\left(\frac{\eta(\tau_2)}{c^2_s(\tau_2)}\right)\nonumber\\
  &&\times\int \frac{d^{3}{\bf k}_1}{(2\pi)^3} \int \frac{d^{3}{\bf k}_2}{(2\pi)^3} \int \frac{d^{3}{\bf k}_3}{(2\pi)^3} \int \frac{d^{3}{\bf k}_4}{(2\pi)^3} \int \frac{d^{3}{\bf k}_5}{(2\pi)^3} \int \frac{d^{3}{\bf k}_6}{(2\pi)^3}\nonumber\\
  &&\times \delta^3\bigg({\bf k}_1+{\bf k}_2+{\bf k}_3\bigg) \delta^3\bigg({\bf k}_4+{\bf k}_5+{\bf k}_6\bigg)\nonumber\\
  &&\times \langle \hat{\zeta}_{\bf p}\hat{\zeta}_{-{\bf p}}\hat{\zeta}^{'}_{{\bf k}_1}(\tau_1)\hat{\zeta}_{{\bf k}_2}(\tau_1)\hat{\zeta}_{{\bf k}_3}(\tau_1)\hat{\zeta}^{'}_{{\bf k}_4}(\tau_2)\hat{\zeta}_{{\bf k}_5}(\tau_2)\hat{\zeta}_{{\bf k}_6}(\tau_2)\rangle^{\dagger},\eea\bea
 \langle\hat{\zeta}_{\bf p}\hat{\zeta}_{-{\bf p}}\rangle_{(1,1)}&=& \frac{M^4_{ pl}}{4}\int^{0}_{-\infty}d\tau_1\frac{a^2(\tau_1)}{c^2_s(\tau_1)}\epsilon(\tau_1)\partial_{\tau_1}\left(\frac{\eta(\tau_1)}{c^2_s(\tau_1)}\right)\;\int^{0}_{-\infty}d\tau_2\;\frac{a^2(\tau_2)}{c^2_s(\tau_2)}\epsilon(\tau_2)\partial_{\tau_2}\left(\frac{\eta(\tau_2)}{c^2_s(\tau_2)}\right)\nonumber\\
  &&\times\int \frac{d^{3}{\bf k}_1}{(2\pi)^3} \int \frac{d^{3}{\bf k}_2}{(2\pi)^3} \int \frac{d^{3}{\bf k}_3}{(2\pi)^3} \int \frac{d^{3}{\bf k}_4}{(2\pi)^3} \int \frac{d^{3}{\bf k}_5}{(2\pi)^3} \int \frac{d^{3}{\bf k}_6}{(2\pi)^3}\nonumber\\
  &&\times \delta^3\bigg({\bf k}_1+{\bf k}_2+{\bf k}_3\bigg) \delta^3\bigg({\bf k}_4+{\bf k}_5+{\bf k}_6\bigg)\nonumber\\
  &&\times \langle \hat{\zeta}^{'}_{{\bf k}_1}(\tau_1)\hat{\zeta}_{{\bf k}_2}(\tau_1)\hat{\zeta}_{{\bf k}_3}(\tau_1)\hat{\zeta}_{\bf p}\hat{\zeta}_{-{\bf p}}\hat{\zeta}^{'}_{{\bf k}_4}(\tau_2)\hat{\zeta}_{{\bf k}_5}(\tau_2)\hat{\zeta}_{{\bf k}_6}(\tau_2)\rangle. \eea
\end{widetext}
\textcolor{black}{The computation of the one-loop contributions after performing all possible Wick contractions and with the above-mentioned momentum integrals can be worked out using the following expression: }
\bea \label{oneloopform}
\langle\hat{\zeta}_{\bf p}\hat{\zeta}_{\bf -p}\rangle_{\textbf{One-loop}}= \langle\hat{\zeta}_{\bf p}\hat{\zeta}_{\bf -p}\rangle_{(1,1)} + 2{\rm Re}\big(\langle\hat{\zeta}_{\bf p}\hat{\zeta}_{\bf -p}\rangle_{(0,2)}\big).
\eea
\textcolor{black}{To compute the above list of integrals explicitly, focusing on the temporal integrals, we make use of a crucial fact about both, the second slow-roll parameter $\eta$ and the effective sound speed $c_{s}$. Both of these parameters remain constant throughout the SRI, USR, and SRII, except at the moment of sharp transitions $\tau_{s_{n}}$ and $\tau_{e_{n}}$. To properly evaluate the integral at the boundaries of the USR, we write the temporal derivative term as follows}:
\bea \partial_{\tau}\left(\frac{\eta(\tau)}{c^2_s(\tau)}\right)\approx\frac{\Delta \eta(\tau)}{c^2_s(\tau)}\bigg(\delta(\tau-\tau_{e_{n}})-\delta(\tau-\tau_{s_{n}})\bigg).\eea
\textcolor{black}{where the Dirac delta contributes at the conformal time boundaries $\tau_{s_{n}}$ and $\tau_{e_{n}}$.
With the above equation, we can write the two-point correlators in the USR in the following manner}:
\bea &&\int^{0}_{-\infty}d\tau\;\frac{1}{c^2_s(\tau)}\partial_{\tau}\left(\frac{\eta(\tau)}{c^2_s(\tau)}\right)\; {\cal K}_{1}(\tau)
=\bigg(\frac{\Delta \eta(\tau_{e_{n}})}{c^2_s}\; {\cal K}_{1}(\tau=\tau_{e_{n}})-\frac{\Delta \eta(\tau_{s_{n}})}{c^2_s}\; {\cal K}_{1}(\tau=\tau_{s_{n}})\bigg)-\underbrace{\int^{0}_{-\infty}d\tau\;\left(\frac{\eta(\tau)}{c^2_s(\tau)}\right)\; {\cal K}^{'}_{1}(\tau)}_{\approx 0}\nonumber\\&&\quad\quad\quad\quad\quad\quad\quad\quad\quad\quad\quad\quad\quad\quad\approx\bigg(\frac{\Delta \eta(\tau_{e_{n}})}{c^2_s}\; {\cal K}_{1}(\tau_{e_{n}})-\frac{\Delta \eta(\tau_{s_{n}})}{c^2_s}\; {\cal K}_{1}(\tau_{s_{n}})\bigg),
\eea
\textcolor{black}{where the kernel ${\cal K}_{1}(\tau)$ contains the remaining momentum integrals. Similarly, for product of two such temporal derivatives in an integral we can write}:
\bea
&&\int^{0}_{-\infty}d\tau_1\;\int^{0}_{-\infty}d\tau_2\;\frac{1}{c^2_s(\tau_1)}\frac{1}{c^2_s(\tau_2)}\partial_{\tau_1}\left(\frac{\eta(\tau_1)}{c^2_s(\tau_1)}\right)\;\partial_{\tau_2}\left(\frac{\eta(\tau_2)}{c^2_s(\tau_2)}\right)\; {\cal K}_{2}(\tau_1,\tau_2)\nonumber\\
&&=\int^{0}_{-\infty}d\tau_2\frac{1}{c^2_s(\tau_2)}\partial_{\tau_2}\left(\frac{\eta(\tau_2)}{c^2_s(\tau_2)}\right)\;\bigg(\frac{\Delta \eta(\tau_e)}{c^4_s}\; {\cal K}_{2}(\tau_1=\tau_{e_{n}},\tau_2)-\frac{\Delta \eta(\tau_{s_{n}})}{c^4_s}\; {\cal K}_{2}(\tau_1=\tau_{s_{n}},\tau_2)\bigg)\nonumber\\
&&\quad\quad-\underbrace{\int^{0}_{-\infty}d\tau_1\;\int^{0}_{-\infty}d\tau_2\;\frac{1}{c^2_s(\tau_1)}\frac{1}{c^2_s(\tau_2)}\left(\frac{\eta(\tau_1)}{c^2_s(\tau_1)}\right)\;\partial_{\tau_2}\left(\frac{\eta(\tau_2)}{c^2_s(\tau_2)}\right)\; \partial_{\tau_1}{\cal K}_{2}(\tau_1,\tau_2)}_{\approx 0}\\&&\approx \bigg(\frac{\Delta \eta(\tau_{e_{n}})}{c^8_s}\; {\cal K}_{2}(\tau_1=\tau_{e_{n}},\tau_2=\tau_{e_{n}})-\frac{\Delta \eta(\tau_s)}{c^8_s}\; {\cal K}_{2}(\tau_1=\tau_{s_{n}},\tau_2=\tau_{s_{n}})\bigg)\nonumber\\&&-\underbrace{\int^{0}_{-\infty}d\tau_2\left(\frac{\eta(\tau_2)}{c^4_s(\tau_2)}\right)\;\bigg(\frac{\Delta \eta(\tau_{e_{n}})}{c^4_s}\; \partial_{\tau_2}{\cal K}_{2}(\tau_1=\tau_e,\tau_2)-\frac{\Delta \eta(\tau_{s_{n}})}{c^4_s}\; \partial_{\tau_2}{\cal K}_{2}(\tau_1=\tau_{s_{n}},\tau_2)\bigg)}_{\approx 0}\approx \bigg(\frac{\Delta \eta(\tau_{e_{n}})}{c^8_s}\; {\cal K}_{2}(\tau_{e_{n}})-\frac{\Delta \eta(\tau_{s_{n}})}{c^8_s}\; {\cal K}_{2}(\tau_{s_{n}})\bigg)\nonumber,
\eea 
\textcolor{black}{which includes a different kernel labelled as ${\cal K}_{2}(\tau_{1},\tau_{2})$. The results from the eqn.(\ref{oneloopform}) after performing the momentum integral and the temporal integral as explained above, are now written as follows: }

\bea \label{sr1correct}
\bigg[\Delta^{2}_{\zeta,\textbf{One-loop}}(k)\bigg]_{\textbf{SR}_{1}} &=& \bigg[\Delta^{2}_{\zeta,\textbf{Tree}}(k)\bigg]_{\textbf{SR}_{1}}^{2}
\times \left(1-\frac{2}{15\pi^{2}}\frac{1}{c_{s}^{2}k_{*}^{2}}\left(1-\frac{1}{c_{s}^{2}}\right)\epsilon\right)\times \left(c_{\textbf{SR}_{1}}-\frac{4}{3}{\cal J}_{\textbf{SR}_{1}}(\tau_{s_{1}})\right),
\\ \label{usrcorrect}
    \bigg[\Delta^{2}_{\zeta,\textbf{One-loop}}(k)\bigg]_{\textbf{USR}_{n}} &=&  \frac{1}{4}\bigg[\Delta^{2}_{\zeta,\textbf{Tree}}(k)\bigg]_{\textbf{SR}_{1}}^{2}
    \times \bigg\{\bigg[\bigg(\frac{\Delta\eta(\tau_{e_{n}})}{\tilde{c}^{4}_{s}}\bigg)^{2}{\cal J}_{\textbf{USR}_{n}}^{(1)}(\tau_{e_{n}}) - \left(\frac{\Delta\eta(\tau_{s_{n}})}{\tilde{c}^{4}_{s}}\right)^{2}{\cal J}_{\textbf{USR}_{n}}^{(1)}(\tau_{s_{n}})\bigg] \nonumber \\ 
    &\quad \quad +& 2 \bigg[\bigg(\frac{\Delta\eta(\tau_{e_{n}})}{\tilde{c}^{4}_{s}}\bigg) {\cal J}_{\textbf{USR}_{n}}^{(2)}(\tau_{e_{n}}) - \bigg(\frac{\Delta\eta(\tau_{e_{n}})}{\tilde{c}^{4}_{s}}\bigg) {\cal J}_{\textbf{USR}_{n}}^{(2)}(\tau_{s_{n}})\bigg]-c_{\textbf{USR}_{n}} \bigg\}\\
     \label{sr2correct}
\bigg[\Delta^{2}_{\zeta,\textbf{One-loop}}(k)\bigg]_{\textbf{SR}_{n+1}} &=& \bigg[\Delta^{2}_{\zeta,\textbf{Tree}}(k)\bigg]_{\textbf{SR}_{1}}^{2}
\times\left(1-\frac{2}{15\pi^{2}}\frac{1}{c_{s}^{2}k_{*}^{2}}\left(1-\frac{1}{c_{s}^{2}}\right)\epsilon\right)
\times \bigg(c_{\textbf{SR}_{n+1}}+{\cal J}_{\textbf{SR}_{n+1}}(\tau_{e_{n}})\bigg).
\eea
Notice the regularization scheme-dependent parameters $c_{\textbf{SR}_{1}},\;c_{\textbf{USR}_{n}}$ and $c_{\textbf{SR}_{n+1}}$, which are computed at the late time $\tau \to 0$, and fixed during the renormalization procedure.
Furthermore, these expressions require the evaluation of the loop integral terms at different conformal times namely, ${\cal J}_{\textbf{SR}_{1}}(\tau_{s_{1}})$ during $\tau < \tau_{s_1}$,  ${\cal J}_{\textbf{USR}_{n}1}(\tau_{s_{n}})$ and ${\cal J}_{\textbf{USR}_{n}2}(\tau_{s_{n}})$, during sharp transition $\tau = \tau_{s_n}$, ${\cal J}_{\textbf{USR}_{n}1}(\tau_{e_{n}})$ and ${\cal J}_{\textbf{USR}_{n}2}(\tau_{e_{n}})$ during $\tau = \tau_{e_n}$, and ${\cal J}_{\textbf{SR}_{n+1}}(\tau_{e_{n}})$ during the conformal time interval $\tau_{e_{n}} \leq \tau < \tau_{s_{n+1}}$. 
Now the integrals in the above expressions can be written as:
\bea
{\cal J}_{\textbf{SR}_{1}}(\tau_{s_{1}}) &=& \int_{k_{*}}^{k_{s_1}}\frac{dk}{k}(1+k^2c_s ^2 \tau^2)\nonumber\\
&=& \ln{\frac{k_{s_1}}{k_{*}}} + \frac{1}{2}\bigg(\frac{k_{s_1}}{k_{*}}\bigg)^{2} -\frac{1}{2}, \\
{\cal J}_{\textbf{USR}_{n}}^{(1)}(\tau_{s_{n}}) &=& \int_{k_{s_{n}}}^{k_{e_{n}}}\frac{dk}{k}\bigg|\alpha^{(2n)}_{\bf k}\left(1+ikc_s\tau\right)e^{-ikc_s\tau}-\beta^{(2n)}_{\bf k}\left(1-ikc_s\tau\right)e^{ikc_s\tau}\bigg|^2\nonumber\\
&\approx &\ln\left(\frac{k_{e_{n}}}{k_{s_{n}}}\right) + \frac{1}{2} \bigg(\frac{k_{e_n}}{k_{s_n}}\bigg)^2 - \frac{1}{2} + \cdots,\quad\quad \eea\bea
{\cal J}_{\textbf{USR}_{n}}^{(1)} (\tau_{e_{n}}) &=& \left(\frac{k_{e_{n}}}{k_{s_{n}}}\right)^{6} \int_{k_{s_{n}}}^{k_{e_{n}}}\frac{dk}{k}\bigg|\alpha^{(2n)}_{\bf k}\left(1+ikc_s\tau\right)e^{-ikc_s\tau}-\beta^{(2n)}_{\bf k}\left(1-ikc_s\tau\right)e^{ikc_s\tau}\bigg|^2\nonumber\\
&\approx&\left(\frac{k_{e_{n}}}{k_{s_{n}}}\right)^{6}{\cal J}_{\textbf{USR}_{n}}^{(1)}(\tau_{s_{n}}),\quad\quad \\
{\cal J}_{\textbf{USR}_{n}}^{(2)}(\tau_{s_{n}}) &=& \int_{k_{s_n}}^{k_{e_n}} d\ln{k} \bigg(\frac{d}{d\ln{k}} \bigg|\alpha^{(2n)}_{\bf k}\left(1+ikc_s\tau\right)e^{-ikc_s\tau}-\beta^{(2n)}_{\bf k}\left(1-ikc_s\tau\right)e^{ikc_s\tau}\bigg|^2 \bigg) \nonumber \\
&=& \bigg[\bigg|\alpha^{(2n)}_{\bf k}\left(1+ikc_s\tau\right)e^{-ikc_s\tau}-\beta^{(2n)}_{\bf k}\left(1-ikc_s\tau\right)e^{ikc_s\tau}\bigg|^2 \bigg]_{k_{s_n}}^{k_{e_n}}, \\
{\cal J}_{\textbf{USR}_{n}}^{(2)}(\tau_{e_{n}}) &=& \left(\frac{k_{e_{n}}}{k_{s_{n}}}\right)^{6} 
{\cal J}_{\textbf{USR}_{n}}^{(2)}(\tau_{e_{n}}), \eea\bea
\label{insrnres}
{\cal J}_{\textbf{SR}_{n+
1}}(\tau_{e_{n}}) &=& \left(\frac{k_{e_{n}}}{k_{s_{n}}}\right)^{6}\int_{k_{e_{n}}}^{k_{s_{n+1}}}\frac{dk}{k}\bigg|\alpha^{(2n+1)}_{\bf k}\left(1+ikc_s\tau\right)e^{-ikc_s\tau}-\beta^{(2n+1)}_{\bf k}\left(1-ikc_s\tau\right)e^{ikc_s\tau}\bigg|^2 \nonumber \\
&\approx & \bigg\{\ln\left(\frac{k_{s_{n+1}}}{k_{e_{n}}}\right) - \frac{27}{32}\bigg[1-\bigg(\frac{k_{e_n}}{k_{s_{n+1}}}\bigg)^{12}\bigg]\bigg\}.
\eea

The dots at the end of the USR$_n ^{(1)}$ integrals represent the oscillatory terms that are suppressed. More details about the computation of these integrals can be found in \cite{Choudhury:2023rks, Choudhury:2023vuj,Choudhury:2023jlt}. Now, the integrals mentioned above get evaluated under the late-time limit. This limit becomes necessary to remove the harmful quadratic UV divergences, while the only ones that remain are the suspected logarithmic IR divergences which are impossible to eradicate while working with quantum field theory in a de-Sitter background. In the next subsection, we are going to present the two schemes of renormalization to remove the harmful UV divergences, namely the late-time renormalization and the wavefunction/adiabatic renormalization. Although the IR divergences can not be removed, they can be smoothened with the help of power spectrum renormalization which is also described in the following subsection. Before proceeding further, we need to simplify the one-loop corrected result of the USR phases by excluding certain heavily suppressed contributions. Therefore, we rewrite the one-loop result at the USR phases represented by eqn.(\ref{usrcorrect}) as :
\bea
\bigg[\Delta^{2}_{\zeta,\textbf{One-loop}}(k)\bigg]_{\textbf{USR}_{n}} &=&  \frac{1}{4}\bigg[\Delta^{2}_{\zeta,\textbf{Tree}}(k)\bigg]_{\textbf{SR}_{1}}^{2} \times \bigg[{\cal T}^{(n)}_{1}+{\cal T}^{(n)}_{2}-  c_{\textbf{USR}_{n}}\bigg], \nonumber \\
&=& \frac{1}{4}\bigg[\Delta^{2}_{\zeta,\textbf{Tree}}(k)\bigg]_{\textbf{SR}_{1}}^{2} \times \bigg[{\cal T}^{(n)}_{1} \bigg(1+ \frac{{\cal T}^{(n)}_{2}}{{\cal T}^{(n)}_{1}}\bigg)- c_{\textbf{USR}_{n}}\bigg],
\eea
where the terms ${\cal T}^{(n)}_{1}$ and ${\cal T}^{(n)}_{2}$ are given by :

\bea
{\cal T}^{(n)}_{1} &=& \bigg[\bigg(\frac{\Delta\eta(\tau_{e_{n}})}{\tilde{c}^{4}_{s}}\bigg)^{2}{\cal J}_{\textbf{USR}_{n}}^{(1)}(\tau_{e_{n}}) - \left(\frac{\Delta\eta(\tau_{s_{n}})}{\tilde{c}^{4}_{s}}\right)^{2}{\cal J}_{\textbf{USR}_{n}}^{(1)}(\tau_{s_{n}})\bigg]. \nonumber \\
{\cal T}^{(n)}_{2} &=& 2 \bigg[\bigg(\frac{\Delta\eta(\tau_{e_{n}})}{\tilde{c}^{4}_{s}}\bigg) {\cal J}_{\textbf{USR}_{n}}^{(2)}(\tau_{e_{n}}) - \bigg(\frac{\Delta\eta(\tau_{e_{n}})}{\tilde{c}^{4}_{s}}\bigg) {\cal J}_{\textbf{USR}_{n}}^{(2)}(\tau_{s_{n}})\bigg].
\eea

Here, the ratio ${\cal T}^{(n)}_{2} / {\cal T}^{(n)}_{1}\ll 1$, as ${\cal T}^{(n)}_{2}$ represents highly suppressed contributions, and so ${\cal T}^{(n)}_{1}\left(1 + {\cal T}^{(n)}_{2} / {\cal T}^{(n)}_{1}\right) \approx {\cal T}^{(n)}_{1}$. Therefore, we can neglect the contributions coming from ${\cal T}^{(n)}_{2}$ to write the final form of the one-loop corrected expression for the USR$_n$ phase as well as the other phases : 

\bea
    \label{Sr1again}
        \bigg[\Delta^{2}_{\zeta,\textbf{One-loop}}(k)\bigg]_{\textbf{SR}_{1}} &=& \bigg[\Delta^{2}_{\zeta,\textbf{Tree}}(k)\bigg]_{\textbf{SR}_{1}}^{2}
        \times \left(1-\frac{2}{15\pi^{2}}\frac{1}{c_{s}^{2}k_{*}^{2}}\left(1-\frac{1}{c_{s}^{2}}\right)\epsilon\right)\times \left(c_{\textbf{SR}_{1}}-\frac{4}{3}{\cal J}_{\textbf{SR}_{1}}(\tau_{s_{1}})\right), \\
    \label{usragain}
        \bigg[\Delta^{2}_{\zeta,\textbf{One-loop}}(k)\bigg]_{\textbf{USR}_{n}} &=&  \frac{1}{4}\bigg[\Delta^{2}_{\zeta,\textbf{Tree}}(k)\bigg]_{\textbf{SR}_{1}}^{2}
        \bigg\{\bigg[\bigg(\frac{\Delta\eta(\tau_{e_{n}})}{\tilde{c}^{4}_{s}}\bigg)^{2}{\cal J}_{\textbf{USR}_{n}}(\tau_{e_{n}}) - \left(\frac{\Delta\eta(\tau_{s_{n}})}{\tilde{c}^{4}_{s}}\right)^{2}{\cal J}_{\textbf{USR}_{n}}(\tau_{s_{n}})\bigg] - c_{\textbf{USR}_{n}}\bigg\}, \;\;\quad \\
    \label{sr2again}
        \bigg[\Delta^{2}_{\zeta,\textbf{One-loop}}(k)\bigg]_{\textbf{SR}_{n+1}} &=& \bigg[\Delta^{2}_{\zeta,\textbf{Tree}}(k)\bigg]_{\textbf{SR}_{1}}^{2}
        \times\left(1-\frac{2}{15\pi^{2}}\frac{1}{c_{s}^{2}k_{*}^{2}}\left(1-\frac{1}{c_{s}^{2}}\right)\epsilon\right)
        \times \bigg(c_{\textbf{SR}_{n+1}}+{\cal J}_{\textbf{SR}_{n+1}}(\tau_{e_{n}})\bigg).
\eea

Here we have also replaced the integral ${\cal J}_{\textbf{USR}_{n}}^{(1)}$ in the USR$_n$ phase by ${\cal J}_{\textbf{USR}_{n}}$ since it does not make much sense to distinguish these two as the effect of ${\cal J}_{\textbf{USR}_{n}}^{(2)}$ has been neglected.
Below is a representative of the loop diagrams for the contributions from eqn.(\ref{Sr1again}), eqn.(\ref{usragain}), and eqn.(\ref{sr2again})
\begin{equation} \label{unrenormloop}
\begin{tikzpicture}[baseline={([yshift=-3.5ex]current bounding box.center)},very thick]
  
  \def\radius{1}
  \scalebox{1}{\draw[green,ultra thick] (0,\radius) circle (\radius);
  \draw[green,ultra thick] (4.5*\radius,0) circle (\radius);}

  \draw[black, very thick] (-1.5*\radius,0) -- (0,0);
  \draw[blue,fill=blue] (0,0) circle (.5ex);
  \draw[black, very thick] (0,0)  -- (1.5*\radius,0);
  \node at (2*\radius,0) {+};
  \draw[black, very thick] (2.5*\radius,0) -- (3.5*\radius,0); 
  \draw[blue,fill=blue] (3.5*\radius,0) circle (.5ex);
  \draw[blue,fill=blue] (5.5*\radius,0) circle (.5ex);
  \draw[black, very thick] (5.5*\radius,0) -- (6.5*\radius,0);
  

\end{tikzpicture}\quad = \quad \text{One-loop unrenormalized contributions},
\end{equation}
where the RHS of the above equation equals the following sum:
\bea
\text{One-loop unrenormalized contributions} =  \bigg[\Delta^{2}_{\zeta,\textbf{One-loop}}(k)\bigg]_{\textbf{SR}_{1}} + \bigg[\Delta^{2}_{\zeta,\textbf{One-loop}}(k)\bigg]_{\textbf{USR}_{n}} + \bigg[\Delta^{2}_{\zeta,\textbf{One-loop}}(k)\bigg]_{\textbf{SR}_{n+1}}.\quad
\eea
The equation (\ref{unrenormloop}) gives a diagrammatic representation of the total unrenormalized one-loop contributions to the scalar power spectrum.  

\subsection{Renormalization using standard technique in Quantum Field Theory}

\textcolor{black}{In this section, our prime objective is to establish the computation in a more commonly known language so that one can make an easier connection with the standard renormalization techniques applicable within the framework of Quantum Field Theory. Instead of using many such techniques in this section, we will further solely concentrate on the technique where the counter-terms are introduced to subtract the divergences appearing at the level of the unrenormalized/bare action. Finally, this will give rise to the renormalized form of the action where all possible harmful divergences, such as quadratic UV divergence particularly can be completely removed and logarithmic IR divergences can be smoothened after successfully performing this process. }

\textcolor{black}{To serve this purpose let us first write down the expression for the third-order perturbed bare action for the comoving curvature perturbation, which are given by the following expression:}
\bea
\label{baction}
         S_{\zeta,{\bf B}}^{(3)} &=& M^{2}_{p}\int d\tau\;d^3x\;\bigg [\left({\bf D}_1\right)_{\bf B}\; \zeta^{\prime} {^2}_{\bf B} \zeta_{\bf B} + \left({\bf D}_2\right)_{\bf B}\;(\partial_i \zeta_{\bf B})^2 \zeta_{\bf B}  -  \left({\bf D}_3\right)_{\bf B}\;\zeta^{\prime}_{\bf B} (\partial_i \zeta_{\bf B}) \bigg (\partial_i \partial ^{-2}\bigg(\frac{\epsilon \zeta^{\prime}_{\bf B}}{c_s ^2}\bigg)\bigg) \nonumber \\ 
        && \quad \quad \quad \quad \quad \quad \quad  - \left({\bf D}_4\right)_{\bf B}\;\left(\zeta^{\prime} {^3}_{\bf B}+\zeta^{\prime}_{\bf B}(\partial_i \zeta_{\bf B})^2 \right) +  \left({\bf D}_5\right)_{\bf B}\;\zeta_{\bf B} \bigg(\partial_i \partial_j \partial^{-2}\bigg (\frac{\epsilon \zeta^{\prime}_{\bf B}}{c_s ^2}\bigg)\bigg)^2 + \underbrace{ \left({\bf D}_6\right)_{\bf B}\zeta^{\prime}_{\bf B} \zeta^2_{\bf B}}_{\textbf{Dominant term}}+.....\bigg],
   \eea 
   \textcolor{black}{where we define the most important EFT bare (unrenormalized) coupling para $\left({\bf D}_i\right)_{\bf B}\forall i=1,2,\cdots,6$ by the following expressions:}
   \bea  \left({\bf D}_1\right)_{\bf B}&=&\bigg(3(c_s ^2 -1)\epsilon + \epsilon ^2 - \frac{\epsilon ^3}{2}\bigg )a^2,\\
    \left({\bf D}_2\right)_{\bf B}&=&\frac{\epsilon}{c_s ^2}\bigg(\epsilon - 2s +1 -c_s ^2 \bigg)a^2,\\
    \left({\bf D}_3\right)_{\bf B}&=&\frac{2 \epsilon}{c_s ^2}a^2,\\ 
    \left({\bf D}_4\right)_{\bf B}&=&\frac{\epsilon}{aH}\bigg (1-\frac{1}{c_s ^2}\bigg)a^2,\\ 
    \left({\bf D}_5\right)_{\bf B}&=&\frac{\epsilon}{2}a^2,\\ 
    \left({\bf D}_6\right)_{\bf B}&=&\frac{\epsilon}{2c_s ^2}\partial_{\tau} \bigg(\frac{\eta}{c_s ^2}\bigg)a^2.\eea
\textcolor{black}{It is important to note that here the above-mentioned bare action contains the previously mentioned harmful quadratic UV divergence as well as the less harmful logarithmic IR divergence. Apart from that some power law structure of the divergences may appear which turns out to be extremely suppressed in the present computation. Here it is important to note that $\zeta_{\bf B}$ corresponds to the unrenormalized/ bare part of the gauge invariant comoving curvature perturbation in this present context of discussion. The subscript {\bf B} is explicitly used to indicate the bare contributions to avoid any further confusion. In the present context of discussion, by introducing counterterms at the level of action one can able to completely remove the contribution of the quadratic UV divergence in the one-loop computation of the two-point cosmological correlation function of the gauge invariant comoving scalar curvature perturbation and its associated 1PI effective action for the corresponding two-point amplitude. It will be more clear in the next two subsections that the present commonly known technique of renormalization in Quantum Field Theory is exactly equivalent to the late time and the adiabatic/ wave function renormalization scheme that we are going to explicitly perform in the next section. In each of the following two subsections, in the two different contexts of the renormalization schemes we are going to explicitly point out how one can determine the individual or the combinational form of the counter terms introduced in the present context at the level of the third order perturbed action for the comoving scalar curvature perturbation. However, with the present computation one cannot able to completely remove the effect of the logarithmic IR divergence contribution, rather with the help of the present computation one can able to soften the behaviour of such less harmful divergence in the present context of the discussion. }

\textcolor{black}{To construct the renormalized version of the third order action for the comoving scalar curvature perturbation we need to follow the following rescaling ansatz of the gauge invariant modes which are extremely helpful to establish the connection among the renormalized, unrenormalized/ bare, and counter-term contribution, which is described by the following expression:}
\bea \zeta_{\bf R}=\zeta_{\bf B}-\zeta_{\bf C}=\sqrt{{\cal Z}^{\rm IR}_{n}}\times \zeta_{\bf B}\quad\quad\quad{\rm where}\quad\quad\quad {\cal Z}^{\rm IR}_{n}:=\left(1+\delta_{{\cal Z}^{\rm IR}_{n}}\right).\eea
\textcolor{black}{Here the superscripts, {\bf R}, {\bf B}, and {\bf C} are used to characterize the renormalized,  bare, and counter-term contributions. Utilizing the above-mentioned fact one can further write down the bare contribution in terms of the counter-term contribution by the following expression:}
\bea \zeta_{\bf B}=\left(\frac{1}{1-\sqrt{{\cal Z}^{\rm IR}_{n}}}\right)\times \zeta_{\bf C}=\left(\frac{1}{1-\sqrt{\left(1+\delta_{{\cal Z}^{\rm IR}_{n}}\right)}}\right)\times \zeta_{\bf C}.\eea
\textcolor{black}{Here it is important to note that the quantity, ${\cal Z}^{\rm IR}_{n}$ (more precisely $\delta_{{\cal Z}^{\rm IR}_{n}}$) is commonly known as the counter-term which we need to explicitly determine by implementing physical renormalization condition in the present computation. In the latter half of the paper, when we discuss the power spectrum renormalization scheme there, we will explicitly show in terms of one-loop corrected 1PI renormalized effective action, which captures the information of the two-point amplitude or in terms of one-loop corrected renormalized power spectrum the behaviour of the logarithmic IR divergence can be softened at the corresponding order of the computation. For this reason, we have explicitly used the superscript `IR' in the nomenclature of the counter-term to distinguish the effects of UV and IR divergences in this context. Additionally, we have introduced a subscript `n' to specifically quantify in which number out of MST we are interested in this computation. It will be more feasible and better understood once we look at the detailed computations of the power spectrum renormalization scheme.} 

\textcolor{black}{Our further job is to convert the expression for the second-order and third-order unrenormalized/bare action in terms of the renormalized version of the action utilizing the newly introduced rescaled renormalized version of the gauge invariant scalar curvature perturbation. This can be easily done by following the steps mentioned below:}
\begin{enumerate}
    \item \underline{\bf Renormalized coupling parameters:}\\ \\
   \textcolor{black}{ With the help of the previously mentioned ansatz the renormalized coupling parameters as appearing in the third-order perturbed action can be expressed in terms of the bare and counter-term contributions by the following expressions:}
    \bea \left({\bf D}_1\right)_{\bf R}&=&\left({\bf D}_1\right)_{\bf B}-\left({\bf D}_1\right)_{\bf C}={\cal Z}_{{\bf D}_1}\times \left({\bf D}_1\right)_{\bf B}\quad\quad\quad{\rm where}\quad\quad\quad {\cal Z}_{{\bf D}_1}:=\left(1+\delta_{{\cal Z}_{{\bf D}_1}}\right),\\
        \left({\bf D}_2\right)_{\bf R}&=&\left({\bf D}_2\right)_{\bf B}-\left({\bf D}_2\right)_{\bf C}={\cal Z}_{{\bf D}_2}\times \left({\bf D}_2\right)_{\bf B}\quad\quad\quad{\rm where}\quad\quad\quad {\cal Z}_{{\bf D}_2}:=\left(1+\delta_{{\cal Z}_{{\bf D}_2}}\right),\\
        \left({\bf D}_3\right)_{\bf R}&=&\left({\bf D}_3\right)_{\bf B}-\left({\bf D}_3\right)_{\bf C}={\cal Z}_{{\bf D}_3}\times \left({\bf D}_3\right)_{\bf B}\quad\quad\quad{\rm where}\quad\quad\quad {\cal Z}_{{\bf D}_3}:=\left(1+\delta_{{\cal Z}_{{\bf D}_3}}\right),\\
        \left({\bf D}_4\right)_{\bf R}&=&\left({\bf D}_4\right)_{\bf B}-\left({\bf D}_4\right)_{\bf C}={\cal Z}_{{\bf D}_4}\times \left({\bf D}_4\right)_{\bf B}\quad\quad\quad{\rm where}\quad\quad\quad {\cal Z}_{{\bf D}_4}:=\left(1+\delta_{{\cal Z}_{{\bf D}_4}}\right),\\
        \left({\bf D}_5\right)_{\bf R}&=&\left({\bf D}_5\right)_{\bf B}-\left({\bf D}_5\right)_{\bf C}={\cal Z}_{{\bf D}_5}\times \left({\bf D}_5\right)_{\bf B}\quad\quad\quad{\rm where}\quad\quad\quad {\cal Z}_{{\bf D}_5}:=\left(1+\delta_{{\cal Z}_{{\bf D}_5}}\right),\\
        \left({\bf D}_6\right)_{\bf R}&=&\left({\bf D}_6\right)_{\bf B}-\left({\bf D}_6\right)_{\bf C}={\cal Z}_{{\bf D}_6}\times \left({\bf D}_6\right)_{\bf B}\quad\quad\quad{\rm where}\quad\quad\quad {\cal Z}_{{\bf D}_6}:=\left(1+\delta_{{\cal Z}_{{\bf D}_6}}\right).\eea
    \textcolor{black}{Here the multiplicative factors ${\cal Z}_{{\bf D}_i}\forall i=1,2,\cdots,6$, (more precisely $\delta_{{\cal Z}_{{\bf D}_i}}\forall i=1,2,\cdots,6$) represent the counter-terms of each of the individual bare coupling parameters $ \left({\bf D}_i\right)_{\bf B}\forall i=1,2,\cdots,6$, which helps us to easily define the renormalized coupling parameters $ \left({\bf D}_i\right)_{\bf R}\forall i=1,2,\cdots,6$ with the corresponding presented scaling ansatz.}


    
    \textcolor{black}{Further utilizing the above-mentioned fact one can further write down the bare contribution in terms of the counter-term contribution by the following expression:}
\bea \left({\bf D}_1\right)_{\bf B}&=&\left(\frac{1}{1-{\cal Z}_{{\bf D}_1}}\right)\times \left({\bf D}_1\right)_{\bf C}=-\frac{1}{\delta_{{\cal Z}_{{\bf D}_1}}}\times \left({\bf D}_1\right)_{\bf C},\\
\left({\bf D}_2\right)_{\bf B}&=&\left(\frac{1}{1-{\cal Z}_{{\bf D}_2}}\right)\times \left({\bf D}_2\right)_{\bf C}=-\frac{1}{\delta_{{\cal Z}_{{\bf D}_2}}}\times \left({\bf D}_2\right)_{\bf C},\\
  \left({\bf D}_3\right)_{\bf B}&=&\left(\frac{1}{1-{\cal Z}_{{\bf D}_3}}\right)\times \left({\bf D}_3\right)_{\bf C}=-\frac{1}{\delta_{{\cal Z}_{{\bf D}_3}}}\times \left({\bf D}_3\right)_{\bf C},\\
  \left({\bf D}_4\right)_{\bf B}&=&\left(\frac{1}{1-{\cal Z}_{{\bf D}_4}}\right)\times \left({\bf D}_4\right)_{\bf C}=-\frac{1}{\delta_{{\cal Z}_{{\bf D}_4}}}\times \left({\bf D}_4\right)_{\bf C},\\
  \left({\bf D}_5\right)_{\bf B}&=&\left(\frac{1}{1-{\cal Z}_{{\bf D}_5}}\right)\times \left({\bf D}_5\right)_{\bf C}=-\frac{1}{\delta_{{\cal Z}_{{\bf D}_5}}}\times \left({\bf D}_5\right)_{\bf C},\\
  \left({\bf D}_6\right)_{\bf B}&=&\left(\frac{1}{1-{\cal Z}_{{\bf D}_6}}\right)\times \left({\bf D}_6\right)_{\bf C}=-\frac{1}{\delta_{{\cal Z}_{{\bf D}_6}}}\times \left({\bf D}_6\right)_{\bf C}.\eea

    \item \underline{\bf Renormalized operator contributions without couplings:}\\ \\
    \textcolor{black}{ With the help of the previously mentioned ansatz the renormalized version of the each of individual operators without the inclusion of the effect of the renormalized coupling parameters can be further expressed in terms of the bare and counter-term contributions by the following expressions:}
     \bea 
     \zeta^{\prime} {^2}_{\bf R} \zeta_{\bf R}&=&\left(\zeta^{\prime}_{\bf B}-\zeta^{\prime}_{\bf C}\right)^2\left(\zeta_{\bf B}-\zeta_{\bf C}\right)\nonumber\\
     &=&\left(1-\left(1-\sqrt{{\cal Z}^{\rm IR}_{n}}\right)\right)^2\left(1-\left(1-\sqrt{{\cal Z}^{\rm IR}_{n}}\right)\right)\times \zeta^{\prime} {^2}_{\bf B} \zeta_{\bf B}\nonumber\\
     &=&\left({\cal Z}^{\rm IR}_{n}\right)^{\frac{3}{2}}\times \zeta^{\prime} {^2}_{\bf B} \zeta_{\bf B},\\
    (\partial_i \zeta_{\bf R})^2 \zeta_{\bf R}&=&\left(\partial_{i}\left(\zeta_{\bf B}-\zeta_{\bf C}\right)\right)^2\left(\zeta_{\bf B}-\zeta_{\bf C}\right)\nonumber\\
     &=&\left(1-\left(1-\sqrt{{\cal Z}^{\rm IR}_{n}}\right)\right)^2\left(1-\left(1-\sqrt{{\cal Z}^{\rm IR}_{n}}\right)\right)\times (\partial_i \zeta_{\bf B})^2 \zeta_{\bf B}\nonumber\\
     &=&\left({\cal Z}^{\rm IR}_{n}\right)^{\frac{3}{2}}\times (\partial_i \zeta_{\bf B})^2 \zeta_{\bf B},\\
     \zeta^{\prime}_{\bf R} (\partial_i \zeta_{\bf R}) \bigg (\partial_i \partial ^{-2}\bigg(\frac{\epsilon \zeta^{\prime}_{\bf R}}{c_s ^2}\bigg)\bigg)&=&\left(\zeta^{\prime}_{\bf B}-\zeta^{\prime}_{\bf C}\right)\left(\partial_{i}\left(\zeta_{\bf B}-\zeta_{\bf C}\right)\right)\bigg (\partial_i \partial ^{-2}\bigg(\frac{\epsilon \left(\zeta^{\prime}_{\bf B}-\zeta^{\prime}_{\bf C}\right)}{c_s ^2}\bigg)\bigg)\nonumber\\
     &=&\left(1-\left(1-\sqrt{{\cal Z}^{\rm IR}_{n}}\right)\right)\left(1-\left(1-\sqrt{{\cal Z}^{\rm IR}_{n}}\right)\right)\left(1-\left(1-\sqrt{{\cal Z}^{\rm IR}_{n}}\right)\right)\nonumber\\
     &&\quad\quad\quad\quad\quad\quad\quad\quad\quad\quad\quad\quad\times \zeta^{\prime}_{\bf B} (\partial_i \zeta_{\bf B}) \bigg (\partial_i \partial ^{-2}\bigg(\frac{\epsilon \zeta^{\prime}_{\bf B}}{c_s ^2}\bigg)\bigg)\nonumber\\
     &=&\left({\cal Z}^{\rm IR}_{n}\right)^{\frac{3}{2}}\times \zeta^{\prime}_{\bf B} (\partial_i \zeta_{\bf B}) \bigg (\partial_i \partial ^{-2}\bigg(\frac{\epsilon \zeta^{\prime}_{\bf B}}{c_s ^2}\bigg)\bigg),\\
     \left(\zeta^{\prime} {^3}_{\bf R}+\zeta^{\prime}_{\bf R}(\partial_i \zeta_{\bf R})^2 \right)&=&\left(\left(\zeta^{\prime}_{\bf B}-\zeta^{\prime}_{\bf C}\right)^3+\left(\zeta^{\prime}_{\bf B}-\zeta^{\prime}_{\bf C}\right)\left(\partial_{i}\left(\zeta_{\bf B}-\zeta_{\bf C}\right)\right)^2\right)\nonumber\\
     &=&\Bigg\{\left(1-\left(1-\sqrt{{\cal Z}^{\rm IR}_{n}}\right)\right)^3\zeta^{\prime} {^3}_{\bf B}\nonumber\\
     &&\quad\quad\quad\quad+\left(1-\left(1-\sqrt{{\cal Z}^{\rm IR}_{n}}\right)\right)\left(1-\left(1-\sqrt{{\cal Z}^{\rm IR}_{n}}\right)\right)^2\zeta^{\prime}_{\bf B}(\partial_i \zeta_{\bf B})^2 \Bigg\}\nonumber\\
     &=&\left({\cal Z}^{\rm IR}_{n}\right)^{\frac{3}{2}}\times \left(\zeta^{\prime} {^3}_{\bf B}+\zeta^{\prime}_{\bf B}(\partial_i \zeta_{\bf B})^2 \right),\eea\bea
     \zeta_{\bf R} \bigg(\partial_i \partial_j \partial^{-2}\bigg (\frac{\epsilon \zeta^{\prime}_{\bf R}}{c_s ^2}\bigg)\bigg)^2&=&\left(\zeta_{\bf B}-\zeta_{\bf C}\right)\bigg (\partial_i\partial_j \partial ^{-2}\bigg(\frac{\epsilon \left(\zeta^{\prime}_{\bf B}-\zeta^{\prime}_{\bf C}\right)}{c_s ^2}\bigg)\bigg)^2\nonumber\\
     &=&\left(1-\left(1-\sqrt{{\cal Z}^{\rm IR}_{n}}\right)\right)\left(1-\left(1-\sqrt{{\cal Z}^{\rm IR}_{n}}\right)\right)^2\nonumber\\
     &&\quad\quad\quad\quad\quad\quad\quad\quad\quad\quad\quad\quad\times \zeta_{\bf B} \bigg(\partial_i \partial_j \partial^{-2}\bigg (\frac{\epsilon \zeta^{\prime}_{\bf B}}{c_s ^2}\bigg)\bigg)^2\nonumber\\
     &=&\left({\cal Z}^{\rm IR}_{n}\right)^{\frac{3}{2}}\times \zeta_{\bf B} \bigg(\partial_i \partial_j \partial^{-2}\bigg (\frac{\epsilon \zeta^{\prime}_{\bf B}}{c_s ^2}\bigg)\bigg)^2,\\
     \zeta^{\prime}_{\bf R} \zeta^2_{\bf R}&=&\left(\zeta^{\prime}_{\bf B}-\zeta^{\prime}_{\bf C}\right)\left(\zeta_{\bf B}-\zeta_{\bf C}\right)^2\nonumber\\
     &=&\left(1-\left(1-\sqrt{{\cal Z}^{\rm IR}_{n}}\right)\right)\left(1-\left(1-\sqrt{{\cal Z}^{\rm IR}_{n}}\right)\right)^2\times \zeta^{\prime}_{\bf B} \zeta {^2}_{\bf B}\nonumber\\
     &=&\left({\cal Z}^{\rm IR}_{n}\right)^{\frac{3}{2}}\times \zeta^{\prime}_{\bf B} \zeta^2_{\bf B}.\eea
\textcolor{black}{      After executing the above-mentioned analysis to renormalize the mentioned form of the operators we found that after renormalization the operators coming from third-order perturbed action scaled with $\left({\cal Z}^{\rm IR}_{n}\right)^{3/2}$ with its bare contribution. }

     \item \underline{\bf Renormalized operator contributions including couplings:}\\ \\
    \textcolor{black}{ Further, the bare contributions including the effect of the couplings can be expressed in terms of the renormalized operators along with the renormalized coupling parameters via the mapping of the above-mentioned scaling relations and considering the contributions of the counter-terms up to some specific order of accuracy in the corresponding power series expansion. Such mapping relations will be extremely helpful to directly convert the bare action in terms of the renormalized version of the action. In the following, we describe such mapping in detail for each of the previously mentioned contributions:}
     \bea \left({\bf D}_1\right)_{\bf R} \zeta^{\prime} {^2}_{\bf R} \zeta_{\bf R}&=&{\cal Z}_{{\bf D}_1}\left({\cal Z}^{\rm IR}_{n}\right)^{\frac{3}{2}}\times \left({\bf D}_1\right)_{\bf B}\zeta^{\prime} {^2}_{\bf B} \zeta_{\bf B}\nonumber\\
     &=&\left(1+\delta_{{\cal Z}_{{\bf D}_1}}\right)\left(1+\frac{3}{2}\delta_{{\cal Z}^{\rm IR}_{n}}+\cdots\right)\times \left({\bf D}_1\right)_{\bf B}\zeta^{\prime} {^2}_{\bf B} \zeta_{\bf B}\nonumber\\
     &=& \left(1+\delta_{{\cal Z}_{{\bf D}_1}}+\frac{3}{2}\delta_{{\cal Z}^{\rm IR}_{n}}+\cdots\right)\times \left({\bf D}_1\right)_{\bf B}\zeta^{\prime} {^2}_{\bf B} \zeta_{\bf B}\nonumber\\
     &=&\left({\bf D}_1\right)_{\bf B}\zeta^{\prime} {^2}_{\bf B} \zeta_{\bf B}-\left({\bf D}_1\right)_{\bf C}\zeta^{\prime} {^2}_{\bf C} \zeta_{\bf C},\\
   \left({\bf D}_2\right)_{\bf R} (\partial_i \zeta_{\bf R})^2 \zeta_{\bf R}&=&{\cal Z}_{{\bf D}_2}\left({\cal Z}^{\rm IR}_{n}\right)^{\frac{3}{2}}\times \left({\bf D}_2\right)_{\bf B}(\partial_i \zeta_{\bf B})^2 \zeta_{\bf B}\nonumber\\
     &=&\left(1+\delta_{{\cal Z}_{{\bf D}_2}}\right)\left(1+\frac{3}{2}\delta_{{\cal Z}^{\rm IR}_{n}}+\cdots\right)\times \left({\bf D}_2\right)_{\bf B}(\partial_i \zeta_{\bf B})^2 \zeta_{\bf B}\nonumber\\
     &=& \left(1+\delta_{{\cal Z}_{{\bf D}_2}}+\frac{3}{2}\delta_{{\cal Z}^{\rm IR}_{n}}+\cdots\right)\times \left({\bf D}_2\right)_{\bf B}(\partial_i \zeta_{\bf B})^2 \zeta_{\bf B}\nonumber\\
     &=&\left({\bf D}_2\right)_{\bf B}(\partial_i \zeta_{\bf B})^2 \zeta_{\bf B}-\left({\bf D}_2\right)_{\bf C}(\partial_i \zeta_{\bf C})^2 \zeta_{\bf C},\\
     \left({\bf D}_3\right)_{\bf R} \zeta^{\prime}_{\bf R} (\partial_i \zeta_{\bf R}) \bigg (\partial_i \partial ^{-2}\bigg(\frac{\epsilon \zeta^{\prime}_{\bf R}}{c_s ^2}\bigg)\bigg)&=&{\cal Z}_{{\bf D}_3}\left({\cal Z}^{\rm IR}_{n}\right)^{\frac{3}{2}}\times \left({\bf D}_3\right)_{\bf B}\zeta^{\prime}_{\bf B} (\partial_i \zeta_{\bf B}) \bigg (\partial_i \partial ^{-2}\bigg(\frac{\epsilon \zeta^{\prime}_{\bf B}}{c_s ^2}\bigg)\bigg)\nonumber\\
     &=&\left(1+\delta_{{\cal Z}_{{\bf D}_3}}\right)\left(1+\frac{3}{2}\delta_{{\cal Z}^{\rm IR}_{n}}+\cdots\right)\times \left({\bf D}_3\right)_{\bf B}(\partial_i \zeta_{\bf B}) \bigg (\partial_i \partial ^{-2}\bigg(\frac{\epsilon \zeta^{\prime}_{\bf B}}{c_s ^2}\bigg)\bigg)\nonumber\\
     &=& \left(1+\delta_{{\cal Z}_{{\bf D}_3}}+\frac{3}{2}\delta_{{\cal Z}^{\rm IR}_{n}}+\cdots\right)\times \left({\bf D}_3\right)_{\bf B}(\partial_i \zeta_{\bf B}) \bigg (\partial_i \partial ^{-2}\bigg(\frac{\epsilon \zeta^{\prime}_{\bf B}}{c_s ^2}\bigg)\bigg)\nonumber\\
     &=&\left({\bf D}_3\right)_{\bf B}\zeta^{\prime}_{\bf B} (\partial_i \zeta_{\bf B}) \bigg (\partial_i \partial ^{-2}\bigg(\frac{\epsilon \zeta^{\prime}_{\bf B}}{c_s ^2}\bigg)\bigg)-\left({\bf D}_3\right)_{\bf C}\zeta^{\prime}_{\bf C} (\partial_i \zeta_{\bf C}) \bigg (\partial_i \partial ^{-2}\bigg(\frac{\epsilon \zeta^{\prime}_{\bf C}}{c_s ^2}\bigg)\bigg),\quad\quad\eea\bea
     \left({\bf D}_4\right)_{\bf R} \left(\zeta^{\prime} {^3}_{\bf R}+\zeta^{\prime}_{\bf R}(\partial_i \zeta_{\bf R})^2 \right)&=&{\cal Z}_{{\bf D}_4}\left({\cal Z}^{\rm IR}_{n}\right)^{\frac{3}{2}}\times \left({\bf D}_4\right)_{\bf B}\left(\zeta^{\prime} {^3}_{\bf B}+\zeta^{\prime}_{\bf B}(\partial_i \zeta_{\bf B})^2 \right)\nonumber\\
     &=&\left(1+\delta_{{\cal Z}_{{\bf D}_4}}\right)\left(1+\frac{3}{2}\delta_{{\cal Z}^{\rm IR}_{n}}+\cdots\right)\times \left({\bf D}_4\right)_{\bf B}\left(\zeta^{\prime} {^3}_{\bf B}+\zeta^{\prime}_{\bf B}(\partial_i \zeta_{\bf B})^2 \right)\nonumber\\
     &=& \left(1+\delta_{{\cal Z}_{{\bf D}_4}}+\frac{3}{2}\delta_{{\cal Z}^{\rm IR}_{n}}+\cdots\right)\times \left({\bf D}_4\right)_{\bf B}\left(\zeta^{\prime} {^3}_{\bf B}+\zeta^{\prime}_{\bf B}(\partial_i \zeta_{\bf B})^2 \right)\nonumber\\
     &=&\left({\bf D}_4\right)_{\bf B}\left(\zeta^{\prime} {^3}_{\bf B}+\zeta^{\prime}_{\bf B}(\partial_i \zeta_{\bf B})^2 \right)-\left({\bf D}_4\right)_{\bf C}\left(\zeta^{\prime} {^3}_{\bf C}+\zeta^{\prime}_{\bf C}(\partial_i \zeta_{\bf C})^2 \right),\quad\quad\\
     \left({\bf D}_5\right)_{\bf R} \zeta_{\bf R} \bigg (\partial_i\partial_j \partial ^{-2}\bigg(\frac{\epsilon \zeta^{\prime}_{\bf R}}{c_s ^2}\bigg)\bigg)^2&=&{\cal Z}_{{\bf D}_5}\left({\cal Z}^{\rm IR}_{n}\right)^{\frac{3}{2}}\times \left({\bf D}_5\right)_{\bf B}\zeta_{\bf B}\bigg (\partial_i\partial_j \partial ^{-2}\bigg(\frac{\epsilon \zeta^{\prime}_{\bf B}}{c_s ^2}\bigg)\bigg)^2\nonumber\\
     &=&\left(1+\delta_{{\cal Z}_{{\bf D}_5}}\right)\left(1+\frac{3}{2}\delta_{{\cal Z}^{\rm IR}_{n}}+\cdots\right)\times \left({\bf D}_5\right)_{\bf B}\zeta_{\bf B}\bigg (\partial_i\partial_j \partial ^{-2}\bigg(\frac{\epsilon \zeta^{\prime}_{\bf B}}{c_s ^2}\bigg)\bigg)^2\nonumber\\
     &=& \left(1+\delta_{{\cal Z}_{{\bf D}_5}}+\frac{3}{2}\delta_{{\cal Z}^{\rm IR}_{n}}+\cdots\right)\times \left({\bf D}_5\right)_{\bf B}\zeta_{\bf B}\bigg (\partial_i\partial_j \partial ^{-2}\bigg(\frac{\epsilon \zeta^{\prime}_{\bf B}}{c_s ^2}\bigg)\bigg)^2\nonumber\\
     &=&\left({\bf D}_5\right)_{\bf B}\zeta_{\bf B}\bigg (\partial_i\partial_j \partial ^{-2}\bigg(\frac{\epsilon \zeta^{\prime}_{\bf B}}{c_s ^2}\bigg)\bigg)^2-\left({\bf D}_5\right)_{\bf C}\zeta_{\bf C}\bigg (\partial_i\partial_j \partial ^{-2}\bigg(\frac{\epsilon \zeta^{\prime}_{\bf C}}{c_s ^2}\bigg)\bigg)^2,\quad\quad\\
     \left({\bf D}_6\right)_{\bf R} \zeta^{\prime}_{\bf R} \zeta^2_{\bf R}&=&{\cal Z}_{{\bf D}_6}\left({\cal Z}^{\rm IR}_{n}\right)^{\frac{3}{2}}\times \left({\bf D}_6\right)_{\bf B}\zeta^{\prime}_{\bf B} \zeta^2_{\bf B}\nonumber\\
     &=&\left(1+\delta_{{\cal Z}_{{\bf D}_6}}\right)\left(1+\frac{3}{2}\delta_{{\cal Z}^{\rm IR}_{n}}+\cdots\right)\times \left({\bf D}_6\right)_{\bf B}\zeta^{\prime}_{\bf B} \zeta^2_{\bf B}\nonumber\\
     &=& \left(1+\delta_{{\cal Z}_{{\bf D}_6}}+\frac{3}{2}\delta_{{\cal Z}^{\rm IR}_{n}}+\cdots\right)\times \left({\bf D}_6\right)_{\bf B}\zeta^{\prime}_{\bf B} \zeta^2_{\bf B}\nonumber\\
     &=&\left({\bf D}_6\right)_{\bf B}\zeta^{\prime}_{\bf B} \zeta^2_{\bf B}-\left({\bf D}_6\right)_{\bf C}\zeta^{\prime}_{\bf C} \zeta^2_{\bf C},\quad\quad
   \eea
   \textcolor{black}{From the above-mentioned analysis, we have found that all the operators in the third-order perturbed action have the universal scaling properties: }
   \bea &&{\cal Z}_{{\bf D}_i}\left({\cal Z}^{\rm IR}_{n}\right)^{\frac{3}{2}}\approx \left(1+\delta_{{\cal Z}_{{\bf D}_i}}+\frac{3}{2}\delta_{{\cal Z}^{\rm IR}_{n}}+\cdots\right)\quad\quad\forall \quad i=1,2,\cdots,6.\eea
  \textcolor{black}{Here the dotted contributions $\cdots$ represent the higher-order terms in the corresponding power-series expansion. For our analysis, we have restricted up to the first-order terms and neglected all the higher-order small effects. This in turn implies that we did the rest of the computation to determine the explicit contributions of the counter-terms in the linear regime of the corresponding expansion.}
\end{enumerate}
\textcolor{black}{Hence keeping all the previously mentioned steps in mind one can finally construct the expression for the renormalized version of the second-order and third-order perturbed action for the gauge invariant comoving curvature perturbation, which are described by the following expressions:}
\bea
\label{raction}
         S_{\zeta,{\bf R}}^{(3)} &=& M^{2}_{p}\int d\tau\;d^3x\;\bigg [\left({\bf D}_1\right)_{\bf R}\; \zeta^{\prime} {^2}_{\bf R} \zeta_{\bf R} + \left({\bf D}_2\right)_{\bf R}\;(\partial_i \zeta_{\bf R})^2 \zeta_{\bf R}  -  \left({\bf D}_3\right)_{\bf R}\;\zeta^{\prime}_{\bf R} (\partial_i \zeta_{\bf R}) \bigg (\partial_i \partial ^{-2}\bigg(\frac{\epsilon \zeta^{\prime}_{\bf R}}{c_s ^2}\bigg)\bigg) \nonumber \\ 
        && \quad \quad \quad \quad \quad \quad \quad  - \left({\bf D}_4\right)_{\bf R}\;\left(\zeta^{\prime} {^3}_{\bf R}+\zeta^{\prime}_{\bf R}(\partial_i \zeta_{\bf R})^2 \right) +  \left({\bf D}_5\right)_{\bf R}\;\zeta_{\bf R} \bigg(\partial_i \partial_j \partial^{-2}\bigg (\frac{\epsilon \zeta^{\prime}_{\bf R}}{c_s ^2}\bigg)\bigg)^2 + \underbrace{ \left({\bf D}_6\right)_{\bf R}\zeta^{\prime}_{\bf R} \zeta^2_{\bf R}}_{\textbf{Dominant term}}+.....\bigg]\nonumber\\
        &=&M^{2}_{p}\int d\tau\;d^3x\;\left({\cal Z}^{\rm IR}_{n}\right)^{\frac{3}{2}}\times\bigg [{\cal Z}_{{\bf D}_1}\left({\bf D}_1\right)_{\bf B}\; \zeta^{\prime} {^2}_{\bf B} \zeta_{\bf B} + {\cal Z}_{{\bf D}_2}\left({\bf D}_2\right)_{\bf B}\;(\partial_i \zeta_{\bf B})^2 \zeta_{\bf B}  -  {\cal Z}_{{\bf D}_3}\left({\bf D}_3\right)_{\bf B}\;\zeta^{\prime}_{\bf B} (\partial_i \zeta_{\bf B}) \bigg (\partial_i \partial ^{-2}\bigg(\frac{\epsilon \zeta^{\prime}_{\bf B}}{c_s ^2}\bigg)\bigg) \nonumber \\ 
        && \quad \quad \quad \quad \quad \quad \quad  - {\cal Z}_{{\bf D}_4}\left({\bf D}_4\right)_{\bf B}\;\left(\zeta^{\prime} {^3}_{\bf B}+\zeta^{\prime}_{\bf B}(\partial_i \zeta_{\bf B})^2 \right) +  {\cal Z}_{{\bf D}_5}\left({\bf D}_5\right)_{\bf B}\;\zeta_{\bf B} \bigg(\partial_i \partial_j \partial^{-2}\bigg (\frac{\epsilon \zeta^{\prime}_{\bf B}}{c_s ^2}\bigg)\bigg)^2 + \underbrace{{\cal Z}_{{\bf D}_6} \left({\bf D}_6\right)_{\bf B}\zeta^{\prime}_{\bf B} \zeta^2_{\bf B}}_{\textbf{Dominant term}}+.....\bigg]\nonumber\\
        &=&M^{2}_{p}\int d\tau\;d^3x\;\bigg [\left(1+\delta_{{\cal Z}_{{\bf D}_1}}+\frac{3}{2}\delta_{{\cal Z}^{\rm IR}_{n}}\right)\left({\bf D}_1\right)_{\bf B}\; \zeta^{\prime} {^2}_{\bf B} \zeta_{\bf B} +\left(1+\delta_{{\cal Z}_{{\bf D}_2}}+\frac{3}{2}\delta_{{\cal Z}^{\rm IR}_{n}}\right)\left({\bf D}_2\right)_{\bf B}\;(\partial_i \zeta_{\bf B})^2 \zeta_{\bf B}\nonumber \\ 
        && \quad \quad \quad \quad \quad \quad \quad    -  \left(1+\delta_{{\cal Z}_{{\bf D}_3}}+\frac{3}{2}\delta_{{\cal Z}^{\rm IR}_{n}}\right)\left({\bf D}_3\right)_{\bf B}\;\zeta^{\prime}_{\bf B} (\partial_i \zeta_{\bf B}) \bigg (\partial_i \partial ^{-2}\bigg(\frac{\epsilon \zeta^{\prime}_{\bf B}}{c_s ^2}\bigg)\bigg) \nonumber \\
        && \quad \quad \quad \quad \quad \quad \quad  - \left(1+\delta_{{\cal Z}_{{\bf D}_4}}+\frac{3}{2}\delta_{{\cal Z}^{\rm IR}_{n}}\right)\left({\bf D}_4\right)_{\bf B}\;\left(\zeta^{\prime} {^3}_{\bf B}+\zeta^{\prime}_{\bf B}(\partial_i \zeta_{\bf B})^2 \right) \nonumber \eea\bea
        && \quad \quad \quad \quad \quad \quad \quad  +  \left(1+\delta_{{\cal Z}_{{\bf D}_5}}+\frac{3}{2}\delta_{{\cal Z}^{\rm IR}_{n}}\right)\left({\bf D}_5\right)_{\bf B}\;\zeta_{\bf B} \bigg(\partial_i \partial_j \partial^{-2}\bigg (\frac{\epsilon \zeta^{\prime}_{\bf B}}{c_s ^2}\bigg)\bigg)^2\nonumber \\ 
        && \quad \quad \quad \quad \quad \quad \quad  + \underbrace{\left(1+\delta_{{\cal Z}_{{\bf D}_6}}+\frac{3}{2}\delta_{{\cal Z}^{\rm IR}_{n}}\right) \left({\bf D}_6\right)_{\bf B}\zeta^{\prime}_{\bf B} \zeta^2_{\bf B}}_{\textbf{Dominant term}}+.....\bigg]\nonumber\\
        &=&M^{2}_{p}\int d\tau\;d^3x\;\bigg [\left({\bf D}_1\right)_{\bf B}\; \zeta^{\prime} {^2}_{\bf B} \zeta_{\bf B} + \left({\bf D}_2\right)_{\bf B}\;(\partial_i \zeta_{\bf B})^2 \zeta_{\bf B}  -  \left({\bf D}_3\right)_{\bf B}\;\zeta^{\prime}_{\bf B} (\partial_i \zeta_{\bf B}) \bigg (\partial_i \partial ^{-2}\bigg(\frac{\epsilon \zeta^{\prime}_{\bf B}}{c_s ^2}\bigg)\bigg) \nonumber \\ 
        && \quad \quad \quad \quad \quad \quad \quad  - \left({\bf D}_4\right)_{\bf B}\;\left(\zeta^{\prime} {^3}_{\bf B}+\zeta^{\prime}_{\bf B}(\partial_i \zeta_{\bf B})^2 \right) +  \left({\bf D}_5\right)_{\bf B}\;\zeta_{\bf B} \bigg(\partial_i \partial_j \partial^{-2}\bigg (\frac{\epsilon \zeta^{\prime}_{\bf B}}{c_s ^2}\bigg)\bigg)^2 + \underbrace{ \left({\bf D}_6\right)_{\bf B}\zeta^{\prime}_{\bf B} \zeta^2_{\bf B}}_{\textbf{Dominant term}}+.....\bigg]\nonumber\\
        &&+M^{2}_{p}\int d\tau\;d^3x\;\bigg [\delta_{{\cal Z}_{{\bf D}_1}}\left({\bf D}_1\right)_{\bf B}\; \zeta^{\prime} {^2}_{\bf B} \zeta_{\bf B} +\delta_{{\cal Z}_{{\bf D}_2}}\left({\bf D}_2\right)_{\bf B}\;(\partial_i \zeta_{\bf B})^2 \zeta_{\bf B}   -  \delta_{{\cal Z}_{{\bf D}_3}}\left({\bf D}_3\right)_{\bf B}\;\zeta^{\prime}_{\bf B} (\partial_i \zeta_{\bf B}) \bigg (\partial_i \partial ^{-2}\bigg(\frac{\epsilon \zeta^{\prime}_{\bf B}}{c_s ^2}\bigg)\bigg) \nonumber \\ 
        && \quad \quad \quad \quad \quad \quad \quad  - \delta_{{\cal Z}_{{\bf D}_4}}\left({\bf D}_4\right)_{\bf B}\;\left(\zeta^{\prime} {^3}_{\bf B}+\zeta^{\prime}_{\bf B}(\partial_i \zeta_{\bf B})^2 \right)  +  \delta_{{\cal Z}_{{\bf D}_5}}\left({\bf D}_5\right)_{\bf B}\;\zeta_{\bf B} \bigg(\partial_i \partial_j \partial^{-2}\bigg (\frac{\epsilon \zeta^{\prime}_{\bf B}}{c_s ^2}\bigg)\bigg)^2\nonumber \eea\bea
        && \quad \quad \quad \quad \quad \quad \quad  + \underbrace{\delta_{{\cal Z}_{{\bf D}_6}} \left({\bf D}_6\right)_{\bf B}\zeta^{\prime}_{\bf B} \zeta^2_{\bf B}}_{\textbf{Dominant term}}+.....\bigg]\nonumber\\
        &&+\frac{3}{2}M^{2}_{p}\int d\tau\;d^3x\;\delta_{{\cal Z}^{\rm IR}_{n}}\times\bigg [\left({\bf D}_1\right)_{\bf B}\; \zeta^{\prime} {^2}_{\bf B} \zeta_{\bf B} +\left({\bf D}_2\right)_{\bf B}\;(\partial_i \zeta_{\bf B})^2 \zeta_{\bf B}   -  \left({\bf D}_3\right)_{\bf B}\;\zeta^{\prime}_{\bf B} (\partial_i \zeta_{\bf B}) \bigg (\partial_i \partial ^{-2}\bigg(\frac{\epsilon \zeta^{\prime}_{\bf B}}{c_s ^2}\bigg)\bigg) \nonumber \\ 
        && \quad \quad \quad \quad \quad \quad \quad\quad \quad \quad \quad \quad  - \left({\bf D}_4\right)_{\bf B}\;\left(\zeta^{\prime} {^3}_{\bf B}+\zeta^{\prime}_{\bf B}(\partial_i \zeta_{\bf B})^2 \right)  +  \left({\bf D}_5\right)_{\bf B}\;\zeta_{\bf B} \bigg(\partial_i \partial_j \partial^{-2}\bigg (\frac{\epsilon \zeta^{\prime}_{\bf B}}{c_s ^2}\bigg)\bigg)^2\nonumber \\ 
        && \quad \quad \quad \quad \quad \quad \quad\quad \quad \quad \quad \quad  + \underbrace{\left({\bf D}_6\right)_{\bf B}\zeta^{\prime}_{\bf B} \zeta^2_{\bf B}}_{\textbf{Dominant term}}+.....\bigg]\nonumber\\
        &=&S_{\zeta,{\bf B}}^{(3)}-S_{\zeta,{\bf C}}^{(3)},
   \eea 
\textcolor{black}{where the bare part of the third-order perturbed action, $S_{\zeta,{\bf B}}^{(3)}$ is explicitly mentioned in the equation (\ref{baction}). After introducing the counter-terms the corresponding contributions in the third-order action can be further described by the following expression:}
\bea
\label{caction}
         S_{\zeta,{\bf C}}^{(3)} &=& M^{2}_{p}\int d\tau\;d^3x\;\bigg [\left({\bf D}_1\right)_{\bf C}\; \zeta^{\prime} {^2}_{\bf C} \zeta_{\bf C} + \left({\bf D}_2\right)_{\bf C}\;(\partial_i \zeta_{\bf C})^2 \zeta_{\bf C}  -  \left({\bf D}_3\right)_{\bf C}\;\zeta^{\prime}_{\bf C} (\partial_i \zeta_{\bf C}) \bigg (\partial_i \partial ^{-2}\bigg(\frac{\epsilon \zeta^{\prime}_{\bf R}}{c_s ^2}\bigg)\bigg) \nonumber \\ 
        && \quad \quad \quad \quad \quad \quad \quad  - \left({\bf D}_4\right)_{\bf C}\;\left(\zeta^{\prime} {^3}_{\bf C}+\zeta^{\prime}_{\bf C}(\partial_i \zeta_{\bf C})^2 \right) +  \left({\bf D}_5\right)_{\bf C}\;\zeta_{\bf C} \bigg(\partial_i \partial_j \partial^{-2}\bigg (\frac{\epsilon \zeta^{\prime}_{\bf C}}{c_s ^2}\bigg)\bigg)^2 + \underbrace{ \left({\bf D}_6\right)_{\bf C}\zeta^{\prime}_{\bf C} \zeta^2_{\bf C}}_{\textbf{Dominant term}}+.....\bigg]\nonumber\\
        &=&\Bigg[S_{\zeta,{\bf C}}^{(3)}\Bigg]_{\bf UV}+\Bigg[S_{\zeta,{\bf C}}^{(3)}\Bigg]_{\bf IR}.
   \eea 
  \textcolor{black}{ Here it is important to note that the counter-term contribution of the third-order action which can able to completely remove the quadratic UV divergence contribution in the one-loop corrected primordial power spectrum is described by the following expression: }
\bea
\label{UVaction}
         \Bigg[S_{\zeta,{\bf C}}^{(3)}\Bigg]_{\bf UV} 
        &=&-M^{2}_{p}\int d\tau\;d^3x\;\bigg [\delta_{{\cal Z}_{{\bf D}_1}}\left({\bf D}_1\right)_{\bf B}\; \zeta^{\prime} {^2}_{\bf B} \zeta_{\bf B} +\delta_{{\cal Z}_{{\bf D}_2}}\left({\bf D}_2\right)_{\bf B}\;(\partial_i \zeta_{\bf B})^2 \zeta_{\bf B}   -  \delta_{{\cal Z}_{{\bf D}_3}}\left({\bf D}_3\right)_{\bf B}\;\zeta^{\prime}_{\bf B} (\partial_i \zeta_{\bf B}) \bigg (\partial_i \partial ^{-2}\bigg(\frac{\epsilon \zeta^{\prime}_{\bf B}}{c_s ^2}\bigg)\bigg) \nonumber \\ 
        && \quad \quad \quad \quad \quad \quad \quad  - \delta_{{\cal Z}_{{\bf D}_4}}\left({\bf D}_4\right)_{\bf B}\;\left(\zeta^{\prime} {^3}_{\bf B}+\zeta^{\prime}_{\bf B}(\partial_i \zeta_{\bf B})^2 \right)  +  \delta_{{\cal Z}_{{\bf D}_5}}\left({\bf D}_5\right)_{\bf B}\;\zeta_{\bf B} \bigg(\partial_i \partial_j \partial^{-2}\bigg (\frac{\epsilon \zeta^{\prime}_{\bf B}}{c_s ^2}\bigg)\bigg)^2\nonumber \\ 
        && \quad \quad \quad \quad \quad \quad \quad  + \underbrace{\delta_{{\cal Z}_{{\bf D}_6}} \left({\bf D}_6\right)_{\bf B}\zeta^{\prime}_{\bf B} \zeta^2_{\bf B}}_{\textbf{Dominant term}}+.....\bigg],\quad\quad\quad\eea
       \textcolor{black}{ Similarly, the counter-term contribution of the third-order action that can able to smoothen the behaviour of logarithmic IR divergence contribution in the one-loop corrected primordial power spectrum is described by the following expression: }
        \bea\label{IRaction}
         \Bigg[S_{\zeta,{\bf C}}^{(3)}\Bigg]_{\bf IR} 
        &=&-\frac{3}{2}M^{2}_{p}\int d\tau\;d^3x\;\delta_{{\cal Z}^{\rm IR}_{n}}\times\bigg [\left({\bf D}_1\right)_{\bf B}\; \zeta^{\prime} {^2}_{\bf B} \zeta_{\bf B} +\left({\bf D}_2\right)_{\bf B}\;(\partial_i \zeta_{\bf B})^2 \zeta_{\bf B}   -  \left({\bf D}_3\right)_{\bf B}\;\zeta^{\prime}_{\bf B} (\partial_i \zeta_{\bf B}) \bigg (\partial_i \partial ^{-2}\bigg(\frac{\epsilon \zeta^{\prime}_{\bf B}}{c_s ^2}\bigg)\bigg) \nonumber \\ 
        && \quad \quad \quad \quad \quad \quad \quad\quad \quad \quad \quad \quad  - \left({\bf D}_4\right)_{\bf B}\;\left(\zeta^{\prime} {^3}_{\bf B}+\zeta^{\prime}_{\bf B}(\partial_i \zeta_{\bf B})^2 \right)  +  \left({\bf D}_5\right)_{\bf B}\;\zeta_{\bf B} \bigg(\partial_i \partial_j \partial^{-2}\bigg (\frac{\epsilon \zeta^{\prime}_{\bf B}}{c_s ^2}\bigg)\bigg)^2\nonumber \\ 
        && \quad \quad \quad \quad \quad \quad \quad\quad \quad \quad \quad \quad  + \underbrace{\left({\bf D}_6\right)_{\bf B}\zeta^{\prime}_{\bf B} \zeta^2_{\bf B}}_{\textbf{Dominant term}}+.....\bigg].\quad\quad\quad
   \eea 
\textcolor{black}{Now with the help of the derived form of the renormalized form of the third-order perturbed action for the gauge invariant comoving scalar curvature perturbation as mentioned in equation (\ref{raction}), our further aim is to compute the explicit expression for the one-loop corrected scalar power spectrum in presence of all the counter-terms introduced previously. In the Quantum Field Theory description, this is nothing but computing the renormalized one-loop 1PI effective action corresponding to the two-point amplitude. To serve this purpose we use the previously mentioned in-in formalism in the present context of the computation, using which we found the following simplified result for each of the individual phases at the one-loop level:}
\bea
    \label{Sr1againA}
        \bigg[\Delta^{2}_{\zeta,\textbf{One-loop}}(k)\bigg]_{\textbf{SR}_{1}} &=& \left(1+\delta_{{\cal Z}^{\rm IR}_{n}}\right)\times\left(1-\frac{2}{15\pi^{2}}\frac{1}{c_{s}^{2}k_{*}^{2}}\left(1-\frac{1}{c_{s}^{2}}\right)\epsilon\right)\nonumber\\
        &&\quad\quad\times \left(\left(\delta_{{\cal Z}_{{\bf D}_1}}+\delta_{{\cal Z}_{{\bf D}_2}}+\delta_{{\cal Z}_{{\bf D}_3}}+\delta_{{\cal Z}_{{\bf D}_4}}+\delta_{{\cal Z}_{{\bf D}_5}}\right)_{\textbf{SR}_{1}}-\frac{4}{3}{\cal J}_{\textbf{SR}_{1}}(\tau_{s_{1}})\bigg[\Delta^{2}_{\zeta,\textbf{Tree}}(k)\bigg]_{\textbf{SR}_{1}}^{2}\right), \\
    \label{usragainB}
        \bigg[\Delta^{2}_{\zeta,\textbf{One-loop}}(k)\bigg]_{\textbf{USR}_{n}} &=&  
        \left(1+\delta_{{\cal Z}^{\rm IR}_{n}}\right)\times\bigg\{\frac{1}{4}\bigg[\Delta^{2}_{\zeta,\textbf{Tree}}(k)\bigg]_{\textbf{SR}_{1}}^{2}\times\bigg[\bigg(\frac{\Delta\eta(\tau_{e_{n}})}{\tilde{c}^{4}_{s}}\bigg)^{2}{\cal J}_{\textbf{USR}_{n}}(\tau_{e_{n}}) - \left(\frac{\Delta\eta(\tau_{s_{n}})}{\tilde{c}^{4}_{s}}\right)^{2}{\cal J}_{\textbf{USR}_{n}}(\tau_{s_{n}})\bigg]\nonumber\\
        &&\quad\quad\quad\quad\quad\quad\quad\quad\quad\quad\quad\quad\quad\quad\quad\quad\quad\quad\quad\quad\quad\quad\quad\quad\quad\quad- \left(\delta_{{\cal Z}_{{\bf D}_6}}\right)_{\textbf{USR}_{n}}\bigg\}, \;\;\quad \\
    \label{sr2againC}
        \bigg[\Delta^{2}_{\zeta,\textbf{One-loop}}(k)\bigg]_{\textbf{SR}_{n+1}} &=& 
\left(1+\delta_{{\cal Z}^{\rm IR}_{n}}\right)\times\left(1-\frac{2}{15\pi^{2}}\frac{1}{c_{s}^{2}k_{*}^{2}}\left(1-\frac{1}{c_{s}^{2}}\right)\epsilon\right)
        \nonumber\\
        &&\quad\quad\times  \bigg(\left(\delta_{{\cal Z}_{{\bf D}_1}}+\delta_{{\cal Z}_{{\bf D}_2}}+\delta_{{\cal Z}_{{\bf D}_3}}+\delta_{{\cal Z}_{{\bf D}_4}}+\delta_{{\cal Z}_{{\bf D}_5}}\right)_{\textbf{SR}_{n+1}}+{\cal J}_{\textbf{SR}_{n+1}}(\tau_{e_{n}})\bigg[\Delta^{2}_{\zeta,\textbf{Tree}}(k)\bigg]_{\textbf{SR}_{1}}^{2}\bigg).\quad\quad
\eea
\textcolor{black}{Here all the loop integral terms at different conformal times namely, ${\cal J}_{\textbf{SR}_{1}}(\tau_{s_{1}})$ during $\tau < \tau_{s_1}$,  ${\cal J}_{\textbf{USR}_{n}1}(\tau_{s_{n}})$ and ${\cal J}_{\textbf{USR}_{n}2}(\tau_{s_{n}})$, during sharp transition $\tau = \tau_{s_n}$, ${\cal J}_{\textbf{USR}_{n}1}(\tau_{e_{n}})$ and ${\cal J}_{\textbf{USR}_{n}2}(\tau_{e_{n}})$ during $\tau = \tau_{e_n}$, and ${\cal J}_{\textbf{SR}_{n+1}}(\tau_{e_{n}})$ during the conformal time interval $\tau_{e_{n}} \leq \tau < \tau_{s_{n+1}}$ are explicitly evaluated in the previous section in detail. Further, including the effects from each of the individual phases appearing in the context of MST construction we get the final following compact form of the one-loop corrected renormalized version of the scalar power spectrum for the $n$-th sharp transition, which is described by:}
\bea
\left[\overline{\Delta_{\zeta,\textbf{EFT}}^{2}(k)}\right]_n &=& \left(1+\delta_{{\cal Z}^{\rm IR}_{n}}\right)\times\bigg\{\bigg[\Delta^{2}_{\zeta,\textbf{Tree}}(k)\bigg]_{\textbf{SR}_{1}}+\left(1-\frac{2}{15\pi^{2}}\frac{1}{c_{s}^{2}k_{*}^{2}}\left(1-\frac{1}{c_{s}^{2}}\right)\epsilon\right)\nonumber\\&&\quad\quad\quad\quad\quad\quad\quad\times \left(\left(\delta_{{\cal Z}_{{\bf D}_1}}+\delta_{{\cal Z}_{{\bf D}_2}}+\delta_{{\cal Z}_{{\bf D}_3}}+\delta_{{\cal Z}_{{\bf D}_4}}+\delta_{{\cal Z}_{{\bf D}_5}}\right)_{\textbf{SR}_{1}}-\frac{4}{3}{\cal J}_{\textbf{SR}_{1}}(\tau_{s_{1}})\bigg[\Delta^{2}_{\zeta,\textbf{Tree}}(k)\bigg]_{\textbf{SR}_{1}}^{2}\right)\nonumber\\
&&\quad\quad\quad\quad\quad+\bigg\{\frac{1}{4}\bigg[\Delta^{2}_{\zeta,\textbf{Tree}}(k)\bigg]_{\textbf{SR}_{1}}^{2}\times\bigg[\bigg(\frac{\Delta\eta(\tau_{e_{n}})}{\tilde{c}^{4}_{s}}\bigg)^{2}{\cal J}_{\textbf{USR}_{n}}(\tau_{e_{n}}) - \left(\frac{\Delta\eta(\tau_{s_{n}})}{\tilde{c}^{4}_{s}}\right)^{2}{\cal J}_{\textbf{USR}_{n}}(\tau_{s_{n}})\bigg]\nonumber\\
        &&\quad\quad\quad\quad\quad\quad\quad\quad\quad\quad\quad\quad\quad\quad\quad\quad\quad\quad\quad\quad\quad\quad\quad\quad\quad\quad- \left(\delta_{{\cal Z}_{{\bf D}_6}}\right)_{\textbf{USR}_{n}}\bigg\}\nonumber\\
&&\quad\quad\quad\quad\quad+\left(1-\frac{2}{15\pi^{2}}\frac{1}{c_{s}^{2}k_{*}^{2}}\left(1-\frac{1}{c_{s}^{2}}\right)\epsilon\right)\nonumber\\
&&\quad\quad\quad\quad\quad\times\bigg(\left(\delta_{{\cal Z}_{{\bf D}_1}}+\delta_{{\cal Z}_{{\bf D}_2}}+\delta_{{\cal Z}_{{\bf D}_3}}+\delta_{{\cal Z}_{{\bf D}_4}}+\delta_{{\cal Z}_{{\bf D}_5}}\right)_{\textbf{SR}_{n+1}}+{\cal J}_{\textbf{SR}_{n+1}}(\tau_{e_{n}})\bigg[\Delta^{2}_{\zeta,\textbf{Tree}}(k)\bigg]_{\textbf{SR}_{1}}^{2}\bigg)\bigg\},\quad\quad
\eea
\textcolor{black}{which can be written further after some readjustment in the following tractable format:}
\bea
\left[\overline{\Delta_{\zeta,\textbf{EFT}}^{2}(k)}\right]_n &=& {\cal Z}^{\rm IR}_{n}{\cal Z}^{\rm UV}_{n}\times \bigg[\Delta^{2}_{\zeta,\textbf{Tree}}(k)\bigg]_{\textbf{SR}_{1}}= {\cal Z}^{\rm IR}_{n}\times \left[\Delta_{\zeta,\textbf{EFT}}^{2}(k)\right]_n,\eea
\textcolor{black}{where IR and UV counter-terms are defined as:}
\bea && {\cal Z}^{\rm IR}_{n}=\left(1+\delta_{{\cal Z}^{\rm IR}_{n}}\right),\\
&& {\cal Z}^{\rm UV}_{n}=\left(1+\delta_{{\cal Z}^{\rm UV}_{n}}\right).\eea
\textcolor{black}{Here we define $\delta_{{\cal Z}^{\rm UV}_{n}}(\delta_{{\cal Z}_{{\bf D}_1}},\delta_{{\cal Z}_{{\bf D}_2}},\delta_{{\cal Z}_{{\bf D}_3}},\delta_{{\cal Z}_{{\bf D}_4}},\delta_{{\cal Z}_{{\bf D}_5}},\delta_{{\cal Z}_{{\bf D}_6}})$ by the following expression:}
\bea &&\delta_{{\cal Z}^{\rm UV}_{n}}(\delta_{{\cal Z}_{{\bf D}_1}},\delta_{{\cal Z}_{{\bf D}_2}},\delta_{{\cal Z}_{{\bf D}_3}},\delta_{{\cal Z}_{{\bf D}_4}},\delta_{{\cal Z}_{{\bf D}_5}},\delta_{{\cal Z}_{{\bf D}_6}})\nonumber\\
&=&\bigg[\Delta^{2}_{\zeta,\textbf{Tree}}(k)\bigg]_{\textbf{SR}_{1}}\times\left(1-\frac{2}{15\pi^{2}}\frac{1}{c_{s}^{2}k_{*}^{2}}\left(1-\frac{1}{c_{s}^{2}}\right)\epsilon\right)\nonumber\\
&&\quad\quad\quad\quad\quad\quad\times\left(\frac{1}{\bigg[\Delta^{2}_{\zeta,\textbf{Tree}}(k)\bigg]_{\textbf{SR}_{1}}^{2}}\times\left(\delta_{{\cal Z}_{{\bf D}_1}}+\delta_{{\cal Z}_{{\bf D}_2}}+\delta_{{\cal Z}_{{\bf D}_3}}+\delta_{{\cal Z}_{{\bf D}_4}}+\delta_{{\cal Z}_{{\bf D}_5}}\right)_{\textbf{SR}_{1}}-\frac{4}{3}{\cal J}_{\textbf{SR}_{1}}(\tau_{s_{1}})\right)\nonumber\\
&&\quad\quad\quad\quad\quad+\frac{1}{4}\bigg[\Delta^{2}_{\zeta,\textbf{Tree}}(k)\bigg]_{\textbf{SR}_{1}}\times\Bigg\{\bigg[\bigg(\frac{\Delta\eta(\tau_{e_{n}})}{\tilde{c}^{4}_{s}}\bigg)^{2}{\cal J}_{\textbf{USR}_{n}}(\tau_{e_{n}}) - \left(\frac{\Delta\eta(\tau_{s_{n}})}{\tilde{c}^{4}_{s}}\right)^{2}{\cal J}_{\textbf{USR}_{n}}(\tau_{s_{n}})\bigg]\nonumber\\
        &&\quad\quad\quad\quad\quad\quad\quad\quad\quad\quad\quad\quad\quad\quad\quad\quad\quad\quad\quad\quad\quad\quad\quad\quad\quad\quad- \frac{4}{\bigg[\Delta^{2}_{\zeta,\textbf{Tree}}(k)\bigg]_{\textbf{SR}_{1}}^{2}}\times\left(\delta_{{\cal Z}_{{\bf D}_6}}\right)_{\textbf{USR}_{n}}\Bigg\}\nonumber\\
&&\quad\quad\quad\quad\quad+\bigg[\Delta^{2}_{\zeta,\textbf{Tree}}(k)\bigg]_{\textbf{SR}_{1}}\times\left(1-\frac{2}{15\pi^{2}}\frac{1}{c_{s}^{2}k_{*}^{2}}\left(1-\frac{1}{c_{s}^{2}}\right)\epsilon\right)\nonumber\\
&&\quad\quad\times\left(\frac{1}{\bigg[\Delta^{2}_{\zeta,\textbf{Tree}}(k)\bigg]_{\textbf{SR}_{1}}^{2}}\times\left(\delta_{{\cal Z}_{{\bf D}_1}}+\delta_{{\cal Z}_{{\bf D}_2}}+\delta_{{\cal Z}_{{\bf D}_3}}+\delta_{{\cal Z}_{{\bf D}_4}}+\delta_{{\cal Z}_{{\bf D}_5}}\right)_{\textbf{SR}_{n+1}}+{\cal J}_{\textbf{SR}_{n+1}}(\tau_{e_{n}})\right),\quad\quad
\eea
\textcolor{black}{Next, we impose the renormalization condition to determine the expression for the counter-terms as appearing in the above-mentioned derived result. This can be understood with the explicit use of the Renormalization Group (RG) flow in the present context of the discussion, which will in terms fix the structure of all the UV and IR sensitive counter-terms appearing in the present computation. This can be technically implemented with the help of the flow equation and corresponding beta functions written for the corresponding 1PI one-loop corrected renormalized two-point amplitude in the Fourier space, which is nothing but representing the renormalized scalar power spectrum in the present context.}

\textcolor{black}{Here in this cosmological setup the Callan–Symanzik equation can be expressed as:}
\bea \frac{d}{d\ln \mu}\Bigg\{\bigg[\Delta^{2}_{\zeta,\textbf{Tree}}(k)\bigg]_{\textbf{SR}_{1}}\Bigg\}=\frac{d}{d\ln \mu}\Bigg\{\frac{\left[\overline{\Delta_{\zeta,\textbf{EFT}}^{2}(k)}\right]_n}{{\cal Z}^{\rm IR}_{n}{\cal Z}^{\rm UV}_{n}}\Bigg\}=0.\eea
\textcolor{black}{Now it is important to note that here the corresponding total differential operator can be further simplified by the following expression:}
\bea \frac{d}{d\ln \mu}=\bigg(\frac{\partial}{\partial\ln \mu}+\sum^{6}_{i=1}\beta_{{\bf D}_i}\frac{\partial}{\partial {\bf D}_i}-\gamma_{\rm IR}-\gamma_{\rm UV}\Bigg),\eea
\textcolor{black}{where the beta functions for the coupling parameters are defined as:}
\bea \beta_{{\bf D}_i}:=\left(\frac{\partial {\bf D}_i}{\partial\ln \mu}\right)\quad\quad\forall\quad\quad i=1,2,\cdots,6.\eea
\textcolor{black}{The individual expressions for the beta functions for each of the coupling parameters can be further simplified in the following form:}
\bea \beta_{{\bf D}_1}:&=&\left(\frac{\partial {\bf D}_1}{\partial\ln \mu}\right)\nonumber\\
&=& 2\epsilon a^2\Bigg[(\epsilon-\eta)\bigg(3(c^2_s-1)+2\epsilon-\frac{3\epsilon^2}{2}\bigg)+\bigg(3(c^2_s-1)+\epsilon-\frac{\epsilon^2}{2}+6\frac{ s c^2_s}{H}\bigg)\Bigg]\left(\frac{d {\cal N}}{d\ln \mu}\right),\\
\beta_{{\bf D}_2}:&=&\left(\frac{\partial {\bf D}_2}{\partial\ln \mu}\right)\nonumber\\
&=&\frac{2\epsilon a^2}{c^2_s}\Bigg[(\epsilon-\eta)\bigg(2\epsilon-2s+1-c^2_s\bigg)+\bigg(\epsilon-2s+1-c^2_s\bigg)\left(1-\frac{s}{H}\right)-\left(\frac{\dot{s}}{H}+\frac{sc^2_s}{H}\right)\Bigg]\left(\frac{d{\cal N}}{d\ln \mu}\right),\\
\beta_{{\bf D}_3}:&=&\left(\frac{\partial {\bf D}_3}{\partial\ln \mu}\right)\nonumber\\
&=&\frac{4\epsilon a^2}{c^2_s}\Bigg(\epsilon-\eta+1-\frac{s}{H}\Bigg)\left(\frac{d{\cal N}}{d\ln \mu}\right),\\
\beta_{{\bf D}_4}:&=&\left(\frac{\partial {\bf D}_4}{\partial\ln \mu}\right)\nonumber\\
&=&\frac{2a\epsilon}{H}\Bigg[(\epsilon-\eta)\left(1-\frac{1}{c^2_s}\right)+\frac{s}{H c^2_s}+\frac{1}{2}(1+\epsilon H)\left(1-\frac{1}{c^2_s}\right)\Bigg]\left(\frac{d{\cal N}}{d\ln \mu}\right),\\
\beta_{{\bf D}_5}:&=&\left(\frac{\partial {\bf D}_5}{\partial\ln \mu}\right)\nonumber\\
&=&\epsilon a^2 \left(\epsilon-\eta+1\right)\left(\frac{d{\cal N}}{d\ln \mu}\right),\\
\beta_{{\bf D}_6}:&=&\left(\frac{\partial {\bf D}_6}{\partial\ln \mu}\right)\nonumber\\
&=&\frac{a^2\epsilon}{c^2_s}\left(\frac{\eta}{c^2_s}\right)^{'}\Bigg[\bigg(\epsilon-\eta+1-\frac{s}{H}\bigg)+\frac{1}{2}\frac{d}{d\ln \mu}\ln \left(\frac{\eta}{c^2_s}\right)^{'}\Bigg]\left(\frac{d{\cal N}}{d\ln \mu}\right).
\eea
\textcolor{black}{Here we use the relations for further simplification purposes}
\bea &&\left(\frac{d\epsilon}{d\ln \mu}\right)=\left(\frac{d\epsilon}{d{\cal N}}\right)\left(\frac{d{\cal N}}{d\ln \mu}\right)=2\epsilon(\epsilon-\eta)\left(\frac{d{\cal N}}{d\ln \mu}\right)\quad\quad\quad{\rm where}\quad \eta=\epsilon-\frac{1}{2\epsilon}\left(\frac{d\epsilon}{d{\cal N}}\right),\quad\\
&&\left(\frac{d\ln a}{d\ln \mu}\right)=\left(\frac{d\ln a}{dt}\right)\left(\frac{dt}{d\ln \mu}\right)=H\left(\frac{dt}{d\ln \mu}\right)=\left(\frac{d {\cal N}}{d\ln \mu}\right)\quad\quad\quad{\rm where}\quad d{\cal N}=H\;dt\quad{\rm with}\quad \left(\frac{d\ln a}{dt}\right)=H,\quad\\
&&\left(\frac{dc_s}{d\ln \mu}\right)=\left(\frac{dc_s}{dt}\right)\left(\frac{dt}{d\ln \mu}\right)=sc_s\left(\frac{dt}{d\ln \mu}\right)=\frac{sc_s}{H}\left(\frac{d {\cal N}}{d\ln \mu}\right)\quad\quad\quad{\rm where}\quad s=\frac{1}{c_s}\left(\frac{dc_s}{dt}\right)=\left(\frac{d\ln c_s}{dt}\right),\quad\\
&&\left(\frac{ds}{d\ln \mu}\right)=\left(\frac{ds}{dt}\right)\left(\frac{dt}{d\ln \mu}\right)=\dot{s}\left(\frac{dt}{d\ln \mu}\right)=\frac{\dot{s}}{H}\left(\frac{d {\cal N}}{d\ln \mu}\right)\quad\\
&&\left(\frac{dH}{d\ln \mu}\right)=\left(\frac{dH}{dt}\right)\left(\frac{dt}{d\ln \mu}\right)=-\epsilon H\left(\frac{d {\cal N}}{d\ln \mu}\right).\eea
\textcolor{black}{Now at the horizon crossing scale, $c_s\mu=c_s k=aH$, and we now evaluate the quantity $\left(\frac{d {\cal N}}{d\ln \mu}\right)$ which appears in all expressions for the beta functions of the coupling parameters of the underlying theory. To serve this purpose let us first express the expression for the first slow-roll parameter in terms of the number of e-foldings, which gives the following result:}
\bea \epsilon=-\frac{\dot{H}}{H^2}=-\frac{d\ln H}{d{\cal N}}.\eea
\textcolor{black}{Utilizing this fact we can compute the following expression:}
\bea \left(\frac{d {\cal N}}{d\ln \mu}\right)=\frac{1}{\displaystyle 1-\left(\epsilon+\frac{s}{H}\right)}\approx \Bigg[1+\left(\epsilon+\frac{s}{H}\right)\Bigg].\eea
\textcolor{black}{Consequently, all the mentioned beta functions can be further simplified by the following expression:}
\bea \beta_{{\bf D}_1}&=& 2\epsilon a^2\Bigg[(\epsilon-\eta)\bigg(3(c^2_s-1)+2\epsilon-\frac{3\epsilon^2}{2}\bigg)+\bigg(3(c^2_s-1)+\epsilon-\frac{\epsilon^2}{2}+6\frac{ s c^2_s}{H}\bigg)\Bigg]\Bigg[1+\left(\epsilon+\frac{s}{H}\right)\Bigg],\\
\beta_{{\bf D}_2}
&=&\frac{2\epsilon a^2}{c^2_s}\Bigg[(\epsilon-\eta)\bigg(2\epsilon-2s+1-c^2_s\bigg)+\bigg(\epsilon-2s+1-c^2_s\bigg)\left(1-\frac{s}{H}\right)-\left(\frac{\dot{s}}{H}+\frac{sc^2_s}{H}\right)\Bigg]\Bigg[1+\left(\epsilon+\frac{s}{H}\right)\Bigg],\\
\beta_{{\bf D}_3}
&=&\frac{4\epsilon a^2}{c^2_s}\Bigg(\epsilon-\eta+1-\frac{s}{H}\Bigg)\Bigg[1+\left(\epsilon+\frac{s}{H}\right)\Bigg],\\
\beta_{{\bf D}_4}
&=&\frac{2a\epsilon}{H}\Bigg[(\epsilon-\eta)\left(1-\frac{1}{c^2_s}\right)+\frac{s}{H c^2_s}+\frac{1}{2}(1+\epsilon H)\left(1-\frac{1}{c^2_s}\right)\Bigg]\Bigg[1+\left(\epsilon+\frac{s}{H}\right)\Bigg],\\
\beta_{{\bf D}_5}
&=&\epsilon a^2 \left(\epsilon-\eta+1\right)\Bigg[1+\left(\epsilon+\frac{s}{H}\right)\Bigg],\\
\beta_{{\bf D}_6}
&=&\frac{a^2\epsilon}{c^2_s}\left(\frac{\eta}{c^2_s}\right)^{'}\Bigg[\bigg(\epsilon-\eta+1-\frac{s}{H}\bigg)+\frac{1}{2}\frac{d}{d\ln \mu}\ln \left(\frac{\eta}{c^2_s}\right)^{'}\Bigg]\Bigg[1+\left(\epsilon+\frac{s}{H}\right)\Bigg].
\eea

\textcolor{black}{Also, we introduce two new parameters which are IR and UV counter-term dependent, and given by the following expression:}
\bea && \gamma_{\rm IR}:=\left(\frac{\partial\ln {\cal Z}^{\rm IR}_{n}}{\partial\ln \mu}\right),\quad\quad\quad
 \gamma_{\rm UV}:=\left(\frac{\partial\ln {\cal Z}^{\rm UV}_{n}}{\partial\ln \mu}\right).\eea
\textcolor{black}{Finally, we get the following simplified version of the Callan–Symanzik equation for the given cosmological setup:}
\bea \bigg(\frac{\partial}{\partial\ln \mu}+\sum^{6}_{i=1}\beta_{{\bf D}_i}\frac{\partial}{\partial {\bf D}_i}-\gamma_{\rm IR}-\gamma_{\rm UV}\Bigg)\overline{\Delta_{\zeta,\textbf{EFT}}^{2}(k)}=0.\eea

\textcolor{black}{Further, one can compute the following flow equations which will be extremely helpful further in determining the IR and UV counter-terms in terms of the renormalization scale in the present context:}
\begin{enumerate}
    \item \underline{\bf Renormalized scalar spectral tilt:}\\ \\
    \textcolor{black}{In terms of the cosmological hierarchical flow the renormalized spectral tilt of the scalar power spectrum can be computed as: }

    \bea 
\left[\overline{n_{\zeta,\textbf{EFT}}(k)-1}\right]_n&=&\frac{d}{d\ln k}\bigg(\ln \left[\overline{\Delta_{\zeta,\textbf{EFT}}^{2}(k)}\right]_n\bigg)\nonumber\\
&=&{\cal Z}^{\rm IR}_{n}{\cal Z}^{\rm UV}_{n}\times \frac{d}{d \ln k}\bigg(\ln \bigg[\Delta^{2}_{\zeta,\textbf{Tree}}(k)\bigg]_{\textbf{SR}_{1}}\bigg)\nonumber\\
&&\quad\quad\quad\quad\quad\quad\quad+\Bigg\{{\cal Z}^{\rm IR}_{n}\underbrace{\bigg(\frac{d{\cal Z}^{\rm UV}_{n}}{d\ln k}\bigg)}_{\neq 0}+{\cal Z}^{\rm UV}_{n}\underbrace{\bigg(\frac{d{\cal Z}^{\rm IR}_{n}}{d\ln k}\bigg)}_{=0}\Bigg\}\times\bigg(\ln \bigg[\Delta^{2}_{\zeta,\textbf{Tree}}(k)\bigg]_{\textbf{SR}_{1}}\bigg)\nonumber\\
&=&{\cal Z}^{\rm IR}_{n}{\cal Z}^{\rm UV}_{n}\times\bigg(\bigg[n_{\zeta,\textbf{Tree}}(k)\bigg]_{{\bf SR}_1}-1\bigg)+{\cal Z}^{\rm IR}_{n}\bigg(\frac{d{\cal Z}^{\rm UV}_{n}}{d\ln k}\bigg)\times\bigg(\ln \bigg[\Delta^{2}_{\zeta,\textbf{Tree}}(k)\bigg]_{\textbf{SR}_{1}}\bigg)\nonumber\\
&=&{\cal Z}^{\rm IR}_{n}\times\Bigg[{\cal Z}^{\rm UV}_{n}\bigg(\bigg[n_{\zeta,\textbf{Tree}}(k)\bigg]_{{\bf SR}_1}-1\bigg)+\bigg(\frac{d{\cal Z}^{\rm UV}_{n}}{d\ln k}\bigg)\bigg(\ln \bigg[\Delta^{2}_{\zeta,\textbf{Tree}}(k)\bigg]_{\textbf{SR}_{1}}\bigg)\Bigg].\eea

\item \underline{\bf Renormalized running of the scalar spectral tilt:}\\ \\
    \textcolor{black}{In terms of the cosmological hierarchical flow the renormalized running of the spectral tilt of the scalar power spectrum can be computed as:}

    \bea 
&&\left[\overline{\alpha_{\zeta,\textbf{EFT}}(k)}\right]_n=\frac{d}{d\ln k}\bigg(\left[\overline{n_{\zeta,\textbf{EFT}}(k)}\right]_n\bigg)\nonumber\\
&=&{\cal Z}^{\rm IR}_{n}{\cal Z}^{\rm UV}_{n}\times\frac{d}{d\ln k}\bigg(\bigg[n_{\zeta,\textbf{Tree}}(k)\bigg]_{{\bf SR}_1}\bigg)+\Bigg\{2{\cal Z}^{\rm IR}_{n}\underbrace{\bigg(\frac{d{\cal Z}^{\rm UV}_{n}}{d\ln k}\bigg)}_{\neq 0}+{\cal Z}^{\rm UV}_{n}\underbrace{\bigg(\frac{d{\cal Z}^{\rm IR}_{n}}{d\ln k}\bigg)}_{=0}\Bigg\}\times\bigg(\bigg[n_{\zeta,\textbf{Tree}}(k)\bigg]_{{\bf SR}_1}-1\bigg)\nonumber\\
&&\quad\quad\quad\quad\quad+\Bigg\{{\cal Z}^{\rm IR}_{n}\underbrace{\bigg(\frac{d^2{\cal Z}^{\rm UV}_{n}}{d\ln k^2}\bigg)}_{\neq 0}+\underbrace{\bigg(\frac{d{\cal Z}^{\rm UV}_{n}}{d\ln k}\bigg)}_{\neq 0}\underbrace{\bigg(\frac{d{\cal Z}^{\rm IR}_{n}}{d\ln k}\bigg)}_{=0}\Bigg\}\times\bigg(\ln \bigg[\Delta^{2}_{\zeta,\textbf{Tree}}(k)\bigg]_{\textbf{SR}_{1}}\bigg)\nonumber\\
&=&{\cal Z}^{\rm IR}_{n}{\cal Z}^{\rm UV}_{n}\times\bigg(\bigg[\alpha_{\zeta,\textbf{Tree}}(k)\bigg]_{{\bf SR}_1}\bigg)+2{\cal Z}^{\rm IR}_{n}\bigg(\frac{d{\cal Z}^{\rm UV}_{n}}{d\ln k}\bigg)\times\bigg(\bigg[n_{\zeta,\textbf{Tree}}(k)\bigg]_{{\bf SR}_1}-1\bigg)\nonumber\\
&&\quad\quad\quad\quad\quad+{\cal Z}^{\rm IR}_{n}\bigg(\frac{d^2{\cal Z}^{\rm UV}_{n}}{d\ln k^2}\bigg)\times\bigg(\ln \bigg[\Delta^{2}_{\zeta,\textbf{Tree}}(k)\bigg]_{\textbf{SR}_{1}}\bigg)\\
&=&{\cal Z}^{\rm IR}_{n}\times\Bigg[{\cal Z}^{\rm UV}_{n}\bigg(\bigg[\alpha_{\zeta,\textbf{Tree}}(k)\bigg]_{{\bf SR}_1}\bigg)+2\bigg(\frac{d{\cal Z}^{\rm UV}_{n}}{d\ln k}\bigg)\bigg(\bigg[n_{\zeta,\textbf{Tree}}(k)\bigg]_{{\bf SR}_1}-1\bigg)+\bigg(\frac{d^2{\cal Z}^{\rm UV}_{n}}{d\ln k^2}\bigg)\bigg(\ln \bigg[\Delta^{2}_{\zeta,\textbf{Tree}}(k)\bigg]_{\textbf{SR}_{1}}\bigg)\Bigg].\nonumber\eea

\item \underline{\bf Renormalized running of the running of scalar spectral tilt:}\\ \\
\textcolor{black}{In terms of the cosmological hierarchical flow the renormalized running of the running of spectral tilt of the scalar power spectrum can be computed as:}

    \bea 
\left[\overline{\beta_{\zeta,\textbf{EFT}}(k)}\right]_n&=&\frac{d}{d\ln k}\bigg(\left[\overline{\alpha_{\zeta,\textbf{EFT}}(k)}\right]_n\bigg)\nonumber\\
&=&{\cal Z}^{\rm IR}_{n}{\cal Z}^{\rm UV}_{n}\times\frac{d}{d\ln k}\bigg(\bigg[\alpha_{\zeta,\textbf{Tree}}(k)\bigg]_{{\bf SR}_1}\bigg)+2{\cal Z}^{\rm IR}_{n}\underbrace{\bigg(\frac{d{\cal Z}^{\rm UV}_{n}}{d\ln k}\bigg)}_{\neq 0}\times\bigg(\bigg[\alpha_{\zeta,\textbf{Tree}}(k)\bigg]_{{\bf SR}_1}\bigg)\nonumber\\
&& \quad\quad\quad\quad\quad+2\Bigg\{\underbrace{\bigg(\frac{d{\cal Z}^{\rm IR}_{n}}{d\ln k}\bigg)}_{=0}\underbrace{\bigg(\frac{d{\cal Z}^{\rm UV}_{n}}{d\ln k}\bigg)}_{\neq 0}+{\cal Z}^{\rm IR}_{n}\underbrace{\bigg(\frac{d^2{\cal Z}^{\rm IR}_{n}}{d\ln k^2}\bigg)}_{\neq 0}\Bigg\}\times\bigg(\bigg[n_{\zeta,\textbf{Tree}}(k)\bigg]_{{\bf SR}_1}-1\bigg)\nonumber\\
&&\quad\quad\quad\quad\quad+\Bigg\{{\cal Z}^{\rm IR}_{n}\underbrace{\bigg(\frac{d^3{\cal Z}^{\rm UV}_{n}}{d\ln k^3}\bigg)}_{\neq 0}+\underbrace{\bigg(\frac{d^2{\cal Z}^{\rm UV}_{n}}{d\ln k^2}\bigg)}_{\neq 0}\underbrace{\bigg(\frac{d{\cal Z}^{\rm IR}_{n}}{d\ln k}\bigg)}_{=0}\Bigg\}\times\bigg(\ln \bigg[\Delta^{2}_{\zeta,\textbf{Tree}}(k)\bigg]_{\textbf{SR}_{1}}\bigg)\nonumber\\
&=&{\cal Z}^{\rm IR}_{n}{\cal Z}^{\rm UV}_{n}\times\bigg(\bigg[\beta_{\zeta,\textbf{Tree}}(k)\bigg]_{{\bf SR}_1}\bigg)+2{\cal Z}^{\rm IR}_{n}\bigg(\frac{d{\cal Z}^{\rm UV}_{n}}{d\ln k}\bigg)\times\bigg(\bigg[\alpha_{\zeta,\textbf{Tree}}(k)\bigg]_{{\bf SR}_1}\bigg)\nonumber\\
&&\quad\quad\quad\quad\quad+3{\cal Z}^{\rm IR}_{n}\bigg(\frac{d^2{\cal Z}^{\rm UV}_{n}}{d\ln k^2}\bigg)\times\bigg(\bigg[n_{\zeta,\textbf{Tree}}(k)\bigg]_{{\bf SR}_1}-1\bigg)+{\cal Z}^{\rm IR}_{n}\bigg(\frac{d^3{\cal Z}^{\rm UV}_{n}}{d\ln k^3}\bigg)\times\bigg(\ln \bigg[\Delta^{2}_{\zeta,\textbf{Tree}}(k)\bigg]_{\textbf{SR}_{1}}\bigg)\nonumber\\
&=&{\cal Z}^{\rm IR}_{n}\times\Bigg[{\cal Z}^{\rm UV}_{n}\bigg(\bigg[\beta_{\zeta,\textbf{Tree}}(k)\bigg]_{{\bf SR}_1}\bigg)+2\bigg(\frac{d{\cal Z}^{\rm UV}_{n}}{d\ln k}\bigg)\bigg(\bigg[\alpha_{\zeta,\textbf{Tree}}(k)\bigg]_{{\bf SR}_1}\bigg)\nonumber\\
&&\quad\quad\quad\quad\quad+3\bigg(\frac{d^2{\cal Z}^{\rm UV}_{n}}{d\ln k^2}\bigg)\bigg(\bigg[n_{\zeta,\textbf{Tree}}(k)\bigg]_{{\bf SR}_1}-1\bigg)+\bigg(\frac{d^3{\cal Z}^{\rm UV}_{n}}{d\ln k^3}\bigg)\bigg(\ln \bigg[\Delta^{2}_{\zeta,\textbf{Tree}}(k)\bigg]_{\textbf{SR}_{1}}\bigg)\Bigg].\eea
\end{enumerate}
\textcolor{black}{The above-mentioned flow equations directly point towards the scale dependence of the two-point amplitude of the primordial power spectrum for the scalar modes in the presence of the IR and UV counter-term effects in each of the individual expressions.} 

\textcolor{black}{Now we impose the renormalization conditions which are going to fix both the structure of the UV and IR sensitive counter-terms. For this, we are now going to utilize the known facts at the CMB pivot scale $k_*$, which restricts us from considering the following necessary constraints that are described in terms of renormalization conditions:}
\begin{enumerate}
    \item \underline{\bf Renormalization condition I:}\\ \\
    \textcolor{black}{The first renormalization condition states that the two-point amplitude of the scalar power spectrum after renormalization has to exactly match the tree-level contribution computed in the first slow-roll phase in our computational setup at the CMB pivot scale $k_*$. In technical language, this statement can be stated as:}
          \bea \left[\overline{\Delta_{\zeta,\textbf{EFT}}^{2}(k_*)}\right]_n &=& \bigg[\Delta^{2}_{\zeta,\textbf{Tree}}(k_*)\bigg]_{\textbf{SR}_{1}}.\eea
         \textcolor{black}{ This condition is not at all ad-hoc and completely justifiable from the physical perspective. We all know that at the CMB map to date, no quantum effects are distinguishable, and for this reason at the corresponding momentum scale, which is the pivot scale such quantum corrections can be easily considered to be absent. CMB shows the outcome of causal phenomena in terms of the distribution of cold and hot spots in the maps and all the a-causal features happening outside the horizon (i.e. in the super-Hubble scale) are immaterial as far as the reliable cosmological observations are concerned. } 
          
     \item  \underline{\bf Renormalization condition II:}\\ \\
    \textcolor{black}{ The second renormalization condition states that the logarithmic derivative of the two-point amplitude of the scalar power spectrum with respect to the momentum scale after renormalization has to exactly match the tree-level contribution computed in the first slow-roll phase in our computational setup at the CMB pivot scale $k_*$. Such logarithmic derivative with respect to the momentum scale basically computes the scale dependence of the two-point amplitude in the context of primordial cosmology, commonly known as the {\it spectral-tilt} or {\it spectral-index} of the scalar power spectrum. In technical language, this statement can be stated as:}
          \bea 
\left[\overline{n_{\zeta,\textbf{EFT}}(k)(k_*)-1}\right]_n&=&\Bigg[\frac{d}{d\ln k}\bigg(\ln \left[\overline{\Delta_{\zeta,\textbf{EFT}}^{2}(k)}\right]_n\bigg)\Bigg]_{k=k_*}=\bigg(\bigg[n_{\zeta,\textbf{Tree}}(k_*)\bigg]_{{\bf SR}_1}-1\bigg).\eea
\textcolor{black}{The above-mentioned condition is perfectly consistent with the previously mentioned condition. Actually, in the present context, it tells us something more about the shape of the scalar power spectrum. Although the tilt is computed in the CMB pivot scale $k_*$, due to the existence of its non-zero value we can further fix the shape of the primordial power spectrum computed for the scalar modes.}

 \item  \underline{\bf Renormalization condition III:}\\ \\
\textcolor{black}{ The third renormalization condition states that the second logarithmic derivative of the two-point amplitude of the scalar power spectrum with respect to the momentum scale, which is basically the running of the scalar spectral tilt after renormalization has to exactly match the tree-level contribution computed in the first slow-roll phase in our computational setup at the CMB pivot scale $k_*$. Such logarithmic double derivative with respect to the momentum scale basically computes the scale dependence of the two-point amplitude in the context of primordial cosmology, commonly known as the {\it running of the spectral-tilt} or {\it running of the spectral-index} of the scalar power spectrum. In technical language, this statement can be stated as:}
          \bea 
\left[\overline{\alpha_{\zeta,\textbf{EFT}}(k_*)}\right]_n=\Bigg[\frac{d}{d\ln k}\bigg(\left[\overline{n_{\zeta,\textbf{EFT}}(k)}\right]_n\bigg)\Bigg]_{k=k_*}=\bigg(\bigg[\alpha_{\zeta,\textbf{Tree}}(k_*)\bigg]_{{\bf SR}_1}\bigg).\eea
\textcolor{black}{The above-mentioned condition is perfectly consistent with the previously mentioned two conditions. Actually, in the present context, it tells us something more about the shape of the scalar power spectrum rather than having a tilt. Although the running of the tilt is computed in the CMB pivot scale $k_*$, due to the existence of its non-zero value we can further fix the shape of the primordial power spectrum computed for the scalar modes in terms of concavity or convexity in the original form of the underlying effective potential or Hubble parameter of the underlying EFT setup. The existence of such running further points towards having some additional features, i.e. inflection point, saddle point, bump/dip in the underlying mathematical structure of the effective potential or in the Hubble parameter as appearing in the present version of the EFT computation.}
\item  \underline{\bf Renormalization condition IV:}\\ \\
\textcolor{black}{The fourth renormalization condition states that the third logarithmic derivative of the two-point amplitude of the scalar power spectrum with respect to the momentum scale, which basically represents the running of the running of scalar spectral tilt after renormalization has to exactly match the tree-level contribution computed in the first slow-roll phase in our computational setup at the CMB pivot scale $k_*$. Such logarithmic triple derivative with respect to the momentum scale basically computes the further minute scale dependence of the two-point amplitude in the context of primordial cosmology, commonly known as the {\it running of the running of spectral-tilt} or {\it running of the running of spectral-index} of the scalar power spectrum. In technical language, this statement can be stated as: }
          \bea 
\left[\overline{\beta_{\zeta,\textbf{EFT}}(k_*)}\right]_n=\Bigg[\frac{d}{d\ln k}\bigg(\left[\overline{\alpha_{\zeta,\textbf{EFT}}(k)}\right]_n\bigg)\Bigg]_{k=k_*}=\bigg(\bigg[\beta_{\zeta,\textbf{Tree}}(k_*)\bigg]_{{\bf SR}_1}\bigg).\eea
\textcolor{black}{The above-mentioned condition is perfectly consistent with the previously mentioned three conditions. Actually, in the present context, it tells us something more about the shape of the scalar power spectrum rather than having a running of the spectral tilt. Although the running of the running of the spectral tilt is computed in the CMB pivot scale $k_*$, due to the existence of its non-zero value we can further fix the shape of the primordial power spectrum very minutely computed for the scalar modes in terms of concavity or convexity as pointed previously in the present version of the EFT computation. }

\end{enumerate}
\textcolor{black}{The immediate consequence of the above-mentioned four renormalization conditions imposes the following further constraints on the properties of the UV and IR sensitive counter-terms, which are appended below point-wise:}
\begin{itemize}
    \item   \underline{\bf Consequence of renormalization condition I:}\\ \\
               \textcolor{black}{ The immediate consequence of the first renormalization condition appears in terms of the following constraint condition:}
                \bea {\cal Z}^{\rm IR}_{n}=  \frac{\small[\overline{\Delta_{\zeta,\textbf{EFT}}^{2}(k_{*})}\small]_{n}}{\small[\Delta_{\zeta,\textbf{EFT}}^{2}(k_{*})\small]_{n}} = \frac{\small[  \Delta_{\zeta,\textbf{Tree}}^{2}(k_{*})\small]_{\textbf{SR}_{1}}}{\small[\Delta_{\zeta,\textbf{EFT}}^{2}(k_{*})\small]_{n}}\quad\quad\quad\Longrightarrow\quad\quad\quad {\cal Z}^{\rm IR}_{n}(k_*){\cal Z}^{\rm UV}_{n}(k_*)=1.\eea
\textcolor{black}{This relation will be extremely helpful to determine the IR counter-term explicitly in this context. However, here it is important to note that from this relation we get the IR counter-term ${\cal Z}^{\rm IR}_{n}$ is the inverse of the UV counter-term ${\cal Z}^{\rm UV}_{n}$ in the present computation. So without fixing the form of the UV counter-term, it is further impossible to fix the form of the IR counter-term. }

    \item \underline{\bf Consequence of renormalization condition II:}\\ \\
\textcolor{black}{The immediate consequence of the second renormalization condition appears in terms of the following constraint conditions:}
                \bea {\cal Z}^{\rm IR}_{n}(k_*){\cal Z}^{\rm UV}_{n}(k_*)=1\quad\quad\quad {\rm and}\quad\quad\quad \bigg(\frac{d{\cal Z}^{\rm UV}_{n}}{d\ln k}\bigg)_{k=k_*}=0.\eea
\textcolor{black}{Careful observation tells us that the direct consequence of condition I is already taken care of in condition II, which means that condition II is more constrained than condition I. This might be helpful in determining the mathematical structure of the UV counter-term due to having an additional constraint appearing in terms of the vanishing logarithmic momentum scale dependent derivative computed at the CMB pivot scale $k_*$. Surprisingly on that scale contribution of such terms is not explicitly visible as well as distinguishable and from this perspective this outcome is completely physically justifiable in the present context of the computation. Once the constrained structure of the UV counter-term is explicitly determined, one can immediately compute the contribution of the IR counter-term and further fix the structure of the renormalized scalar spectrum computed from this EFT setup.}

    \item \underline{\bf Consequence of renormalization condition III:}\\ \\

   \textcolor{black}{ The immediate consequence of the third renormalization condition appears in terms of the following constraint conditions:}
                \bea {\cal Z}^{\rm IR}_{n}(k_*){\cal Z}^{\rm UV}_{n}(k_*)=1,\quad\quad\quad \quad\quad\quad \bigg(\frac{d{\cal Z}^{\rm UV}_{n}}{d\ln k}\bigg)_{k=k_*}=0\quad\quad\quad\quad{\rm and}\quad\quad\quad \bigg(\frac{d^2{\cal Z}^{\rm UV}_{n}}{d\ln k^2}\bigg)_{k=k_*}=0.\eea

               \textcolor{black}{ It gives tighter constraints than the previous two which impose further restrictions on the scale-dependent behaviour of the UV-counter term at the pivot scale.}

    \item \underline{\bf Consequence of renormalization condition IV:}\\ \\
     \textcolor{black}{The immediate consequence of the fourth renormalization condition appears in terms of the following constraint conditions:}
               \bea {\cal Z}^{\rm IR}_{n}(k_*){\cal Z}^{\rm UV}_{n}(k_*)=1,\quad\quad  \bigg(\frac{d{\cal Z}^{\rm UV}_{n}}{d\ln k}\bigg)_{k=k_*}=0\quad\quad\bigg(\frac{d^2{\cal Z}^{\rm UV}_{n}}{d\ln k^2}\bigg)_{k=k_*}=0\quad\quad{\rm and}\quad\quad \bigg(\frac{d^3{\cal Z}^{\rm UV}_{n}}{d\ln k^3}\bigg)_{k=k_*}=0.\quad\eea
\textcolor{black}{It gives further tighter constraints than the previous three which impose further restrictions on the scale-dependent behaviour of the UV-counter term at the pivot scale.}
\end{itemize}

\textcolor{black}{After analyzing the problem in detail we found that we need to very carefully determine the UV counter-term ${\cal Z}^{\rm UV}_{n}$ such that it satisfy the previously obtained sets of constraint conditions. This can be only possible if we correctly remove the contribution of the quadratic divergence contribution. Once this can be done, it will automatically fix the form of the IR counter-term ${\cal Z}^{\rm IR}_{n}$. However, at this level it is extremely difficult to determine the exact form of the UV counter-term ${\cal Z}^{\rm UV}_{n}$ just having the previously mentioned constraints at the CMB pivot scale. The main difficulty at the technical level comes because of the fact that in ${\bf SR}_1$, ${\bf USR}_n$ and ${\bf SR}_{n+1}$ phases counter-terms needs to separately remove the contribution of the quadratic divergences. It seems like easy, but at the technical level it is extremely difficult to pursue just performing the computation we did up to this point. Here comes the importance of the next three sections where the late time renormalization scheme or equivalently the adiabatic renormalization scheme helps us to completely remove the contributions of the quadratic UV divergences from the ${\bf SR}_1$, ${\bf USR}_n$ and ${\bf SR}_{n+1}$ phases individually. Once this is done then we can immediately determine the explicit form of the IR counter-term using the condition, ${\cal Z}^{\rm IR}_{n}(k_*){\cal Z}^{\rm UV}_{n}(k_*)=1$. In the context of power spectrum renormalization we have used this condition along with the quadratic divergence free result obtained from late time renormalization scheme or equivalently the adiabatic renormalization scheme for ${\cal Z}^{\rm UV}_{n}$.}

\textcolor{black}{To avoid any further confusion regarding the schemes of renormalization used or the inter-connection among various tools and techniques used in this paper in the next subsections let us mention the underlying connection among adding counter-term at the level of action (which is a standard approach within the framework of Quantum Field Theory) with the late time, adiabatic/wave function and power spectrum renormalization schemes clearly. We strongly believe such an explanation will be helpful to understand the applicability of the derived results in this paper. Let us provide the explanations in detail which are appended below point-wise:}
\begin{enumerate}
    \item \textcolor{black}{In the present computation complete removal of the harmful quadratic UV divergence is directly associated with the counter-term contribution of the third order perturbed action, which in our computation we have denoted by, $\Bigg[S_{\zeta,{\bf C}}^{(3)}\Bigg]_{\bf UV}$ in the equation (\ref{UVaction}). As an immediate outcome of the computation with this specific part it is possible to explicitly show that the combination of the counter-terms $\delta_{{\cal Z}_{{\bf D}_i}}\forall i=1,2,\cdots,6$ for previously mentioned six operators can be expressed in terms of a cumulative factor. We identify this factor as the counter-term contribution that will completely remove the quadratic UV divergence as appearing in the context of late-time and wave function/adiabatic renormalization schemes, which we will discuss in the later half of this paper. In the next subsections, we will show that the cumulative counter-terms in both schemes can be expressed in terms of certain specific combinations of the counter-terms corresponding to the mentioned six operators in the third-order perturbed action. Such connections will help us to completely remove the quadratic UV divergence from the expression for the one-loop corrected primordial scalar power spectrum.}

    \item \textcolor{black}{On the other hand, the coarse graining and smoothening of the behaviour of the logarithmic IR divergence is directly associated with the counter-term contribution of the third-order perturbed action, which in our computation we have denoted by, $\Bigg[S_{\zeta,{\bf C}}^{(3)}\Bigg]_{\bf IR}$ in the equation (\ref{IRaction}). As an immediate outcome of the computation with this specific part, it is possible to explicitly show that the single counter term $\delta_{{\cal Z}^{\rm IR}_{n}}$. This factor during the computation of 1PI one-loop corrected two-point amplitude will smooth the behaviour of the logarithmic IR divergence by shifting it to the higher order, which corresponds to the higher even loop diagrams appearing in the perturbative expansion. In the power spectrum renormalization scheme, we will discuss this issue in detail in the later half of this paper.}
\end{enumerate}

\subsection{Renormalized scalar power spectrum from MST-EFT}

\subsubsection{Late time renormalization scheme}
In this section, we are going to provide a well-established method to deal with the quadratic UV and IR divergences appearing in the result of the one-loop corrected scalar power spectrum. There are two similar approaches to resolving this: late-time renormalization ($\tau \to 0$), and adiabatic renormalization. 
Let us first begin with a general momentum integral  that reads:
\bea
\label{genI}
{\cal J}_{\rm renG} = \int_{k_{\rm IR}}^{k_{\rm UV}} \frac{dk}{k} (A+B c_s ^2 k^2 \tau^2) + C = \bigg\{A \ln \bigg( \frac{k_{\rm UV}}{k_{\rm IR}}\bigg)+\frac{1}{2}B \tau ^2 c_s ^2 (k_{\rm UV}^2 - k_{\rm IR}^2) + C \bigg\}.
\eea
This integral is a representative which is persistent in phases of SR$_1$, USR$_n$, and SR$_{n+1}$. Also, A and B are the constants that depend on the Bogoliubov coefficients in different phases while $k_{\rm UV}$ and $k_{\rm IR}$ represent the UV and IR momentum cut-off scales respectively which we have still not fixed. C here somewhat plays the role of a counter term which appears during a standard renormalization procedure to tackle the divergences.
Now taking the super-horizon scale limit, we get, 
\bea
\label{latelimit}
\lim_{\tau \rightarrow 0}{\cal J}_{\rm renG} = A \ln{\bigg(\frac{k_{\rm UV}}{k_{\rm IR}}\bigg)}.
\eea
 We fix the value of C to be :
\bea
C= \lim_{\tau \rightarrow 0} \bigg(-\frac{1}{2}B\; \tau^2 c_s ^2 ( k_{\rm UV}^2 - k_{\rm IR}^2 ) \bigg) = 0
\eea
Notice that the value of this counter term vanishes because of implementing the late time scale and not due to the condition $k_{\rm UV} = k_{\rm IR}$, because this condition does not stand for our considerations. Now, we set the momentum cut-off scales to $k_{\rm IR} = k_{s_n}$, and $k_{\rm UV} = k_{e_n}$. 
As mentioned earlier, the logarithmic IR divergences are not harmful, and also cannot be eradicated utilizing the counter term. However, imposing this 
constraint condition on C enables us to smoothen the IR divergences. Now to ensure that there is no overcounting of the momentum modes, we hereby re-evaluate the integral of eqn. (\ref{genI}). We insert an intermediate limit scale $k_{\rm int}$ to evaluate the integral such that the momentum integral is now broken into two intervals. The second part mainly deals with the quadratic UV divergences in the presence of time dependence in the upper limit of the momentum-dependent integration, while the first part addresses the finite contributions. Therefore, now the eqn.(\ref{genI}) can be re-expressed as :

\bea 
{\cal J}_{\rm renG} = \bigg\{\bigg [\int_{k_{\rm IR}}^{k_{\rm int}} + \int_{k_{\rm int}}^{\Lambda a(\tau)/c_s}\bigg] \frac{dk}{k} (A+B c_s ^2 k^2 \tau^2)\bigg\} + C.
\eea

Here the upper limit of integration in the second part has implicit time dependence introduced with the scale $k_{\rm UV}= \Lambda a(\tau)/c_s$. Note that $\Lambda$ here is not the cosmological constant but rather the contribution coming from the UV cut-off scale. In the super-horizon scale limit with $k_{\rm int} \gg k_{\rm IR}$, we see that the factor with $\tau$ dependence i.e. the second term vanishes. Now when we use the relation $a(\tau) / c_s k_{\rm int} = 1/H$. Thus the above integral is re-expressed as :
\bea 
\label{igenen}
{\cal J}_{\rm renG} = A\bigg\{\ln{\bigg(\frac{k_{\rm int}}{k_{\rm IR}}}\bigg) + \ln{\bigg(\frac{\Lambda}{H}}\bigg)\bigg\} + \frac{B}{2} \bigg\{\bigg(\frac{\Lambda}{H}\bigg)^{2} - 1 \bigg\} + C.
\eea
Now that the expression is further simplified, we invoke yet another constraint from adiabatic renormalization at the pivot scale through which we can have an idea about the counter term C that will be given by :
\bea
C(\mu, \Lambda) = A\bigg\{\ln{\bigg(\frac{\mu}{H}}\bigg) - \ln{\bigg(\frac{\Lambda}{H}}\bigg)\bigg\} + \frac{B}{2} \bigg\{\bigg(\frac{\mu}{H}\bigg)^2 - \bigg(\frac{\Lambda}{H}\bigg)^2\bigg\},
\eea
where $\mu$ denotes the renormalization scale associated with the wavefunction renormalization approach. Further, when we substitute the above expression of C in eqn.(\ref{igenen}), we obtain the following expression from the one-loop contribution individually valid in all the phases of inflation mentioned before. 
\bea
{\cal J}_{\rm renG}(\mu) = A \bigg\{\ln{\bigg(\frac{k_{\rm int}}{k_{\rm IR}}}\bigg) + \ln{\bigg(\frac{\mu}{H}}\bigg)\bigg\} + \frac{B}{2} \bigg\{\bigg(\frac{\mu}{H}\bigg)^2 - 1 \bigg\}, 
\eea
Now taking the renormalization scheme to be connected with adiabatic renormalization at $\mu = H$, the above expression for the loop-integral converges to :
\bea
{\cal J}_{\rm renG}(\mu) = {\cal J}_{\rm renG}(H)=A\ln{\bigg(\frac{k_{\rm int}}{k_{\rm IR}}}\bigg).
\eea
Therefore, you can appreciate the fact that this is the same result as obtained from the eqn.(\ref{latelimit}). Moreover, note that the final result is independent of the UV cut-off $\Lambda$ . This analysis points to the similarity of the obtained results of the late-time renormalization and the wavefunction renormalization. This computation also reassures that the results are free from the overcounting of irrelevant momentum modes. Finally, using the late-time renormalization scheme, we can express the integrals in equations (\ref{Sr1again}), (\ref{usragain}) and (\ref{sr2again}) as :
\bea
\lim_{\tau \rightarrow 0}{\cal J}_{\textbf{SR}_{1}}(\tau_{s_{1}}) &=& \ln{\bigg(\frac{k_{s_1}}{k_{*}}}\bigg),  \\
\lim_{\tau \rightarrow 0}{\cal J}_{\textbf{USR}_{n}}(\tau_{s_{n}}) &=& \ln\left(\frac{k_{e_{n}}}{k_{s_{n}}}\right), \\
\lim_{\tau \rightarrow 0}{\cal J}_{\textbf{USR}_{n}}(\tau_{e_{n}})&=&\left(\frac{k_{e_{n}}}{k_{s_{n}}}\right)^{6}{\cal J}_{\textbf{USR}_{n}}(\tau_{s_{n}}), \\
\lim_{\tau \rightarrow 0}{\cal J}_{\textbf{SR}_{n}}(\tau_{e_{n}}) &=& \ln\left(\frac{k_{s_{n+1}}}{k_{e_{n}}}\right)
\eea
Therefore, the total one-loop corrected late-time renormalized scalar power spectrum after the application of these limits becomes:
\bea
\label{late-tpowspec}
\Delta^{2}_{\zeta,\textbf{EFT}}(k) &=& \bigg[\Delta^{2}_{\zeta,\textbf{Tree}}(k)\bigg]_{\textbf{SR}_{1}}
\; \bigg\{ 1+ \bigg[\Delta^{2}_{\zeta,\textbf{Tree}}(k)\bigg]_{\textbf{SR}_{1}} \bigg [ \left(1-\frac{2}{15\pi^{2}}\frac{1}{c_{s}^{2}k_{*}^{2}}\left(1-\frac{1}{c_{s}^{2}}\right)\epsilon\right)\times \left(-\frac{4}{3}\ln{\frac{k_{s_1}}{k_{*}}}\right)\bigg] \nonumber \\
&& \quad \quad + \frac{1}{4} \bigg[\Delta^{2}_{\zeta,\textbf{Tree}}(k)\bigg]_{\textbf{SR}_{1}} \times \sum^{N}_{n=1}\bigg(
     \bigg[\bigg(\frac{\Delta\eta(\tau_{e_{n}})}{\tilde{c}^{4}_{s}}\bigg)^{2}\ln{\bigg(\frac{k_{e_n}}{k_{s_n}}\bigg)^6} - \left(\frac{\Delta\eta(\tau_{s_{n}})}{\tilde{c}^{4}_{s}}\right)^{2}\bigg] \ln{\bigg(\frac{k_{e_n}}{k_{s_n}}\bigg)} \bigg)\nonumber \\
     && \quad \quad  +  
\bigg[\Delta^{2}_{\zeta,\textbf{Tree}}(k)\bigg]_{\textbf{SR}_{1}}\left(1-\frac{2}{15\pi^{2}}\frac{1}{c_{s}^{2}k_{*}^{2}}\left(1-\frac{1}{c_{s}^{2}}\right)\epsilon\right)
\times \sum^{N}_{n=1} 
\bigg(\ln{\bigg(\frac{k_{s_{n+1}}}{k_{e_n}}}\bigg)\bigg)\bigg\}.
\eea
Given below is a diagrammatic representation of the contributions to the total one-loop corrected scalar power spectrum from eqn.(\ref{late-tpowspec}):  
\begin{equation}
\begin{tikzpicture}[baseline={([yshift=-3.5ex]current bounding box.center)},very thick]
  
  \def\radius{1}
  \scalebox{1}{\draw[red,very thick] (0,\radius) circle (\radius);
  \draw[red,very thick] (4.5*\radius,0) circle (\radius);}

  \draw[black, very thick] (-4*\radius,0) -- 
  (-2.5*\radius,0);
  \node at (-2*\radius,0) {+};
  \draw[black, very thick] (-1.5*\radius,0) -- (0,0);
  \draw[blue,fill=blue] (0,0) circle (.5ex);
  \draw[black, very thick] (0,0)  -- (1.5*\radius,0);
  \node at (2*\radius,0) {+};
  \draw[black, very thick] (2.5*\radius,0) -- (3.5*\radius,0); 
  \draw[blue,fill=blue] (3.5*\radius,0) circle (.5ex);
  \draw[blue,fill=blue] (5.5*\radius,0) circle (.5ex);
  \draw[black, very thick] (5.5*\radius,0) -- (6.5*\radius,0);
  

\end{tikzpicture}\quad = \quad {\rm Late-time \,\,renormalized} \,\,\Delta^{2}_{\zeta,\textbf{EFT}}(k),
\end{equation}
where the diagrams represent the tree and only one-loop contributions to the quantity $\Delta^{2}_{\zeta,\textbf{EFT}}(k)$ obtained after the late-time renormalization scheme.  

\textcolor{black}{In the present computation, we have further found the following important facts regarding the connection between the counter-terms appearing in the present context with the counter-terms appearing in the context of renormalization of the bare action as appearing in the context of Quantum Field Theory in quasi de Sitter space with the gauge invariant coming curvature perturbation, which is given by :}
\bea \sum^{6}_{i=1}\delta_{{\cal Z}_{{\bf D}_i}}=C(\mu=H,\Lambda)=\left(1+\delta_{{\cal Z}_{C}}(\mu=H,\Lambda)\right)=-\Bigg\{A\ln{\bigg(\frac{\Lambda}{H}}\bigg) + \frac{B}{2}\bigg(\frac{\Lambda}{H}\bigg)^2\Bigg\},\eea
\textcolor{black}{where $C(\mu=H,\Lambda)$ (more precisely $\delta_{{\cal Z}_{C}}(\mu=H,\Lambda)$) is the counter-term in the late-time renormalization scheme. Utilizing this relation, we get the following relation:}
\bea \left(\sum^{6}_{i=1}\delta_{{\cal Z}_{{\bf D}_i}}-1\right)&=&\left(\delta_{{\cal Z}_{{\bf D}_1}}+\delta_{{\cal Z}_{{\bf D}_2}}+\delta_{{\cal Z}_{{\bf D}_3}}+\delta_{{\cal Z}_{{\bf D}_4}}+\delta_{{\cal Z}_{{\bf D}_5}}+\delta_{{\cal Z}_{{\bf D}_6}}-1\right)\nonumber\\
&=&\delta_{{\cal Z}_{C}}(\mu=H,\Lambda)\nonumber\\
&=&-\Bigg\{1+A\ln{\bigg(\frac{\Lambda}{H}}\bigg) + \frac{B}{2}\bigg(\frac{\Lambda}{H}\bigg)^2\Bigg\}.\eea
\textcolor{black}{From this relation, it is clear how the operator counter terms of the third-order action are connected to the present scheme. Here it is important to note that, the renormalization scale $\mu=H$ is arbitrary due to having an inherent scale dependence in the Hubble parameter. For this reason, this result describes the contributions as well as connections among the counter-terms in a generic fashion. To know more about the explicit contributions and further effects it is further desirable to split the results into the parts which are tagged by the symbols ${\rm SR}_1$, ${\rm USR}_n$ and ${\rm SR}_n$. Here $n$ characterizes the number of MST in the present context of the computation. After splitting the above-mentioned results into the corresponding MST phases we get the following simplified results:}
\bea &&\underline{{\bf SR}_1:}\quad\quad\sum^{6}_{i=1}\delta_{{\cal Z}_{{\bf D}_i}}=C(\mu=H_{s_1},\Lambda)=\left(1+\delta_{{\cal Z}_{C,{\bf SR}_1}}\right)\quad{\rm with}\quad\delta_{{\cal Z}_{{\bf D}_6}}=0,\\
&&\underline{{\bf USR}_n:}\quad\quad\delta_{{\cal Z}_{{\bf D}_6}}=C(\mu=H_{e_n},\Lambda)=\left(1+\delta_{{\cal Z}_{C,{\bf USR}_n}}\right)\quad {\rm with}\quad \sum^{6}_{i=1}\delta_{{\cal Z}_{{\bf D}_i}}=0,\\
&&\underline{{\bf SR}_{n+1}:}\quad\quad\sum^{6}_{i=1}\delta_{{\cal Z}_{{\bf D}_i}}=C(\mu=H_{\rm end},\Lambda)=\left(1+\delta_{{\cal Z}_{C,{\bf SR}_{n+1}}}\right)\quad{\rm with}\quad\delta_{{\cal Z}_{{\bf D}_6}}=0,\eea
\textcolor{black}{using which we get the following simplified relations:}
\bea &&\underline{{\bf SR}_1:}\quad\quad\left(\delta_{{\cal Z}_{{\bf D}_1}}+\delta_{{\cal Z}_{{\bf D}_2}}+\delta_{{\cal Z}_{{\bf D}_3}}+\delta_{{\cal Z}_{{\bf D}_4}}+\delta_{{\cal Z}_{{\bf D}_5}}-1\right)=\delta_{{\cal Z}_{C,{\bf SR}_1}},\\
&&\underline{{\bf USR}_n:}\quad\quad\left(\delta_{{\cal Z}_{{\bf D}_6}}-1\right)=\delta_{{\cal Z}_{C,{\bf USR}_n}},\\
&&\underline{{\bf SR}_{n+1}:}\quad\quad\left(\delta_{{\cal Z}_{{\bf D}_1}}+\delta_{{\cal Z}_{{\bf D}_2}}+\delta_{{\cal Z}_{{\bf D}_3}}+\delta_{{\cal Z}_{{\bf D}_4}}+\delta_{{\cal Z}_{{\bf D}_5}}-1\right)=\delta_{{\cal Z}_{C,{\bf SR}_{n+1}}}.\eea
\textcolor{black}{Here for the consecutive phases we have the following results:}
\bea \underline{{\bf SR}_1:}\quad\quad C(\mu=H_{s_1},\Lambda)&=&\left(1+\delta_{{\cal Z}_{C,{\bf SR}_1}}\right)\nonumber\\
&=&\left(\delta_{{\cal Z}_{{\bf D}_1}}+\delta_{{\cal Z}_{{\bf D}_2}}+\delta_{{\cal Z}_{{\bf D}_3}}+\delta_{{\cal Z}_{{\bf D}_4}}+\delta_{{\cal Z}_{{\bf D}_5}}\right)\nonumber\\
&=&-\Bigg\{A\ln{\bigg(\frac{\Lambda}{H_{s_1}}}\bigg) + \frac{B}{2}\bigg(\frac{\Lambda}{H_{s_1}}\bigg)^2\Bigg\}\nonumber\\
&&{\rm where}\quad\quad A=-\frac{4}{3}B=-\frac{4}{3}\left(1-\frac{2}{15\pi^{2}}\frac{1}{c_{s}^{2}k_{*}^{2}}\left(1-\frac{1}{c_{s}^{2}}\right)\epsilon\right)\bigg[\Delta^{2}_{\zeta,\textbf{Tree}}(k)\bigg]^2_{\textbf{SR}_{1}},\\
\underline{{\bf USR}_n:}\quad\quad C(\mu=H_{e_n},\Lambda)&=&\left(1+\delta_{{\cal Z}_{C,{\bf USR}_n}}\right)\nonumber\\
&=&\delta_{{\cal Z}_{{\bf D}_6}}\nonumber\\
&=&-\Bigg\{A\ln{\bigg(\frac{\Lambda}{H_{e_n}}}\bigg) + \frac{B}{2}\bigg(\frac{\Lambda}{H_{e_n}}\bigg)^2\Bigg\}\nonumber\\
&&{\rm where}\quad\quad A=B=\frac{1}{4}\bigg[\bigg(\frac{\Delta\eta(\tau_{e_{n}})}{\tilde{c}^{4}_{s}}\bigg)^{2}\ln{\bigg(\frac{k_{e_n}}{k_{s_n}}\bigg)^6} - \left(\frac{\Delta\eta(\tau_{s_{n}})}{\tilde{c}^{4}_{s}}\right)^{2}\bigg]\bigg[\Delta^{2}_{\zeta,\textbf{Tree}}(k)\bigg]^2_{\textbf{SR}_{1}},\quad\quad\quad\\
\underline{{\bf SR}_{n+1}:}\quad\quad C(\mu=H_{\rm end},\Lambda)&=&\left(1+\delta_{{\cal Z}_{C,{\bf SR}_{n+1}}}\right)\nonumber\\
&=&\left(\delta_{{\cal Z}_{{\bf D}_1}}+\delta_{{\cal Z}_{{\bf D}_2}}+\delta_{{\cal Z}_{{\bf D}_3}}+\delta_{{\cal Z}_{{\bf D}_4}}+\delta_{{\cal Z}_{{\bf D}_5}}\right)\nonumber\\
&=&-\Bigg\{A\ln{\bigg(\frac{\Lambda}{H_{\rm end}}}\bigg) + \frac{B}{2}\bigg(\frac{\Lambda}{H_{\rm end}}\bigg)^2\Bigg\}\nonumber\\
&&{\rm where}\quad\quad A=B=\left(1-\frac{2}{15\pi^{2}}\frac{1}{c_{s}^{2}k_{*}^{2}}\left(1-\frac{1}{c_{s}^{2}}\right)\epsilon\right)\bigg[\Delta^{2}_{\zeta,\textbf{Tree}}(k)\bigg]^2_{\textbf{SR}_{1}},\eea 
\textcolor{black}{As a result, we have the following expression for the UV divergence-free contribution of the two-point amplitude of the power spectrum:}
\bea
\left[\Delta_{\zeta,\textbf{EFT}}^{2}(k)\right]_n &=& {\cal Z}^{\rm UV}_{n}\times \bigg[\Delta^{2}_{\zeta,\textbf{Tree}}(k)\bigg]_{\textbf{SR}_{1}},\eea
\textcolor{black}{where the factor ${\cal Z}^{\rm UV}_{n}$ is given by for the $n$ th sharp transition by the following expression:}
\bea {\cal Z}^{\rm UV}_{n}&=&\left(1+\delta_{{\cal Z}^{\rm UV}_{n}}\right),\\
&=&\bigg\{ 1+ \bigg[\Delta^{2}_{\zeta,\textbf{Tree}}(k)\bigg]_{\textbf{SR}_{1}} \bigg [ \left(1-\frac{2}{15\pi^{2}}\frac{1}{c_{s}^{2}k_{*}^{2}}\left(1-\frac{1}{c_{s}^{2}}\right)\epsilon\right)\times \left(-\frac{4}{3}\ln{\frac{k_{s_1}}{k_{*}}}\right)\bigg] \nonumber \\
&& \quad \quad + \frac{1}{4} \bigg[\Delta^{2}_{\zeta,\textbf{Tree}}(k)\bigg]_{\textbf{SR}_{1}} \times \bigg(
     \bigg[\bigg(\frac{\Delta\eta(\tau_{e_{n}})}{\tilde{c}^{4}_{s}}\bigg)^{2}\ln{\bigg(\frac{k_{e_n}}{k_{s_n}}\bigg)^6} - \left(\frac{\Delta\eta(\tau_{s_{n}})}{\tilde{c}^{4}_{s}}\right)^{2}\bigg] \ln{\bigg(\frac{k_{e_n}}{k_{s_n}}\bigg)} \bigg)\nonumber \\
     && \quad \quad  +  
\bigg[\Delta^{2}_{\zeta,\textbf{Tree}}(k)\bigg]_{\textbf{SR}_{1}}\left(1-\frac{2}{15\pi^{2}}\frac{1}{c_{s}^{2}k_{*}^{2}}\left(1-\frac{1}{c_{s}^{2}}\right)\epsilon\right)
\times 
\bigg(\ln{\bigg(\frac{k_{s_{n+1}}}{k_{e_n}}}\bigg)\bigg)\bigg\}. \eea
\textcolor{black}{This identification not only helps us to understand the underlying connection between the counter-terms appearing in the bare action (UV sensitive part) and present contexts but also establishes the fact that the quadratic UV divergence can be completely removed in the expression for ${\cal Z}^{\rm UV}_{n}$. Since the connection is now established this identification will be further helpful to make a bridge between the present late-time scheme with the standard renormalization technique available within the framework of Quantum Field Theory of quasi de Sitter space. Most importantly, the late-time scheme helps to exactly fix the mathematical structure of the quantity ${\cal Z}^{\rm UV}_{n}$ in terms of which the total power spectrum for the scalar modes is now determined after renormalization. Only the problem is that by knowing the structure of ${\cal Z}^{\rm UV}_{n}$ one can compute the IR-counter term ${\cal Z}^{\rm IR}_{n}$ using the previously mentioned constraint appearing at the CMB pivot scale.}

\subsubsection{Adiabatic or Wave function renormalization scheme}

In principle, the quadratic UV divergences can be sold at a late time scale. However, we provide a systematic approach for the renormalization scheme, enabling the elimination of these UV divergences with the inclusion of a counter-term. We thereby steer our resolution boat towards the adiabatic renormalization which when applied helps to smoothen the large fluctuations at the USR and SR phases. To understand this idea in-depth, check refs.(\cite{Durrer:2009ii,Wang:2015zfa,PhysRevD.9.341,Finelli:2007fr,Marozzi:2011da,Boyanovsky:2005sh,PhysRevD.10.3905,PhysRevD.35.2955}). Adiabatic subtraction can lead to significant reduction in the weight of the UV modes. This procedure renormalizes the divergent quantities in curved spacetime by considering an adiabatic vacuum, and thereby subtracting the expectation value of such divergent quantities. We adopt a minimal subtraction approach that is the addition of a counter term in the underlying theory to expel the UV divergent contribution from the short-range UV moves that are pronounced in the subhorizon scale ($-k c_s\tau \gg 1)$. This is because the impact of adiabatic subtraction on the scalar power spectrum is less impactful after the horizon exit \cite{Durrer:2009ii}. This is why adiabatic renormalization can remove the UV contributions but not the IR divergences.

The regularization in adiabatic renormalization is built upon a WKB-like adiabatic expansion of the field modes, which when expanded up to $m$th order gives: 
\bea 
 \chi_{k}^{(m)} &\equiv& ah_{k}^{(m)} \sim \frac{1}{\sqrt{2 {\cal W}_{k}^{(m)}(\tau)}} \exp{\bigg(-i \int_{\tau_0}^{\tau}{\cal W}_{k}^{(m)}(\tau')d\tau \bigg)}, \\
 \chi_{k}^{(m)^{*}} &\equiv& ah_{k}^{(m)^{*}} \sim \frac{1}{\sqrt{2 {\cal W}_{k}^{(m)}(\tau)}} \exp{\bigg(i \int_{\tau_0}^{\tau}{\cal W}_{k}^{(m)}(\tau')d\tau \bigg)},
\eea
where $a$ is the scale factor and ${\cal W}_{k}^{m}$ is given by :
\bea
{\cal W}_{k}^{(m)} \equiv \omega_{k}^{(0)} + \omega_{k}^{(1)}+\omega_{k}^{(2)}+ \cdots + \omega_{k}^{(m)}.
\eea
The superscripts in $\omega_{k}$ denote the adiabatic order, which is the number of time derivatives it contains. More on the structure of the $\omega$'s can be found here \cite{Ferreiro:2023uvr}. An important thing to note here is that in the present context, the WKB approximation helps us to construct a regularized wave function over the adiabatic limit. This in turn plays a role in removing the UV divergent contributions from short-range modes. To this effect, we consider that the UV divergent contributions appear as mth power in the mode functions in the adiabatic limit. Further, it is crucial to note that the adiabatic renormalization scheme directly renormalizes the comoving curvature perturbation modes within the adiabatic limit of cosmological perturbations. Therefore, we can express the mode solutions for the SR$_1$, USR$_n$, and SR$_{n+1}$ phases as :
\bea
\label{genWKB}
\zeta_{\bf k, \rm SR_1}^{(m)}(\tau) &=& -\bigg[\frac{c_s H \tau}{2 M_p \sqrt{\epsilon} \sqrt{{\cal W}_{k}^{(m)}(\tau)}}\bigg] \bigg[\alpha_{\bf k}^{(1)} \exp{\bigg(-i \int_{\tau_{0}}^{\tau} d \tau^{'} {\cal W}_{k}^{(m)}(\tau ^{'})}\bigg) + \beta_{\bf k}^{(1)} \exp{\bigg(i \int_{\tau_{0}}^{\tau} d \tau^{'} {\cal W}_{k}^{(m)}(\tau ^{'})}\bigg)\bigg], \\
\zeta_{\bf k, \rm USR_{n}}^{(m)}(\tau) &=& -\bigg[\frac{c_s H \tau}{2 M_p \sqrt{\epsilon} \sqrt{{\cal W}_{k}^{(m)}(\tau)}}\bigg] \bigg(\frac{\tau_{0}}{\tau}\bigg)^{3} \bigg[\alpha_{\bf k}^{(2n)} \exp{\bigg(-i \int_{\tau_{0}}^{\tau}d \tau^{'} {\cal W}_{k}^{(m)}(\tau ^{'})}\bigg) \nonumber \\ 
 && \quad \quad \quad \quad \quad \quad \quad \quad \quad \quad \quad \quad \quad \quad \quad \quad \quad \quad + \beta_{\bf k}^{(2n)} \exp{\bigg(i \int_{\tau_{0}}^{\tau}d \tau^{'} {\cal W}_{k}^{(m)}(\tau ^{'})}\bigg)\bigg], \\
\zeta_{\bf k, \rm SR_{n+1}}^{(m)}(\tau) &=& -\bigg[\frac{c_s H \tau}{2 M_p \sqrt{\epsilon} \sqrt{{\cal W}_{k}^{(m)}(\tau)}}\bigg] \bigg(\frac{\tau_{0}}{\tau}\bigg)^{3} \bigg[\alpha_{\bf k}^{(2n+1)} \exp{\bigg(-i \int_{\tau_{0}}^{\tau}d \tau^{'} {\cal W}_{k}^{(m)}(\tau ^{'})}\bigg) \nonumber \\
&& \quad \quad \quad \quad \quad \quad \quad \quad \quad \quad \quad \quad \quad \quad \quad \quad \quad \quad + \beta_{\bf k}^{(2n+1)} \exp{\bigg(i \int_{\tau_{0}}^{\tau}d \tau^{'} {\cal W}_{k}^{(m)}(\tau ^{'})}\bigg)\bigg]. 
\eea
From the expressions above, $n$ represents the number of sharp transitions while m represents the order of WKB mode. Additionally, we have written the general $m$th order mode function for the SR$_1$ phase without fixing the initial Bunch Davies condition for the corresponding Bogoliubov coefficients. Here ${\cal W}_{k}^{(m)}$ denotes the characteristics function representative of the conformal time dependency factor within adiabatic regularization that is defined for the $m$th order as :
\bea
{\cal W}^{(m)}(\tau) = \sqrt{\bigg(c_s^2 k^2 - \frac{z^{''}}{z}\bigg) - \frac{1}{2}\bigg[\frac{{\cal W}_{k}^{(m-2)''}(\tau)}{{\cal W}_{k}^{m-2}(\tau)} - \frac{3}{2}\bigg(\frac{{\cal W}_{k}^{(m-2)''}(\tau)}{{\cal W}_{k}^{m-2}(\tau)}\bigg)^2\bigg]}, \quad \quad {\rm where} \quad \frac{z''}{z}\approx \frac{2}{\tau ^2}
\eea

It is to be noted that due to the adiabatic limit set in the scalar modes, there are no drastic modifications to the Bogoliubov coefficients in the USR$_n$ phases which is justified from the perspective of the validity of adiabatic regularization for the $m$th mode. Through meticulous computation, we will show that the final result is independent of the dynamics of the Bogoliubov coefficients, as it does not affect the short-range UV modes. Imagine a scenario where there are significant changes to the structure of the underlying vacuum state defined by the Bogoliubov coefficients when shifted from the initial Bunch Davies condition. It will directly challenge the utility of the adiabatic regularization approach which is already an established scheme. Here, one can appreciate the beauty of working in the Quantum field theory of Curved Quasi de-sitter space-time which maintains the form of the underlying shifted quantum vacuum state even in the adiabatic limit of regularization. Note that these discussions are valid for SR$_1$, USR$_n$, as well as SR$_{n+1}$ phases.
As mentioned earlier, the UV divergences manifest during the SR$_1$, USR$_n$, as well as SR$_{n+1}$ phases of inflation as quadratic terms. So, we would need a correction not more than order two to remove these divergences.
Hence we fix our discussion with $m=2$. With this above discussion, the WKB modes under the adiabatic limit can be expressed as :
\bea
\label{2orderWKB}
\zeta_{\bf k, \rm SR_1}^{(2)}(\tau) &=& -\bigg[\frac{c_s H \tau}{2 M_p \sqrt{\epsilon} \sqrt{{\cal W}_{k}^{(2)}(\tau)}}\bigg] \bigg[\alpha_{\bf k}^{(1)} \exp{\bigg(-i \int_{\tau_{0}}^{\tau} d \tau^{'} {\cal W}_{k}^{(2)}(\tau ^{'})}\bigg) + \beta_{\bf k}^{(1)} \exp{\bigg(i \int_{\tau_{0}}^{\tau} d \tau^{'} {\cal W}_{k}^{(2)}(\tau ^{'})}\bigg)\bigg], \\
\zeta_{\bf k, \rm USR_{n}}^{(2)}(\tau) &=& -\bigg[\frac{c_s H \tau}{2 M_p \sqrt{\epsilon} \sqrt{{\cal W}_{k}^{(2)}(\tau)}}\bigg] \bigg(\frac{\tau_{0}}{\tau}\bigg)^{3} \bigg[\alpha_{\bf k}^{(2n)} \exp{\bigg(-i \int_{\tau_{0}}^{\tau}d \tau^{'} {\cal W}_{k}^{(2)}(\tau ^{'})}\bigg) \nonumber \\ 
 && \quad \quad \quad \quad \quad \quad \quad \quad \quad \quad \quad \quad \quad \quad \quad \quad \quad \quad + \beta_{\bf k}^{(2n)} \exp{\bigg(i \int_{\tau_{0}}^{\tau}d \tau^{'} {\cal W}_{k}^{(2)}(\tau ^{'})}\bigg)\bigg], \\
\zeta_{\bf k, \rm SR_{n+1}}^{(2)}(\tau) &=& -\bigg[\frac{c_s H \tau}{2 M_p \sqrt{\epsilon} \sqrt{{\cal W}_{k}^{(2)}(\tau)}}\bigg] \bigg(\frac{\tau_{0}}{\tau}\bigg)^{3} \bigg[\alpha_{\bf k}^{(2n+1)} \exp{\bigg(-i \int_{\tau_{0}}^{\tau}d \tau^{'} {\cal W}_{k}^{(2)}(\tau ^{'})}\bigg) \nonumber \\
&& \quad \quad \quad \quad \quad \quad \quad \quad \quad \quad \quad \quad \quad \quad \quad \quad \quad \quad + \beta_{\bf k}^{(2n+1)} \exp{\bigg(i \int_{\tau_{0}}^{\tau}d \tau^{'} {\cal W}_{k}^{(2)}(\tau ^{'})}\bigg)\bigg]. 
\eea

Here the second-order characteristic function takes the form :
\bea
{\cal W}_{k}^{(2)} (\tau) = \sqrt{c_s ^2k^2 - \frac{2}{\tau ^2}} \approx c_s k.
\eea
The above expression uses the limit $-kc_s \tau \rightarrow  \infty$ to express the contribution from the short-range UV modes only. Therefore, with this result, the second-order WKB mode functions of the comoving curvature perturbation are further simplified to :
\bea
\label{simplewkb2}
\zeta_{\bf k, \rm SR_1}^{(2)}(\tau) &\approx & -\bigg[\frac{c_s H \tau}{2 M_p \sqrt{\epsilon} \sqrt{c_sk}}\bigg] \bigg[\alpha_{\bf k}^{(1)} \exp{\bigg(-ic_sk(\tau - \tau_{0})\bigg)} + \beta_{\bf k}^{(1)} \exp{\bigg(ic_sk(\tau - \tau_{0})\bigg)} \bigg]\\
\zeta_{\bf k, \rm USR_n}^{(2)}(\tau) &\approx & -\bigg[\frac{c_{s}H \tau}{2 M_p \sqrt{\epsilon} \sqrt{c_sk}}\bigg] \bigg(\frac{\tau_{0}}{\tau}\bigg)^{3} \bigg[\alpha_{\bf k}^{(2n)}\exp{\bigg(-ic_{s}k(\tau - \tau_{0})\bigg)} + \beta_{\bf k}^{(2n)} \exp{\bigg(ic_sk(\tau - \tau_{0})\bigg)} \bigg], \\
\zeta_{\bf k, \rm SR_{n+1}}^{(2)}(\tau) &\approx& -\bigg[\frac{c_s H \tau}{2 M_p \sqrt{\epsilon} \sqrt{c_sk}}\bigg] \bigg(\frac{\tau_{0}}{\tau}\bigg)^{3} 
\bigg[\alpha_{\bf k}^{(2n+1)} \exp{\bigg(-ic_sk(\tau - \tau_{0})\bigg)} + \beta_{\bf k}^{(2n+1)} \exp{\bigg(ic_sk(\tau - \tau_{0})\bigg)} \bigg], 
\eea
Now we move on to the main part of this scheme which is computing the expression for the counter terms of these periods of inflation. We introduce the following counter terms dependent on the adiabatic renormalization scheme-dependent parameters as seen before i.e. $c_{\rm SR_{1}}(\mu, \mu_0)$, and $c_{\rm USR_{n}}(\mu, \mu_0)$ and $c_{\rm SR_{n+1}}(\mu,\mu_0)$ for the respective periods of inflation SR$_{1}$, USR$_{n}$, and SR$_{n+1}$. 
\bea
\label{count1}
{\cal Z}_{\bf \zeta, \rm SR_{1}}^{\rm UV} (\mu, \mu_0) &=& \bigg[\Delta^{2}_{\zeta,\textbf{Tree}}(k)\bigg]_{\textbf{SR}_{1}}^{2} \times c_{\rm SR_{1}}(\mu, \mu_0), \\
\label{count2}
{\cal Z}_{\bf \zeta, \rm USR_{n}}^{\rm UV} (\mu, \mu_0) &=& \frac{1}{4}\bigg[\Delta^{2}_{\zeta,\textbf{Tree}}(k)\bigg]_{\textbf{SR}_{1}}^{2} \times c_{\rm USR_{n}}(\mu, \mu_0), \\
\label{count3}
{\cal Z}_{\bf \zeta, \rm SR_{n+1}}^{\rm UV} (\mu, \mu_0) &=& \bigg[\Delta^{2}_{\zeta,\textbf{Tree}}(k)\bigg]_{\textbf{SR}_{1}}^{2} \times c_{\rm SR_{n+1}}(\mu, \mu_0). 
\eea
Here $\mu$ represents the renormalization scale of the underlying quantum field theory in the quasi-de sitter background. Also, $\mu_0$ represents the renormalization scale at the conformal time taken as a reference to perform the adiabatic regularization scheme. These scales can be taken according to convenience as long as it is justified under the current context. Let us now evaluate the renormalization scheme-dependent parameters explicitly :
\bea
c_{\rm SR_{1}}(\mu, \mu_0) &=& \int_{\mu_0}^{\mu}dk\;k^2 c_s^2 \frac{\tau ^2}{k} = \frac{c_s^2}{2}(\mu ^2 - \mu_0 ^2 ) \tau ^2 = \frac{1}{2}\bigg[\bigg(\frac{\mu}{\mu_0}\bigg)^2 - 1 \bigg], \\
c_{\rm USR_{n}}(\mu, \mu_0) &=& \bigg[\left(\frac{\Delta\eta(\tau_{0})}{\tilde{c}^{4}_{s}}\right)^{2}\bigg(\frac{\mu}{\mu_0}\bigg)^6  - \left(\frac{\Delta\eta(\tau_{0})}{\tilde{c}^{4}_{s}}\right)^{2}\bigg] \int_{\mu_0}^{\mu} dk \; c_s ^2 k^2 \frac{\tau ^2}{k} \nonumber \\
&=& \frac{1}{2} \bigg[\left(\frac{\Delta\eta(\tau_{0})}{\tilde{c}^{4}_{s}}\right)^{2}\bigg(\frac{\mu}{\mu_0}\bigg)^6  - \left(\frac{\Delta\eta(\tau_{0})}{\tilde{c}^{4}_{s}}\right)^{2} \bigg]\bigg[\bigg(\frac{\mu}{\mu_0}\bigg)^2 - 1\bigg],\\
c_{\rm SR_{n+1}}(\mu, \mu_0) &=& \int_{\mu_0}^{\mu} dk \; k^2 c_s ^2 \frac{\tau ^2}{k} + C_1 = \frac{1}{2}\bigg[\bigg(\frac{\mu}{\mu_0}\bigg)^2 - 1 \bigg] + C_1.
\eea
The above integrals used the relation of $-c_s \tau = 1/\mu_0$. In addition, the term $C_1$ has been chosen such that to address the second term in eqn.(\ref{insrnres}). This term represents a highly suppressed contribution in the one-loop corrected scalar power spectrum. Therefore, going forward, we will drop this term as its exclusion will have an infinitesimally negligible impact on the resulting scalar power spectrum. Substituting these results obtained into the counter-terms of eqn.(\ref{count1}), eqn.(\ref{count2}), and (\ref{count3}), we obtain the following expressions : \vspace{-0.3em}
\bea
\label{revcount1}
{\cal Z}_{\bf \zeta, \rm SR_{1}}^{\rm UV} (\mu, \mu_0) &=& \bigg[\Delta^{2}_{\zeta,\textbf{Tree}}(k)\bigg]_{\textbf{SR}_{1}}^{2} \times \frac{1}{2}\bigg[\bigg(\frac{\mu}{\mu_0}\bigg)^2 - 1 \bigg], \\
\label{revcount2}
{\cal Z}_{\bf \zeta, \rm USR_{n}}^{\rm UV} (\mu, \mu_0) &=& \frac{1}{8}\bigg[\Delta^{2}_{\zeta,\textbf{Tree}}(k)\bigg]_{\textbf{SR}_{1}}^{2} \times \bigg[\left(\frac{\Delta\eta(\tau_{0})}{\tilde{c}^{4}_{s}}\right)^{2}\bigg(\frac{\mu}{\mu_0}\bigg)^6  - \left(\frac{\Delta\eta(\tau_{0})}{\tilde{c}^{4}_{s}}\right)^{2} \bigg]\bigg[\bigg(\frac{\mu}{\mu_0}\bigg)^2 - 1\bigg], \\
\label{revcount3}
{\cal Z}_{\bf \zeta, \rm SR_{n+1}}^{\rm UV} (\mu, \mu_0) &=& \bigg[\Delta^{2}_{\zeta,\textbf{Tree}}(k)\bigg]_{\textbf{SR}_{1}}^{2} \times \frac{1}{2}\bigg[\bigg\{\bigg(\frac{\mu}{\mu_0}\bigg)^2 - 1 \bigg\}  \bigg].
\eea        

Therefore, the total one-loop corrected adiabatically renormalized scalar power spectrum can be written as :

\bea
\label{adiasr1}
\bigg[\Delta^{2}_{\zeta,\textbf{One-loop}}(k,\mu,\mu_0)\bigg]_{\textbf{SR}_{1}} &=& \bigg[\Delta^{2}_{\zeta,\textbf{Tree}}(k)\bigg]_{\textbf{SR}_{1}}^{2}
\times \bigg[ \left(1-\frac{2}{15\pi^{2}}\frac{1}{c_{s}^{2}k_{*}^{2}}\left(1-\frac{1}{c_{s}^{2}}\right)\epsilon\right) \nonumber \\
&& \quad \quad \quad \quad \quad \bigg(\frac{1}{2}\bigg\{ \bigg(\frac{\mu}{\mu_0}\bigg)^2 - \bigg(\frac{k_{s_1}}{k_{*}}\bigg)^2 \bigg\} - \frac{4}{3} \ln\bigg({\frac{k_{s_1}}{k_{*}}}\bigg) \bigg)\bigg], \\
\label{adiausrn}
\bigg[\Delta^{2}_{\zeta,\textbf{One-loop}}(k,\mu,\mu_0)\bigg]_{\textbf{USR}_{n}} &=& \frac{1}{4}\bigg[\Delta^{2}_{\zeta,\textbf{Tree}}(k)\bigg]_{\textbf{SR}_{1}}^{2} \nonumber \\
&& \quad \times \bigg\{\bigg[\bigg(\frac{\Delta\eta(\tau_{e_{n}})}{\tilde{c}^{4}_{s}}\bigg)^{2}\bigg(\frac{k_{e_n}}{k_{s_n}}\bigg)^6  - \left(\frac{\Delta\eta(\tau_{s_{n}})}{\tilde{c}^{4}_{s}}\right)^{2}\bigg]
\bigg[\ln\bigg({\frac{k_{e_n}}{k_{s_n}}}\bigg) + \frac{1}{2}\bigg\{ \bigg(\frac{k_{e_n}}{k_{s_n}}\bigg)^2 - 1 \bigg\}\bigg] \nonumber \\
&& \quad \quad \quad  - \frac{1}{2}\bigg[\bigg(\frac{\Delta\eta(\tau_{e_{n}})}{\tilde{c}^{4}_{s}}\bigg)^{2} \bigg(\frac{\mu}{\mu_0}\bigg)^6 - \left(\frac{\Delta\eta(\tau_{s_{n}})}{\tilde{c}^{4}_{s}}\right)^{2}\bigg] \bigg[\bigg(\frac{\mu}{\mu_0}\bigg)^2 -1\bigg] \bigg\}, \\
\label{adisrnrest}
\bigg[\Delta^{2}_{\zeta,\textbf{One-loop}}(k,\mu,\mu_0)\bigg]_{\textbf{SR}_{n+1}} &=& \bigg[\Delta^{2}_{\zeta,\textbf{Tree}}(k)\bigg]_{\textbf{SR}_{1}}^{2} \times \bigg[ \left(1-\frac{2}{15\pi^{2}}\frac{1}{c_{s}^{2}k_{*}^{2}}\left(1-\frac{1}{c_{s}^{2}}\right)\epsilon\right) \bigg(\frac{1}{2}\bigg[\bigg(\frac{\mu}{\mu_0}\bigg)^2 - 1 \bigg] \nonumber \\ 
&& \quad \quad \quad \quad \quad  + \bigg(\ln{\bigg(\frac{k_{s_{n+1}}}{k_{e_n}}}\bigg)\bigg)\bigg) \bigg].
\eea

As you can see from the above expressions for the scalar power spectrum of the comoving curvature perturbation, the UV divergences will be removed upon settling for some fixed renormalization scale fixed by $\mu$, and corresponding reference scale $\mu_0$. However, the fate of the IR divergences remains to be questioned. Consequently, it results in an IR-sensitive but UV-protected quantum field theory of curved space-time. This sort of IR nature is conclusive evidence of the validation of perturbative approximations, which must always be obeyed. You can always assume some arbitrariness in the renormalization scale but the perturbativity is maintained at all points of time if the scale is fixed in the neighborhood of the UV cut-off. Thus, we would like to present the renormalization scales taken for our purpose as well as remind you of the cut-off scales initially chosen to evaluate the integrals. 

\begin{enumerate}
    \item \textbf{\underline{Catalogue of Cut-off scales:}} \\
    \begin{align*}
\mathbf{\underline{For\;SR_1}:} & \quad \Lambda_{\rm UV} = k_{s_1}, \quad  \Lambda_{\rm IR} = k_{*}, \\
\mathbf{\underline{For\;USR_n}:} & \quad \Lambda_{\rm UV} = k_{e_n}, \quad  \Lambda_{\rm IR} = k_{s_n}, \\
\mathbf{\underline{For\;SR_{n+1}}:} & \quad \Lambda_{\rm UV} = k_{s_{n+1}}, \quad \Lambda_{\rm IR} = k_{e_n}.
\end{align*}

\item \textbf{\underline{Catalogue of Renormalization scales and necessary constraints:}} \\
\begin{align*}
\mathbf{\underline{For\;SR_1}:} & \quad \mu = k_{s_1} = \Lambda_{\rm UV}^{\rm SR_1}, \quad  \mu_0 = k_{*} = \Lambda_{\rm IR}^{\rm SR_1}, \\
\mathbf{\underline{For\;USR_n}:} & \quad \mu = k_{e_n} =  \Lambda_{\rm UV}^{\rm USR_n}, \quad  \mu_0 = k_{s_n} =  \Lambda_{\rm IR}^{\rm USR_n}, \\
\mathbf{\underline{For\;SR_{n+1}}:} & \quad \mu = \mu_0 = k_{\rm end}.
\end{align*}
\end{enumerate}

After applying the above renormalization scales, the expressions of eqn.(\ref{adiasr1}), (\ref{adiausrn}), and (\ref{adisrnrest}) respectively take the form : 
\bea
\bigg[\Delta^{2}_{\zeta,\textbf{One-loop}}(k)\bigg]_{\textbf{SR}_{1}} &=& \bigg[\Delta^{2}_{\zeta,\textbf{Tree}}(k)\bigg]_{\textbf{SR}_{1}}^{2}
\times \bigg[ \left(1-\frac{2}{15\pi^{2}}\frac{1}{c_{s}^{2}k_{*}^{2}}\left(1-\frac{1}{c_{s}^{2}}\right)\epsilon\right) \bigg( - \frac{4}{3} \ln\bigg({\frac{k_{s_1}}{k_{*}}}\bigg) \bigg)\bigg], \\
\bigg[\Delta^{2}_{\zeta,\textbf{One-loop}}(k)\bigg]_{\textbf{USR}_{n}} &=& \frac{1}{4}\bigg[\Delta^{2}_{\zeta,\textbf{Tree}}(k)\bigg]_{\textbf{SR}_{1}}^{2} \nonumber \\
&& \quad \times \bigg\{\bigg[\bigg(\frac{\Delta\eta(\tau_{e_{n}})}{\tilde{c}^{4}_{s}}\bigg)^{2}\bigg(\frac{k_{e_n}}{k_{s_n}}\bigg)^6  - \left(\frac{\Delta\eta(\tau_{s_{n}})}{\tilde{c}^{4}_{s}}\right)^{2}\bigg]
\bigg[\ln\bigg({\frac{k_{e_n}}{k_{s_n}}}\bigg) + \frac{1}{2}\bigg\{ \bigg(\frac{k_{e_n}}{k_{s_n}}\bigg)^2 - 1 \bigg\}\bigg] \nonumber \\
&& \quad \quad \quad  - \frac{1}{2}\bigg[\bigg(\frac{\Delta\eta(\tau_{e_{n}})}{\tilde{c}^{4}_{s}}\bigg)^{2} \bigg(\frac{k_{e_n}}{k_{s_n}}\bigg)^6 - \left(\frac{\Delta\eta(\tau_{s_{n}})}{\tilde{c}^{4}_{s}}\right)^{2}\bigg] \bigg[\bigg(\frac{k_{e_n}}{k_{s_n}}\bigg)^2 -1\bigg] \bigg\} \nonumber \eea\bea
&=& \frac{1}{4} \bigg[\Delta^{2}_{\zeta,\textbf{Tree}}(k)\bigg]_{\textbf{SR}_{1}}^{2} \bigg\{\bigg[\bigg(\frac{\Delta\eta(\tau_{e_{n}})}{\tilde{c}^{4}_{s}}\bigg)^{2}\bigg(\frac{k_{e_n}}{k_{s_n}}\bigg)^6  - \left(\frac{\Delta\eta(\tau_{s_{n}})}{\tilde{c}^{4}_{s}}\right)^{2}\bigg] \bigg(\ln{\frac{k_{e_n}}{k_{s_n}}}\bigg)  \nonumber \\
&& + \frac{1}{2} \bigg[\bigg(\frac{\Delta\eta(\tau_{e_{n}})}{\tilde{c}^{4}_{s}}\bigg)^{2}\bigg(\frac{k_{e_n}}{k_{s_n}}\bigg)^6  - \left(\frac{\Delta\eta(\tau_{s_{n}})}{\tilde{c}^{4}_{s}}\right)^{2}\bigg] \nonumber \\ && \quad \quad \quad  - \frac{1}{2}\bigg[\bigg(\frac{\Delta\eta(\tau_{e_{n}})}{\tilde{c}^{4}_{s}}\bigg)^{2}\bigg(\frac{k_{e_n}}{k_{s_n}}\bigg)^6  - \left(\frac{\Delta\eta(\tau_{s_{n}})}{\tilde{c}^{4}_{s}}\right)^{2}\bigg] \bigg\}\nonumber \\
&=& \frac{1}{4}\bigg[\Delta^{2}_{\zeta,\textbf{Tree}}(k)\bigg]_{\textbf{SR}_{1}}^{2} \bigg[\bigg(\frac{\Delta\eta(\tau_{e_{n}})}{\tilde{c}^{4}_{s}}\bigg)^{2}\bigg(\frac{k_{e_n}}{k_{s_n}}\bigg)^6  - \left(\frac{\Delta\eta(\tau_{s_{n}})}{\tilde{c}^{4}_{s}}\right)^{2}\bigg] \bigg(\ln{\frac{k_{e_n}}{k_{s_n}}}\bigg), \eea\bea
\bigg[\Delta^{2}_{\zeta,\textbf{One-loop}}(k)\bigg]_{\textbf{SR}_{n+1}} &=& \bigg[\Delta^{2}_{\zeta,\textbf{Tree}}(k)\bigg]_{\textbf{SR}_{1}}^{2} \times \bigg[ \left(1-\frac{2}{15\pi^{2}}\frac{1}{c_{s}^{2}k_{*}^{2}}\left(1-\frac{1}{c_{s}^{2}}\right)\epsilon\right) \bigg(\frac{1}{2}[1^2 - 1 ] \nonumber \\ 
&& \quad \quad \quad \quad \quad  + \ln{\bigg(\frac{k_{s_{n+1}}}{k_{e_n}}}\bigg)\bigg) \bigg] \nonumber \\
&=& \bigg[\Delta^{2}_{\zeta,\textbf{Tree}}(k)\bigg]_{\textbf{SR}_{1}}^{2} \times \bigg[ \left(1-\frac{2}{15\pi^{2}}\frac{1}{c_{s}^{2}k_{*}^{2}}\left(1-\frac{1}{c_{s}^{2}}\right)\epsilon\right) \bigg( \ln{\bigg(\frac{k_{s_{n+1}}}{k_{e_n}}}\bigg)\bigg) \bigg].
\eea
Thus the total one-loop corrected adiabatic renormalized power spectrum under these renormalization scales converges to :

\bea
\label{adiapowspec}
\Delta^{2}_{\zeta,\textbf{Total}}(k) &=& \bigg[\Delta^{2}_{\zeta,\textbf{Tree}}(k)\bigg]_{\textbf{SR}_{1}}
\; \bigg\{ 1+ \bigg[\Delta^{2}_{\zeta,\textbf{Tree}}(k)\bigg]_{\textbf{SR}_{1}} \bigg [ \left(1-\frac{2}{15\pi^{2}}\frac{1}{c_{s}^{2}k_{*}^{2}}\left(1-\frac{1}{c_{s}^{2}}\right)\epsilon\right)\times \left(-\frac{4}{3}\ln{\frac{k_{s_1}}{k_{*}}}\right)\bigg] \nonumber \\
&& \quad \quad + \frac{1}{4} \bigg[\Delta^{2}_{\zeta,\textbf{Tree}}(k)\bigg]_{\textbf{SR}_{1}} \times \sum^{N}_{n=1}\bigg(
     \bigg[\bigg(\frac{\Delta\eta(\tau_{e_{n}})}{\tilde{c}^{4}_{s}}\bigg)^{2}\ln{\bigg(\frac{k_{e_n}}{k_{s_n}}\bigg)^6} - \left(\frac{\Delta\eta(\tau_{s_{n}})}{\tilde{c}^{4}_{s}}\right)^{2}\bigg] \ln{\bigg(\frac{k_{e_n}}{k_{s_n}}\bigg)} \bigg)\nonumber \\
     && \quad \quad \quad +  
\bigg[\Delta^{2}_{\zeta,\textbf{Tree}}(k)\bigg]_{\textbf{SR}_{1}}\left(1-\frac{2}{15\pi^{2}}\frac{1}{c_{s}^{2}k_{*}^{2}}\left(1-\frac{1}{c_{s}^{2}}\right)\epsilon\right)
\times \sum^{N}_{n=1} 
\bigg(\ln{\bigg(\frac{k_{s_{n+1}}}{k_{e_n}}}\bigg)\bigg)\bigg\}.
\eea

As you can already appreciate from the above results, the quadratic UV divergences have been removed and the logarithmic IR divergences sustained the procedure which we are going to smoothen in our next subsection which discusses the power spectrum renormalization. The common factor that comes out of the whole expression for the total power spectrum somewhat dictates the order of its amplitude. This is the tree-level contribution of SR$_1$ given as :
\bea
 \bigg[\Delta^{2}_{\zeta,{\bf Tree}}(k)\bigg]_{\textbf{SR}_{1}} = \displaystyle{ \bigg[\frac{H^2}{8\pi ^2 M_{p}^2 \epsilon c_s}\bigg]_{*}}\left[1+(k/k_{s_{1}})^{2}\right] \xrightarrow{\rm Super-horizon \; scale \; k\ll k_{s_1}} \displaystyle{ \bigg[\frac{H^2}{8\pi ^2 M_{p}^2 \epsilon c_s}\bigg]_{*}}.
\eea
Now carefully notice that the final result of the total one loop corrected renormalized power spectrum obtained from the adiabatic renormalization in eqn.(\ref{adiapowspec}) is exactly the same as the final result obtained from the late-time renormalization scheme as in eqn.(\ref{late-tpowspec}). This suggests the theory should be renormalization scheme dependent. We will speculate on this matter and some of the other renormalization schemes in the discussions to follow.  However, before proceeding further, we have to design certain quantities for convenience and consistency.
\bea
\label{Utot}
U&=& U_{\textbf{SR}_{1}} + U_{\textbf{SR}_{\textbf{rest}}},\\
\label{Usr1}
U_{\textbf{SR}_{1}} &=&  -\frac{4}{3}\bigg[\Delta^{2}_{\zeta,\textbf{Tree}}(k)\bigg]_{\textbf{SR}_{1}}^{2}\times\left(1-\frac{2}{15\pi^{2}}\frac{1}{c_{s}^{2}k_{*}^{2}}\left(1-\frac{1}{c_{s}^{2}}\right)\epsilon\right)\ln\left(\frac{k_{s_1}}{k_{*}}\right),
\\
\label{Usrrest}
U_{\textbf{SR}_{\textbf{rest}}} &=& \sum^{N}_{n=1}U_{n} = \bigg[\Delta^{2}_{\zeta,\textbf{Tree}}(k)\bigg]_{\textbf{SR}_{1}}^{2}\times\sum^{N}_{n=1}\left(1-\frac{2}{15\pi^{2}}\frac{1}{c_{s}^{2}k_{*}^{2}}\left(1-\frac{1}{c_{s}^{2}}\right)\epsilon\right)\ln\left(\frac{k_{s_{n+1}}}{k_{e_{n}}}\right),
\\
\label{V}
V &=& \sum^{N}_{n=1}V_{n} = \frac{1}{4}\bigg[\Delta^{2}_{\zeta,\textbf{Tree}}(k)\bigg]_{\textbf{SR}_{1}}^{2}\times\sum^{N}_{n=1}\bigg(\frac{\left(\Delta\eta(\tau_{e_{n}})\right)^{2}}{\tilde{c}^{8}_{s}}\left(\frac{k_{e_{n}}}{k_{s_{n}}}\right)^{6} - \frac{\left(\Delta\eta(\tau_{s_{n}})\right)^{2}}{\tilde{c}^{8}_{s}}\bigg )\ln\left(\frac{k_{e_{n}}}{k_{s_{n}}}\right).
\eea

Here $U$ represents the total loop contributions from SR$_1$ and SR$_{n+1}$ periods of inflation, while $V$ represents the total loop contributions from the USR$_{n}$ periods of inflation. We have written these above expressions for any general $n$ number of sharp transitions which for our case we have considered as $n=6$.  Therefore, now the total one-loop corrected scalar power spectrum can be re-expressed as :

\bea
 \label{unrenormps}
\Delta^{2}_{\zeta, {\textbf{EFT}}}(k) = \bigg[\Delta^{2}_{\zeta, {\textbf{Tree}}}(k)\bigg]_{\textbf{SR}_{1}}\bigg\{1+U+V\bigg\}.
\eea

Given below is a diagrammatic representation of the contributions to the total one-loop corrected scalar power spectrum from eqn.(\ref{unrenormps}):
\begin{equation}
\begin{tikzpicture}[baseline={([yshift=-3.5ex]current bounding box.center)},very thick]
  
  \def\radius{1}
  \scalebox{1}{\draw[cyan,very thick] (0,\radius) circle (\radius);
  \draw[cyan,very thick] (4.5*\radius,0) circle (\radius);}

 \draw[black, very thick] (-4*\radius,0) -- (-2.5*\radius,0);
 \node at (-1.7*\radius,0){+};
  \draw[black, very thick] (-1*\radius,0) -- (0,0);
  \draw[blue,fill=blue] (0,0) circle (.5ex);
  \draw[black, very thick] (0,0)  -- (1*\radius,0);
  \node at (2*\radius,0) {+};
  \draw[black, very thick] (2.5*\radius,0) -- (3.5*\radius,0); 
  \draw[blue,fill=blue] (3.5*\radius,0) circle (.5ex);
  \draw[blue,fill=blue] (5.5*\radius,0) circle (.5ex);
  \draw[black, very thick] (5.5*\radius,0) -- (6.5*\radius,0);
  
\end{tikzpicture}  \quad = \quad {\rm Adiabatically \,\,renormalized} \,\,\Delta^{2}_{\zeta, \textbf{EFT}}(k),
\end{equation}
where the diagrams represent the only one-loop contributions to the quantity $\Delta^{2}_{\zeta, \textbf{EFT}}(k)$ obtained after the adiabatic renormalization scheme.

Therefore, if you compare the loop diagrams for both adiabatic and the late-time renormalization, you get the exact same results as depicted below :
\begin{equation}
\begin{tikzpicture}[baseline={([yshift=-3.5ex]current bounding box.center)},very thick]
  
  \def\radius{1}
  \scalebox{1}{\draw[red,very thick] (-0.6*\radius,0.75*\radius) circle (0.75*\radius);
  \draw[red,very thick] (2.45*\radius,0) circle (0.75*\radius);\draw[cyan,very thick](8.4*\radius,0.75*\radius) circle(0.75*\radius);\draw[cyan,very thick](11.55*\radius,0) circle(0.75*\radius);}

   \draw[black, very thick] (-3*\radius,0) -- (-2.2*\radius,0);
   \node at (-1.8*\radius,0) {+};
  \draw[black, very thick] (-1.4*\radius,0) -- (-0.6*\radius,0);
  \draw[blue,fill=blue] (-0.6*\radius,0) circle (.5ex);
  \draw[black, very thick] (-0.6*\radius,0)  -- (0.2*\radius,0);
  \node at (0.6*\radius,0) {+};
  \draw[black, very thick] (0.9*\radius,0) -- (1.7*\radius,0); 
  \draw[blue,fill=blue] (1.7*\radius,0) circle (.5ex);
  \draw[blue,fill=blue] (3.2*\radius,0) circle (.5ex);
  \draw[black, very thick] (3.2*\radius,0) -- (4*\radius,0);
  \node at (5*\radius,0) {=};
   \draw[black, very thick] (6*\radius,0) -- (6.8*\radius,0);
   \node at (7.2*\radius,0) {+};
   \draw[black, very thick] (7.6*\radius,0) -- (8.4*\radius,0);
  \draw[blue,fill=blue] (8.4*\radius,0) circle (.5ex);
  \draw[black, very thick] (8.4*\radius,0)  -- (9.2*\radius,0);
  \node at (9.6*\radius,0) {+};
  \draw[black, very thick] (10*\radius,0) -- (10.8*\radius,0); 
  \draw[blue,fill=blue] (10.8*\radius,0) circle (.5ex);
  \draw[blue,fill=blue] (12.3*\radius,0) circle (.5ex);
  \draw[black, very thick] (12.3*\radius,0) -- (13.1*\radius,0);
   
\end{tikzpicture}  
\end{equation}

\textcolor{black}{In the present computation, we have further found the following important facts regarding the connection between the counter-terms appearing in the present context with the counter-terms appearing in the context of renormalization of the bare action as appearing in the context of Quantum Field Theory in quasi de Sitter space with the gauge invariant coming curvature perturbation, which is given by :}
\bea &&\underline{{\bf SR}_1:}\quad\quad\sum^{6}_{i=1}\delta_{{\cal Z}_{{\bf D}_i}}={\cal Z}_{\bf \zeta, \rm SR_{1}}^{\rm UV}=\left(1+\delta_{{\cal Z}_{\bf \zeta, \rm SR_{1}}^{\rm UV}}\right)\quad{\rm with}\quad\delta_{{\cal Z}_{{\bf D}_6}}=0,\\
&&\underline{{\bf USR}_n:}\quad\quad\delta_{{\cal Z}_{{\bf D}_6}}={\cal Z}_{\bf \zeta, \rm USR_{n}}^{\rm UV}=\left(1+\delta_{{\cal Z}_{\bf \zeta, \rm USR_{n}}^{\rm UV}}\right)\quad {\rm with}\quad \sum^{6}_{i=1}\delta_{{\cal Z}_{{\bf D}_i}}=0,\\
&&\underline{{\bf SR}_{n+1}:}\quad\quad\sum^{6}_{i=1}\delta_{{\cal Z}_{{\bf D}_i}}={\cal Z}_{\bf \zeta, \rm SR_{n+1}}^{\rm UV}=\left(1+\delta_{{\cal Z}_{\bf \zeta, \rm SR_{n+1}}^{\rm UV}}\right)\quad{\rm with}\quad\delta_{{\cal Z}_{{\bf D}_6}}=0,\eea
\textcolor{black}{using which we get the following simplified relations:}
\bea &&\underline{{\bf SR}_1:}\quad\quad\left(\delta_{{\cal Z}_{{\bf D}_1}}+\delta_{{\cal Z}_{{\bf D}_2}}+\delta_{{\cal Z}_{{\bf D}_3}}+\delta_{{\cal Z}_{{\bf D}_4}}+\delta_{{\cal Z}_{{\bf D}_5}}-1\right)=\delta_{{\cal Z}_{\bf \zeta, \rm SR_{1}}^{\rm UV}},\\
&&\underline{{\bf USR}_n:}\quad\quad\left(\delta_{{\cal Z}_{{\bf D}_6}}-1\right)=\delta_{{\cal Z}_{\bf \zeta, \rm USR_{n}}^{\rm UV}},\\
&&\underline{{\bf SR}_{n+1}:}\quad\quad\left(\delta_{{\cal Z}_{{\bf D}_1}}+\delta_{{\cal Z}_{{\bf D}_2}}+\delta_{{\cal Z}_{{\bf D}_3}}+\delta_{{\cal Z}_{{\bf D}_4}}+\delta_{{\cal Z}_{{\bf D}_5}}-1\right)=\delta_{{\cal Z}_{\bf \zeta, \rm SR_{n+1}}^{\rm UV}}.\eea
\textcolor{black}{Here for the consecutive phases we have the following results:}
\bea \underline{{\bf SR}_1:}\quad\quad {\cal Z}_{\bf \zeta, \rm SR_{1}}^{\rm UV}&=&\left(1+\delta_{{\cal Z}_{\bf \zeta, \rm SR_{1}}^{\rm UV}}\right)\nonumber\\
&=&\left(\delta_{{\cal Z}_{{\bf D}_1}}+\delta_{{\cal Z}_{{\bf D}_2}}+\delta_{{\cal Z}_{{\bf D}_3}}+\delta_{{\cal Z}_{{\bf D}_4}}+\delta_{{\cal Z}_{{\bf D}_5}}\right)\nonumber\\
&=&\left(1-\frac{2}{15\pi^{2}}\frac{1}{c_{s}^{2}k_{*}^{2}}\left(1-\frac{1}{c_{s}^{2}}\right)\epsilon\right)\bigg[\Delta^{2}_{\zeta,\textbf{Tree}}(k)\bigg]^2_{\textbf{SR}_{1}}\nonumber\\
&&\quad\quad\quad\quad\quad\quad\quad\quad\quad\quad\quad\quad\times\Bigg\{\frac{1}{2}\bigg[\bigg(\frac{\mu}{\mu_0}\bigg)^2 - 1 \bigg]\Bigg\}_{\mu=k_{s_1},\mu_0=k_*}
\nonumber\\
&=&\left(1-\frac{2}{15\pi^{2}}\frac{1}{c_{s}^{2}k_{*}^{2}}\left(1-\frac{1}{c_{s}^{2}}\right)\epsilon\right)\bigg[\Delta^{2}_{\zeta,\textbf{Tree}}(k)\bigg]^2_{\textbf{SR}_{1}}\nonumber\\
&&\quad\quad\quad\quad\quad\quad\quad\quad\quad\quad\quad\quad\times\Bigg\{\frac{1}{2}\bigg[\bigg(\frac{k_{s_1}}{k_*}\bigg)^2 - 1 \bigg]\Bigg\},\\
\underline{{\bf USR}_n:}\quad\quad {\cal Z}_{\bf \zeta, \rm USR_{n}}^{\rm UV}&=&\left(1+\delta_{{\cal Z}_{\bf \zeta, \rm USR_{n}}^{\rm UV}}\right)\nonumber\\
&=&\delta_{{\cal Z}_{{\bf D}_6}}\nonumber\\
&=&\frac{1}{4}\bigg[\bigg(\frac{\Delta\eta(\tau_{e_{n}})}{\tilde{c}^{4}_{s}}\bigg)^{2}\ln{\bigg(\frac{k_{e_n}}{k_{s_n}}\bigg)^6} - \left(\frac{\Delta\eta(\tau_{s_{n}})}{\tilde{c}^{4}_{s}}\right)^{2}\bigg]\bigg[\Delta^{2}_{\zeta,\textbf{Tree}}(k)\bigg]^2_{\textbf{SR}_{1}}\nonumber\\
&&\quad\quad\quad\quad\quad\quad\quad\quad\quad\quad\quad\quad\times\bigg\{
\frac{1}{2}\bigg[\bigg(\frac{\mu}{\mu_0}\bigg)^2 - 1 \bigg]\Bigg\}_{\mu=k_{e_n},\mu_0=k_{s_n}}\nonumber\\
&=&\frac{1}{4}\bigg[\bigg(\frac{\Delta\eta(\tau_{e_{n}})}{\tilde{c}^{4}_{s}}\bigg)^{2}\ln{\bigg(\frac{k_{e_n}}{k_{s_n}}\bigg)^6} - \left(\frac{\Delta\eta(\tau_{s_{n}})}{\tilde{c}^{4}_{s}}\right)^{2}\bigg]\bigg[\Delta^{2}_{\zeta,\textbf{Tree}}(k)\bigg]^2_{\textbf{SR}_{1}}\nonumber\\
&&\quad\quad\quad\quad\quad\quad\quad\quad\quad\quad\quad\quad\times\bigg\{
\frac{1}{2}\bigg[\bigg(\frac{k_{e_n}}{k_{s_n}}\bigg)^2 - 1 \bigg]\Bigg\},\quad\quad\quad\eea\bea
\underline{{\bf SR}_{n+1}:}\quad\quad {\cal Z}_{\bf \zeta, \rm SR_{n+1}}^{\rm UV}&=&\left(1+\delta_{{\cal Z}_{\bf \zeta, \rm SR_{n+1}}^{\rm UV}}\right)\nonumber\\
&=&\left(\delta_{{\cal Z}_{{\bf D}_1}}+\delta_{{\cal Z}_{{\bf D}_2}}+\delta_{{\cal Z}_{{\bf D}_3}}+\delta_{{\cal Z}_{{\bf D}_4}}+\delta_{{\cal Z}_{{\bf D}_5}}\right)\nonumber\\
&=&\left(1-\frac{2}{15\pi^{2}}\frac{1}{c_{s}^{2}k_{*}^{2}}\left(1-\frac{1}{c_{s}^{2}}\right)\epsilon\right)\bigg[\Delta^{2}_{\zeta,\textbf{Tree}}(k)\bigg]^2_{\textbf{SR}_{1}}\nonumber\\
&&\quad\quad\quad\quad\quad\quad\quad\quad\quad\quad\quad\quad\times\Bigg\{\frac{1}{2}\bigg[\bigg(\frac{\mu}{\mu_0}\bigg)^2 - 1 \bigg]\Bigg\}_{\mu=\mu_0=k_{\rm end}}\nonumber\\
&=&0,\eea 
\textcolor{black}{As a result, we have the following expression for the UV divergence-free contribution of the two-point amplitude of the power spectrum: }
\bea
\left[\Delta_{\zeta,\textbf{EFT}}^{2}(k)\right]_n &=& {\cal Z}^{\rm UV}_{n}\times \bigg[\Delta^{2}_{\zeta,\textbf{Tree}}(k)\bigg]_{\textbf{SR}_{1}},\eea
\textcolor{black}{where the factor ${\cal Z}^{\rm UV}_{n}$ is given by for the $n$ th sharp transition by the following expression:}
\bea {\cal Z}^{\rm UV}_{n}&=&\left(1+\delta_{{\cal Z}^{\rm UV}_{n}}\right),\\
&=&\bigg\{ 1+ \bigg[\Delta^{2}_{\zeta,\textbf{Tree}}(k)\bigg]_{\textbf{SR}_{1}} \bigg [ \left(1-\frac{2}{15\pi^{2}}\frac{1}{c_{s}^{2}k_{*}^{2}}\left(1-\frac{1}{c_{s}^{2}}\right)\epsilon\right)\times \left(-\frac{4}{3}\ln{\frac{k_{s_1}}{k_{*}}}\right)\bigg] \nonumber \\
&& \quad \quad + \frac{1}{4} \bigg[\Delta^{2}_{\zeta,\textbf{Tree}}(k)\bigg]_{\textbf{SR}_{1}} \times \bigg(
     \bigg[\bigg(\frac{\Delta\eta(\tau_{e_{n}})}{\tilde{c}^{4}_{s}}\bigg)^{2}\ln{\bigg(\frac{k_{e_n}}{k_{s_n}}\bigg)^6} - \left(\frac{\Delta\eta(\tau_{s_{n}})}{\tilde{c}^{4}_{s}}\right)^{2}\bigg] \ln{\bigg(\frac{k_{e_n}}{k_{s_n}}\bigg)} \bigg)\nonumber \\
     && \quad \quad  +  
\bigg[\Delta^{2}_{\zeta,\textbf{Tree}}(k)\bigg]_{\textbf{SR}_{1}}\left(1-\frac{2}{15\pi^{2}}\frac{1}{c_{s}^{2}k_{*}^{2}}\left(1-\frac{1}{c_{s}^{2}}\right)\epsilon\right)
\times 
\bigg(\ln{\bigg(\frac{k_{s_{n+1}}}{k_{e_n}}}\bigg)\bigg)\bigg\}. \eea
\textcolor{black}{This identification not only helps us to understand the underlying connection between the counter-terms appearing in the bare action (UV sensitive part) and present contexts but also establishes the fact that the quadratic UV divergence can be completely removed in the expression for ${\cal Z}^{\rm UV}_{n}$. Since the connection is now established this identification will be further helpful to make a bridge between the present adiabatic/wave function scheme with the standard renormalization technique available within the framework of Quantum Field Theory of quasi de Sitter space. Most importantly, the adiabatic/wave function scheme helps to exactly fix the mathematical structure of the quantity ${\cal Z}^{\rm UV}_{n}$ in terms of which the total power spectrum for the scalar modes is now determined after renormalization. Only the problem is that by knowing the structure of ${\cal Z}^{\rm UV}_{n}$ one can compute the IR-counter term ${\cal Z}^{\rm IR}_{n}$ using the previously mentioned constraint appearing at the CMB pivot scale.}

\textcolor{black}{Also, we have found from our computations that late-time and adiabatic renormalization schemes give rise to the unique structure of the term ${\cal Z}^{\rm UV}_{n}$ after removal of the quadratic divergence contribution. For this reason, the final result turns out to be scheme-independent and we have now the renormalized one-loop spectrum where the one-loop effect is described by the logarithmic IR divergent contributions. Now since the structure of the the term ${\cal Z}^{\rm UV}_{n}$ is uniquely determined, one can use the previously derived constraint at the CMB pivot scale to further determine the IR-counter term ${\cal Z}^{\rm IR}_{n}$. With the help of the computation performed in the next section, i.e. implementing the power spectrum renormalization we are going to fix the precise structure of the IR counter-term. The results derived in the next section are actually the continuation of the undetermined results that we have computed with the help of standard techniques of renormalization as appearing in the context of Quantum Field Theory.}

\subsubsection{Power spectrum renormalization scheme} 

Hereafter, we employ the power spectrum renormalization scheme, which ultimately saves us from logarithmic IR divergences by suppressing them and making them more controllable. This scheme uses a counter-term, determined by invoking a renormalization condition at the pivot scale $k_{*}$, to give us the renormalized version of the scalar power spectrum. The procedure gets performed for each interval involving a sharp transition, which in the present context implies $6$ times. The final step involves the implementation of the resummation procedure to obtain a finite physical output. Here we will present the power spectrum renormalization as done in the condition with an arbitrary $N$ number of sharp transitions. Let us now dive into this very procedure by rescaling the tree-level contribution to the power spectrum as :
\bea \label{rescale}
\small[\Delta_{\zeta,\textbf{Tree}}^{2}(k)\small]_{\textbf{SR}_{1}} = N \times \small[\widetilde  \Delta_{\zeta,\textbf{Tree}}^{2}(k)\small]_{\textbf{SR}_{1}}.
\eea
This is done for calculation convenience and does not affect the overall structure of the result. The above equation enables us to write the renormalized version of the scalar power spectrum when considering only a single sharp transition:
\bea \label{neft}
\small[\Delta_{\zeta,\textbf{EFT}}^{2}(k)\small]_{n} = \small[\widetilde  \Delta_{\zeta,\textbf{Tree}}^{2}(k)\small]_{\textbf{SR}_{1}}\{1 + W_{n} + NV_{n}\},
\eea
where $W_{n}= U_{\rm SR_1} + NU_{n}$, and the variable $n$ is used to keep track of the number of times renormalization takes place, which in the present context goes from $n = 1\;{\rm to}\;6$ totaling the $N=6$ MSTs.
Using this, the expression for the renormalized version of the scalar power spectrum for $n$ sharp transitions combined turns out to:
\bea
\Delta_{\zeta,\textbf{EFT}}^{2}(k) = \sum_{n=1}^{N}\; \small[\Delta_{\zeta,\textbf{EFT}}^{2}(k)\small]_{n}.
\eea
Upon utilizing the rescaling introduced earlier in eqn.(\ref{rescale}) we can modify the above equation as follows:
\bea
\Delta_{\zeta,\textbf{EFT}}^{2}(k) = \small[\widetilde  \Delta_{\zeta,\textbf{Tree}}^{2}(k)\small]_{\textbf{SR}_{1}}\sum_{n=1}^{N}\;\{1 + W_{n} + NV_{n}\}.
\eea
\textcolor{black}{In the present context of discussion, the renormalized version of the 1PI correlation function for any arbitrary $m$-point amplitude computed within the framework of EFT of Single Field Inflation can be expressed by the following simplified expression:}
\bea \overline{\Gamma_{{\bf EFT}}[\zeta]}=\sum^{\infty}_{m=2}\frac{i}{m!}\int\prod^{m}_{j=1}d^4x_j\,\overline{\Gamma^{(m)}_{\zeta,{\bf EFT}}(x_i)}\,\zeta(x_j),\eea
\textcolor{black}{where it is important to note that in the Fourier space such $m$-point renormalized amplitude, $\overline{\Gamma^{(m)}_{\zeta,{\bf EFT}}(x_j)}$ can be further expressed as:}
\bea \overline{\Gamma^{(m)}_{\zeta,{\bf EFT}}(x_j)}:=\int \frac{d^4k_j}{(2\pi)^4}\, e^{ik_j. x_j}\,\overline{\Gamma^{(m)}_{\zeta,{\bf EFT}}(k_j,\mu,\mu_0)}\times (2\pi)^4\delta^4\left(\sum^{m}_{j=1}k_j\right)\quad\quad\forall\quad j=1,2,\cdots,m.\eea
\textcolor{black}{where $\mu$ and $\mu_0$ represent the renormalization scale and reference scale respectively. The explicit values of these scales are mentioned below precisely for the two renormalization schemes, late time and adiabatic method:}
\begin{enumerate}
    \item  \underline{\bf For late time scheme:}

\bea \mathbf{\underline{For\;SR_1}:} & \quad \mu = H_{s_1}, \quad  \mu_0 = \Lambda, \\
\mathbf{\underline{For\;USR_n}:} & \quad \mu = H_{e_n}, \quad  \mu_0 = \Lambda, \\
\mathbf{\underline{For\;SR_{n+1}}:} & \quad \mu = H_{\rm end}, \quad  \mu_0 = \Lambda.\eea
    \item \underline{\bf For adiabatic/wave function scheme:}
    
    \bea \mathbf{\underline{For\;SR_1}:} & \quad \mu = k_{s_1} = \Lambda_{\rm UV}^{\rm SR_1}, \quad  \mu_0 = k_{*} = \Lambda_{\rm IR}^{\rm SR_1}, \\
\mathbf{\underline{For\;USR_n}:} & \quad \mu = k_{e_n} =  \Lambda_{\rm IR}^{\rm USR_n}, \quad  \mu_0 = k_{s_n} =  \Lambda_{\rm IR}^{\rm USR_n}, \\
\mathbf{\underline{For\;SR_{n+1}}:} & \quad \mu = \mu_0 = k_{\rm end}.\eea
\end{enumerate}
\textcolor{black}{Further, in terms of the 1PI effective action, any general $m$-point renormalized amplitude in the Fourier space can be expressed by the following expression:}
\bea \overline{\Gamma^{(m)}_{\zeta,{\bf EFT}}(k_1,k_2,k_3,\cdots,k_m,\mu,\mu_0)}=\left({\cal Z}^{\rm IR}_{n}\right)^{\frac{m}{2}}\Gamma^{(m)}_{\zeta,{\bf EFT}}(k_1,k_2,k_3,\cdots,k_m)\eea
\textcolor{black}{We can now perform the renormalization by introducing a counter-term for each of the $n$th contributions corresponding to a single sharp transition and combining them. The counter-term or the renormalization factor, denoted by ${\cal Z}^{\rm IR}_{n}$ from hereon. This statement can be translated in terms of $m$-point renormalized cosmological correlation function which can be further written in terms of the unrenormalized/ bare contribution by the following simplified expression:}
\bea \overline{\langle \zeta_{\bf k_1}\zeta_{\bf k_2}\zeta_{\bf k_3}\cdots\cdots\zeta_{\bf k_m}\rangle_{\zeta,{\bf EFT}}}=\left({\cal Z}^{\rm IR}_{n}\right)^{\frac{m}{2}}\langle \zeta_{\bf k_1}\zeta_{\bf k_2}\zeta_{\bf k_3}\cdots\cdots\zeta_{\bf k_m}\rangle_{\zeta,{\bf EFT}}.\eea
\textcolor{black}{This statement is established at the level of $m$-point correlation function would be easily realized in terms of the connecting relationship among the renormalized, unrenormalized/ bare, and counter-term contribution in the expression for the gauge invariant comoving scalar curvature perturbation, which is described by the following expression:}
\bea \zeta^{\bf R}_{\bf k}=\zeta^{\bf B}_{\bf k}-\zeta^{\bf C}_{\bf k}=\sqrt{{\cal Z}^{\rm IR}_{n}}\times \zeta^{\bf B}_{\bf k}.\eea
\textcolor{black}{Here the superscripts, {\bf R}, {\bf B}, and {\bf C} are used to characterize the renormalized,  bare, and counter-term contributions. Utilizing the above-mentioned fact one can further write down the bare contribution in terms of the counter-term contribution by the following expression:}
\bea \zeta^{\bf B}_{\bf k}=\left(\frac{1}{1-\sqrt{{\cal Z}^{\rm IR}_{n}}}\right)\times \zeta^{\bf C}_{\bf k}.\eea
\textcolor{black}{Now, since we are interested in the renormalization of the scalar power spectrum, henceforth we restrict our discussion by fixing $m=2$, which describes the two-point amplitude of the cosmological correlation function in the Fourier space. After fixing this the corresponding 1PI effective action for the two-point amplitude can be further expressed as:}
\bea \overline{\Gamma^{(2)}_{\zeta,{\bf EFT}}(k_1,k_2,\mu,\mu_0)}={\cal Z}^{\rm IR}_{n}\times\Gamma^{(2)}_{\zeta,{\bf EFT}}(k_1,k_2),\eea
\textcolor{black}{which can be further translated at the level of the two-point renormalized cosmological correlation function of the comoving scalar curvature perturbation as:}
\bea \overline{\langle \zeta_{\bf k_1}\zeta_{\bf k_2}\rangle}=\langle \zeta^{\bf R}_{\bf k_1}\zeta^{\bf R}_{\bf k_2}\rangle&=&\langle\left(\zeta^{\bf B}_{\bf k}-\zeta^{\bf C}_{\bf k}\right)\left(\zeta^{\bf B}_{\bf k}-\zeta^{\bf C}_{\bf k}\right)\rangle\nonumber\\
&=&\langle \zeta^{\bf B}_{\bf k_1}\zeta^{\bf B}_{\bf k_2}\rangle+\langle \zeta^{\bf C}_{\bf k_1}\zeta^{\bf C}_{\bf k_2}\rangle-\langle \zeta^{\bf B}_{\bf k_1}\zeta^{\bf C}_{\bf k_2}\rangle-\langle \zeta^{\bf C}_{\bf k_1}\zeta^{\bf B}_{\bf k_2}\rangle\nonumber\\
&=&\langle \zeta^{\bf B}_{\bf k_1}\zeta^{\bf B}_{\bf k_2}\rangle+\langle \zeta^{\bf C}_{\bf k_1}\zeta^{\bf C}_{\bf k_2}\rangle-2\langle \zeta^{\bf B}_{\bf k_1}\zeta^{\bf C}_{\bf k_2}\rangle\nonumber\\
&=&\langle \zeta^{\bf B}_{\bf k_1}\zeta^{\bf B}_{\bf k_2}\rangle+\left(1-\sqrt{{\cal Z}^{\rm IR}_{n}}\right)^2\langle \zeta^{\bf B}_{\bf k_1}\zeta^{\bf B}_{\bf k_2}\rangle-2\left(1-\sqrt{{\cal Z}^{\rm IR}_{n}}\right)\langle \zeta^{\bf B}_{\bf k_1}\zeta^{\bf B}_{\bf k_2}\rangle\nonumber\\
&=&\Bigg[1-2\left(1-\sqrt{{\cal Z}^{\rm IR}_{n}}\right)+\left(1-\sqrt{{\cal Z}^{\rm IR}_{n}}\right)^2\Bigg]\times \langle \zeta^{\bf B}_{\bf k_1}\zeta^{\bf B}_{\bf k_2}\rangle\nonumber\\
&=&{\cal Z}^{\rm IR}_{n}\times \langle \zeta^{\bf B}_{\bf k_1}\zeta^{\bf B}_{\bf k_2}\rangle.\label{corr}\eea
\textcolor{black}{In terms of the sole counter-term contribution one can also write down the expression for the two-point renormalized cosmological correlation function of the comoving scalar curvature perturbation as:}
\bea \overline{\langle \zeta_{\bf k_1}\zeta_{\bf k_2}\rangle}=\langle \zeta^{\bf R}_{\bf k_1}\zeta^{\bf R}_{\bf k_2}\rangle&=& \Bigg\{\frac{{\cal Z}^{\rm IR}_{n}}{\left(1-{\cal Z}^{\rm IR}_{n}\right)^2}\Bigg\}\times \langle \zeta^{\bf C}_{\bf k_1}\zeta^{\bf C}_{\bf k_2}\rangle.\label{count}\eea
\textcolor{black}{This further implies that comparing equation (\ref{corr}) and equation (\ref{count}), we get the following connecting relationship between the bare part and counter-part of the two-point cosmological correlation function of the comoving scalar curvature perturbation, which is given by the following simplified expression:}
\bea  \langle \zeta^{\bf B}_{\bf k_1}\zeta^{\bf B}_{\bf k_2}\rangle=\Bigg\{\frac{1}{\left(1-{\cal Z}^{\rm IR}_{n}\right)^2}\Bigg\}\times \langle \zeta^{\bf C}_{\bf k_1}\zeta^{\bf C}_{\bf k_2}\rangle.\eea
\textcolor{black}{Here to achieve the above-mentioned results we have used the following facts explicitly for the computation:}
\bea \langle \zeta^{\bf B}_{\bf k_1}\zeta^{\bf C}_{\bf k_2}\rangle=\langle \zeta^{\bf C}_{\bf k_1}\zeta^{\bf B}_{\bf k_2}\rangle=\left(1-\sqrt{{\cal Z}^{\rm IR}_{n}}\right)\langle \zeta^{\bf B}_{\bf k_1}\zeta^{\bf B}_{\bf k_2}\rangle.\eea
\textcolor{black}{Further, we use the following facts to write down the expressions for the scalar power spectrum from the two-point correlation function:}
\bea &&\label{r1}\overline{\langle \zeta_{\bf k_1}\zeta_{\bf k_2}\rangle}=\langle \zeta^{\bf R}_{\bf k_1}\zeta^{\bf R}_{\bf k_2}\rangle=(2\pi)^3\delta^3\left({\bf k_1}+{\bf k_2}\right)\frac{2\pi^2}{k^3_1}\small[\overline{\Delta_{\zeta,\textbf{EFT}}^{2}(k)}\small]_{n},\\
&&\label{r2}\langle \zeta^{\bf B}_{\bf k_1}\zeta^{\bf B}_{\bf k_2}\rangle=(2\pi)^3\delta^3\left({\bf k_1}+{\bf k_2}\right)\frac{2\pi^2}{k^3_1}\small[\Delta_{\zeta,\textbf{EFT}}^{2}(k)\small]_{n}.\eea
\textcolor{black}{Now using equation (\ref{r1}) and (\ref{r2}) in equation (\ref{corr}) and fixing $k=k_1$ to avoid any further confusion due to nomenclature the renormalized version of the $n$th contribution of the scalar power spectrum consisting of a single sharp transition is governed by the following simplified expression:}
\bea \label{zeft}
\small[\overline{\Delta_{\zeta,\textbf{EFT}}^{2}(k)}\small]_{n} = {\cal Z}^{\rm IR}_{n}\times\;\small[\Delta_{\zeta,\textbf{EFT}}^{2}(k)\small]_{n},
\eea
which results in a softening of the logarithmic IR divergences through the counter-term multiplication. This counter-term has its value determined via the following renormalization condition fixed at the pivot scale $k_{*}$:
\bea
\small[\overline{\Delta_{\zeta,\textbf{EFT}}^{2}(k_{*})}\small]_{n} = \small[\widetilde  \Delta_{\zeta,\textbf{Tree}}^{2}(k_{*})\small]_{\textbf{SR}_{1}},
\eea
As a result of the above renormalization condition, we determine the form of the $n$th counter-term after following the definitions in Eqs.(\ref{neft}), and (\ref{zeft}):
\bea
{\cal Z}^{\rm IR}_{n} =  \frac{\small[\overline{\Delta_{\zeta,\textbf{EFT}}^{2}(k_{*})}\small]_{n}}{\small[\Delta_{\zeta,\textbf{EFT}}^{2}(k_{*})\small]_{n}} = \frac{\small[\widetilde  \Delta_{\zeta,\textbf{Tree}}^{2}(k_{*})\small]_{\textbf{SR}_{1}}}{\small[\Delta_{\zeta,\textbf{EFT}}^{2}(k_{*})\small]_{n}} = \frac{1}{1 + W_{n,*} + NV_{n,*}} \approx (1-W_{n,*}-NV_{n,*}+\cdots),
\eea
where the series is truncated after just expanding up to the first order and neglecting other higher-order contributions. This determination of the counter-term finally enables the writing of the combined renormalized one-loop corrected scalar power spectrum coming from contributions across all the $n$ sharp transitions:
\bea
\overline{\Delta_{\zeta,\textbf{EFT}}^{2}(k)} = \small[\Delta_{\zeta,\textbf{Tree}}^{2}(k)\small]_{\textbf{SR}_{1}}\bigg\{1 + \frac{1}{N}\sum_{n=1}^{N}{\cal Q}_{n,\textbf{EFT}}\bigg\},
\eea
with a new EFT dependent term, ${\cal Q}_{n,\textbf{EFT}}$, related to the contribution from $n$th transition. The structure of this new term is as follows:
\bea \label{Qneft}
{\cal Q}_{n,\textbf{EFT}} = \frac{-N\small[\Delta_{\zeta,\textbf{Tree}}^{2}(k)\small]_{\textbf{SR}_{1}}}{\small[\Delta_{\zeta,\textbf{Tree}}^{2}(k_{*})\small]_{\textbf{SR}_{1}}}\{W_{n,*}^{2} + N^{2}V_{n,*}^{2} + \cdots \}.
\eea

The importance of the term ${\cal Q}_{n,\textbf{EFT}}$ will become more apparent as we will venture, in the upcoming sections, towards performing the final necessary procedure of resummation over the logarithmic contributions to obtain a physically relevant output. The scheme presented in this section can be applied to a general $N$ number of sharp transitions. For the analysis of this work, we have fixed $N=6$, which proves to be the lowest number allowed to resolve the horizon problem. The resulting eqn.(\ref{Qneft}) shows that the first order logarithmic terms contained inside the factors $W_{n}$ and $V_{n}$ have been promoted to give quadratic dependence over the final renormalized one-loop corrected scalar power spectrum. This is a crucial result to bear in mind as now the logarithmic divergent terms have been further softened by becoming more sub-leading than before. The diagrammatic representation of the loop contributions included within eqn.(\ref{Qneft}) is shown below:
\vspace{-6em}
\begin{equation}\label{twoloop}
\begin{tikzpicture}[baseline={([yshift=-3.5ex]current bounding box.center)},very thick]
  
  \def\radius{0.76}
  \scalebox{0.5}{
  \draw[red,very thick] (3*\radius,0) circle (\radius);
  \draw[red,very thick] (5*\radius,0) circle (\radius);
  \draw[red,very thick] (13*\radius,\radius) circle (\radius);
  \draw[red,very thick] (13*\radius,3*\radius) circle (\radius);
  \draw[red,very thick] (21*\radius,0) circle (\radius);}
  \draw[black, very thick] (0,0) -- (\radius,0); 
  \draw[blue,fill=blue] (\radius,0) circle (.3ex);
  \draw[blue,fill=blue] (2*\radius,0) circle (.3ex);
  \draw[blue,fill=blue] (3*\radius,0) circle (.3ex);
  \draw[black, very thick] (3*\radius,0) -- (4*\radius,0);
  \node at (4.5*\radius,0) {+};
  \draw[black, very thick] (5*\radius,0) -- (6.5*\radius,0);
  \draw[blue,fill=blue] (6.5*\radius,0) circle (.3ex);
  \draw[blue,fill=blue] (6.5*\radius,\radius) circle (.3ex);
  \draw[black, very thick] (6.5*\radius,0) -- (8*\radius,0);
  \node at (8.5*\radius,0) {+};
  \draw[black, very thick] (9*\radius,0) -- (10*\radius,0);
  \draw[blue,fill=blue] (10*\radius,0) circle (.3ex);
  \draw[red, very thick] (10*\radius,0) -- (11*\radius,0);
  \draw[blue,fill=blue] (11*\radius,0) circle (.3ex);
  \draw[black, very thick] (11*\radius,0) -- (12*\radius,0);
  \node at (15*\radius,0) {= \;\text{ Two-loop contributions,}};
\end{tikzpicture}
\end{equation}
\begin{equation}\label{fourloop}
    \begin{tikzpicture}[baseline={([yshift=-.5ex]current bounding box.center)},very thick]
    \draw [line width=1pt] (-14.5,0)-- (-14,0);
    \draw[blue,fill=blue] (-14,0) circle (.3ex);
    \draw [red,line width=0.8pt] (-14,0)-- (-13.5,0.5);
    \draw[blue,fill=blue] (-13.5,0.5) circle (.3ex);
    \draw [red,line width=0.8pt] (-14,0)-- (-13.5,-0.5);
    \draw[blue,fill=blue] (-13.5,-0.5) circle (.3ex);
    \draw [red,line width=0.8pt] (-13.5,0.5)-- (-13.5,-0.5);
    \draw [red,line width=0.8pt] (-13.5,0.5)-- (-12.5,0.5);
    \draw [red,line width=0.8pt] (-12.5,0.5)-- (-12.5,-0.5);
    \draw [red,line width=0.8pt] (-13.5,0)-- (-12.5,0);
    \draw[blue,fill=blue] (-13.5,0) circle (.3ex);
    \draw[blue,fill=blue] (-12.5,0) circle (.3ex);
    \draw[blue,fill=blue] (-12.5,0.5) circle (.3ex);
    \draw [red,line width=0.8pt] (-13.5,-0.5)-- (-12.5,-0.5);
    \draw[blue,fill=blue] (-12.5,-0.5) circle (.3ex);
    \draw [red,line width=0.8pt] (-12.5,0.5)-- (-12,0);
    \draw[blue,fill=blue] (-12,0) circle (.3ex);
    \draw[blue,fill=blue] (-12.5,-0.5) circle (.3ex);
    \draw [red,line width=0.8pt] (-12.5,-0.5)-- (-12,0);
    \draw [black,line width=0.8pt] (-12,0)-- (-11.5,0);

    \node at (-11,0) {+};

    \draw [line width=1pt] (-10.5,0)-- (-10,0);
    \draw[blue,fill=blue] (-10,0) circle (.3ex);
    \draw [red,line width=0.8pt] (-10,0)-- (-9.5,0.5);
    \draw [red,line width=0.8pt] (-10,0)-- (-9.5,-0.5);
    \draw [red,line width=0.8pt] (-9.5,0.5)-- (-8.5,0.5);
    \draw [red,line width=0.8pt] (-9.5,-0.5)-- (-8.5,-0.5);
    \draw [red,line width=0.8pt] (-8.5,0.5)-- (-8,0);
    \draw [red,line width=0.8pt] (-8.5,-0.5)-- (-8,0);
    \draw [line width=1pt] (-8,0)-- (-7.5,0);
    \draw [red,line width=0.8pt] (-9.5,0.5)-- (-9,0);
    \draw [red,line width=0.8pt] (-9.5,-0.5)-- (-9,0);
    \draw [red,line width=0.8pt] (-8.5,0.5)-- (-9,0);
    \draw [red,line width=0.8pt] (-8.5,-0.5)-- (-9,0);
    \draw[blue,fill=blue] (-9.5,0.5) circle (.3ex);
    \draw[blue,fill=blue] (-9.5,-0.5) circle (.3ex);
    \draw[blue,fill=blue] (-8.5,0.5) circle (.3ex);
    \draw[blue,fill=blue] (-8.5,-0.5) circle (.3ex);
    \draw[blue,fill=blue] (-9,0) circle (.3ex);
    \draw[blue,fill=blue] (-8,0) circle (.3ex);

    \node at (-7,0) {+};

    \draw [line width=1pt] (-6.5,0)-- (-6,0);
    \draw[blue,fill=blue] (-6,0) circle (.3ex);
    \draw [red,line width=0.8pt] (-6,0)-- (-5.5,0.5);
    \draw[blue,fill=blue] (-5.5,0.5) circle (.3ex);
    \draw [red,line width=0.8pt] (-6,0)-- (-5.5,-0.5);
    \draw[blue,fill=blue] (-5.5,-0.5) circle (.3ex);
    \draw [red,line width=0.8pt] (-5.5,0.5)-- (-5.5,-0.5);
    \draw [red,line width=0.8pt] (-5.5,0.5)-- (-4.5,0.5);
    \draw [red,line width=0.8pt] (-4.5,0.5)-- (-4.5,-0.5);
    \draw [red,line width=0.8pt] (-5,0.5)-- (-5,-0.5);
    \draw [red,line width=0.8pt] (-5.5,-0.5)-- (-4.5,-0.5);
    \draw [red,line width=0.8pt] (-4.5,0.5)-- (-4,0);
    \draw [red,line width=0.8pt] (-4.5,-0.5)-- (-4,0);
    \draw [line width=0.8pt] (-4,-0)-- (-3.5,0);
    \draw[blue,fill=blue] (-5,0.5) circle (.3ex);
    \draw[blue,fill=blue] (-5,-0.5) circle (.3ex);
    \draw[blue,fill=blue] (-4.5,0.5) circle (.3ex);
    \draw[blue,fill=blue] (-4.5,-0.5) circle (.3ex);
    \draw[blue,fill=blue] (-4,0) circle (.3ex);

    \node at (-3,0) {+ $\cdots$};
\end{tikzpicture} = \quad \text{Four-loop contributions,}
\end{equation}
the above equations Eqs.(\ref{twoloop},\ref{fourloop}) represent the two-loop and four-loop contributions present inside the term ${\cal Q}_{n,\textbf{EFT}}$. Notice that there can be many more four-loop and higher-order loop contributions present, and we have shown only a few leading order terms; these are more sub-leading than the two-loop contributions coming from the terms $W_{n,*}^{2}$ and $V_{n,*}^{2}$ and hence are denoted inside eqn.(\ref{Qneft}) by using ellipses ($\cdots$).

\subsubsection{More on other renormalization schemes: Is the renormalized result scheme dependent?}

Thus far, we have looked into the intricacies of the renormalization schemes such as late time and adiabatic renormalization. We have been able to completely handle the harmful UV divergences through these two methods. However, it becomes mandatory to mention that although the final momentum-dependent integrals from the one-loop corrected power spectrum yield the same results for both these above methods, the explicit form of the counter term is different. Regardless, there is no effect on the overall conclusion derived from these schemes. They both sell out the UV divergences, while coarse-graining the IR divergences through the smoothening. Now when the UV divergences are removed, we smoothened the logarithmic IR divergences with the Power Spectrum Renormalization scheme. However, there might be issues and drawbacks of some of the renormalization procedures and therefore, the other approaches to renormalization must be explored. With the light of this discussion, we would like to mention the other renormalization schemes that find frequent use in the quantum field theory of curved space-time: The minimal subtraction (MS) scheme and the related modified
minimal subtraction (MS) scheme \cite{PhysRevD.8.3497, THOOFT1973455, collins_1984}, On-Shell scheme \cite{article}, Bogoliubov-Parasiuk-Hepp-Zimmermann (BPHZ) scheme \cite{PhysRev.75.1736,Kraus:1997bi,Piguet:1986ug,Zimmermann:1968mu,Zimmermann:1969jj}, Bogoliubov-Parasiuk-Hepp-Zimmermann-Lowenstein (BPHZL) scheme, \cite{Lowenstein:1975rg,Lowenstein:1975ps}, Hadamard renormalization\cite{PhysRevD.107.025004}, Dimensional Renormalization (DR) scheme\cite{Coquereaux:1979eq,Belusca-Maito:2020ala,Binetruy:1980xn}, Algebraic Renormalization (AR) scheme \cite{Adler:1969er,Batalin:1981jr,Becchi:1973gu,tHooft:1972tcz,Piguet:1995er}, and so on. These schemes are all equally important and an extensive analysis should be performed with each of these to check for the consistency of the derived results.  In our paper, we have employed only three of the renormalization schemes, abstaining from exploring additional approaches. This decision was made to stay aligned with our primary motivation and prevent diversion from the original purpose we set out to address i.e. the impact of the EoS parameter $w$. Since we already found from two renormalization techniques that the results are consistent, we do not expect drastic differences to be observed for outcomes from other approaches. However, we still encourage everyone to seek various methods mentioned to check for consistency or departure. Of the above-mentioned schemes, we seek to inspect the Dimensional regularization scheme in future work which is crucial in the discussion of the present context. Thereby, the necessity for renormalization stems from the fact that there are largely pronounced divergent contributions originating from the USR phases that need to be regulated. Hence the strong constraints of renormalization and resummation ensure that the USR phase is not prolonged, but also produces sufficient amplitude for the generation of viable PBH abundance ($f_{\rm PBH} \sim {\cal O}(10^{-3}-1$). Renormalization is also quintessential in the context of strong No-go constraints of PBH masses for single sharp transitions. As a result of strong constraints coming from renormalization, the production of large mass PBHs ${\cal O}( M_{\odot})$ is forbidden as the conditions for inflation are not met in such a scenario i.e. specifically the number of e-foldings to end inflation is not satisfied i.e. (${\cal N}_{\rm Total} \sim {\cal O} (20-25)$). To overcome such barriers, we proposed an EFT set-up in \cite{Bhattacharya:2023ysp}  whereby the solution to the above-mentioned problem was sought out with $6$ sharp transitions. The number of sharp transitions chosen for the construction of this EFT set-up is also a consequence of the strong renormalization procedure. This sets a minimum of $6$ transitions for our case based on the scales that we had set. This number is the minimum number of transitions that you would require to solve the horizon problem. However, you can take any number greater than $6$ which not only solves the horizon problem but also matches the condition of $\Omega_{k}=0$. 
There is but an alternative approach to investigate this via the smooth transitions as mentioned earlier. The authors in \cite{Riotto:2023gpm,Firouzjahi:2023ahg,Firouzjahi:2023aum} claim that single smooth transitions can offer a solution to evade the No-go theorem on PBHs mass to generate heavy mass PBHs while also explain the production of SIGWs that are consistent with the observational data. In light of this context, we might design a set-up of multiple smooth transitions to examine this scenario in the scope of our future work. We like to highly emphasize the importance of renormalization scheme and the dependency of the results on the scheme adopted. Since late time and adiabatic renormalization provide the same results, we state that there will not be much deviation between different schemes. If the results of the renormalized loop-corrected power spectrum are heavily dependent on the procedure for renormalization, its implications are disturbing and unphysical challenging the very nature of regularization and renormalization schemes.

\subsection{Dynamical Renormalization Group (DRG) resummed  scalar power spectrum from MST-EFT}

This section is aimed towards highlighting the importance of performing the Dynamical Renormalization Group (DRG) procedure \cite{Chen:2016nrs,Baumann:2019ghk,Boyanovsky:1998aa,Boyanovsky:2001ty,Boyanovsky:2003ui,Burgess:2009bs,Dias:2012qy,Chaykov:2022zro,Chaykov:2022pwd}. The basic idea is to effectively capture the quantum effects by completely taking care of the logarithmically divergent contributions. This requires summing over such terms in an infinite series coming from all the higher-order loop corrections in a perturbative computation involving scalar modes. However, summing itself is not warranted unless the terms in the series follow stringent convergence criteria on the horizon crossing and the super-horizon scales. When the convergence is established, this approach of DRG resummation extracts for us the late time behaviour without forcing us to know the explicit behaviour of the terms involved in the perturbative expansion. 

The initial idea behind the Renormalization Group (RG) resummation technique was to bring the momentum-dependent contributions into the scale-dependent running couplings of the underlying theory. The DRG resummation method is an improvement on this initial approach. By extending the running energy scales to a much broader range, the improvement allows to stay within the small coupling regime where the perturbative approximations remain intact within the EFT framework of inflation. In the cosmological sense concerning our work, the DRG resummation in the late-time limit enables us to focus on the study of the cosmological $\beta$ functions. These $\beta$ functions are none other than the well-known features of the scalar power spectrum characterized by the tilt, running, and running of running of the tilt as a function of the energy scales. These are physically relevant quantities whose finite presence confirms the departure from scale invariant nature of the primordial scalar power spectrum. To summarize, the DRG resummation approach justifies our summing over the secular time and momentum-dependent terms in the perturbation expansion, at the horizon crossing and superhorizon scales, and provides us with the knowledge of the entire series comprising of the logarithmically divergent terms without worrying about individual behaviour of such terms involved. 

Now after presenting the general idea behind the DRG approach, we can perform the resummation of all higher orders of logarithmic contributions in a safe manner. The following gives us the resummed version of the power spectrum associated with the scalar modes: 
\bea \label{DRG}
\overline{\overline{\Delta_{\zeta,\textbf{EFT}}^{2}(k)}} &=& \bigg[\Delta_{\zeta,\textbf{Tree}}^{2}(k)\bigg]_{\textbf{SR}_{1}}\bigg(1 + {\cal Q}_{\textbf{EFT}} + \frac{1}{2!}{\cal Q}_{\textbf{EFT}}^{2} + \cdots \bigg)\times \bigg\{1+{\cal O}\bigg[\Delta_{\zeta,\textbf{Tree}}^{2}(k_{*})\bigg]^{2}_{\textbf{SR}_{1}}\bigg\} \nonumber\\
&=& \bigg[\Delta_{\zeta,\textbf{Tree}}^{2}(k)\bigg]_{\textbf{SR}_{1}}\exp{\left({\cal Q}_{\textbf{EFT}}\right)}\times \bigg\{1+{\cal O}\bigg[\Delta_{\zeta,\textbf{Tree}}^{2}(k_{*})\bigg]^{2}_{\textbf{SR}_{1}}\bigg\},
\eea
In terms of the diagrammatic contributions, the DRG resummed version of the scalar power spectrum can be restated as follows:
\vspace{-1em}
\bea
\overline{\overline{\Delta_{\zeta,\textbf{EFT}}^{2}(k)}} = \begin{tikzpicture}[baseline={([yshift=-.5ex]current bounding box.center)},very thick]
\draw [line width=1pt] (0,0)-- (1,0);
\end{tikzpicture}\times\exp{({\cal F})}\times \bigg\{1+{\cal O}\bigg[\Delta_{\zeta,\textbf{Tree}}^{2}(k_{*})\bigg]^{2}_{\textbf{SR}_{1}}\bigg\},
\eea
which includes the tree-level diagram, \begin{tikzpicture}[baseline={([yshift=-.5ex]current bounding box.center)},very thick]
\draw [line width=1pt] (0,0)-- (1,0);
\end{tikzpicture} $\equiv \bigg[\Delta_{\zeta,\textbf{Tree}}^{2}(k)\bigg]_{\textbf{SR}_{1}}$, as well as the other Feynman diagrammatic loop representations denoted by ${\cal F}$ which is summed as follows : \vspace{-1.5em}
\begin{equation}
    \sum {\cal F} =  
\begin{tikzpicture}[baseline={([yshift=-5.5ex]current bounding box.center)},very thick]

  \def\radius{0.5}
  \scalebox{0.5}{
  \draw[red, ultra thick] (3*\radius,0) circle (\radius);
  \draw[red,ultra thick] (5*\radius,0) circle (\radius);
  \draw[red,ultra thick] (13*\radius,\radius) circle (\radius);
  \draw[red,ultra thick] (13*\radius,3*\radius) circle (\radius);
  \draw[red,ultra thick] (21*\radius,0) circle (\radius);}

  \draw[black, very thick] (0,0) -- (\radius,0); 
  \draw[blue,fill=blue] (\radius,0) circle (.3ex);
  \draw[blue,fill=blue] (2*\radius,0) circle (.3ex);
  \draw[blue,fill=blue] (3*\radius,0) circle (.3ex);
  \draw[black, very thick] (3*\radius,0) -- (4*\radius,0);
  \node at (4.5*\radius,0) {+};
  \draw[black, very thick] (5*\radius,0) -- (6.5*\radius,0);
  \draw[blue,fill=blue] (6.5*\radius,0,0) circle (.3ex);
  \draw[blue,fill=blue] (6.5*\radius,\radius) circle (.3ex);
  \draw[black, very thick] (6.5*\radius,0,0) -- (8*\radius,0);
  \node at (8.5*\radius,0) {+};
  \draw[black, very thick] (9*\radius,0) -- (10*\radius,0);
  \draw[blue,fill=blue] (10*\radius,0,0) circle (.3ex);
  \draw[red, ultra thick] (10*\radius,0) -- (11*\radius,0);
  \draw[blue,fill=blue] (11*\radius,0,0) circle (.3ex);
  \draw[black, very thick] (11*\radius,0) -- (12*\radius,0);
  \node at (12.5*\radius,0) {+};
  \draw[black, very thick](13*\radius,0) --(14*\radius,0);
  \draw[blue,fill=blue](14*\radius,0,0) circle (.3ex);
  \draw[red,  thick](14*\radius,0) --(14.5*\radius,\radius);
  \draw[blue,fill=blue](14.5*\radius,\radius) circle (.3ex);
\draw[red, thick](14.5*\radius,\radius) --(15.5*\radius,\radius);
\draw[blue,fill=blue](15.5*\radius,\radius) circle (.3ex);
\draw[red, thick](14*\radius,0) -- (14.5*\radius,-\radius);\draw[blue,fill=blue](14.5*\radius,-\radius) circle (.3ex);
\draw[red,  thick](14.5*\radius,\radius) -- (14.5*\radius,-\radius);
\draw[red, thick](14.5*\radius,-\radius) -- (15.5*\radius,-\radius);
\draw[blue,fill=blue](15.5*\radius,-\radius) circle (.3ex);
\draw[blue,fill=blue](14.5*\radius,0) circle (.3ex);
\draw[red, thick](14.5*\radius,0) -- (15.5*\radius,0);
\draw[red,  thick](15.5*\radius,\radius) -- (15.5*\radius,-\radius);
\draw[blue,fill=blue](15.5*\radius,0) circle (.3ex);
\draw[red, thick](15.5*\radius,\radius) -- (16*\radius,0);
\draw[blue,fill=blue](16*\radius,0) circle (.3ex);
\draw[red, thick](16*\radius,0) -- (15.5*\radius,-\radius);
\draw[black, very thick](16*\radius,0) -- (17*\radius,0);
\node at (17.5*\radius,0) {+};
\draw[black, very thick](18*\radius,0) -- (19*\radius,0);
\draw[blue,fill=blue](19*\radius,0) circle (.3ex);
\draw[red, thick](19*\radius,0) -- (19.5*\radius,\radius);
\draw[red, thick](19*\radius,0) -- (19.5*\radius,-\radius);
\draw[blue,fill=blue](19.5*\radius,\radius) circle (.3ex);
 \draw[blue,fill=blue](19.5*\radius,-\radius) circle (.3ex); 
 \draw[red, thick](19.5*\radius,\radius) -- (20.5*\radius,\radius);
 \draw[red, thick](19.5*\radius,-\radius) -- (20.5*\radius,-\radius);
 \draw[blue,fill=blue](20.5*\radius,\radius) circle (.3ex);
 \draw[blue,fill=blue](20.5*\radius,-\radius) circle (.3ex);
 \draw[red, thick](20.5*\radius,\radius) -- (21*\radius,0);
 \draw[red, thick](20.5*\radius,-\radius) -- (21*\radius,0);
\draw[blue,fill=blue](21*\radius,0) circle (.3ex);
\draw[black,  thick](21*\radius,0) -- (22*\radius,0); 
\draw[red, thick](19.5*\radius,\radius) -- (20.5*\radius,-\radius);
\draw[red, thick](20.5*\radius,\radius) -- (19.5*\radius,-\radius);
\node at (22.5*\radius,0) {+};
\draw[black, thick](23*\radius,0) -- (24*\radius,0);
\draw[red, thick](24*\radius,0) -- (24.5*\radius,\radius);
\draw[red, thick](24*\radius,0) -- (24.5*\radius,-\radius);
\draw[blue,fill=blue](24*\radius,0) circle (.3ex);
\draw[blue,fill=blue](24.5*\radius,\radius) circle (.3ex);
\draw[blue,fill=blue](24.5*\radius,-\radius) circle (.3ex);
\draw[red, thick](24.5*\radius,\radius) -- (25.5*\radius,\radius);
\draw[red,  thick](24.5*\radius,-\radius) -- (25.5*\radius,-\radius);
\draw[red,  thick](25.5*\radius,\radius) -- (26*\radius,0);
\draw[red, thick](26*\radius,0) -- (25.5*\radius,-\radius);
\draw[blue,fill=blue](25.5*\radius,\radius) circle (.3ex);
\draw[blue,fill=blue](26*\radius,0) circle (.3ex);
\draw[blue,fill=blue](25.5*\radius,-\radius) circle (.3ex);
\draw[red,  thick](24.5*\radius,\radius) -- (24.5*\radius,-\radius);
\draw[red, thick](25*\radius,\radius) -- (25*\radius,-\radius);
\draw[red, thick](25.5*\radius,\radius) -- (25.5*\radius,-\radius);
\draw[blue,fill=blue](25*\radius,\radius) circle (.3ex);
\draw[blue,fill=blue](25*\radius,-\radius) circle (.3ex);
\draw[black, thick](26*\radius,0) -- (27*\radius,0); 
\node at (28*\radius,0) {+};
\node at (29*\radius,0) {...};

\end{tikzpicture}
\end{equation}
The above equation results after subjecting to the strict convergence criteria in terms of the quantity satisfying $|{\cal Q}_{\textbf{EFT}}| \ll 1$.  It is to be noted carefully that the above formula is written in presence of a single sharp transition $(n=1)$ and the DRG resummed version as treated above is valid for the logarithmic IR divergent contributions present in the term ${\cal Q}_{\textbf{EFT}}$ which comes from each phase consisting of a sharp transition in the momentum space. Such logarithmic contributions are softened further by performing the DRG procedure when compared with the renormalized scalar power spectrum version discussed previously. Since, in the present context, we are working with multiple sharp transitions, we can equivalently add the respective resummed versions coming for each sharp transition scenario to obtain the final resummed version of the one-loop corrected scalar power spectrum. Each contributing term in the infinite series resummation resembles the function of an even-order quantum loop correction present in a perturbative series and, amazingly, without the explicit calculation of any such higher-order terms we can package their effects to give a finite sum in form of an exponential function.

Let us now visualize this softening using the one-loop results from the functions $U$ and $V$ defined previously in Eqs.(\ref{Utot},\ref{V}). We will work out the upcoming analysis of this softening using the term ${\cal Q}_{n,\textbf{EFT}}$ from the $n$th sharp transition phase. For the time being, we neglect the function $U$, since the factor $(k_{e_{n}}/k_{s_{n}})^{6}\ln{(k_{e_{n}}/k_{s_{n}})}$ in $V$ dominates comparatively. These approximations lead to the following result for the function.
\bea
{\cal Q}_{n,\textbf{EFT}} \approx -\mathbf{S}(k)\times\small[V_{n,*}^{2}+\cdots\small] = -\mathbf{S}(k)\times\small[\chi_{n,*}({\cal T}^{(n)}_{1,*})^{2}+\cdots\small]  
\eea
where some new functions are introduced as follows:  
\bea
\mathbf{S}(k) &=& \frac{\bigg[\Delta_{\zeta,\textbf{Tree}}^{2}(k)\bigg]_{\textbf{SR}_{1}}}{\bigg[\Delta_{\zeta,\textbf{Tree}}^{2}(k_{*})\bigg]_{\textbf{SR}_{1}}}, \\
{\cal T}^{(n)}_{1,*} &\approx& \frac{1}{4}\bigg[\Delta_{\zeta,\textbf{Tree}}^{2}(k_{*})\bigg]_{\textbf{SR}_{1}}\bigg(\frac{\Delta\eta(\tau_{e_{n}})}{\tilde{c}^{4}_{s}}\bigg)^{2}\bigg(\frac{k_{e_n}}{k_{s_n}}\bigg)^{6}\times\ln{\left(\frac{k_{e_{n}}}{k_{s_{n}}}\right)}, \\
\chi_{n,*} &=& 1 + \frac{{\cal T}^{(n)}_{2,*}}{{\cal T}^{(n)}_{1,*}},
\eea 
and a further argument to simplify the term ${\cal T}^{(n)}_{1,*}$ is used: \bea\displaystyle{\left(\Delta\eta(\tau_{e_{n}})\right)^{2}\bigg(\frac{k_{e_n}}{k_{s_n}}\bigg)^{6}} \gg \left(\Delta\eta(\tau_{s_{n}})\right)^{2} \nonumber \eea. As a result of the above approximations, we are led to the final simplified form of the term: 
\bea
{\cal Q}_{n,\textbf{EFT}} \approx -\delta_{n,*}\ln^{2}{\left(\frac{k_{e_{n}}}{k_{s_{n}}}\right)},
\eea
which uses a factor defined as: 
\bea
\delta_{n,*} = \frac{\chi_{n,*}}{4}\bigg[\Delta_{\zeta,\textbf{Tree}}^{2}(k_{*})\bigg]_{\textbf{SR}_{1}}\bigg(\frac{\Delta\eta(\tau_{e_{n}})}{\tilde{c}^{4}_{s}}\bigg)^{2}\bigg(\frac{k_{e_n}}{k_{s_n}}\bigg)^{6} \ll 1.
\eea
where the factor $\delta_{n,*}$ in principle takes different values corresponding to each $n$th contribution from the sharp transition phase.  
A closer look at the above equation shows why the DRG resummed result further softens the divergent logarithmic IR contributions in the final exponentiated result as in eqn.(\ref{DRG}). After implementing the above discussed approximations, we present a simplified version of the previous formula in eqn.(\ref{DRG}) for a single sharp transition $(n=1)$:
\bea \label{simpleDRG}
\overline{\overline{\Delta_{\zeta,\textbf{EFT}}^{2}(k)}} \approx \bigg[\Delta_{\zeta,\textbf{Tree}}^{2}(k)\bigg]_{\textbf{SR}_{1}}\left(\frac{k_{e_{1}}}{k_{s_{1}}}\right)^{-2.3\delta_{*}}\times\bigg\{1+{\cal O}\bigg[\Delta_{\zeta,\textbf{Tree}}^{2}(k_{*})\bigg]^{2}_{\textbf{SR}_{1}}\bigg\}
\eea
Returning to the present context of the discussion, in our MST setup, we have $n$ number of sharp transitions, and depending on the position of a single USR phase, we would expect its corresponding UV divergences removed and IR contributions softened terms in ${\cal Q}_{n,\textbf{EFT}}$. However, since the present scenario also considers a background with a constant EoS parameter $w$, it is not sure whether the features present in the final quantum loop corrected renormalized DRG resummed scalar power spectrum amplitude remain preserved in the later part of the induced gravitational wave spectrum. We will observe from section \ref{s5} that only when $w=1/3$ do we obtain a similar constant amplitude feature across all frequency ranges. To ensure that no violation of the perturbative approximations occurs across all sharp transitions, we suggest using phenomenological coarse-graining factors. 

The coarse-graining factors equal the number of total sharp transitions present and must be tuned specific to the individual phase. But, one must understand that increasing the dependency on such coarse-graining factors decreases the reliability of predictions from the underlying setup. Our current setup requires the use of five such coarse-graining factors, except for the first sharp transition, and their behaviour differs based on the parameter $w$. In general, the final DRG resummed version of the renormalized scalar power spectrum corresponding to the $n$th transition modifies to give the following: 
\bea \label{drgfinal}
\overline{\overline{\Delta_{\zeta,\textbf{EFT}}^{2}(k)}} &=& \bigg[\Delta_{\zeta,\textbf{Tree}}^{2}(k)\bigg]_{\textbf{SR}_{1}} \bigg[\left({\cal Q}_{1,\textbf{EFT}} + \frac{1}{2!}{\cal Q}^{2}_{1,\textbf{EFT}} + \cdots\right) +  \Theta(k-k_{s_{2}})g_{2}(w)\left({\cal Q}_{2,\textbf{EFT}} + \frac{1}{2!}{\cal Q}^{2}_{2,\textbf{EFT}} + \cdots \right) \nonumber\\
&+& \Theta(k-k_{s_{3}})g_{3}(w)\left({\cal Q}_{3,\textbf{EFT}} + \frac{1}{2!}{\cal Q}^{2}_{3,\textbf{EFT}} + \cdots\right) + \cdots + \Theta(k-k_{s_{n}})g_{n}(w)\left({\cal Q}_{n,\textbf{EFT}} + \frac{1}{2!}{\cal Q}^{2}_{n,\textbf{EFT}} + \cdots\right) \bigg]\nonumber\\
&& \quad\quad\quad\quad\quad\quad\quad\quad\quad\quad\quad\quad\quad\quad\quad\quad\quad\quad\quad\quad\quad\quad\quad \times \bigg\{1+{\cal O}\bigg[\Delta_{\zeta,\textbf{Tree}}^{2}(k_{*})\bigg]^{2}_{\textbf{SR}_{1}}\bigg\}, \nonumber\\
&=& \bigg[\Delta_{\zeta,\textbf{Tree}}^{2}(k)\bigg]_{\textbf{SR}_{1}} \left[\exp{\left({\cal Q}_{1,\textbf{EFT}}\right)} + \sum_{n=2}^{N}\Theta(k-k_{s_{n}})g_{n}(w)\exp{\left({\cal Q}_{n,\textbf{EFT}}\right)} \right]\times \bigg\{1+{\cal O}\bigg[\Delta_{\zeta,\textbf{Tree}}^{2}(k_{*})\bigg]^{2}_{\textbf{SR}_{1}}\bigg\},\quad\quad \\
&=& \begin{tikzpicture}[baseline={([yshift=-.5ex]current bounding box.center)},very thick]
\draw [line width=1pt] (0,0)-- (1,0);
\end{tikzpicture}\times \left[\exp{({\cal F}_{1})} + \sum_{n=2}^{N}\Theta(k-k_{s_{n}})g_{n}(w)\exp{({\cal F}_{n})} \right] \nonumber \\ 
&& \quad \quad \quad \quad \quad \quad \quad \quad \quad \quad \quad \quad \quad \quad \quad \quad \quad \quad \quad \quad \quad \quad \quad \times \bigg\{1+{\cal O}\bigg[\Delta_{\zeta,\textbf{Tree}}^{2}(k_{*})\bigg]^{2}_{\textbf{SR}_{1}}\bigg\},\quad\quad
\eea
where we have again used the diagrammatic version for the tree-level contribution, \begin{tikzpicture}[baseline={([yshift=-.5ex]current bounding box.center)},very thick]
\draw [line width=1pt] (0,0)-- (1,0);
\end{tikzpicture} $\equiv \bigg[\Delta_{\zeta,\textbf{Tree}}^{2}(k)\bigg]_{\textbf{SR}_{1}}$, and for each contribution ${\cal Q}_{n,\textbf{EFT}}$, we have the remaining diagrammatic relations ${\cal F}$ as follows: \vspace{-1.5em}
\begin{equation}
    \sum {\cal F}_n =  
\begin{tikzpicture}[baseline={([yshift=-5.5ex]current bounding box.center)},very thick]

  \def\radius{0.5}
  \scalebox{0.5}{
  \draw[red, ultra thick] (3*\radius,0) circle (\radius);
  \draw[red,ultra thick] (5*\radius,0) circle (\radius);
  \draw[red,ultra thick] (13*\radius,\radius) circle (\radius);
  \draw[red,ultra thick] (13*\radius,3*\radius) circle (\radius);
  \draw[red,ultra thick] (21*\radius,0) circle (\radius);}

  \draw[black, very thick] (0,0) -- (\radius,0); 
  \draw[blue,fill=blue] (\radius,0) circle (.3ex);
  \draw[blue,fill=blue] (2*\radius,0) circle (.3ex);
  \draw[blue,fill=blue] (3*\radius,0) circle (.3ex);
  \draw[black, very thick] (3*\radius,0) -- (4*\radius,0);
  \node at (4.5*\radius,0) {+};
  \draw[black, very thick] (5*\radius,0) -- (6.5*\radius,0);
  \draw[blue,fill=blue] (6.5*\radius,0,0) circle (.3ex);
  \draw[blue,fill=blue] (6.5*\radius,\radius) circle (.3ex);
  \draw[black, very thick] (6.5*\radius,0,0) -- (8*\radius,0);
  \node at (8.5*\radius,0) {+};
  \draw[black, very thick] (9*\radius,0) -- (10*\radius,0);
  \draw[blue,fill=blue] (10*\radius,0,0) circle (.3ex);
  \draw[red, ultra thick] (10*\radius,0) -- (11*\radius,0);
  \draw[blue,fill=blue] (11*\radius,0,0) circle (.3ex);
  \draw[black, very thick] (11*\radius,0) -- (12*\radius,0);
  \node at (12.5*\radius,0) {+};
  \draw[black, very thick](13*\radius,0) --(14*\radius,0);
  \draw[blue,fill=blue](14*\radius,0,0) circle (.3ex);
  \draw[red,  thick](14*\radius,0) --(14.5*\radius,\radius);
  \draw[blue,fill=blue](14.5*\radius,\radius) circle (.3ex);
\draw[red, thick](14.5*\radius,\radius) --(15.5*\radius,\radius);
\draw[blue,fill=blue](15.5*\radius,\radius) circle (.3ex);
\draw[red, thick](14*\radius,0) -- (14.5*\radius,-\radius);\draw[blue,fill=blue](14.5*\radius,-\radius) circle (.3ex);
\draw[red,  thick](14.5*\radius,\radius) -- (14.5*\radius,-\radius);
\draw[red, thick](14.5*\radius,-\radius) -- (15.5*\radius,-\radius);
\draw[blue,fill=blue](15.5*\radius,-\radius) circle (.3ex);
\draw[blue,fill=blue](14.5*\radius,0) circle (.3ex);
\draw[red, thick](14.5*\radius,0) -- (15.5*\radius,0);
\draw[red,  thick](15.5*\radius,\radius) -- (15.5*\radius,-\radius);
\draw[blue,fill=blue](15.5*\radius,0) circle (.3ex);
\draw[red, thick](15.5*\radius,\radius) -- (16*\radius,0);
\draw[blue,fill=blue](16*\radius,0) circle (.3ex);
\draw[red, thick](16*\radius,0) -- (15.5*\radius,-\radius);
\draw[black, very thick](16*\radius,0) -- (17*\radius,0);
\node at (17.5*\radius,0) {+};
\draw[black, very thick](18*\radius,0) -- (19*\radius,0);
\draw[blue,fill=blue](19*\radius,0) circle (.3ex);
\draw[red, thick](19*\radius,0) -- (19.5*\radius,\radius);
\draw[red, thick](19*\radius,0) -- (19.5*\radius,-\radius);
\draw[blue,fill=blue](19.5*\radius,\radius) circle (.3ex);
 \draw[blue,fill=blue](19.5*\radius,-\radius) circle (.3ex); 
 \draw[red, thick](19.5*\radius,\radius) -- (20.5*\radius,\radius);
 \draw[red, thick](19.5*\radius,-\radius) -- (20.5*\radius,-\radius);
 \draw[blue,fill=blue](20.5*\radius,\radius) circle (.3ex);
 \draw[blue,fill=blue](20.5*\radius,-\radius) circle (.3ex);
 \draw[red, thick](20.5*\radius,\radius) -- (21*\radius,0);
 \draw[red, thick](20.5*\radius,-\radius) -- (21*\radius,0);
\draw[blue,fill=blue](21*\radius,0) circle (.3ex);
\draw[black,  thick](21*\radius,0) -- (22*\radius,0); 
\draw[red, thick](19.5*\radius,\radius) -- (20.5*\radius,-\radius);
\draw[red, thick](20.5*\radius,\radius) -- (19.5*\radius,-\radius);
\node at (22.5*\radius,0) {+}; 
\draw[black, thick](23*\radius,0) -- (24*\radius,0);
\draw[red, thick](24*\radius,0) -- (24.5*\radius,\radius);
\draw[red, thick](24*\radius,0) -- (24.5*\radius,-\radius);
\draw[blue,fill=blue](24*\radius,0) circle (.3ex);
\draw[blue,fill=blue](24.5*\radius,\radius) circle (.3ex);
\draw[blue,fill=blue](24.5*\radius,-\radius) circle (.3ex);
\draw[red, thick](24.5*\radius,\radius) -- (25.5*\radius,\radius);
\draw[red,  thick](24.5*\radius,-\radius) -- (25.5*\radius,-\radius);
\draw[red,  thick](25.5*\radius,\radius) -- (26*\radius,0);
\draw[red, thick](26*\radius,0) -- (25.5*\radius,-\radius);
\draw[blue,fill=blue](25.5*\radius,\radius) circle (.3ex);
\draw[blue,fill=blue](26*\radius,0) circle (.3ex);
\draw[blue,fill=blue](25.5*\radius,-\radius) circle (.3ex);
\draw[red,  thick](24.5*\radius,\radius) -- (24.5*\radius,-\radius);
\draw[red, thick](25*\radius,\radius) -- (25*\radius,-\radius);
\draw[red, thick](25.5*\radius,\radius) -- (25.5*\radius,-\radius);
\draw[blue,fill=blue](25*\radius,\radius) circle (.3ex);
\draw[blue,fill=blue](25*\radius,-\radius) circle (.3ex);
\draw[black, thick](26*\radius,0) -- (27*\radius,0); 
\node at (28*\radius,0) {+ $\cdots$};
\end{tikzpicture}
\end{equation}
where the factors $g_{n}(w)$ perform the coarse-graining operation of the amplitude to maintain the perturbative approximations across all wavenumber ranges. Note that the notation used in the above equation represents a modified version of a similar quantity from eqn.(\ref{simpleDRG}) for a single $(n=1)$ transition and hence the notation used will refer to the present modified version in the future discussions.The first equality demonstrates that we can add the DRG resummed results coming from each phase of sharp transition under the influence of a Heaviside theta function $\Theta(k-k_{s_{n}})$ acting at the scale of the $n$th transition $k_{s_{n}}$. Notice the $n$ and $w$ dependence in the factors $g_{n}(w)$ for each contribution of the sharp transition phase from the total of $N$ such phases. Just as done in eqn.(\ref{simpleDRG}), for the case of $n>1$ number of transitions, we can write the following simplified version:
\bea \label{simpleDRGn}
\overline{\overline{\Delta_{\zeta,\textbf{EFT}}^{2}(k)}} &\approx& \bigg[\Delta_{\zeta,\textbf{Tree}}^{2}(k)\bigg]_{\textbf{SR}_{1}} \bigg[\left(\frac{k_{e_{1}}}{k_{s_{1}}}\right)^{-2.3\delta_{1,*}} + \Theta(k-k_{s_{2}})g_{2}(w)\left(\frac{k_{e_{2}}}{k_{s_{2}}}\right)^{-2.3\delta_{2,*}} + \cdots \nonumber\\
&& \quad\quad\quad\quad\quad\quad\quad\quad\quad \cdots + \Theta(k-k_{s_{n}})g_{n}(w)\left(\frac{k_{e_{n}}}{k_{s_{n}}}\right)^{-2.3\delta_{n,*}} \bigg]
\times \bigg\{1+{\cal O}\bigg[\Delta_{\zeta,\textbf{Tree}}^{2}(k_{*})\bigg]^{2}_{\textbf{SR}_{1}}\bigg\},\nonumber\\
&=& \bigg[\Delta_{\zeta,\textbf{Tree}}^{2}(k)\bigg]_{\textbf{SR}_{1}} \bigg[\left(\frac{k_{e_{1}}}{k_{s_{1}}}\right)^{-2.3\delta_{1,*}} + \sum_{n=1}^{N}\Theta(k-k_{s_{n}})g_{n}(w)\left(\frac{k_{e_{n}}}{k_{s_{n}}}\right)^{-2.3\delta_{n,*}}\bigg] \nonumber \\ 
&& \quad \quad \quad \quad \quad \quad \quad \quad \quad \quad \quad \quad \quad \quad \quad \quad \quad \quad \quad \quad \quad \quad \quad \quad \quad \quad  \times
\bigg\{1+{\cal O}\bigg[\Delta_{\zeta,\textbf{Tree}}^{2}(k_{*})\bigg]^{2}_{\textbf{SR}_{1}}\bigg\}.\quad\quad
\eea

\subsubsection{Comment on the EoS dependent coarse-graining involved in DRG}
\label{tab1x}
\begin{table}[H]
\centering
\begin{tabular}{|l|l|l|}

\hline\hline
\multicolumn{3}{|l|}{\normalsize \textbf{Behaviour of the EoS dependent coarse-graining $g_{n}(w)$}} \\

\hline

\textbf{EoS parameter} & \textbf{Coarse-graining} & \textbf{Condition on $g_{n}(w)$} \\
\hline
$w=1/3$ & Zero & $g_{n}(w=1/3)=1$  \\  \hline

$1/3 \leq w \leq 1$ & Negative & $g_{n+1}(w) < g_{n}(w)$                                           \\ \hline
$0< w < 1/3$ & Positive & $g_{n+1}(w) > g_{n}(w)$                                          \\ \hline
$-0.05 < w \leq 0$ & Positive & $g_{n+1}(w) > g_{n}(w)$                                          \\ \hline
$-0.55 \leq w \leq -0.05$ & Positive & $g_{n+1}(w) > g_{n}(w)$                                       \\ \hline\hline

\end{tabular}
\caption{Nature of the coarse-graining and its values for various EoS parameters.}
\end{table}

Here we will stress the nature of the phenomenological coarse-graining parameters depending on the background EoS $w$, while keeping our focus on four specific cases involving $w=1/3\;,1/3 < w \leq 1\;,0 < w < 1/3\;,{\rm and}\;w \leq 0$ with two subcases. We proceed to demonstrate each case in the order mentioned here. Table \ref{tab1x} shows the signature of the coarse-graining factor which when applied to the power spectrum for various scenarios of $w$ brings it to a level necessary to maintain perturbativity and get an optimum amplitude
best for obtaining a viable PBHs mass fraction and generating SIGWs consistent with the NANOGrav 15 signal. The details of the corresponding discussions and its physical outcomes are appended below point-wise:

\begin{enumerate}
    \item \underline{$ w=1/3:$} \\ \\
        This case is argued here to be the most favourable scenario in terms of satisfying the perturbative approximations and giving us results closest to the findings from experiments, as will become clear later in section \ref{s5}. The important point to understand for this case is that no coarse-graining is required when working with $w=1/3$. This implies that the final eqn.(\ref{drgfinal}) transforms for this case in the form:
        \bea \label{finaldrgps1} \overline{\overline{\Delta_{\zeta,\textbf{EFT}}^{2}(k)}} = \bigg[\Delta_{\zeta,\textbf{Tree}}^{2}(k)\bigg]_{\textbf{SR}_{1}} \left[\exp{\left({\cal Q}_{1,\textbf{EFT}}\right)} + \sum_{n=2}^{N}\Theta(k-k_{s_{n}})\exp{\left({\cal Q}_{n,\textbf{EFT}}\right)} \right]\times \bigg\{1+{\cal O}\bigg[\Delta_{\zeta,\textbf{Tree}}^{2}(k_{*})\bigg]^{2}_{\textbf{SR}_{1}}\bigg\},
        \quad\quad \eea 
        where in the above we set $g_{n}(w=1/3)=1$ for each coarse-graining factor involved for the $n$ number of sharp transitions. This symbolizes no coarse-graining and any deviation from the value $g_{n}(w=1/3)=1$, for each $n$, will signal finite coarse-graining features. The simplified version of the above equation is then written as:
        \bea
            \overline{\overline{\Delta_{\zeta,\textbf{EFT}}^{2}(k)}} = \bigg[\Delta_{\zeta,\textbf{Tree}}^{2}(k)\bigg]_{\textbf{SR}_{1}} \bigg[\left(\frac{k_{e_{1}}}{k_{s_{1}}}\right)^{-2.3\delta_{1,*}} + \sum_{n=1}^{N}\Theta(k-k_{s_{n}})\left(\frac{k_{e_{n}}}{k_{s_{n}}}\right)^{-2.3\delta_{n,*}}\bigg]\times
            \bigg\{1+{\cal O}\bigg[\Delta_{\zeta,\textbf{Tree}}^{2}(k_{*})\bigg]^{2}_{\textbf{SR}_{1}}\bigg\}.
        \quad\quad \eea
        Any theory that has too much coarse-graining involved in it becomes to an extent unreliable. In this case, there is an absence of any kind of coarse-graining. This particular change for $w=1/3$ matches with results from this case where the constant scalar power spectrum amplitude property is preserved in the generated SIGWs, as we will see through our results. We conclude that $w=1/3$ is the best scenario to make a reliable prediction from the theoretical side, since no coarse-graining is required to explain the observationally relevant phenomenon of GW experiments and confront the overproduction issue successfully as seen in the results section \ref{s5}. 
    \item \underline{$1/3 < w \leq 1:$} \\ \\
        In this scenario, we observe a different feature needed for the coarse-graining factors as the number of sharp transitions keeps increasing for larger wavenumbers. Here, we will observe that the peak amplitude keeps increasing, and thus, it might lead to breaking the perturbative approximations and incurring overproduction of the resulting PBHs. To avoid such situations, the coarse-graining factors, $g_{n}(w;1/3 < w \leq 1)$, must keep decreasing in magnitude in order to suppress the overall peak amplitude of the $n$th transition. In particular, it should follow the fashion where, $g_{n+1}(w) < g_{n}(w)$ for $1/3 < w \leq 1$ and $n=1\;{\rm to}\;N$ with $N$ total transitions. The last peak suffers the largest negative coarse-graining effect. It might also happen that one cannot obtain the desired PBH masses with having a sizeable abundance denoted by the parameter, $f_{\rm PBH} \in (10^{-3},1)$. Hence, this sort of ``negative'' coarse-graining is required for this case of $w$ values. This interval of $w$ displays a large fine-tuning of the coarse-graining parameters as the transitions increase, and as a result, the Universe conditions generated due to this scenario are ruled out.  
    \item \underline{$0 < w < 1/3:$} \\ \\
        In this scenario, the feature opposite to what we observed in the previous case is required. Here, the peak amplitude keeps decreasing, so no harm gets done to the perturbative approximations within each sharp transition phase, and the overproduction does not appear as an issue. However, if we wish to explain signatures from our resulting SIGW spectrum in the regime probed by the NANOGrav-15 data and other GW experiments detecting at larger frequencies, $f \in {\cal O}(10^{-5}-10^{5}){\rm Hz}$, then peak amplitude within such a regime is not substantial. Therefore, a ``positive'' coarse-graining feature is required such that, $g_{n+1}(w) > g_{n}(w)$ for $0 < w < 1/3$ and $n=1\;{\rm to}\;N$ with $N$ total transitions. The last peak suffers the largest positive coarse-graining effect, while the first peak amplitude requires the least amount to satisfy the NANOGrav-15 data.  
    \item \underline{$-0.05 < w \leq 0:$} \\ \\
        This scenario presents an interesting case. We will show from the results in section \ref{s5} that the peak amplitude, when $-0.05 < w < 0$, is surprisingly just greater than the $w=0$ condition. Hence, during such conditions we must have $g_{n}(-0.05 < w < 0) < g_{n}(w=0)$ for each of the $n=1\;{\rm to}\;N$ transitions in comparison. This case again comes under ``positive'' coarse-graining, as the peak amplitudes tend to decrease quickly with an increasing number of transitions, making the last peak require more of this fine-tuning than the first peak if getting observational signatures is the primary goal. The perturbativity in this case is still maintained since the amplitude is much less comparatively but it is not much of a desirable scenario as far as viewed from the lenses of the experimental signatures.  
    \item \underline{$-0.55 \leq w \leq -0.05:$} \\ \\
        For this last remaining interval, as allowed by the approximations of linearity discussed in section \ref{s4}, the conditions are overall similar to the case (c). The peak amplitudes keep decreasing evermore drastically in magnitude with increasing transitions and thus require a large amount of fine-tuning as we go lower in $w$. This also describes ``positive'' coarse-graining and lowering the values of $w$ decreases the reliability of the predictions. 
\end{enumerate}

\begin{figure*}
\includegraphics[width=18.5cm,height=12.5cm] {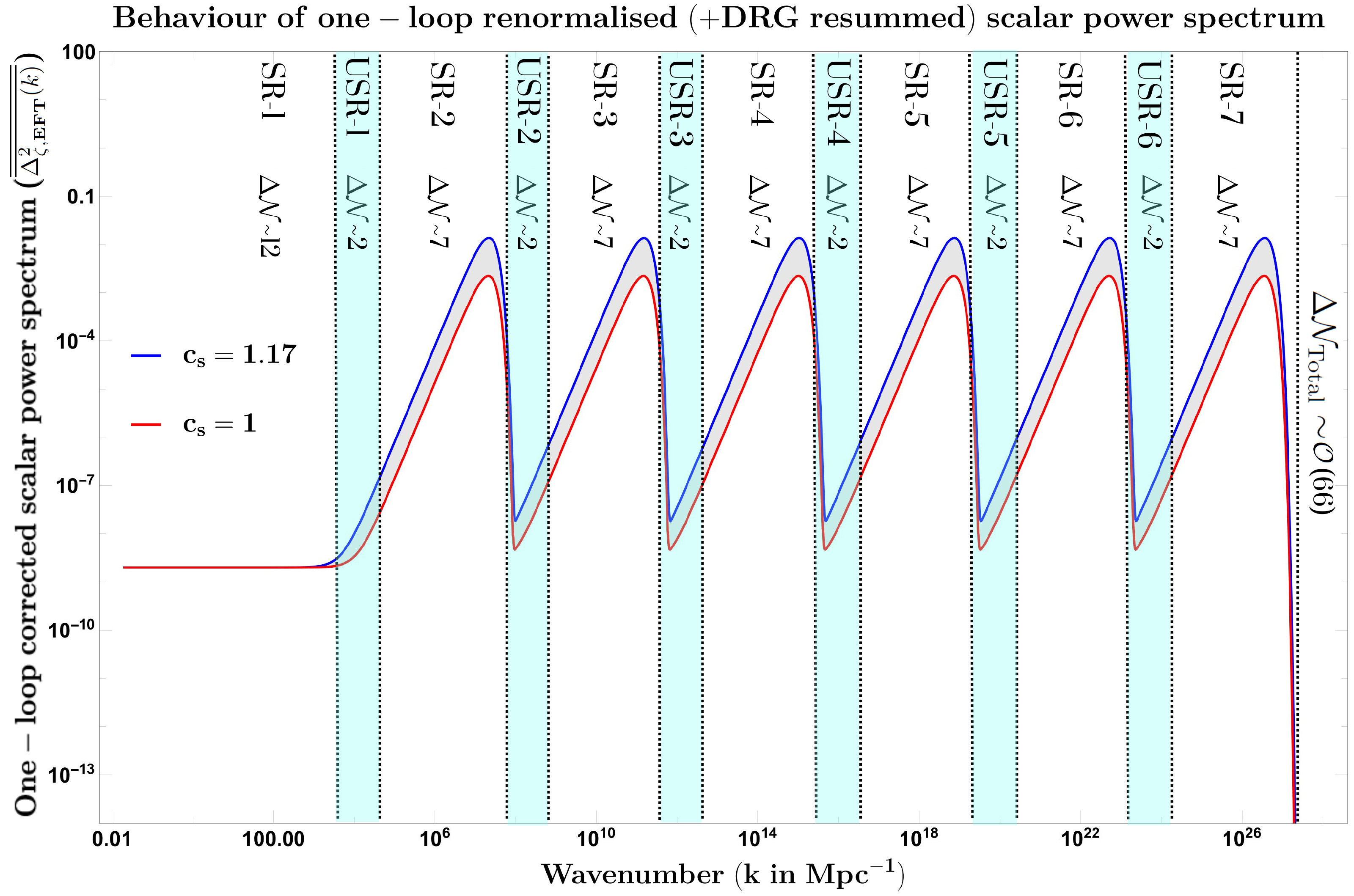}
        \label{w4}
       
    	\caption[Optional caption for list of figures]{The one-loop corrected renormalized and DRG-resummed scalar power spectrum with $c_s =1$ represented by red line and $c_s = 1.17$ by blue line. The ash shaded region displays the allowed parameter space from the theory satisfying $1 \leq c_{s} \leq 1.17$.} 
    	\label{Psplot}
    \end{figure*}

With the final result obtained from eqn.(\ref{finaldrgps1}), we present this one-loop corrected renormalized and DRG resumed scalar power spectrum with the momentum scales for two specific values of $c_s =1\;{\rm and}\;1.17$ in fig.(\ref{Psplot}). This plot has zero coarse-graining factors incorporated in it. As is evident from the graph, there are $6$ sharp transition scales that correspond to the scale of PBH formation. The pivot scale has been set to $k_{*} = 0.02$Mpc$^{-1}$. The USR intervals account for a total number of foldings as $\Delta {\cal N}_{\rm USR_{n}} \approx {\cal O}(2)$ due to strong renormalization and DRG resummation constraints which also does not allow the amplitude of the scalar power spectrum to shoot beyond violating the perturbativity constraints. The individual SR$_{n+1}$ regions cover an e-folding of $\Delta {\cal N}_{\rm SR_{n+1}} \approx {\cal O}(7)$ except for the SR$_1$ phase which accounts for $\Delta {\cal N}_{\rm SR_1}\approx {\cal O}(12)$. 

\subsubsection{Equivalence between DRG resummation and $\delta {\cal N}$ formalism}

    \begin{figure*}[ht!]
    	\centering
   {
   \includegraphics[width=18.5cm,height=12.5cm] {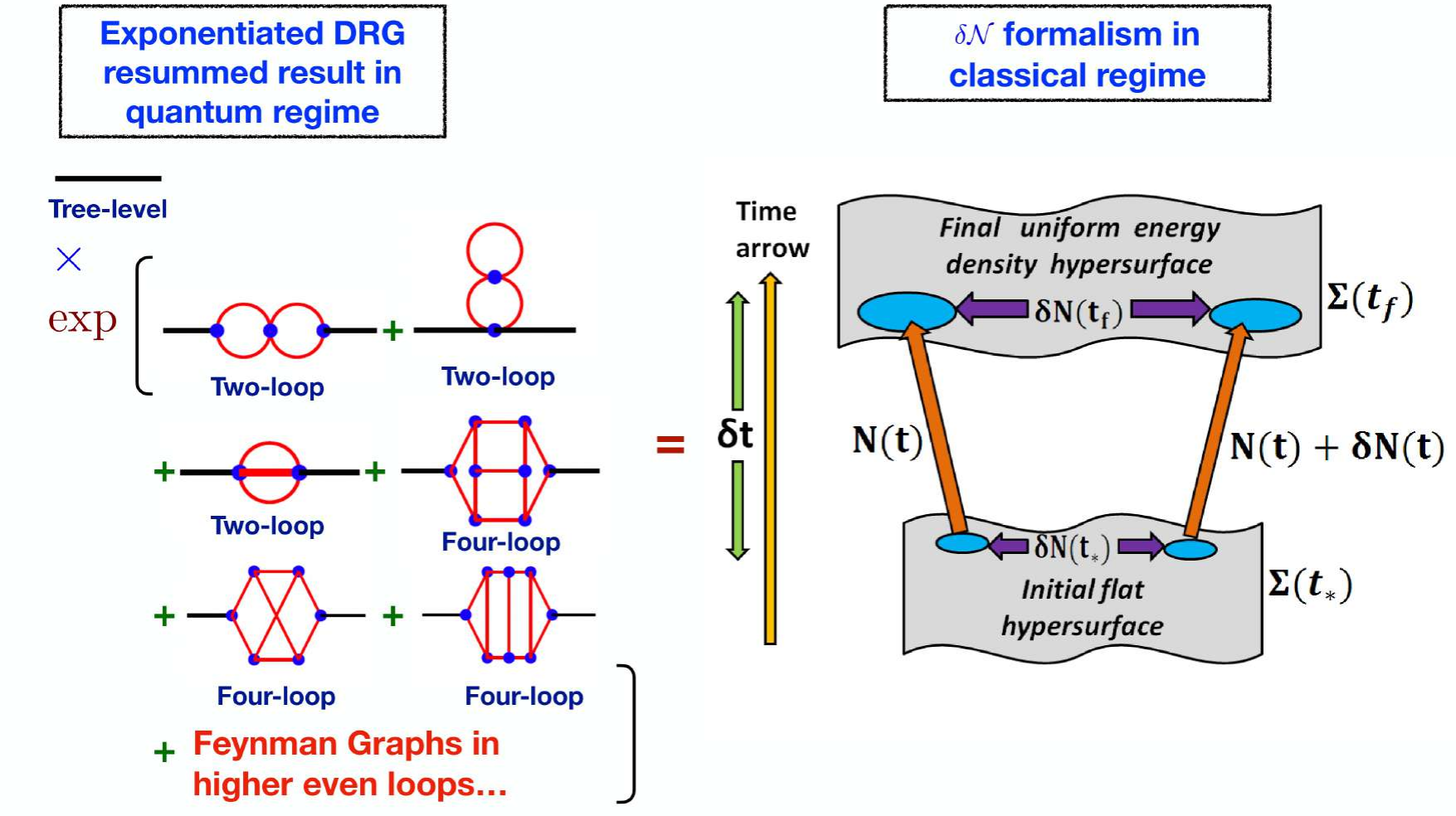}
    }
    	\caption[Optional caption for list of figures]{Representative diagram highlighting the equivalence between the DRG resummation approach and the $\delta{\cal N}$ formalism. } 
    	\label{DRGDelN}
    \end{figure*}
The separate universe approach or the $\delta {\cal N}$ formalism to perturbation theory is an excellent method to determine the time dependence of the cosmological $n$-point correlation functions \cite{Dias:2012qy, Burgess:2009bs, Burgess:2014eoa, Burgess:2015ajz, Chaykov:2022zro, Chaykov:2022pwd, Jackson:2023obv}. Such correlations are evaluated at late times and become our observables. To accurately determine these observables, one requires a set of initial conditions that require knowledge of the subhorizon quantum fluctuations, which later become classical upon horizon crossing. This information is difficult to obtain via the separate universe approach as it works best on large scales. The assertions made from this approach for the cosmological correlators evaluated at late times, $k/aH \ll 1$, highlight that such correlators can be written in terms of their value at an earlier horizon-crossing time and involve specific coefficients along with it. It is later shown in \cite{Dias:2012qy} that the mentioned coefficients carry information about the late-time divergent IR modifications to the necessary correlators at their lowest order. The dynamics of such coefficients can be determined most reliably by using the framework of quantum field theory, specifically, the Callan-Symanzik equations. We want to point out that the coefficients discussed before carrying the IR divergences are shown as analogous to form factors appearing in QCD calculations \cite{Dias:2012qy}. Hence, resummation becomes necessary to account for the infinite divergent contributions from the late time limit in the correlators and present a finite result. Using the renormalization group equations (RGE), one can package the effects from large scales inside the mentioned coefficients, and the initial conditions to solve such equations get generated by the same correlators at horizon-crossing time. Ultimately, the dynamical renormalization group (DRG) analysis leads to an all-order reconstruction of the correlation functions, earlier required from the separate universe approach, purely from quantum field theoretic methods. The DRG resummation allows us to shift the focus on the cosmological $\beta$ functions, namely the spectral tilt, running and running of the running of spectral tilt with the momentum scales, which are the actual physical quantities and have already become free from the one-loop IR divergences at pivot scale after power-spectrum renormalization. From our work, matching the results obtained through DRG resummation with the observational results increases the trustability of the $\delta{\cal N}$ approach in the superhorizon scales as a consequence of this equivalence.

Eqn.(\ref{DRG}) shows a similar treatment of the logarithmic IR divergences discussed in the above paragraph. Each term in the series expansion before participating in resummation into the final exponentiated form, represents the contribution from all even-order loop correction terms, starting with ${\cal Q}_{\textbf{EFT}}$ which resembles a two-loop contribution, then ${\cal Q}_{\textbf{EFT}}^{2}$ representing a four-loop contribution, and so on. This behaviour shows that DRG resummation allows for an all-order reconstruction of the correlations by knowing information about the lowest-order terms in the perturbative expansion. Ultimately, we obtain a much-softened version of the logarithmic IR divergent contributions compared to the softening done by the power spectrum renormalization back in Eqn.(\ref{Qneft}). A diagram illustrating this established equivalence between the two approaches (DRG and $\delta{\cal N}$) is presented in fig.(\ref{DRGDelN}).

\section{Scalar Induced Gravitational Waves with general equation of state ($w$-SIGW) with MST}
\label{s3}

The purpose of this section is to present a review of the theory of scalar-induced gravitational waves. Induced GWs are generated at the second-order through mode couplings of the first-order perturbations to the FLRW metric in the cosmological perturbation theory. This effect generates a significant enhancement in the observed GW spectrum and there must be a large scalar perturbation compared to the fluctuations in the CMB. What we are interested in exploring is the situation where the end of inflation is accompanied by a state where the equation of state is unknown with a value $w$. The induced GW generated in the early universe can exist with the modes deep inside the horizon in this background of general $w$ where they are primarily sourced. 
We write the energy density spectrum of the SIGWs for a general equation of state and propagation speed as:
\bea
\label{Energydensity}
\Omega_{\rm{GW},0}h^2 = 1.62 \times 10^{-5}\;\bigg[\frac{\Omega_{r,0}h^2}{4.18 \times 10^{-5}}\bigg] \bigg[\frac{g_{*}(T_c)}{106.75}\bigg]\bigg[\frac{g_{*,s}(T_c)}{106.75}\bigg]^{-4/3}\Omega_{\rm GW,c}.
\eea
Here $\Omega_{r,0}h^2$ represents the radiation energy density as observed today, and $g_{*},g_{*,s}$ are the energy and entropy effective degree of freedom. $\Omega_{\rm {GW},c}$ denote the fractional energy density of the SIGWs evaluated at a conformal time, in the radiation-dominated universe, when the GW density fraction does not vary.\\

The SIGW spectrum for the scales $k\geq k_{*}$ can be expressed as \cite{Domenech:2021ztg}:
\bea
\label{omegagw}
\Omega_{\rm {GW},c}= \bigg(\frac{k}{k_{*}}\bigg)^{-2b}\int_{0}^{\infty}dv \int_{|{1-v}|}^{1+v} du \; {\cal T}(u,v,b,c_s) \;\;\overline{\overline{\Del_{\zeta,{\bf EFT}}^{2}(ku)}} \times \overline{\overline{\Del_{\zeta,{\bf EFT}}^{2}(kv)}},
\eea
where $b$ is an EoS dependent parameter defined to be as $b=(1-3w)/(1+3w)$. Also, as mentioned before, $k_{*}$ is the usual pivot scale (CMB). Notice the appearance of a general factor of $(k/k_{*})^{-2b}$ in the eqn.(\ref{omegagw}). which disappears, should you plug in the value of EoS for a radiation-dominated era, i.e. $w=1/3$, in the expression. Moreover, throughout this paper, we have performed our analysis considering the pivot scale ($k_{*}$) instead of the reheating scale that is employed in other works of similar kind \cite{Domenech:2020ers, Domenech:2021ztg}. Its implications may look trivial but are fairly significant. A major concern with reheating is that there is an assumption of the equilibrium scales that lie between the GeV and TeV scales. The underlying microphysics and proper thermalization behind this reheating are not completely known. So, it cannot said to be considered completely reliable for precise calculations. To shed some light, you can refer to the previous works performed in thermalization here \cite{Banerjee:2021lqu,Choudhury:2021tuu,Choudhury:2020yaa}. In most cases in the literature, people try to incorporate reheating in an ad-hoc fashion mostly from phenomenology. We are not against the inclusion of such a phenomenological approach but it comes with its demerits, specifically introducing model-dependency into the picture. Thus, you need to be extra careful while performing the computation even when thermalization has been achieved and the system is in equilibrium. Now since the theory of reheating is not well established, we conduct our analysis through normalization w.r.t the pivot scale ($k_{*}$) as it is quite popular from other works and observations such as \cite{Planck:2018jri,Planck:2018vyg,Planck:2015sxf,Choudhury:2015pqa,Choudhury:2014kma,Choudhury:2014sua}. Furthermore, in our previous works \cite{Choudhury:2023vuj, Choudhury:2023jlt, Choudhury:2023rks, Bhattacharya:2023ysp}, considering a radiation-dominated era, we have obtained accurate results with the pivot scale for our estimation.

Now, the transfer function in the Eqn.(\ref{omegagw}) for the general case of constant values of the EoS $w$ and propagation speed $c_{s}$ is given by \cite{Domenech:2021ztg}:
\bea
\label{Transfer func}
{\cal T}(u,v,b,c_s)&=&{\cal N}(b,c_s) \bigg[\frac{4v^2 - (1-u^2+v^2)^2}{4u^2v^2}\bigg]^2 \abs{1-y^2}^b \times \bigg\{\big(P_{b}^{b}(y) + \frac{b+2}{b+1}P_{b+2}^{-b}(y)\big)^2 \Theta(c_s(u+v)-1) \nonumber \\
&& \quad \quad \quad \quad + \frac{4}{\pi^2}\bigg[\big(Q_b ^{-b}(y) + \frac{b+2}{b+1}Q_{b+2}^{-b}(y)\big)^2 \; \Theta(c_s (u+v)-1) \nonumber \\
&& \quad \quad \quad \quad  + \big( {\cal Q}_{b}^{-b}(-y)+ 2\frac{b+2}{b+1}{\cal Q}_{b+2}^{-b}(-y) \; \Theta(1-c_s(u+v))\bigg]\bigg\}.
\eea
A propagation speed dependent parameter $y$ has been introduced for simplicity and given as :
\bea
y\equiv 1- \frac{1-c_s^2(u-v)^2}{2c_s^2uv}.
\eea
The expression also contains a normalized EoS dependent parameter given as :
\bea
{\cal N}(b,c_s)&\equiv& \frac{4^{2b}}{3c_{s}^{4}}\Gamma \bigg(b+\frac{3}{2}\bigg)^4
\bigg(\frac{b+2}{2b+3}\bigg)^2 (1+b)^{-2(1+b)}. \eea
The transfer function in the expression for the SIGW spectrum is expressed in terms of associated Legendre functions $P_{\nu}^{\mu}(x)$ and $Q_{\nu}^{\mu}(x)$ which are given by : 
\bea
P_{\nu}^{\mu}(x) &=& \bigg
[\frac{1+x}{1-x}\bigg]^{\mu /2} \; {\bf F}\bigg(v+1,-v;1-\mu;\frac{1}{2}(1-x) \bigg) ,\nonumber \\
Q_{\nu}^{\mu}(x) &=& \frac{\pi}{2\sin{(\mu\pi)}} \bigg[\cos{(\mu \pi)}\bigg(\frac{1+x}{1-x}\bigg)^{\mu /2} {\bf F}\bigg(v+1,-v;1-\mu;\frac{1}{2}(1-x)\bigg) \nonumber\\
&& \quad \quad \quad \quad \quad \quad \quad - \frac{\Gamma (v+\mu +1)}{\Gamma (v-\mu +1)}\bigg(\frac{1-x}{1+x}\bigg)^{\mu /2} {\bf F}\bigg(v+1,-v;1+\mu;\frac{1}{2}(1-x)\bigg )\bigg].
\eea
which includes the new notation where ${\bf F}(a,b;c;x) = F(a,b;c;x)/\Gamma(c)$. Here $F(a,b;c;x)$ is the Gauss's Hypergeometric function that is scaled with $\Gamma(c)$ to obtain the new function ${\bf F}(a,b;c;x)$. These above functions are known as Ferrer's functions and their validity extends to $\abs{x}<1$. However, when $\abs{x}>1$, we have the Olver's function given by :

\bea
\label{calQ}
{\cal Q}_{\nu}^{\mu}(x) &=& \frac{\pi}{2 \sin{\mu \pi}\Gamma(\nu + \mu +1)}\bigg[\bigg(\frac{x+1}{x-1}\bigg)^{\mu /2}{\bf F}\bigg(v+1,-v;1-\mu;\frac{1}{2}-\frac{1}{2}x \bigg)\nonumber \\
&& \quad \quad \quad \quad \quad \quad \quad \quad \quad \quad \quad \quad  -\frac{\Gamma (v+\mu +1)}{\Gamma (v-\mu +1)}\bigg(\frac{x-1}{x+1}\bigg)^{\mu /2}{\bf F}\bigg(v+1,-v;1-\mu;\frac{1}{2}(1-x)\bigg) \bigg].
\eea
Notice however that $P_{\nu}^{\mu}(x)$ is only valid for $\abs{x}<1$, and not for $\abs{x}>1$. This is because of the nature of Bessel's functions from which it is derived. This function $P_{\nu}^{\mu}(x)$ converges for $\abs{x}<1$ i.e. valid within the interval $(-1,1)$. It is not well-behaved for $\abs{x}>1$. The mathematical details concerning the detailed properties of these functions can be found in \cite{Domenech:2021ztg}.  Now that we have provided the definitions that are crucial and relevant for the discussions in this paper, we turn to check different scenarios of the equation of state and the form of the transfer function that comes out as a result.

\subsection{Radiation dominated case ($w=1/3$)}
In the case of radiation domination, we have $b=0$. In that case, the transfer function reads as:

\bea
{\cal T}_{\rm RD} (u,v,c_s, w=1/3)= \frac{y^2}{3c_s ^4}\bigg(\frac{4v^2 - (1-u^2 +v^2)^2}{4u^2v^2}\bigg)^2 \times \bigg[\frac{\pi ^2 y^2}{4}\Theta[c_s(u+v)-1]+\bigg(1-\frac{1}{2}y \ln \bigg|\frac{1+y}{1-y}\bigg|\bigg)^2\bigg].
\eea

So we can see that the general transfer function reduces to the above expression which is consistent with the results obtained in \cite{Kohri:2018awv}.

\subsection{Kination (Stiff fluid) dominated case ($w=1$)}

 A typical instance where this kind of scenario occurs is the quintessential inflation. The transfer function in this case is given by :

\bea
{\cal T}_{\rm KD}(u,v,c_s,w=1)&=&\frac{4}{3 \pi c_s^4 \abs{1-y^2}}\bigg[\frac{4v^2 - (1-u^2 +v^2 )^2}{4u^2v^2}\bigg]^2 \nonumber \\
&& \times \bigg[(1+3y^2)\; \Theta[c_s(u+v)-1]+\big(1-3y^2+3y\sqrt{\abs{1-y^2}}\big)^2\; \Theta[1-c_s(u+v)]\bigg].
\eea
The transfer function obtained here is consistent with results from \cite{Domenech:2019quo}.
 From our analysis presented in the following section using the MST set-up, we have found that this type of scenario is not favorable when considering the observational constraints from NANOGrav-15 data. However, from a general physical perspective, kination domination should be investigated as it might prove to be of importance in other contexts and so should not be ruled out. 

\subsection{Soft Fluid Dominated case ($w=1/9$)}
This is a case with $b=1/2$ and can be seen because of a scalar field rolling down an exponential potential  \cite{Granda:2019jqy}. The transfer function in this case is given by :
\bea
{\cal T}_{\rm SFD}(u,v,c_s,w=1/9)&=& \frac{2^8}{3^8 \pi c_s^4}\bigg[\frac{4v^2 - (1-u^2+v^2)^2}{4u^2v^2}\bigg] \times \bigg[(4+45y^2)\; \Theta [c_s(u+v)-1]\nonumber \\
&& \quad \quad \quad \quad +\big(y(3-10y^2)+(2+10y^2)\sqrt{\abs{1-y^2}}\big)^2 \Theta[1-c_s(u+v)]\bigg].
\eea
Again, this can be considered theoretically but by imposing constraints from the NANOGrav-15 data disallows it as a suitable scenario.

\subsection{Matter dominated case ($w=0$)}

Here we discuss the pressure-less fluid for which $w=0$. Induced gravitational waves generated from the radiation-dominated era have been thoroughly explored but recent studies have seen an increased attention to GW signal from the early Matter-dominated era \cite{Pearce:2023kxp}. A scalar field oscillating coherently around the bottom of a potential can be categorized in this section. 
Now, The transfer function for $w=0$ reduces to :
\bea
{\cal T}_{\rm MD}(u,v,c_s,w=0)&=& \frac{3^3 5^2}{2^{14}c_s ^4}\bigg[\frac{4v^2-(1-u^2+v^2)^2}{4u^2v^2}\bigg]^2  \times \bigg[\frac{\pi ^2}{4}(1-y^2)^2 (1+3y^2)^2 \Theta[c_s (u+v)-1]\nonumber \\
&& \quad \quad \quad \quad +\bigg(y(1-3y^2)-1/2(1+2y^2-3y^4)\ln{\abs{\frac{1+y}{1-y}}}\bigg)^2 \bigg].
\eea

\subsection{Negative EoS dominated case ($w<0$)}
This is an interesting case to consider. Here we find that the analytical calculations for the kernel to obtain induced GW spectrum becomes complicated. This is due to modified Bessel's functions that appear as solutions of the gravitational potential and tensor modes at linear order in Cosmological Perturbation theory. Hence, it is not wise to comment on anything here without substantial evidence since this particular scenario is a bit tricky. There is a healthy discussion in the sections to follow where we have analyzed the different cases of EoS. For the time being, we present the transfer function by considering the case of $w=-1/9$ (for convenient reasons \footnote{Upon analysis, at this value, it is still possible to explain the PBH mass range obtained from the NANOGrav-15 data. Diving below this negative mark steers the results away from expectations. Detailed analysis of this has been done in the section \ref{s5} to follow)}). 

\bea
{\cal T}_{\rm NEoS}(u,v,c_s,w=-1/9) &=& \frac{5^2 7^2}{2^8 3^9 c_s ^4}\bigg(\frac{4v^2 -(1-u^2+v^2)^2}{4u^2 v^2}\bigg)^2 \times \bigg[\frac{225 \pi^2}{4}(1-y^2)^2 (1+y^2-2y^4)^2 \;\Theta [c_s(u+v)-1] \nonumber \\ 
&& \quad \quad \quad \quad \quad \quad + \bigg(y(9+35y^2 - 30y^4)+\frac{15}{2}(1-3y^4+2y^6)\ln{\abs{\frac{1+y}{1-y}}\bigg)^2}\bigg].
\eea

\section{Primordial Black Hole Mass Fraction from MST} \label{s4}

In this section, we will brief the theory that leads to calculating the PBH abundance using the Press-Schechter Formalism. As mentioned, PBH formation occurs when the curvature perturbations gravitationally collapse upon horizon re-entry. Precisely, the formation will occur on those Hubble patches for which the overdensities created during inflation exceed a specific threshold value $\delta_{\rm th}$. Now as per Carr's criterion ($c_s ^2 =1)$ \cite{1975ApJ...201....1C}, the variation of $\delta_{\rm th}$ w.r.t the EoS $w$ is given by :
\bea
\label{deltath}
\delta_{\rm th} = \frac{3(1+w)}{5+3w}.
\eea
For background with a general equation of state $w$, the dependence of the mass of PBHs corresponding to their respective comoving scale of formation is \cite{Alabidi:2013lya}: 
\bea
\label{mpbh}
M_{\rm PBH} = 1.13 \times 10^{15} \times \bigg[\frac{\gamma}{0.2}\bigg]\bigg[\frac{g_{*}}{106.75}\bigg]^{-1/6}\bigg[\frac{k_{*}}{k_{\rm s}}\bigg]^{\frac{3(1+w)}{1+3w}} M_{\odot}.
\eea
where $M_{\odot}$ represents the solar mass and  $k_{\rm s}$ denotes the wavenumber corresponding to PBH formation. As mentioned in section \ref{s1}, we will base our estimations on the pivot scale $k_{*}$ rather than the reheating scale since it provides a model-independent approach.
At this point, it is crucial to mention that we conduct our analysis by working in the linear regime, which is characterized by the assumption of a mostly linear behaviour for the density contrast in the superhorizon scales and which leads to the calculation of the threshold to lie in the interval, $2/5 \leq \delta_{\rm th} \leq 2/3$ \cite{Musco:2020jjb}.    So, using the eqn.(\ref{deltath}), the corresponding range of values of $w$ in this region is $-5/9 \leq w \leq 1/3$. The linear regime involves the following approximation between the density contrast and the comoving curvature perturbation on the superhorizon scale:
\bea
\delta(t,\mathbf{x}) \cong \frac{2(1+w)}{5+3w}\left(\frac{k}{aH}\right)^{2}\nabla^{2}\zeta(k).
\eea
From here, we move on to write the expression for the variance of the density perturbations smoothed over the scale of mass M,
\bea
\sigma_{\rm M_{\rm PBH}} ^2 = \bigg[\frac{2(1+w)}{5+3w}\bigg]^2 \int \frac{dk}{k} \; (k
R)^4 \; W^2(kR) \; \overline{\overline{\triangle ^2 _{\zeta, \rm \bf EFT}(k)}}.\eea
Here, $W(kR)$ is the Gaussian window function given by $\exp{(-k^2 R^2 /4)}$, used as a smoothing function over the scales of PBH formation $R=1/(\Tilde{c_s} k_{\rm PBH})$.
The term $\Tilde{c_s}$ is the effective sound speed parameter at the transition scale from SR to USR phase and USR to the succeeding SR phase. The specific parameterization for this sound speed is detailed in section \ref{s2}. In addition, $\overline{\overline{\triangle ^2 _{\zeta, \rm \bf EFT}(k)}}$ is the one-loop corrected renormalized DRG resummed scalar power spectrum. Previously in \cite{Bhattacharya:2023ysp}, with the MST set-up, we had calculated the PBH mass fraction for the case of the radiation-dominated era with $w=1/3$. However, now we proceed to include the $w$ dependence on the current PBH mass fraction, which reads as \cite{Sasaki:2018dmp}:
\bea
\beta_{\rm M_{\rm PBH}} = \gamma \times \frac{\sigma_{\rm M_{\rm PBH}}}{\sqrt{2\pi}\delta_{\rm th}}\exp{\bigg(\frac{-\delta_{\rm th}^2}{2\sigma_{\rm M_{\rm PBH}}^2}\bigg)},
\eea
where $\gamma \sim 0.2$ is the effective factor of collapse and helps define the fraction of mass inside the Hubble Horizon which collapses to form PBHs.
It is made clear from the definition that the $w$ dependence in the above equation gets encoded in the variance. In addition, it is also important to clarify that the above result for the mass fraction assumes our use of the Gaussian statistics for the density perturbations and we omit any presence of non-gaussianities and non-linearities while writing these expressions. We will consider them in the scope of future work. Now, we move on to write the present day fractional abundance of the PBHs,
\bea
f_{\rm PBH} \equiv \frac{\Omega_{\rm PBH}}{\Omega_{\rm CDM}}= 1.68\times 10^8 \times \bigg(\frac{\gamma}{0.2}\bigg)^{0.5} \times \bigg(\frac{g_{*}}{106.75}\bigg)^{-0.25} \times \left(M_{\rm PBH}\right)^{-\frac{6w}{3(1+w)}}\times \beta_{M_{\rm PBH}},\eea
where $g_{*}$ represents the relativistic degrees of freedom. Another relation helpful for our analysis is the conversion between the frequency $f$(Hz) of the GW that are also sensitive to the NANOGrav 15 signal and the related wavenumber $k(\rm Mpc^{-1}$):
\bea
f = \frac{k}{2\pi} \simeq 1.6\times 10^{-9}{\rm Hz}\left(\frac{k}{10^{6}{\rm Mpc^{-1}}}\right).
\eea
In the section to follow, we have discussed within an MST setup, how the abundance $f_{\rm PBH}$ is affected by the variation of EoS, and our analysis will prove to be of vital importance as it will reinforce the already existing idea about which of the cases of EoS in section \ref{s3} is more favorable than the other.

\section{Numerical outcomes and Discussions} \label{s5}

    \begin{figure*}[hb!]
    	\centering
   {
   \includegraphics[width=18.5cm,height=12.5cm] {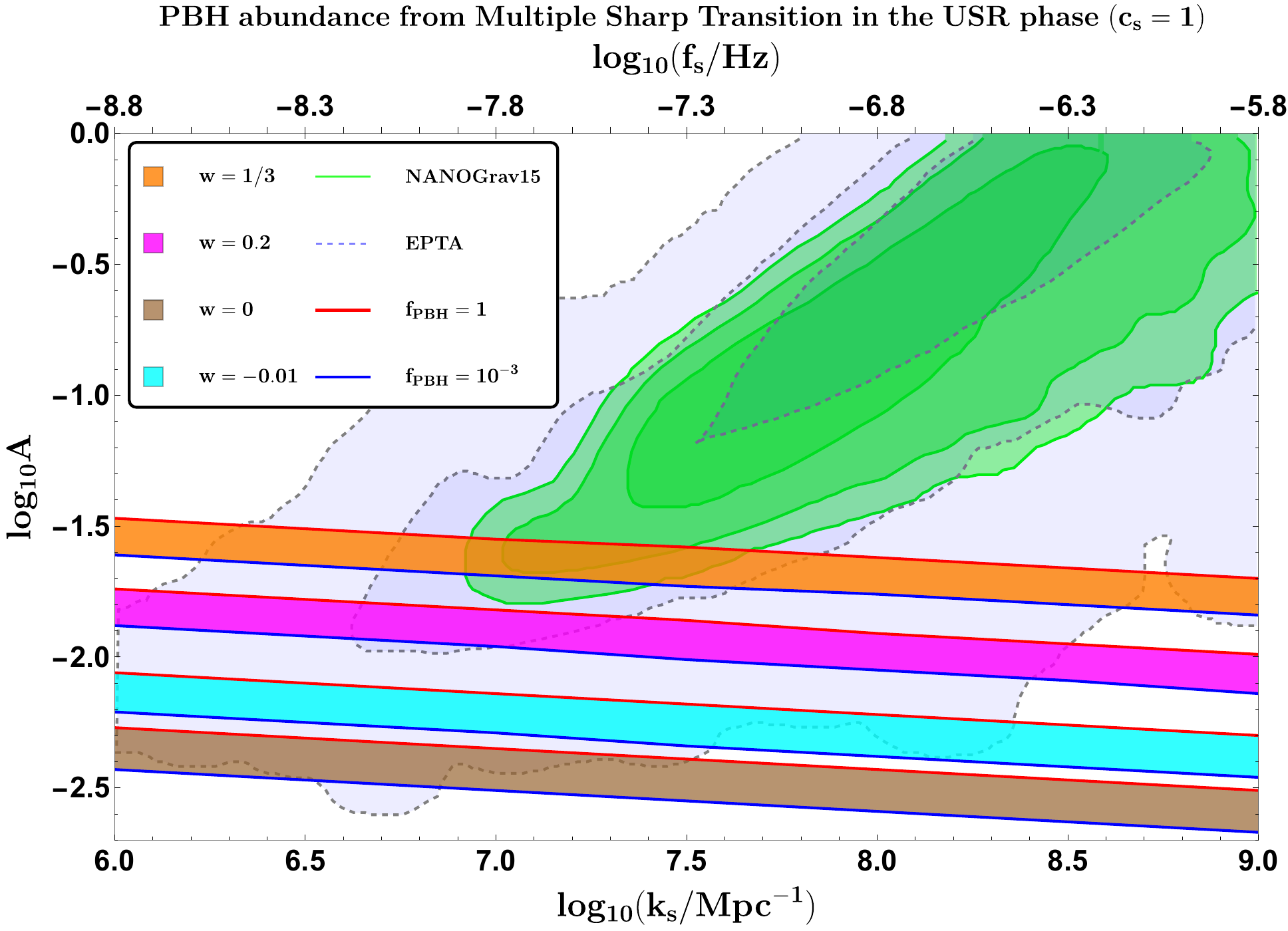}
    }
    	\caption[Optional caption for list of figures]{Figure depicts the variation of amplitude $A$ of the one-loop renormalized and DRG-resummed scalar power spectrum with the transition wavenumber, which avoids PBH overproduction in the MST setup. The effective sound speed parameter is fixed to take $c_{s}=1$. Red and blue lines enclose the region of sizeable abundance $f_{\rm PBH} \in (1,10^{-3})$. Orange, Magenta, Cyan, and Brown colours denote the specific regions related to the EoS parameter $w \in \{1/3,0.2,0,-0.01\}$. The posteriors for the NANOGrav 15 and EPTA data, represented in the green and light-blue colors, respectively, are taken from \cite{Franciolini:2023pbf}.} 
    	\label{overprodcs1}
    \end{figure*}

In this section, we will examine the outcomes related to the diverse challenges we initially aimed to tackle. One of the critical issues we seek to address is the PBH overproduction. PBHs are formed as the curvature perturbations gravitationally collapse upon re-entering the horizon. Now when the perturbations are too large, it could lead to a huge number of PBHs formation even exceeding the present dark matter abundance. This raises a concern because PBHs are thought to be possible candidates for relic dark matter. The abundance of PBH becoming more than dark matter implies that all of the dark matter that we see today is made of PBHs which is against the observational imprints. Therefore, we simply have to refine our models in such a way that this overproduction issue is dealt with. One way to address this issue is by looking through the lens of the EoS parameter, which we have studied in this paper. Another approach is to incorporate non-linearity and non-Gaussianity \cite{Choudhury:2023fwk,Franciolini:2023pbf,Franciolini:2023wun,LISACosmologyWorkingGroup:2023njw, Inui:2023qsd,Chang:2023aba, Gorji:2023ziy,Li:2023xtl, Li:2023qua,Firouzjahi:2023xke}. In addition to these, we also have yet another method to resolve the subject issue which is the introduction of a spectator field \cite{Gorji:2023sil,Ota:2022xni}. The ultimate solution, which should also encompass a comprehensive analysis, must take into account linearities, non-linearities, non-Gaussianities, and the Equation of State (EoS) parameter. We plan to delve into the latter aspect in our upcoming research work. 

    \begin{figure*}[htb!]
    	\centering
   {
   \includegraphics[width=18.5cm,height=12.5cm] {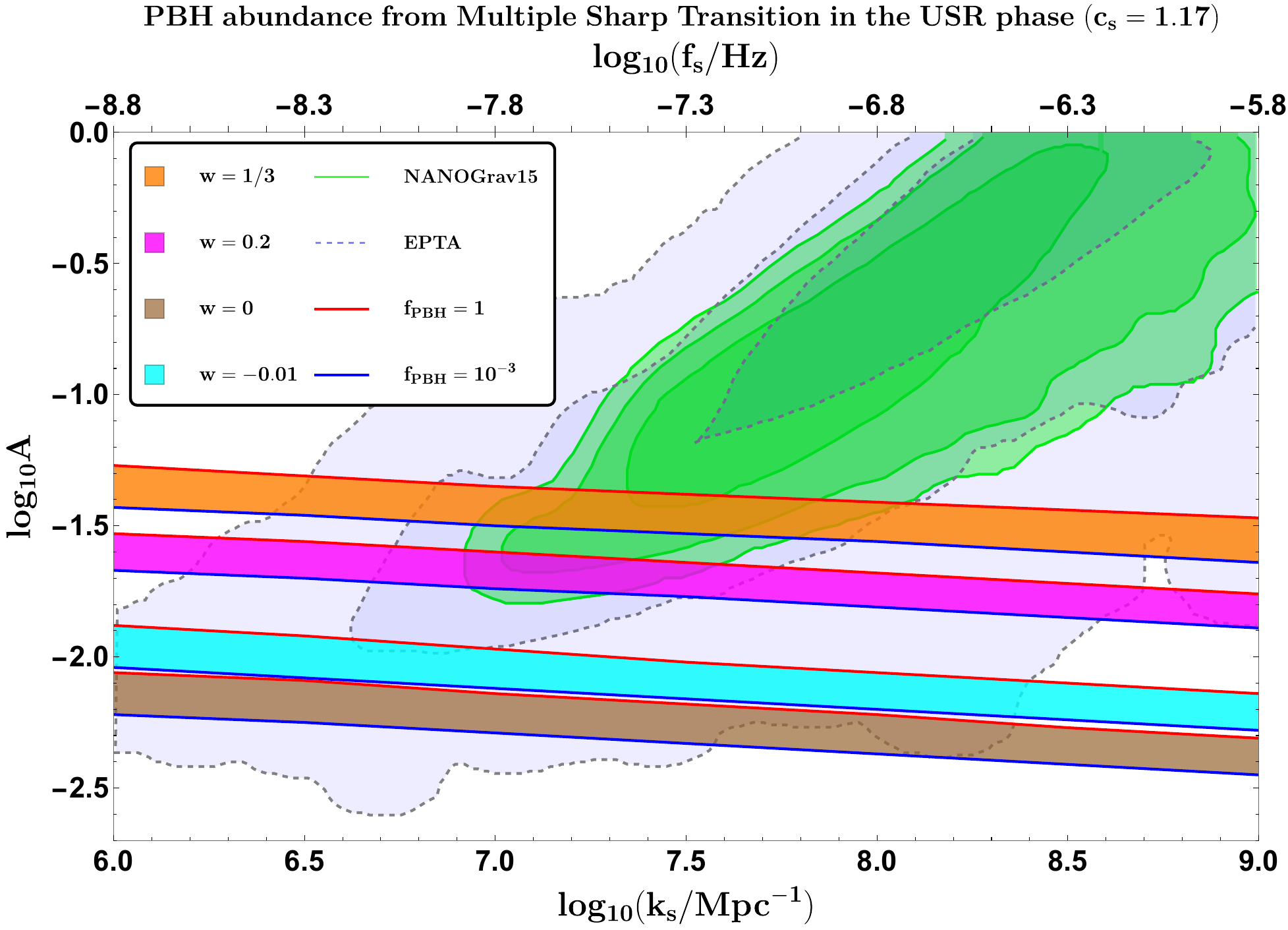}
    }
    	\caption[Optional caption for list of figures]{Figure depicts the variation of amplitude $A$ of the one-loop renormalized and DRG-resummed scalar power spectrum with the transition wavenumber, which avoids PBH overproduction in the MST setup. The effective sound speed parameter is fixed to take $c_{s}=1.17$. Red and blue lines enclose the region of sizeable abundance $f_{\rm PBH} \in (1,10^{-3})$. Orange, Magenta, Cyan, and Brown colours denote the specific regions related to the EoS parameter $w \in \{1/3,0.2,0,-0.01\}$. The posteriors for the NANOGrav 15 and EPTA data, represented in the green and light-blue colors, respectively, are taken from \cite{Franciolini:2023pbf}.} 
    	\label{overprodcs117}
    \end{figure*}

The highlight of this paper is the dependence of the results on the equation of state parameter i.e. which value of $w$ will be more favored than the others. With this in mind, we begin with our results that discuss the effects of having an arbitrary but instantaneous constant EoS parameter $w$ (i.e. constant at a particular slice of time) on the abundance of the PBH, which are formed corresponding to the frequencies that are also sensitive to the NANOGrav 15 signal. Specifically, using the Eqn.(\ref{mpbh}), we vary $k_{s}$ within the range of values that lie inside the NANOGrav-15 frequency spectrum i.e. within, and
compute the range of amplitude that provides us with the suitable PBH abundance for different $w$'s and thereafter tally with the NANOGrav and EPTA data curves. We shall follow this up with the analysis of the obtained values of abundance with the microlensing experiments \footnote{Experiments such as Subaru HSC, OGLE, and MACHO/EROS provide for the study of the nature of dark matter via the gravitational microlensing phenomenon when focusing on their potential galactic sources. They currently offer constraints on the abundance and mass of PBHs as potential dark matter candidates, which include a planet-size mass range of $M_{\rm PBH} \sim {\cal O}(10^{-6}-10^{-2})M_{\odot}$ also consistent with the allowed range from the recent NANOGrav results.} \cite{Niikura:2017zjd,Niikura:2019kqi,EROS-2:2006ryy}. 
Finally, we will present the results of the SIGW spectra for the specific benchmark values of EoS $w$ parameter where such spectra are confronted with the recent NANOGrav 15 and EPTA signal, along with its signature in other ground and space-based experiments \cite{LISA:2017pwj, Kawamura:2011zz, Punturo:2010zz, Reitze:2019iox, Crowder:2005nr, LIGOScientific:2014pky, VIRGO:2014yos, KAGRA:2018plz}.

    \begin{figure*}[htb!]
    	\centering
    \subfigure[]{
      	\includegraphics[width=8.5cm,height=7.5cm] {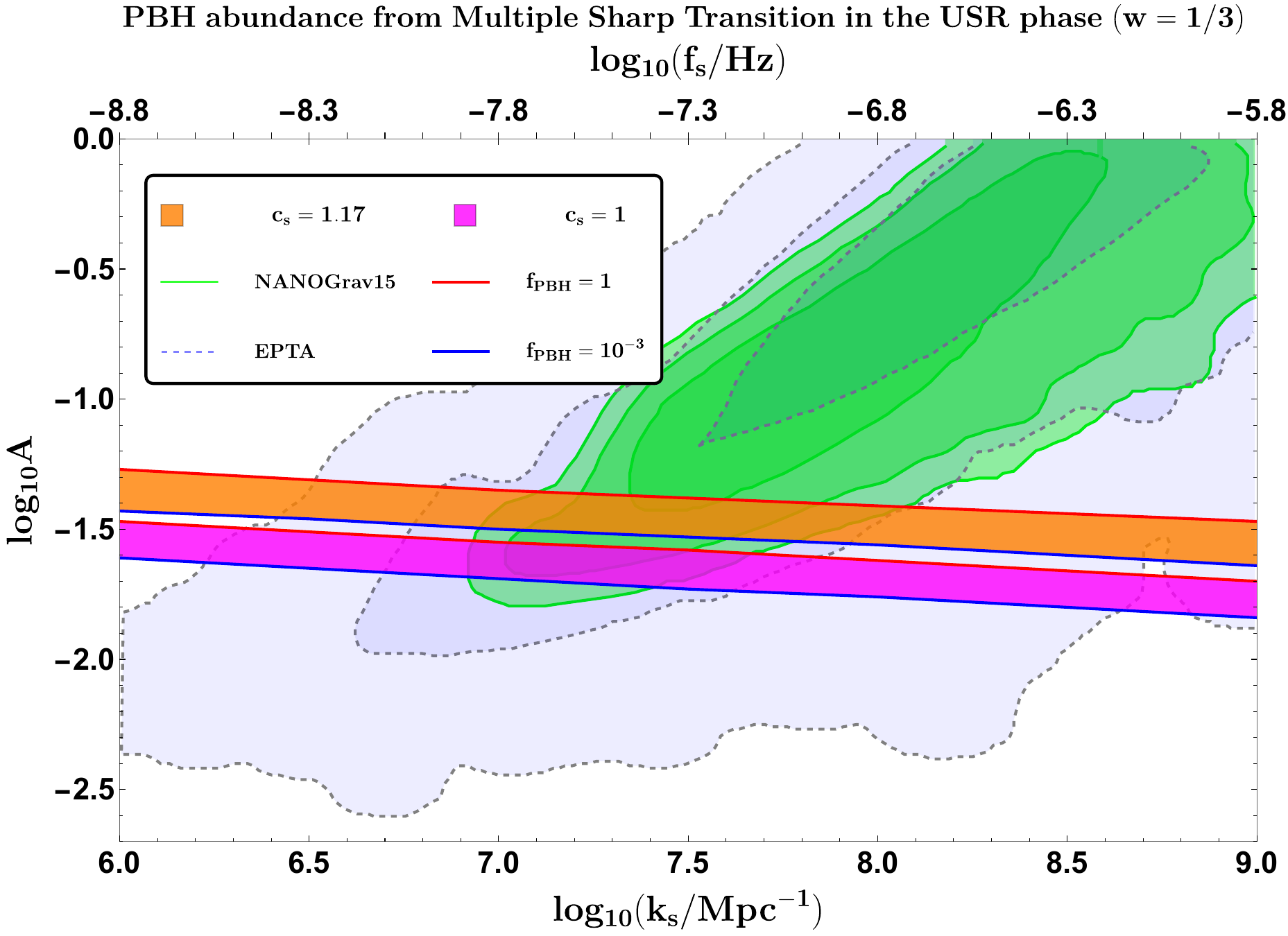}
        \label{w1}
    }
    \subfigure[]{
        \includegraphics[width=8.5cm,height=7.5cm] {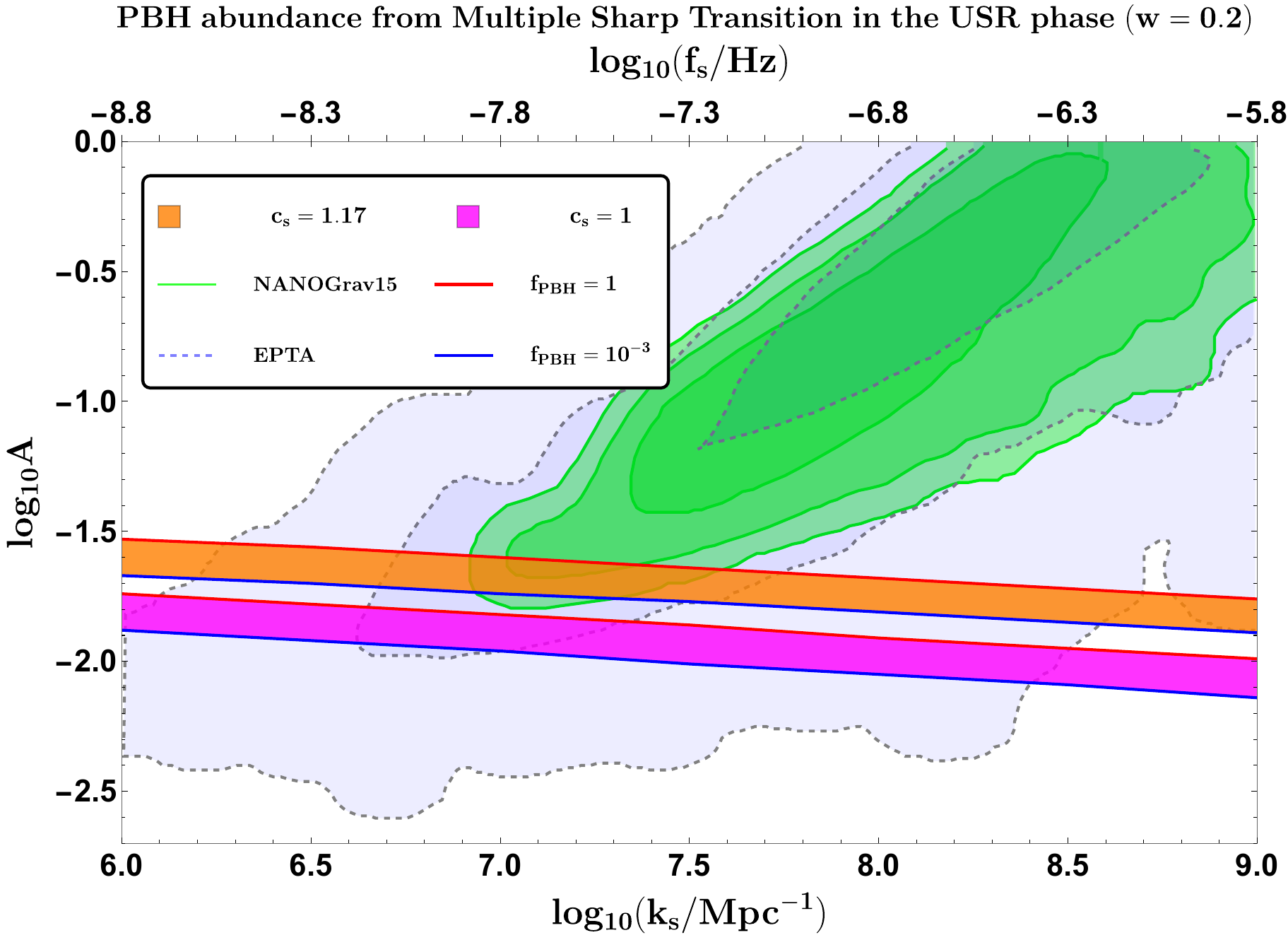}
        \label{w2}
       }
        \subfigure[]{
        \includegraphics[width=8.5cm,height=7.5cm] {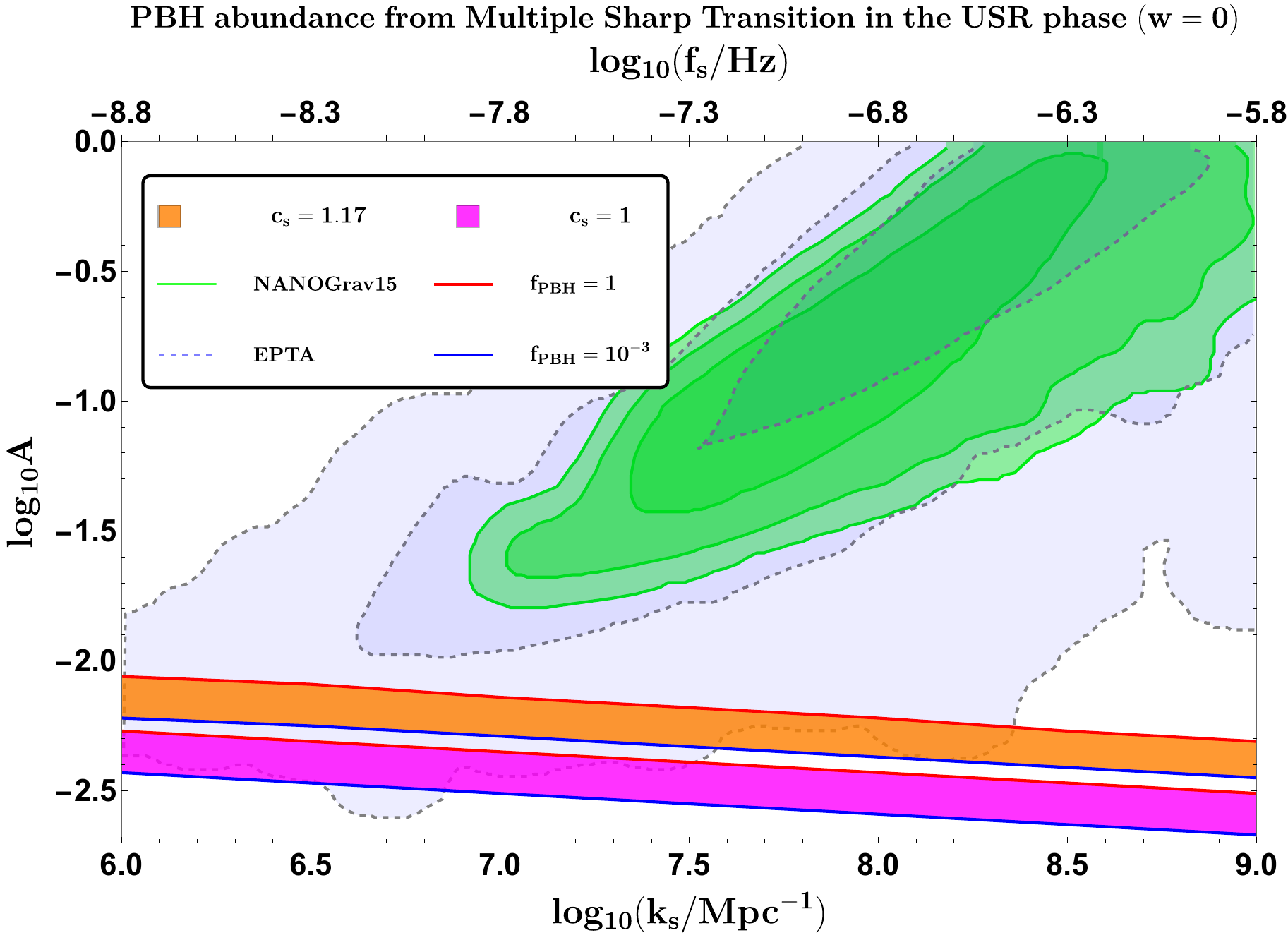}
        \label{w3}
       }
        \subfigure[]{
        \includegraphics[width=8.5cm,height=7.5cm] {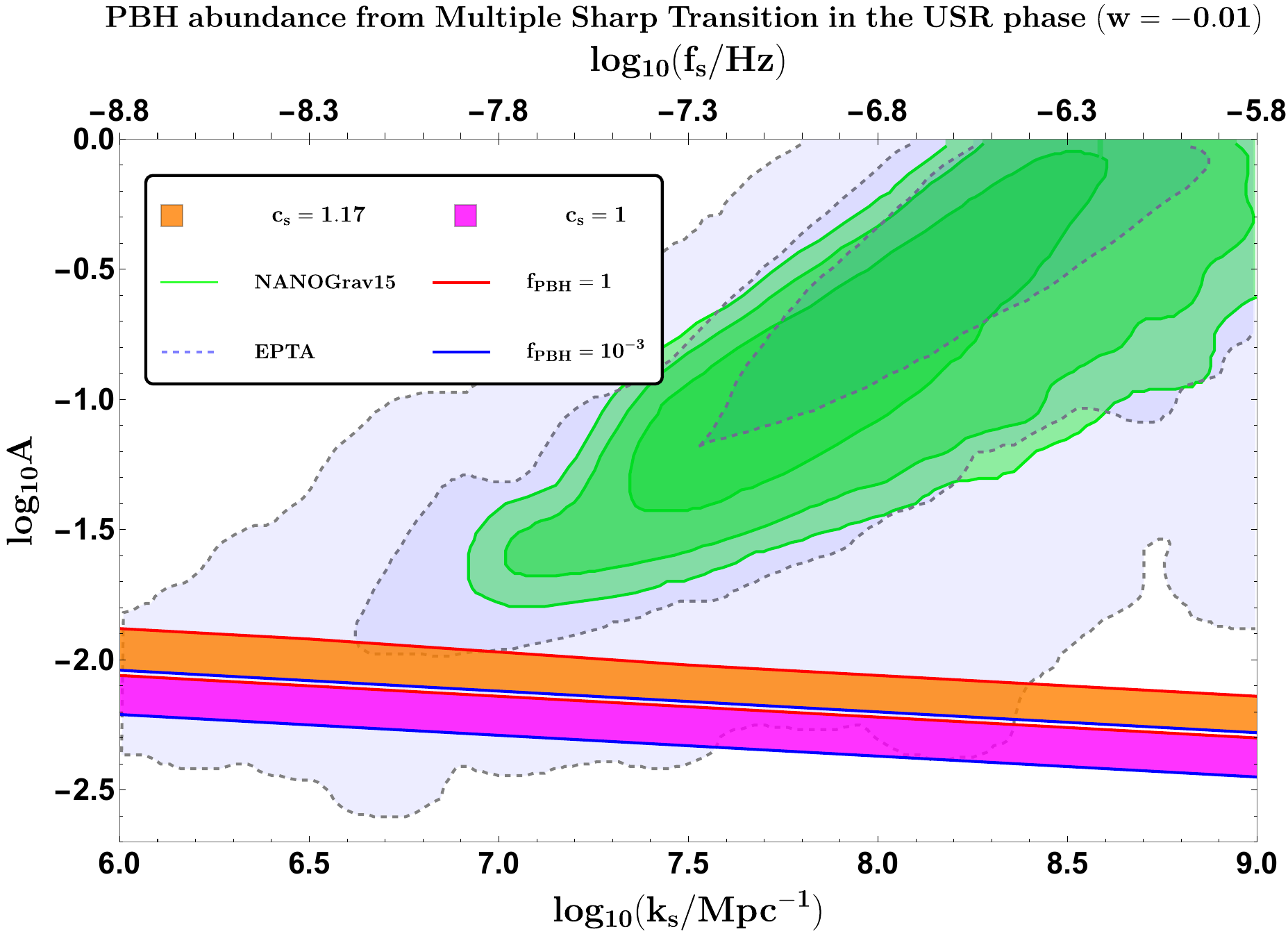}
        \label{w4}
       }
    	\caption[Optional caption for list of figures]{Comparative behaviour of the amplitude $A$ of the one-loop renormalized and DRG-resummed scalar power spectrum with changing transition wavenumber for the two cases, $c_{s}=1$ in magenta and $c_{s}=1.17$ in orange, for each value in the benchmark set, $w \in \{1/3,0.2,0,-0.01\}$.} 
    	\label{overprodcompare}
    \end{figure*}

The fig.(\ref{overprodcs1}) shows how the amplitude of the one-loop corrected renormalized and DRG-resummed scalar power spectrum behaves with increasing frequency in the interval sensitive to the PTA experiments, namely the NANOGrav 15 and EPTA signals. The effective sound speed is fixed here to take $c_{s}=1$. In this figure, we have chosen a set of benchmark values for the EoS parameter, $w \in \{1/3,0.2,0,-0.01\}$, which are allowed by the approximation where the linearities in the density contrast dominate on super-horizon scales. The values taken within the linear range are to some extent arbitrary but there's a caveat as you will notice further down the discussion. Now, the said approximation leads to the allowed interval for the threshold density contrast, i.e., $2/5 \leq \delta_{\rm th} \leq 2/3$, and consequently to the interval for the EoS, $-0.55 \leq w \leq 1/3$, from Eqn.(\ref{deltath}). Notice here that in principle, $\delta_{\rm th}$ can take values between $1/3 < \delta_{\rm th} < 1$, but the range of $\delta_{\rm th}$ taken for our analysis is crucial since the linear approximation in the cosmological perturbation theory is maintained in this range. Outside this interval requires the incorporation of non-linear approximation to develop the perturbation theory. Let us now look into fig.\ref{overprodcs1}. Here, we notice the case of $w=1/3$ highly preferred for the effective sound speed parameter $c_{s}=1$ scenario. The corresponding orange band agrees to $2\sigma$ from the NANOGrav 15 signal. As we get lower in $w$, at $w=0.2$ represented by the magenta band, we find ourselves just outside the $3\sigma$ region from NANOGrav 15 while still well inside the $2\sigma$ region of the EPTA signal. Interestingly, towards the values close to $w=0$, we find that the $w=0$ case in brown is further removed where the amplitude is $A < {\cal O}(10^{-2})$ as compared to the case of $w=-0.01$ in cyan colour. The reason for this will become clearer when we explore the shape of the SIGW spectrum and the amplitude of the total power spectrum for these same benchmark values of $w$ in detail.

    \begin{figure*}[ht!]
    	\centering
   {
   \includegraphics[width=18.5cm,height=12.5cm] {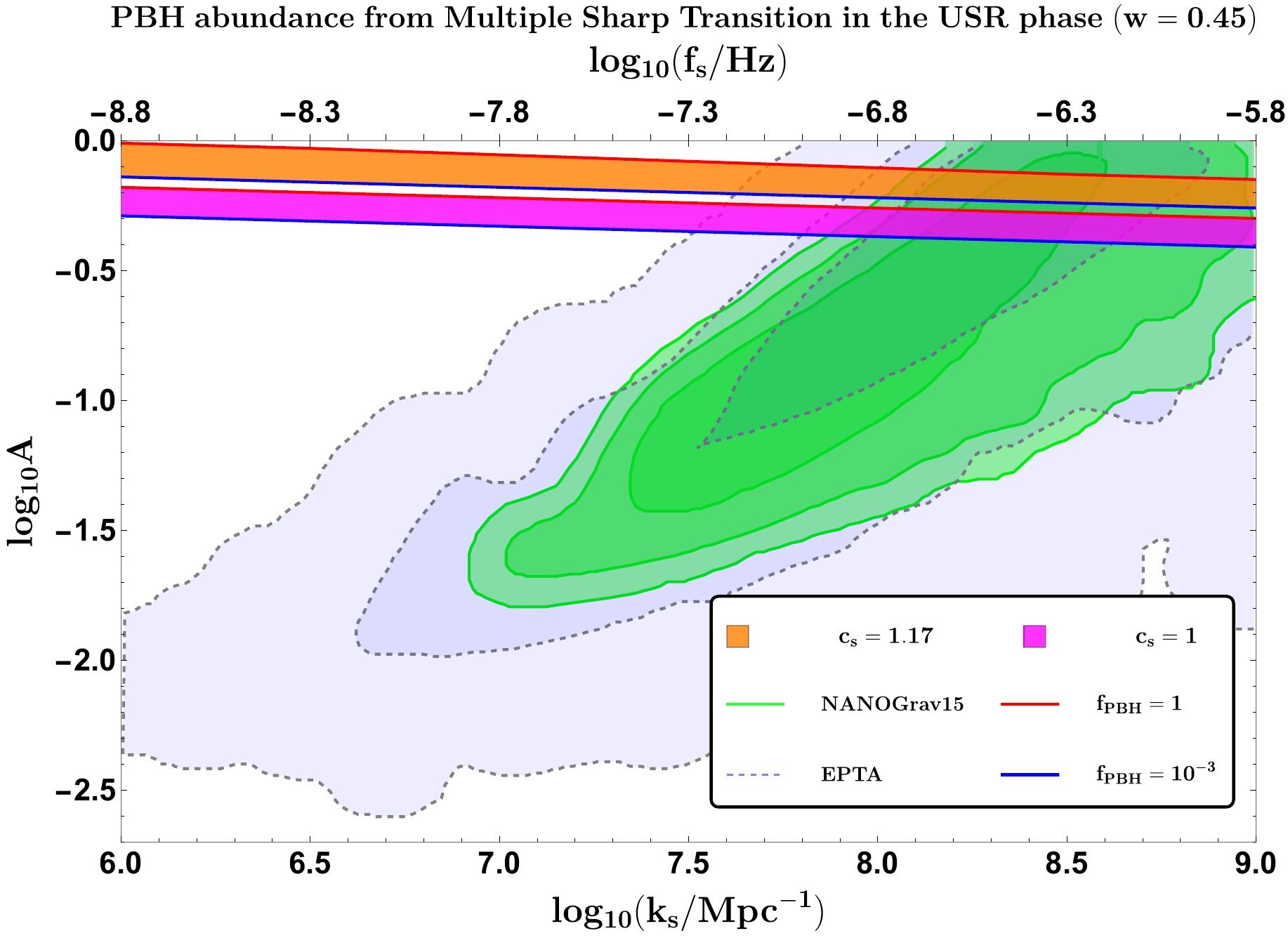}
    }
    	\caption[Optional caption for list of figures]{Figure shows change in the amplitude of the one-loop renormalized and DRG-resummed scalar power spectrum with changing transition wavenumber for the two cases, $c_{s}=1$ in magenta and $c_{s}=1.17$ in orange, for $w=0.45$. The red and blue lines enclose the region corresponding to sizeable abundance $f_{\rm PBH} \in (1,10^{-3})$.} 
    	\label{w5}
    \end{figure*}

The fig.(\ref{overprodcs117}) shows another case for the amplitude behaviour of the one-loop corrected renormalized and DRG-resummed scalar power spectrum, but this time with the effective sound speed fixed at $c_{s}=1.17$. Notice the apparent enhancement in the power spectrum amplitude needed to achieve a sizeable abundance of the PBHs, relative to the case of similar values of EoS $w$ but when the sound speed was fixed earlier to take $c_{s}=1$ in fig.(\ref{overprodcs1}). Now, for the case of $w=1/3$, the PBH abundance obtained lies well within the 1$\sigma$ contour of NANOGrav-15, showcasing yet again the fact that $w=1/3$ is the most favored condition. The overall bands are shifted above as compared to $c_s=1$. This plot re-enforces the idea that slightly increasing the value from $c_{s}=1$ with $c_s=1.17$ presents a more favorable scenario to avoid PBH overproduction. 

    \begin{figure*}[htb!]
    	\centering
    \subfigure[]{
      	\includegraphics[width=8.5cm,height=7.5cm] {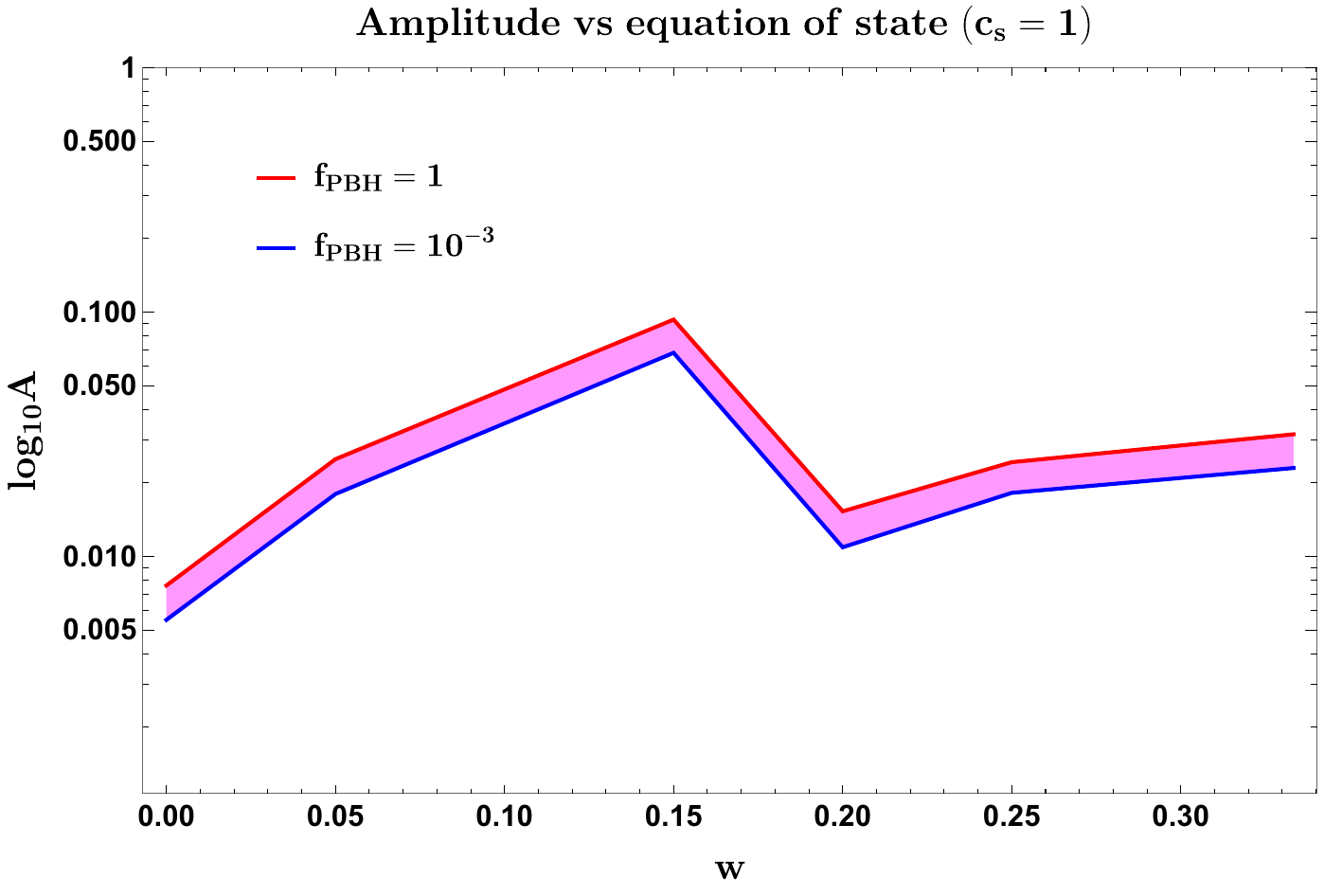}
        \label{ampvswc1}
    }
    \subfigure[]{
        \includegraphics[width=8.5cm,height=7.5cm] {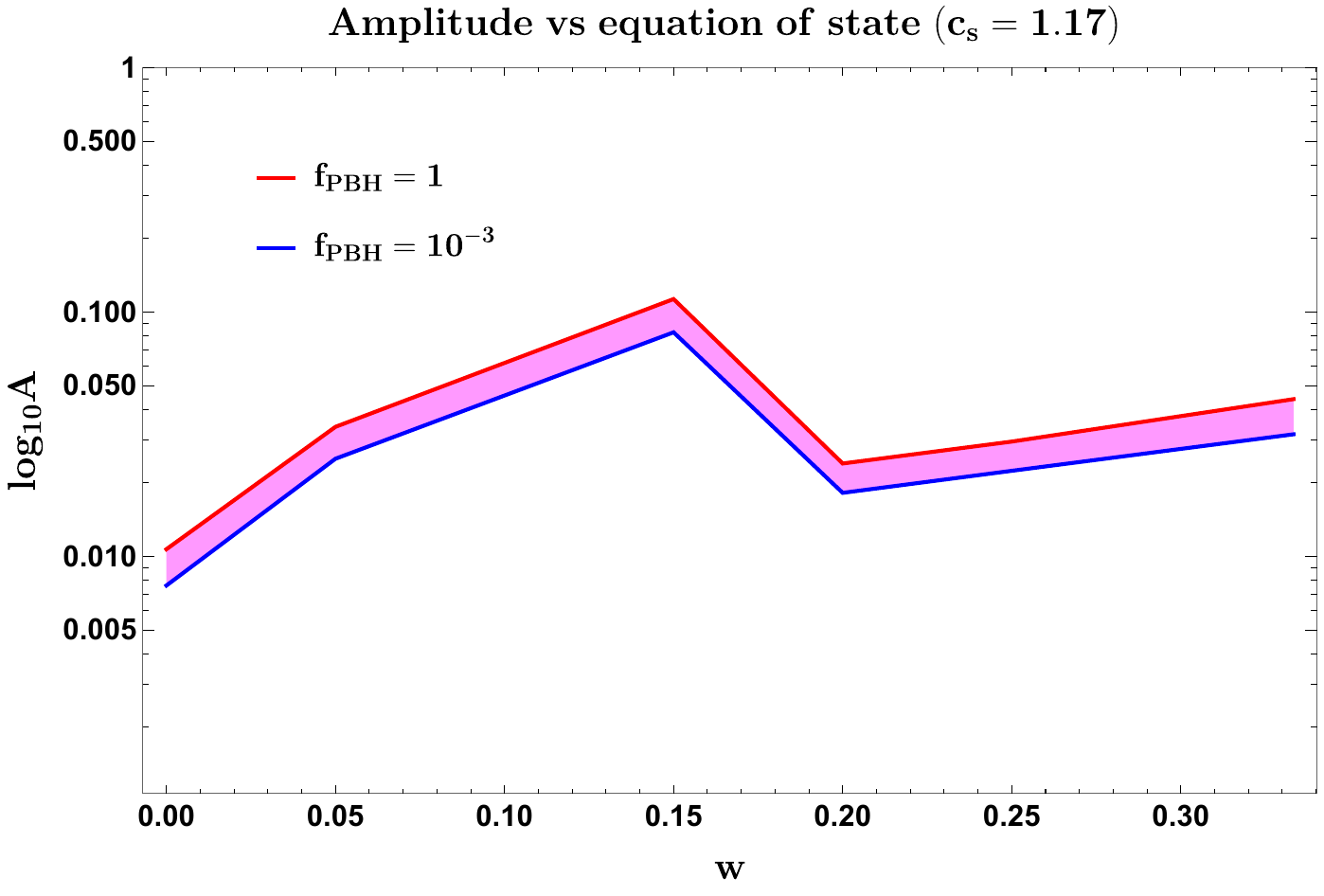}
        \label{ampvswc117}
       }
    	\caption[Optional caption for list of figures]{Behaviour of the amplitude $A$ of the one-loop renormalized and DRG-resummed scalar power spectrum with changing EoS $w$ when the transition wavenumber is fixed to satisfy $k_{\rm PBH} \sim {\cal O}(10^{7}{\rm Mpc^{-1}})$. The red and blue solid lines enclose the allowed region in magenta where the $f_{\rm PBH} \in (1,10^{-3})$ is satisfied.} 
    	\label{ampvsw}
    \end{figure*}

Thereafter, fig.(\ref{overprodcompare}) provides a comparative analysis for the behaviour of the scalar power spectrum amplitude between different values of the effective sound speed while keeping the $w$ value fixed for each panel. On an individual basis, we can recognize that an elevated value of $c_{s}=1.17$ is a more favourable condition for effectively explaining overproduction. This allows for a sufficient amplitude in the USR phase. At this point, let us make it clear that our original approximation of working in the linear regime, where the interval, $2/5 \leq \delta_{\rm th} \leq 2/3$, holds for the density contrast, gives $w=1/3$ to be its maximum value. However, if we try to increase $w$ beyond this interval, one is left to ponder over the current overproduction issue at hand. In that case, fig.(\ref{w5}) provides the answer where we increase the value to $w=0.45$ and find that we can elevate the amplitude to even larger values such that perturbativity comes to the verge of breaking. Overproduction can still be avoided in such a case, but the outcome of such considerations does not generate useful conclusions since production of high mass PBHs is not possible considering highly suppressed amplitude at this NANOGrav-15 scale. Hence, we conclude that increasing $w \geq 0.45$ can lead to problems with perturbation theory and we would have already exceeded the allowed bound set by the linear regime approximation. The above analysis reinforces the idea that $w=1/3$ is the best possible scenario to keep the theoretical constraints intact while producing better results satisfying observational parameters with ease.
Within the effective field theory, it is visible that one can not reach the kination domination, except probably for some cases like the quintessential inflation where this possibility may arise with the inclusion of a bump that acts as a catalyst. However, in all other scenarios, this is almost close to impossible.
Till now, we have witnessed the changing behaviour of the quantum loop renormalized DRG-resummed scalar power spectrum amplitude with the momentum values which can help resolve overproduction, but the trend in the amplitude is not always linear concerning changing EoS parameter $w$. fig.(\ref{ampvsw}) focuses on this matter. Here we direct our attention towards the range, $0 \leq w \leq 1/3$, which reveals that the ascent of the power spectrum amplitude with increasing $w$ is not always monotonic. There exists a regime near $w \in (0.1,0.2)$ where the amplitude shows a significant relative increase such that we can obtain a sizeable abundance of the PBHs, $f_{\rm PBH} \in (1,10^{-3})$. We have previously noticed from fig.\ref{overprodcompare} that after reaching values near $w \sim 0$, the overall amplitude is quite low to lie within the displayed sensitivities of either PTA posteriors. The case of values near $w = 0$ is interesting since the amplitude for negative values, $w \sim -0.05$, is higher than $w=0$. However, upon descending further below, for instance, $w \leq -0.1$, we get heavily suppressed amplitudes, but those results are not illuminating from the perspective of discussing the overproduction problem. Hence, we do not analyze the behaviour for the rest of the remaining interval of allowed values lying in the range, $-0.55 \leq w \leq -0.05$. Further insights into this behaviour will be expounded upon during the discussion on the SIGW spectrum.

    \begin{figure*}[htb!]
    	\centering
    \subfigure[]{
      	\includegraphics[width=8.5cm,height=7.5cm] {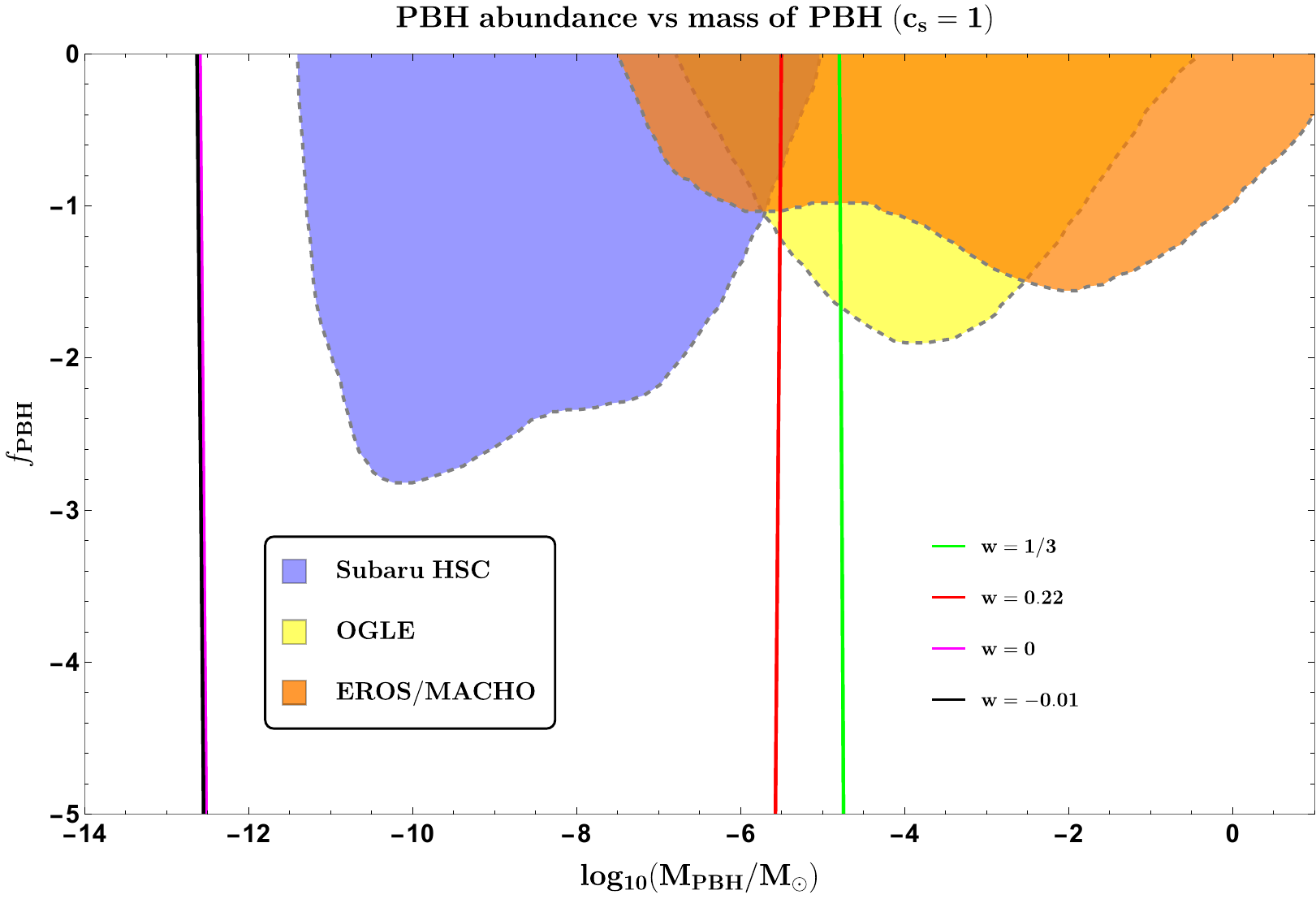}
        \label{microw1}
    }
    \subfigure[]{
        \includegraphics[width=8.5cm,height=7.5cm] {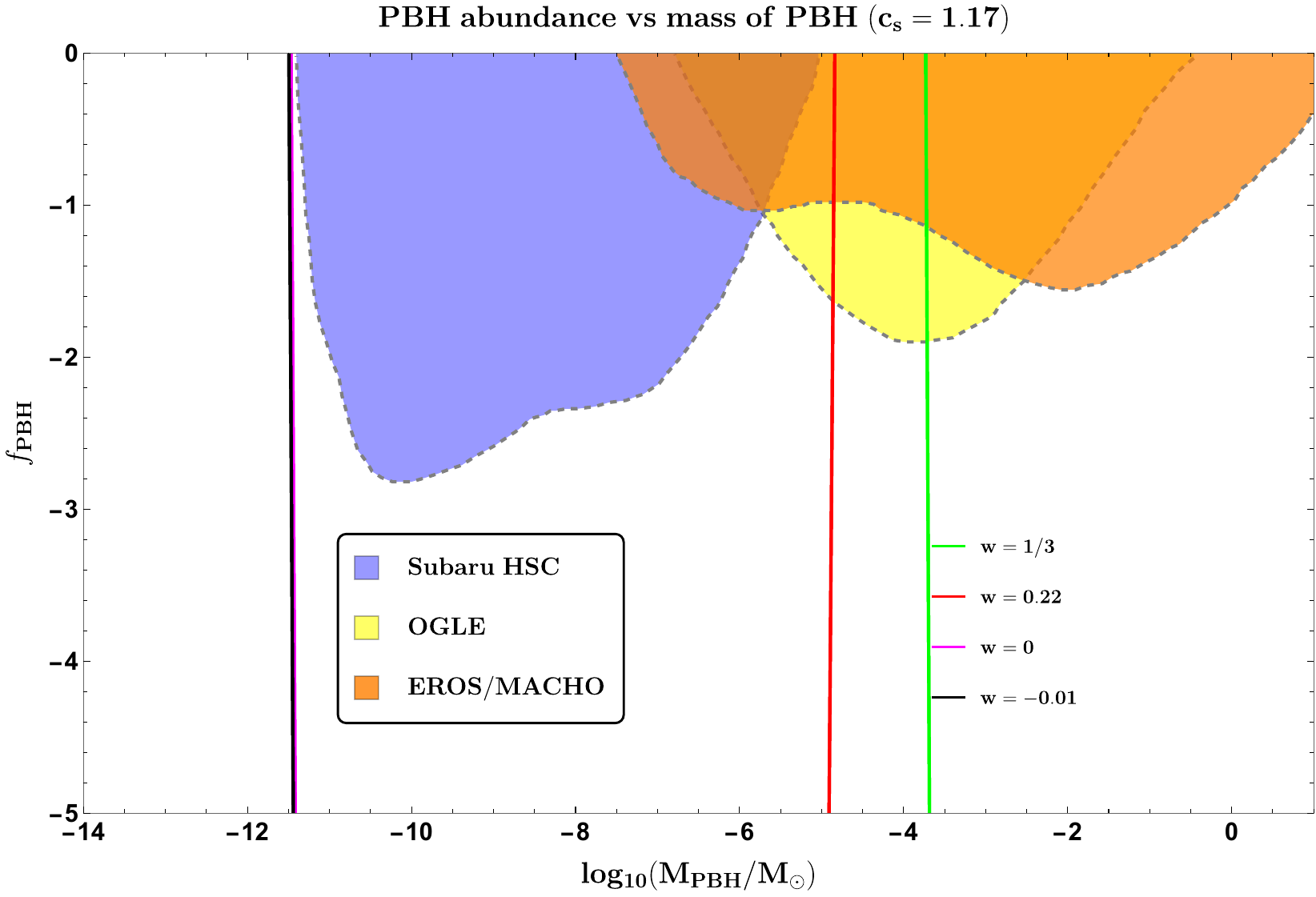}
        \label{microw2}
       }
    	\caption[Optional caption for list of figures]{The behaviour of PBH abundance as a function of its mass. The effective sound speed $c_{s}$ is fixed to have $c_{s}=1$ in the left panel and $c_{s}=1.17$ in the right panel. In blue, yellow and orange, we highlight the regions excluded by the microlensing experiments Subaru HSC, OGLE, and EROS/MACHO, respectively. Green and red lines highlight the abundance change when EoS takes on values $w=1/3$ and $w=0.22$, respectively. For these values, masses obtained are found within the range $M_{\rm PBH} \sim {\cal O}(10^{-6}-10^{-3})M_{\odot}$. The magenta and black colours represent the abundance behaviour for $w=0$ and $w=-0.01$, respectively. PBH masses corresponding to EoS values $w < 0$ have ranges where $M_{\rm PBH} \lesssim {\cal O}(10^{-13})M_{\odot}$, and therefore lie beyond the sensitivities of the microlensing experiments as shown in the plot.} 
    	\label{microlens}
    \end{figure*}

Figures \ref{ampvswc1} and \ref{ampvswc117} represent the variation in the amplitude of the final scalar power spectrum with arbitrary EoS parameter $w$ for two different propagation speeds $c_s=1$ and $c_s=1.17$. We take a moment to point again that there is an absence of heavy difference between the two cases of $c_s$, albeit an elevated peak amplitude is observed for the case of $c_s=1.17$, indicating that a slight increase in its value from $c_{s}=1$ offers a more favourable scenario for inflation.
Regardless, this discussion highlights the reassurance of a more favourable EoS scenario. With this in mind, let us dive deeper into the intricacies. The curves in these plots are a result of varying the PBH formation scales within the range sensitive specifically to NANOGrav-15 and trying to estimate amplitudes such that the PBH abundance lies in the sizeable range $f_{\rm PBH} \in (1,10^{-3})$. As one can visualize from the nature of both plots, there is no specific trend to the change in amplitude with the EoS parameter. Initially, the peak amplitude of the scalar power spectrum increases, reaches a maximum at some value, and then takes a dip. Thereafter, the amplitude starts to rise again as $w$ increases. It is expected that the amplitude will keep on increasing with a further increase in $w$, and challenge the perturbativity constraints as $w \rightarrow 1$ i.e. the case of kination dominated era. Here we have not depicted the case of $w \le 0$ in the graph, but from our analysis, we found that as we go below $w=0$ till $w\sim-0.05$, we find a steady increase in the peak amplitude, which again suffers the safe fate of decline as you descend further below $w \sim -0.05$. Even though all the conditions of perturbativity are maintained at each transition, the desired result of obtaining the high mass PBHs can not be achieved with such low values of $w$. In addition, notice that the analysis is done for some preferred range of $w$, and although there might be an expected behaviour of this curve based on the underlying theory, it is left at your discretion to try for even lower values of $w$ to check the pattern that comes out of it.

       \begin{figure*}[htb!]
    	\centering
    \subfigure[]{
      	\includegraphics[width=8.5cm,height=7.5cm] {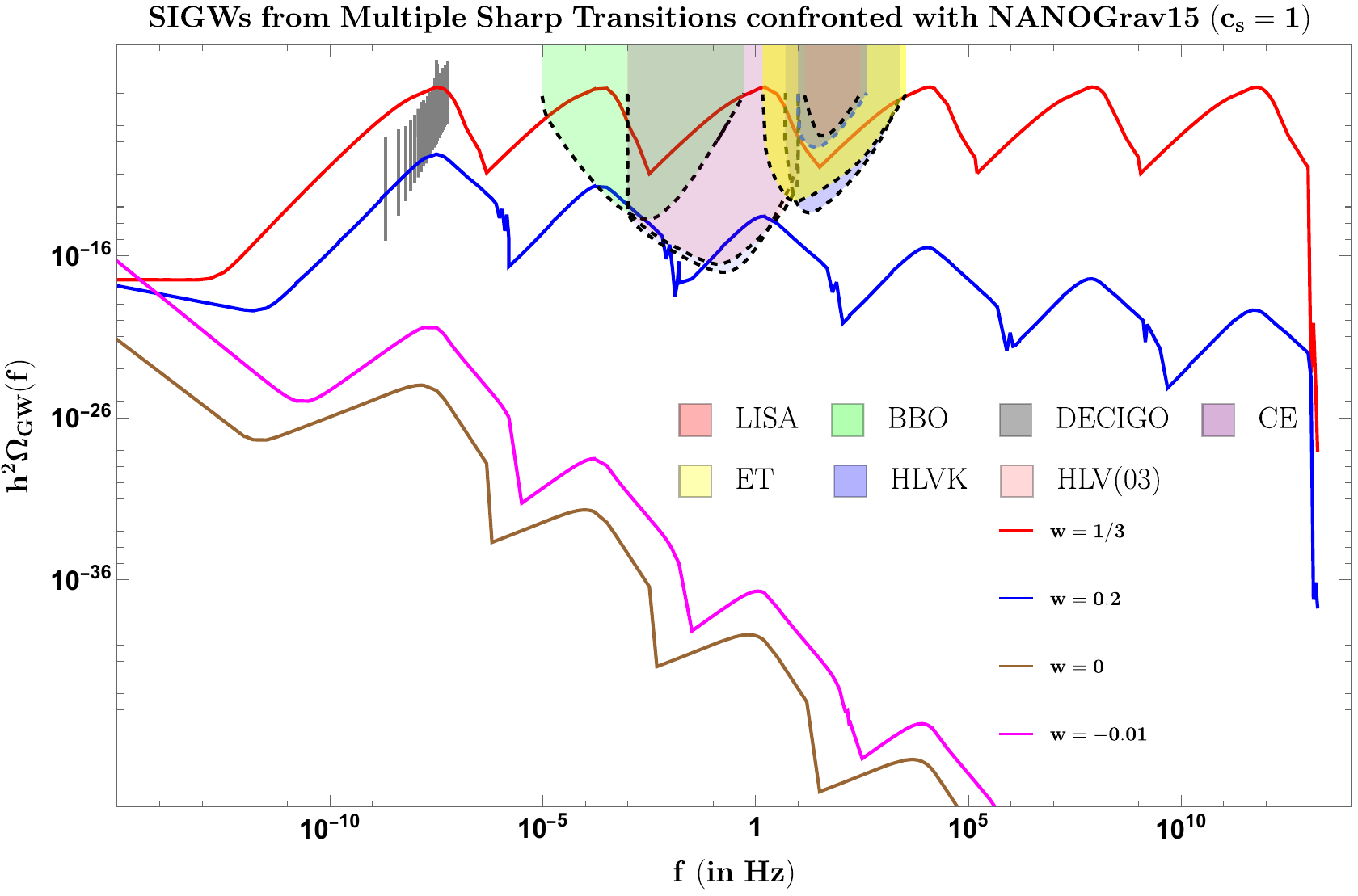}
        \label{SIGWc1NANOa}
    }
    \subfigure[]{
        \includegraphics[width=8.5cm,height=7.5cm] {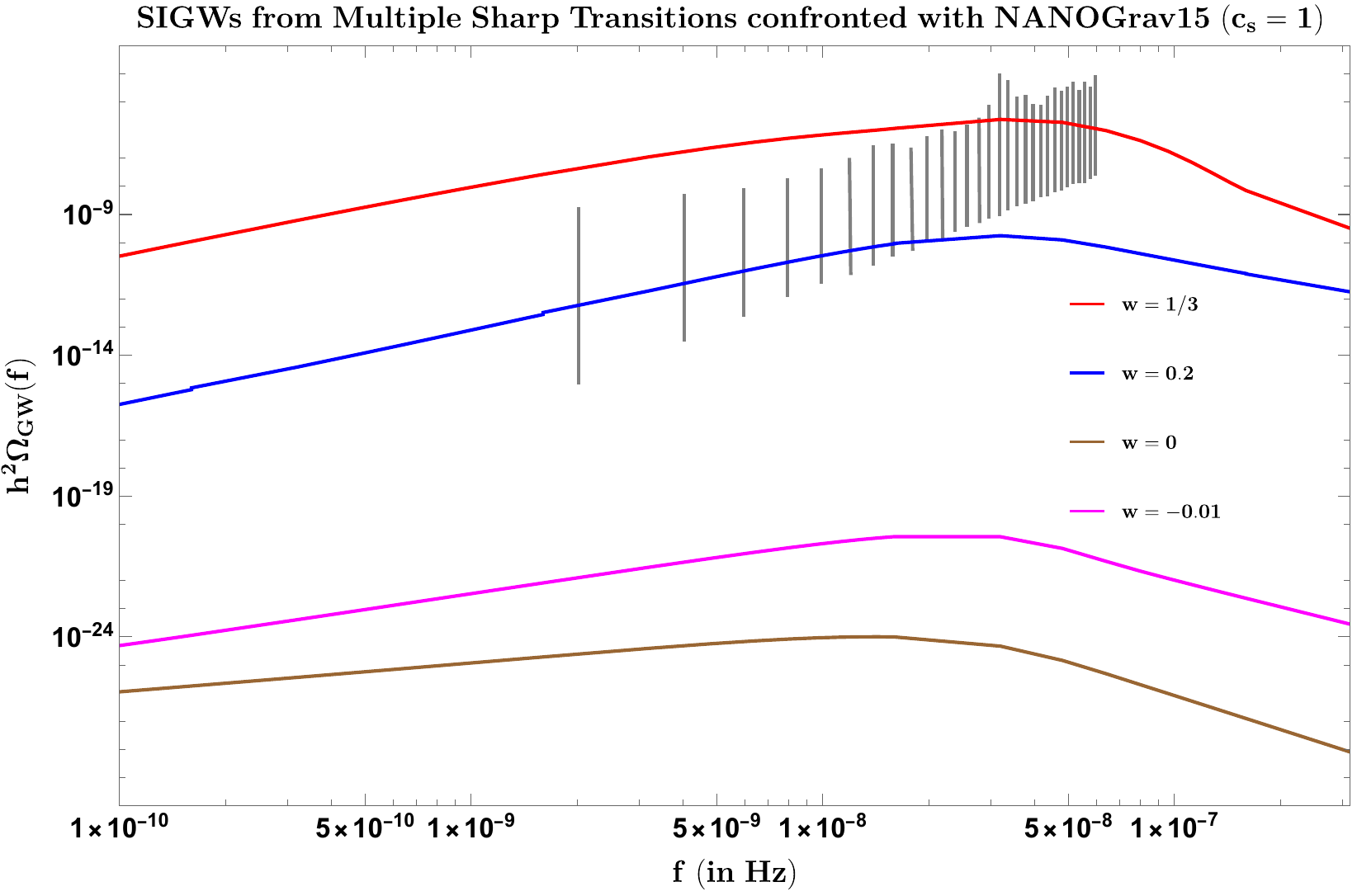}
        \label{SIGWc1NANOb}
       }
    \subfigure[]{
      	\includegraphics[width=8.5cm,height=7.5cm] {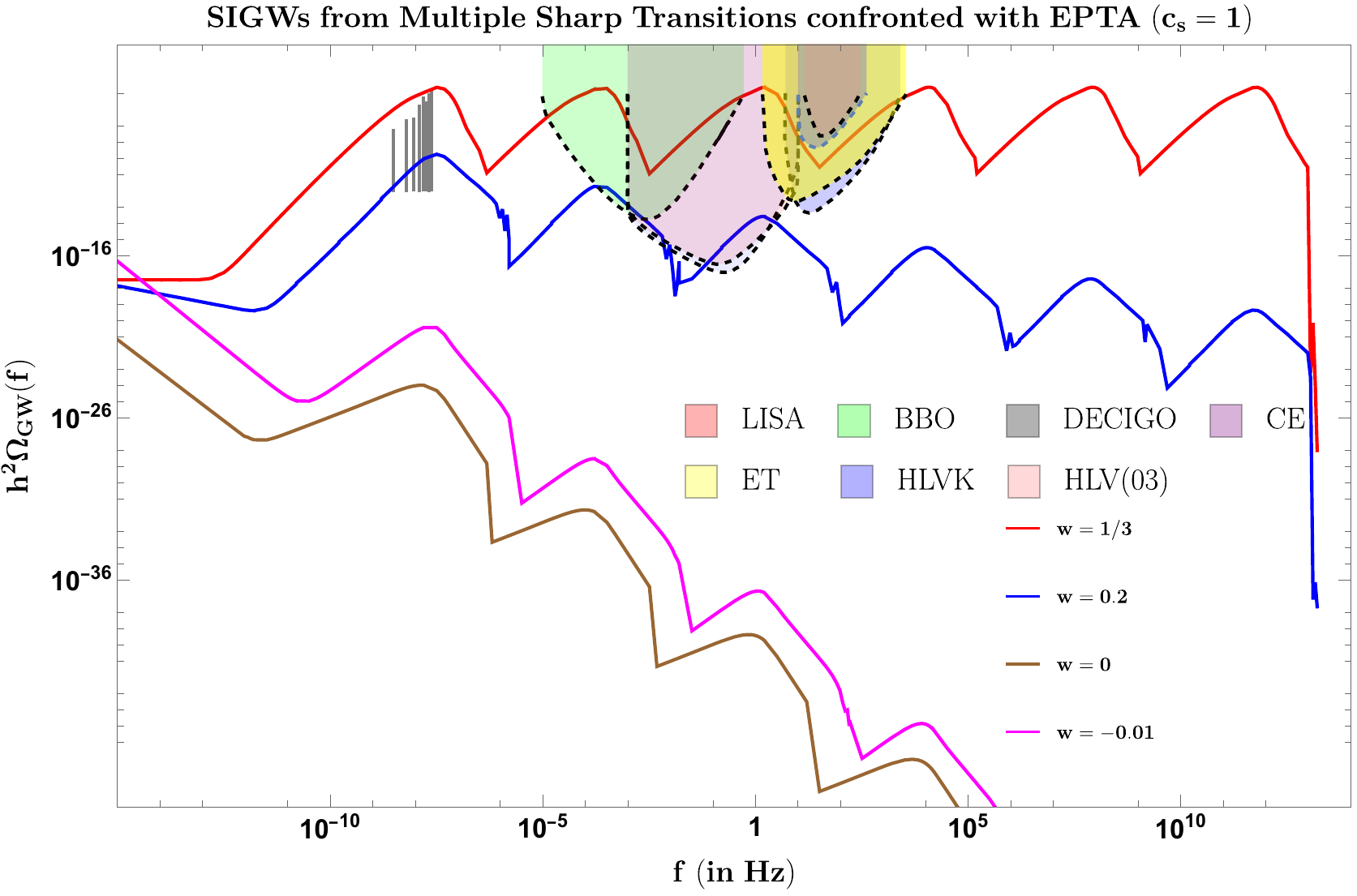}
        \label{SIGWc1EPTAa}
    }
    \subfigure[]{
        \includegraphics[width=8.5cm,height=7.5cm] {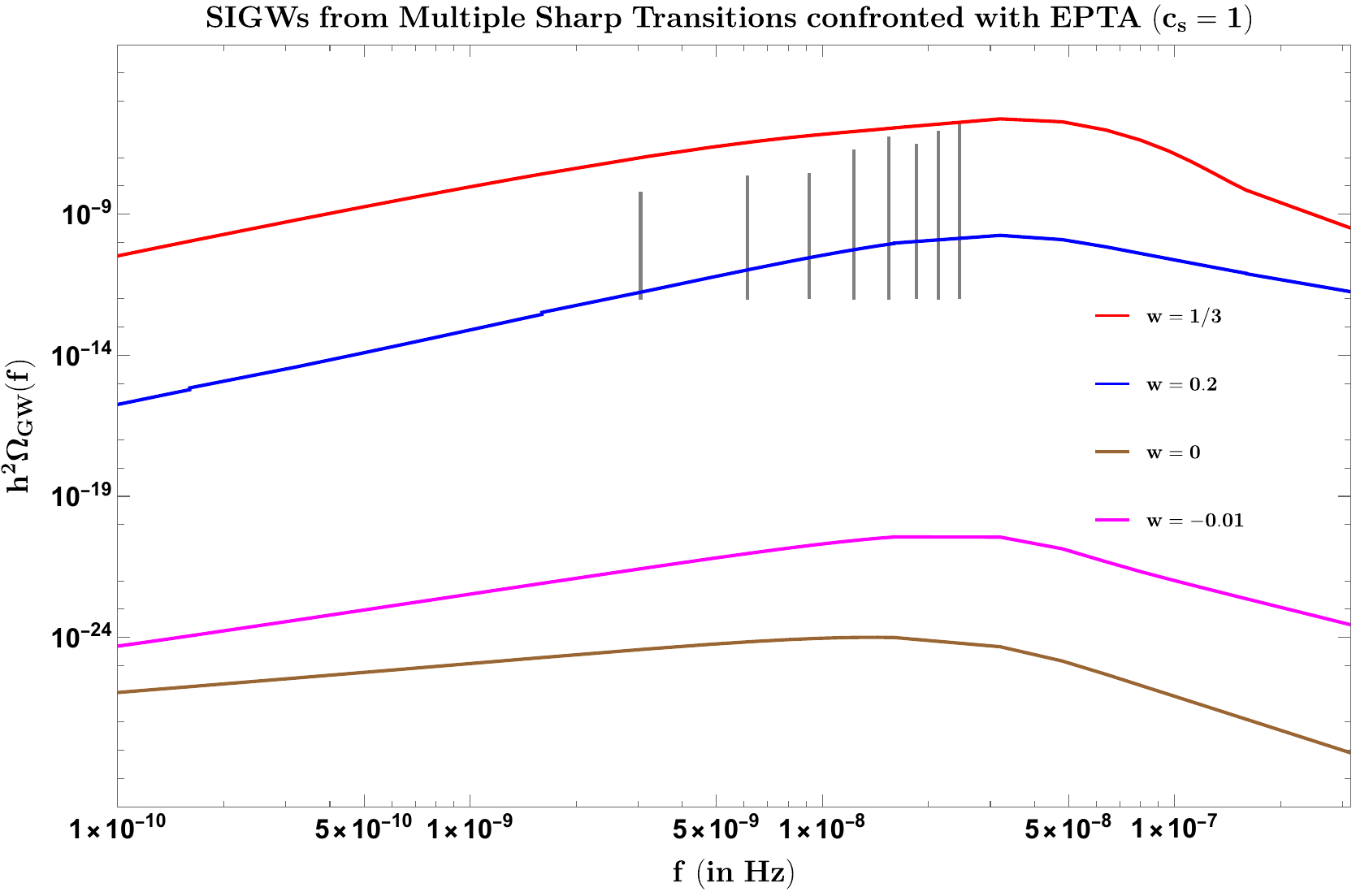}
        \label{SIGWc1EPTAb}
       }
    	\caption[Optional caption for list of figures]{The spectrum of SIGW as a function of its frequency. All the panels feature the value of effective sound speed $c_{s}=1$. The left panel (both top and bottom) shows the complete spectrum, which covers frequencies sensitive to the data from NANOGrav-15, EPTA, and the ground and space-based experiments, which include LISA, DECIGO, BBO, Einstein Telescope (ET), Cosmic Explorer (CE), the HLVK network (aLIGO in Livingstone and Hanford, aVIRGO, and KAGRA), and HLV (O3). The right panel (both top and bottom) focuses primarily on the frequencies involving the NANOGrav 15 and EPTA signals. Red, blue, magenta and brown represent the spectrum corresponding to the EoS values $w \in \{1/3,0.2,0,-0.01\}$, respectively.} 
    	\label{SIGWc1ab}
    \end{figure*}
\begin{figure*}[htb!]
    	\centering
    \subfigure[]{
      	\includegraphics[width=8.5cm,height=7.5cm] {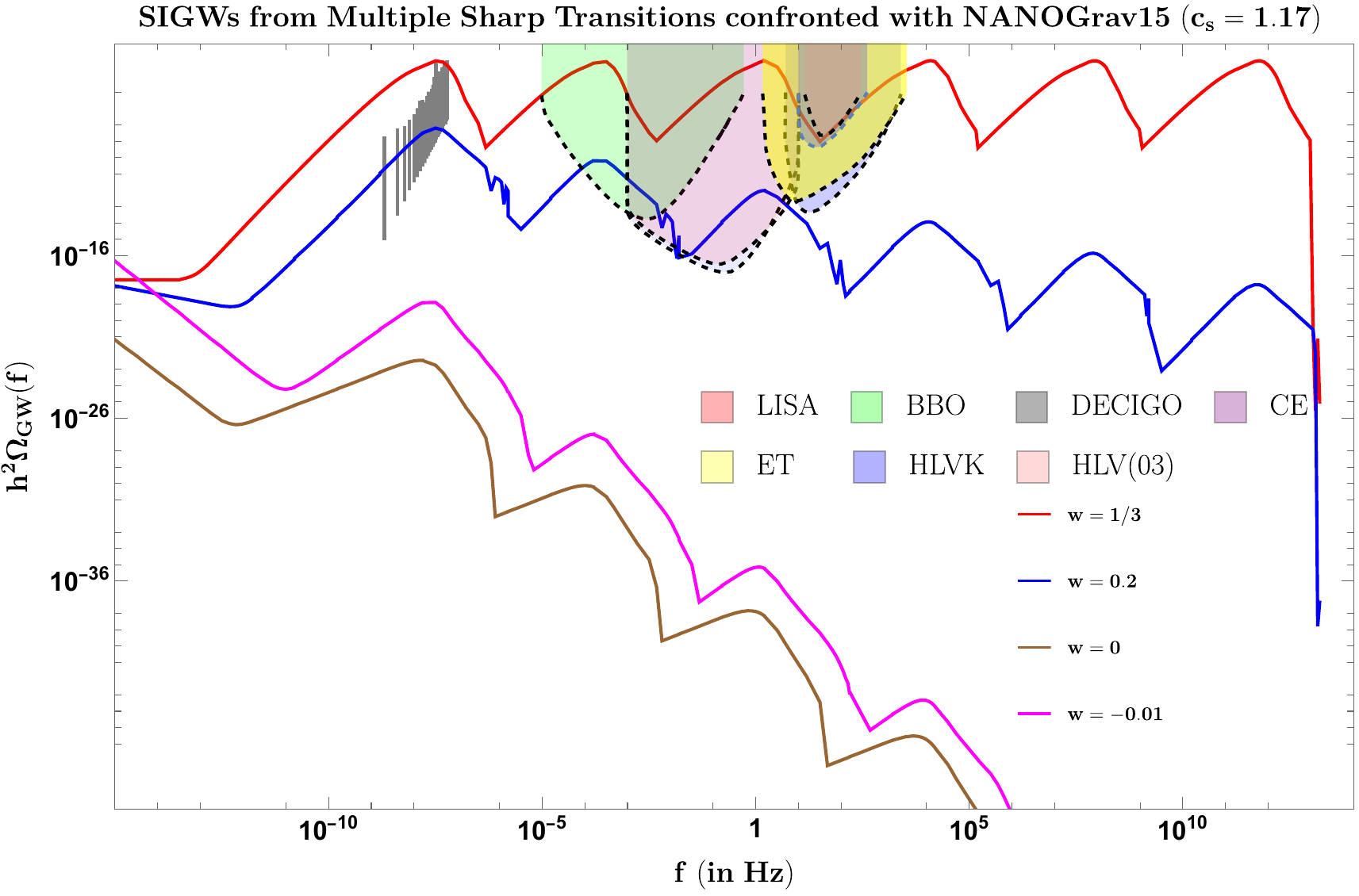}
        \label{SIGWc117NANOa.pdf}
    }
    \subfigure[]{
        \includegraphics[width=8.5cm,height=7.5cm] {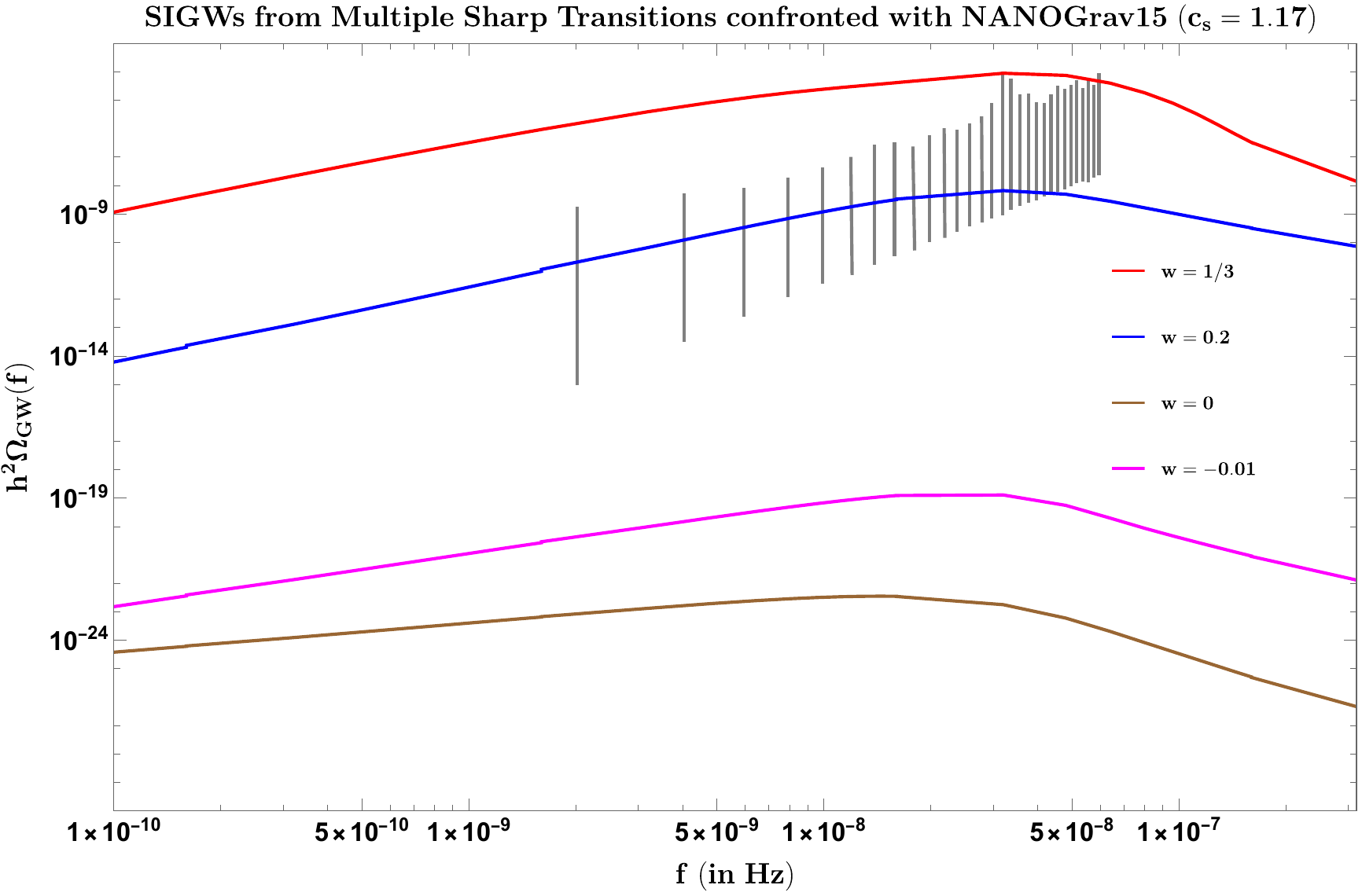}
        \label{SIGWc117NANOb}
       }
    \subfigure[]{
      	\includegraphics[width=8.5cm,height=7.5cm] {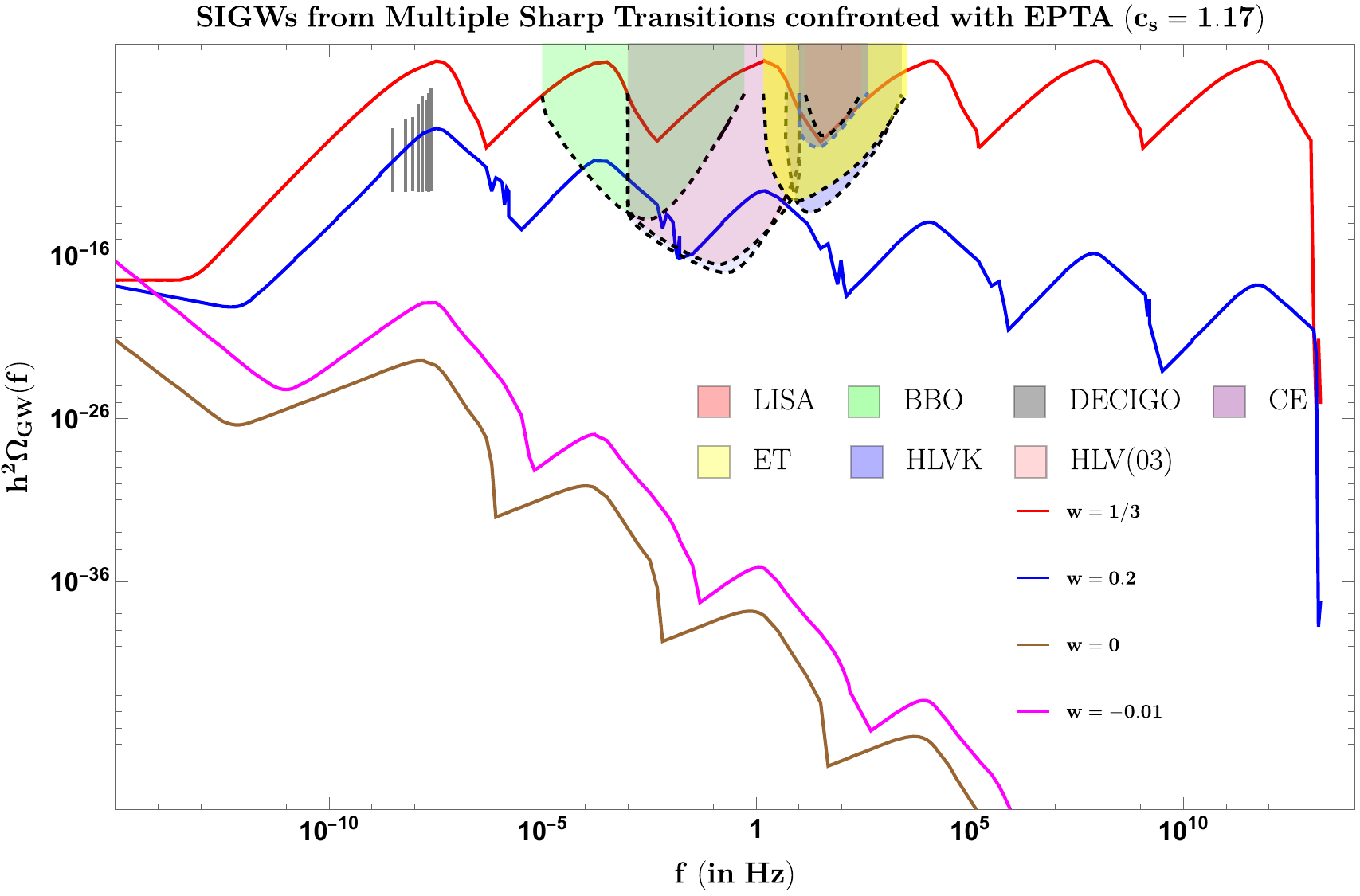}
        \label{SIGWc117EPTAa}
    }
    \subfigure[]{
        \includegraphics[width=8.5cm,height=7.5cm] {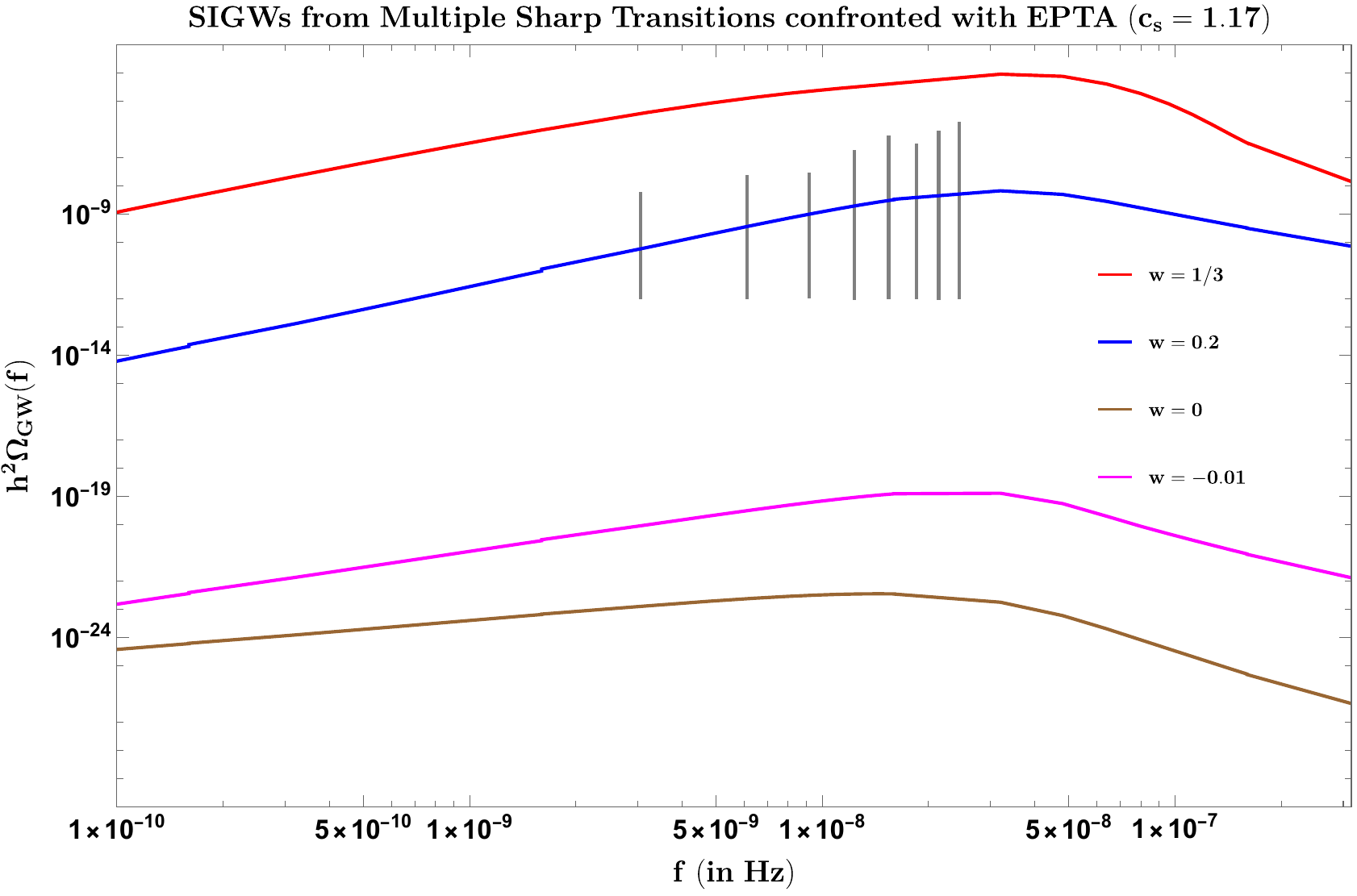}
        \label{SIGWc117EPTAb}
       }
    	\caption[Optional caption for list of figures]{The spectrum of SIGW as a function of its frequency. All the panels feature the value of effective sound speed $c_{s}=1.17$. The left panel (both top and bottom) shows the complete spectrum, which covers frequencies sensitive to the data from NANOGrav-15, EPTA, and the ground and space-based experiments, which include LISA, DECIGO, BBO, Einstein Telescope (ET), Cosmic Explorer (CE), the HLVK network (aLIGO in Livingstone and Hanford, aVIRGO, and KAGRA), and HLV (O3). The right panel (both top and bottom) focuses primarily on the frequencies involving the NANOGrav 15 and EPTA signals. Red, blue, magenta and brown represent the spectrum corresponding to the EoS values $w \in \{1/3,0.2,0,-0.01\}$, respectively.} 
    	\label{SIGWc117ab}
    \end{figure*}
\begin{figure*}
\includegraphics[width=19.5cm,height=12.5cm] {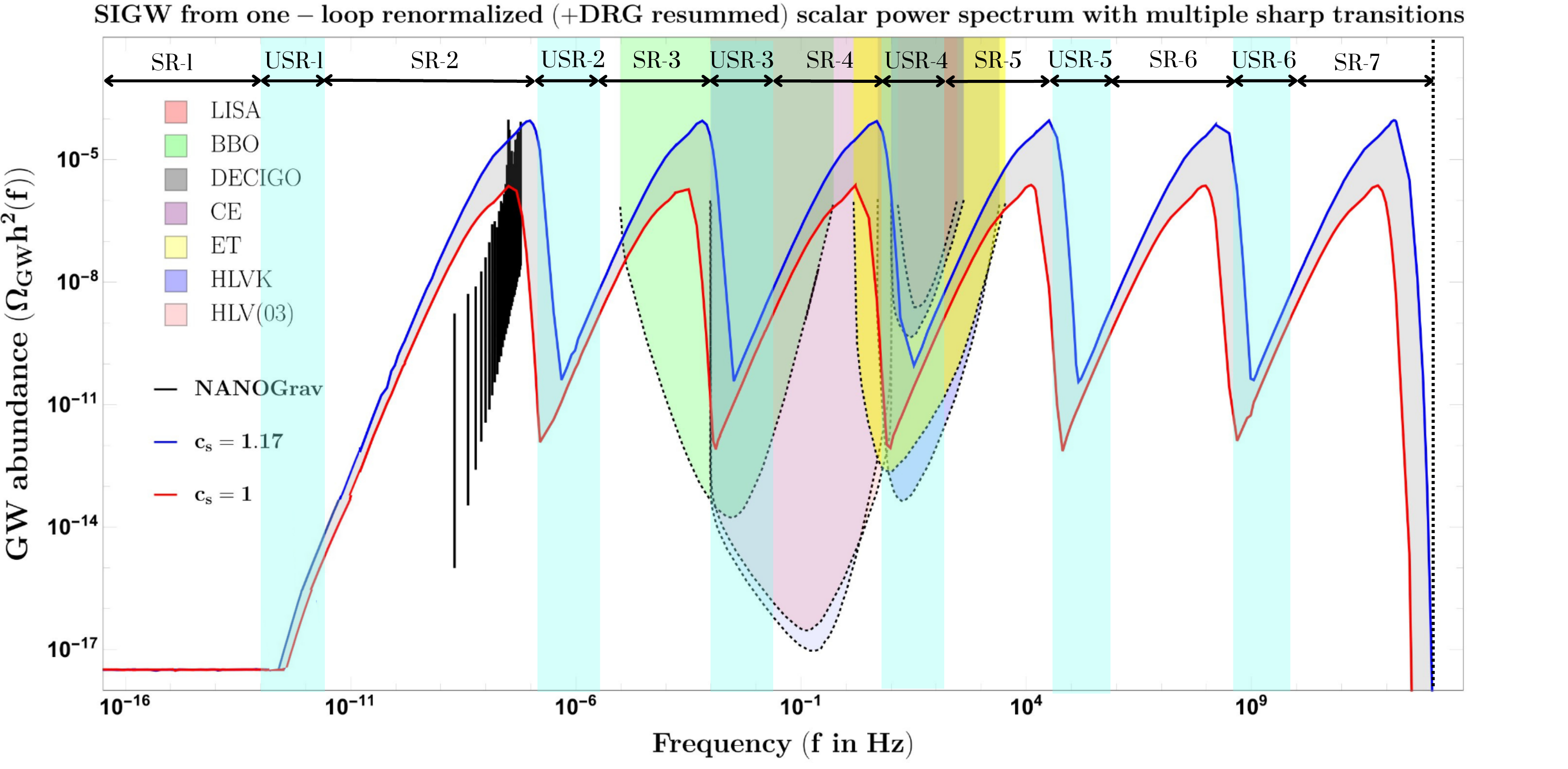}
        \label{w4}
       
    	\caption[Optional caption for list of figures]{SIGW spectrum as a function of the frequency. The spectrum is generated with the fixed EoS condition $w=1/3$. The red and blue lines represent the effective sound speed values $c_{s}=1\;{\rm and }\;1.17$, respectively. The ash-coloured shaded region represents the allowed parameter space for the SIGW signal where $1 \leq c_{s} \leq 1.17$ is satisfied. The first peak is shown to align with the NANOGrav-15 signal and the complete spectrum contains its signature within the sensitivities of the several ground and space-based GW experiments, which include LISA, DECIGO, BBO, Einstein Telescope (ET), Cosmic Explorer (CE), the HLVK network (aLIGO in Livingstone and Hanford, aVIRGO, and KAGRA), and HLV (O3).} 
    	\label{Gw13plot}
    \end{figure*}

Now we turn our attention to the fig.(\ref{microlens}), which demonstrates the change in abundance of PBH as a function of its mass. The abundance is extremely sensitive to the range of masses where one can expect to have a reasonable estimate of $f_{\rm PBH} \in (1,10^{-3})$. In both the panels, we show the variation for the same set of EoS values, $w \in \{1/3,0.22,0,-0.01\}$, and found that for higher values of EoS, $w \gtrsim 0.2$, we can expect the mass range to fall within the region $M_{\rm PBH} \sim {\cal O}(10^{-6}-10^{-3})M_{\odot}$ that is also agreeable for the frequencies related to the NANOGrav-15 signal. For such values of $w$, the constraints coming from the set of microlensing experiments as shown in the plot still allow for a sizeable abundance. As we go below $w=0$, we obtain masses wherefore the abundance lies outside the regime of sensitivity of the experiments shown. The black and magenta lines become almost coincident but with black $(w=-0.01)$ at the left of magenta $(w=0)$. We do not go further below in $w$ as this will only lead to extremely lower values of the PBH mass, $M_{\rm PBH} \lesssim {\cal O}(10^{-13})M_{\odot}$. From the above fig.(\ref{microlens}), we again confirm the benefits of a slightly increased value of $c_{s}$ by obtaining a larger mass range with $c_{s}=1.17$ that allows for a sizeable abundance of PBHs.

Let us now turn to fig.(\ref{SIGWc1ab}), which depicts the variation of the amplitude of various SIGW spectra generated from the one-loop renormalized and DRG-resummed scalar power spectrum for some benchmark values of the EoS parameter, $w \in \{1/3,0.2,0,-0.01\}$. The results get further confronted with the NANOGrav-15 and EPTA data with the effective sound speed fixed at $c_{s}=1$. fig.(\ref{SIGWc1ab}) illustrates the advantage of having $w=1/3$ where the resulting spectra agree most closely with the signals from NANOGrav-15 and EPTA. The remarkable feature of constant peak amplitude is also on display for the $w=1/3$ case. For the value of $w \sim 0.2$, we find the tail region of the generated spectra to align with both the NANOGrav-15 and EPTA signals. As we go lower in $w$, the resulting spectra are extremely low in amplitude, which helps to exclude the region of $w \lesssim 0$ as having any potential to explain the PTA signals. The case of amplitude for $w=-0.01$ being higher than $w=0$ is crucial to observe and which gets translated into the amplitude features previously observed in fig.(\ref{overprodcs1}) when discussing overproduction. The change in tilt of the spectrum after achieving peak amplitude is also an interesting feature to note in the plots for $w \gtrsim 0.2$. The fig.(\ref{SIGWc117ab}) depicts a similar variation of the SIGW amplitude with the frequency when considering the benchmark values, $w \in \{1/3,0.2,0,-0.01\}$, but this time the effective sound speed takes at value $c_{s}=1.17$. This change in $c_{s}$ is reflected by an enhancement in the SIGW amplitude across the complete frequency range for each EoS value in the benchmark set mentioned before. Notice that for $w=1/3$, the regime near the peak amplitude of the generated spectrum shows the closest approximation to the NANOGrav-15 signal by keeping within the parameter space allowed by $1 \leq c_{s} \leq 1.17$. Values of $w \sim 0.2$ are also compatible with both the NANOGrav-15 and EPTA signals, and going any below will not suffice to explain the desired features. Lastly, the cumulative features, like the behaviour of the spectrum tilt before and after reaching their peak values for each EoS $w$ scenario, and the overall features across the complete frequency range covering all the listed GW experiments, remain the same except for the same order of magnitude enhancement in the amplitude of the spectrum for $c_{s}=1.17$ in fig.(\ref{SIGWc117ab}) when compared with fig.(\ref{SIGWc1ab}) for $c_{s}=1$. To elaborate on the case of $w=1/3$ and its advantages from the perspective of the SIGW spectrum, we show the plot in fig.(\ref{Gw13plot}). The details concerning the behaviour of the said spectrum following each interval of SR and USR, the corresponding signature in the sensitivities of the various GW experiments, NANOGrav-15 and other higher-frequency experiments,  along with a closer look at the effects brought by the change in effective sound speed within the interval $1 \leq c_{s} \leq 1.17$ is evident from the plot and matches with the results as discussed previously for the $w=1/3$ scenario.

    \begin{figure}[ht!]
    	\centering
   {
   \includegraphics[width=18.5cm,height=12.5cm] {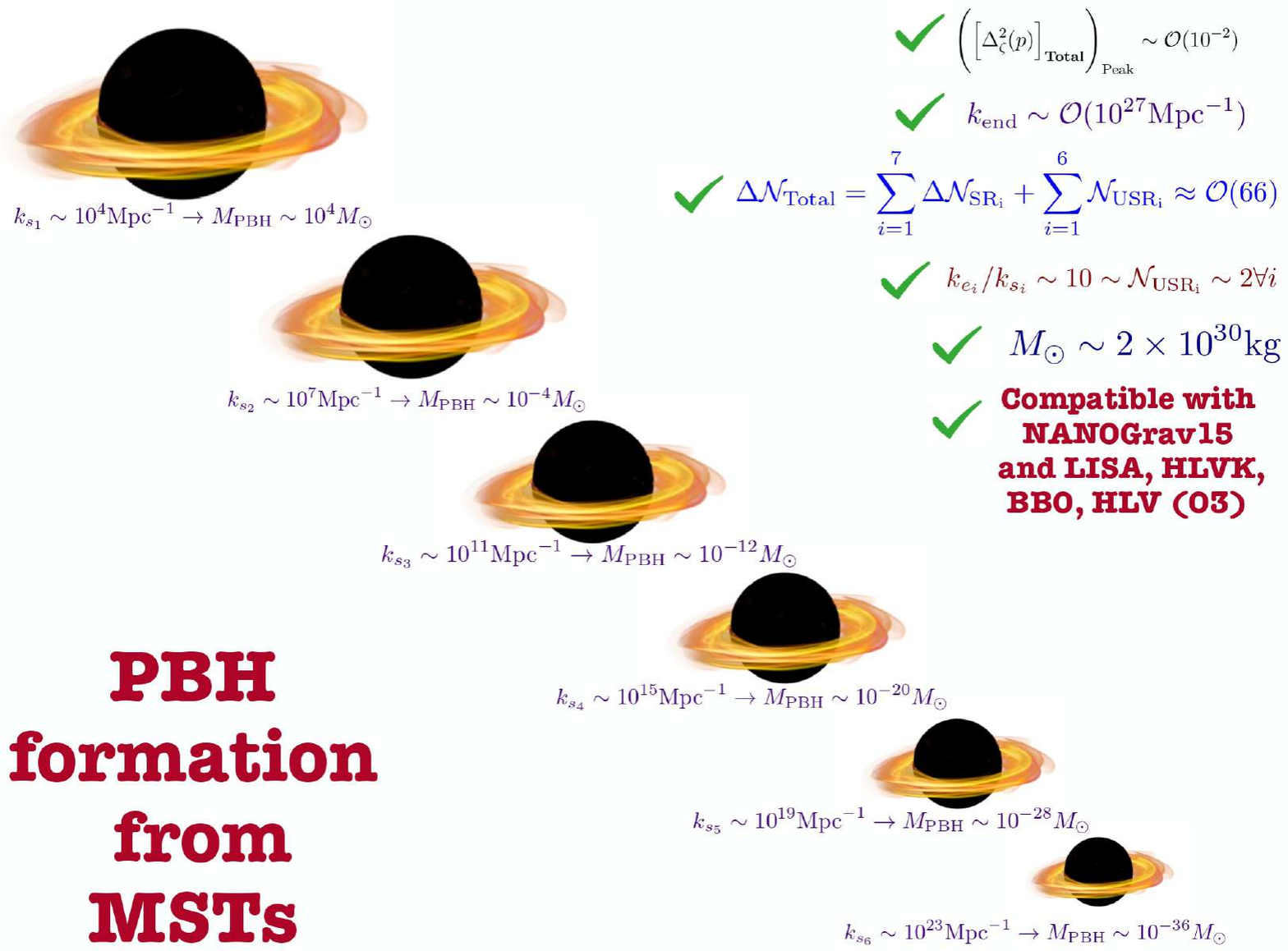}
    }
    	\caption[Optional caption for list of figures]{Representative figure depicting a spectrum of PBHs generated from the MST setup. The mass of each PBH corresponds to the position of the respective transition wavenumber $k_{s_{n}}$ with $n$ denoting the number of sharp transitions, which equals $6$ in the figure shown.} 
    	\label{pbhmst}
    \end{figure}

\section{Illuminating Conceptual Notions: Comparative discussion between Sharp and Smooth Transitions}\label{s6}

The uniqueness of MSTs is that we utilize this setup for its capability to overcome the one-loop constraints by integrating multiple ultra-slow roll periods with sharp transitions for each period. This, however, could be achieved using smooth transitions as prescribed by authors in \cite{Franciolini:2023agm, Taoso:2021uvl, Riotto:2023gpm, Firouzjahi:2023aum, Firouzjahi:2023ahg}. The concept of smooth transitions can also evade the No-go theorem of PBHs in the presence of loop corrections while satisfying the necessary inflation conditions. This approach also leads to the generation of large-mass PBHs and the production of induced Gravitational waves. To generalize these verdicts would require strong support for renormalization and resummation, as performed in this paper. In the MST setup, we anticipate a corresponding potential that has a multiple-step function-like nature, which consequently translates to the sharp behaviour of the second slow roll parameter.
On the contrary, we anticipate a mild potential in the case of smooth transitions characterized by a finite width, which can be realized by including a bump or dip in the inflationary potential. However, we tend to refrain from speculating more on the possible inferences without integrating the renormalization and resummation procedure into smooth transitions. Given this, we plan to dive into the computations involved in smooth transitions with comprehensive renormalization and DRG resummation in our upcoming work on \textit{Multiple Smooth Transitions} (M$\Bar{S}$T), where $\Bar{S}$ stands for smooth.

\section{Conclusion}
\label{s7}

The ambiguity in the exact nature of the equation of state parameter (EoS) $w$ right before the onset of the BBN and its implication on the severe PBH overproduction issue, when paralleled with the observed GW energy density spectrum from the PTA collaboration (including both NANOGrav-15 and EPTA), is the topic of discussion in this paper. The underlying theoretical setup comprises of a single-field inflation model in the EFT framework with multiple sharp transitions (MSTs). The model is used to study the generation of PBHs and the power spectrum of the scalar modes, where the necessary one-loop contributions are taken care of after incorporating the renormalization and resummation techniques. The details behind the implementation of such techniques, including possible renormalization methods like the late-time and wavefunction/adiabatic renormalization and the Dynamical Renormalization Group (DRG) approach, are elaborated in great detail to keep the procedure more transparent. An important fact used throughout the discussions of this paper comes from the results after analyzing the effects of varying effective sound speed values. This analysis provides us with the key interval of $1 \leq c_{s} \leq 1.17$ which helps to achieve the best features in the one-loop renormalized and DRG-resummed scalar power spectrum suited to study PBH formation. To illustrate the scope of the above-mentioned theoretical setup in generating a spectrum of PBH masses ranging from $M_{\rm PBH} \gtrsim {\cal O}(M_{\odot})$ to $M_{\rm PBH} \lesssim {\cal O}(10^{-31}M_{\odot})$ we provide a representative figure fig.(\ref{pbhmst}) for this purpose. The figure depicts in detail the satisfaction of necessary theoretical constraints to observe this spectrum of mass formation while being compatible with the observational signatures. We conducted the analysis of studying the GW spectrum for an arbitrary but constant $w$ by focusing on the scenario of linearities dominating the density contrast fluctuations. Under such conditions, we found that the current allowed interval on the threshold of the density contrast from numerical studies, $2/5 \leq \delta_{\rm th} \leq 2/3$, provides us with the interval on EoS, $-0.55 \leq w \leq 1/3$. Keeping the further analysis of this paper restricted in this interval for $w$, we found after imposing the condition of no overproduction the case of, $0.2 \lesssim w \leq 1/3$, as being the most favourable to obtain a sizeable PBH abundance of $f_{\rm PBH} \in (1,10^{-3})$. The related amplitude $A$ of the one-loop renormalized and DRG-resummed scalar power spectrum is then constrained to values where close statistical agreement of near $2\sigma$ with the NANOGrav-15 and EPTA signals is obtained. Regarding $w \lesssim 0$, where $w=0$ signals a matter dominated epoch, we found the amplitude $A$ to be constrained at much lower values, $A < {\cal O}(10^{-2})$, with no agreement even within $3\sigma$ for the NANOGrav-15 data. We conclude that stronger results in terms of their theoretical and observational agreement follow when we consider the value of $c_{s}=1.17$ rather than $c_{s}=1$. The analysis of the abundance as function of PBH mass tend to show production of large mass PBH, $M_{\rm PBH} \sim {\cal O}(10^{-6}-10^{-3})M_{\odot}$, when adhering to $w \gtrsim 0.2$ under both $c_{s}=1\;{\rm and}\;1.17$, and still qualifying for sizeable abundance after considering the constraints coming from microlensing experiments. For $w \lesssim 0$, mass range turns out with $M_{\rm PBH} \lesssim {\cal O}(10^{-13})M_{\odot}$ which is too low to consider based on the constraints. Next, we also confront the behavior of the amplitude as a function of the parameter $w$. The exact nature of the amplitude variation is not pre-determined, and from our analysis, we witness that to be the case after increasing EoS within the interval, $0 \leq w \leq 1/3$. Our analysis up to this point culminates with the SIGW spectrum as a function of the frequency for various values of the EoS, $w \in \{1/3,0.2,0,-0.01\}$ and both $c_{s}=1\;{\rm and}\;1.17$. The remarkable nature of featuring constant amplitude across all wavenumbers sufficient to generate large mass PBH for $w=1/3$ is also translated to the corresponding SIGW spectra irrespective of considered $c_{s}$ values. The peak regime of the data of GW spectrum from PTA is well-approximated by the first transition phase with $w=1/3$ and up to $w \sim 0.2$, we found notable similarities between the observational data and features of the GW spectrum and power spectrum amplitude from the theory of MSTs. The scenario for SIGW spectra with $w \lesssim 0$ is also seen as consistent with its interpretation from the behaviour of the power spectrum amplitude for the first transition when examining overproduction. For such values of $w$, the resulting amplitude is highly suppressed, which diminishes its value in explaining any observationally relevant data as arising from the study of MSTs. 

We stress again the use of linear approximations in our analysis while using the density contrast $(\delta \equiv \delta\rho/\rho)$ to determine the PBH abundance and SIGW spectra and point out that an accurate study of PBH abundance, under the influence of an arbitrary EoS $w$, remains absent where one needs to incorporate the non-linearities in $\delta$ on the superhorizon scales and the non-Gaussianities developed thereafter in the distribution function of $\delta$. We plan to address such a construction in the near future. What possible changes occur in the present analysis of arbitrary constant $w$ when a sharp transition is replaced by a smooth transition into the USR phase is also a prospective question for future studies which we plan to build on.

\subsection*{Acknowledgement}

SC would like to thank The National Academy of Sciences (NASI), Prayagraj, India for being elected as a member of the academy. SC would like to thank Md. Sami 
for useful discussions. SC would like to express gratitude
for the work-friendly environment of The Thanu Padmanabhan Centre for Cosmology and Science Popularization (CCSP), SGT University, Gurugram, which provided vast support for this research. Finally, we would
like to acknowledge our debt to individuals from various parts of the world for their generous and unwavering
support for research in the natural sciences.

\subsection*{Data availability statement}
No data is available or has been used for this particular submission.

\bibliographystyle{utphys}

\end{document}